\pdfoutput=1
\ifx\inPreamble\undefined \else %%%% 'MAGIC' %%%%%%%%%%%%%%%%%%%%%%%%%%%%%%%%%%%
\newcommand{\artderarbeit}{Master-Thesis}
\newcommand{\thema}{Analyzing the State of Computer Science Research with the DBLP Discovery Dataset}
\setbool{verlaengerung}{false}
\setbool{danksagung}{true}
\newcommand{\autor}{Lennart Küll}
\newcommand{\matrikelnummer}{1549540}
\newcommand{\studiengang}{Informatik}
\newcommand{\schwerpunkt}{Data Analytics}
\newcommand{\lehrstuhl}{Lehrstuhl für Data \& Knowledge Engineering}
\newcommand{\betreuer}{Dr. Terry Ruas\\Jan Philip Wahle M.Sc.}
\newcommand{\prueferA}{Prof. Dr.-Ing. Bela Gipp}
\newcommand{\prueferB}{Dr. Terry Ruas}
\newcommand{\abgabedatum}{10. November 2022}
\newcommand{\ort}{Wuppertal}
\newcommand{\schlagwoerter}{Thesis, Master, Bergische Universität Wuppertal, University of Wuppertal, Computer Science, DBLP, Scientometrics}
\setbool{doppelseitig}{true}			% Einseitiger oder doppelseitiger Druck?

\setbool{linksMarkieren}{true}			% Anklickbare Links im PDF-Dokument markieren?

\setbool{abbildungsverzeichnis}{true} 	% Abbildungsverzeichnis erzeugen?
\setbool{quellcodeverzeichnis}{false}	% Quellcodeverzeichnis erzeugen?
\setbool{tabellenverzeichnis}{true}		% Tabellenverzeichnis erzeugen?
\setbool{symbolverzeichnis}{false}		% Symbolverzeichnis erzeugen?
\setbool{akronymverzeichnis}{false}		% Akronymverzeichnis erzeugen?
\setbool{abkuerzungsverzeichnis}{true}	% Abkürzungsverzeichnis erzeugen?
\setbool{glossar}{true}					% Glossar erzeugen?

\setbool{verzeichnisseImInhaltsverzeichnis}{true} % Verzeichnisse im Inhaltsverzeichnis erwähnen?

\setbool{verzeichnisseZusammenfassen}{false} % Seitenumbruch zwischen den Verzeichnissen deaktivieren?

\newcommand\kapitelTiefeNummerierung{4} 	  % Wie tief soll nummeriert werden?
\newcommand\kapitelTiefeInhaltsverzeichnis{2} % Bis zu welcher Tiefe sollen die Einträge ins Inhaltsverzeichnis?

\setbool{zusatzErklaerung}{false} 		% Hiermit können zusätzliche Erklärungen eingebunden werden (wird im Normalfall nicht benötigt)

\setbool{color}{true} % Farbe für Design-Elemente verwenden (true/false)

\newcommand{\colormodel}{cmyk} % z.B. cmyk, rgb oder gray

\setbool{nichtZitiertInweiterfuehrendeLiteratur}{false}
\setbool{weiterfuehrendeLiteratur}{false}
\newcommand{\explizitesNocite}{
}

\setbool{alleAutorenExplizitNennen}{true} % Steuert die Nennung der Autoren im Literaturverzeichnis
\newcommand{\hauptsprache}{english} % Sprache, in der das Dokument geschrieben ist. Wird auch für die Verzeichnisnamen etc. verwendet
\newcommand{\weitereSprachen}{english, ngerman}% weitere Sprachen, auf die kurzfristig mit \selectlanguage{language} gewechselt wird. (Wenn es mehrere sind werden diese mit Kommata getrennt)

\fi %%%%%%%%%%%%%%%%%%%%%%%%%%%%%%%%%%%%%%%%%%%%%%%%%%%%%%%%%%%%%%%%%%%%%%%%%%%%
\def\inPreamble{}% Flag setzen: wird sind in der Präambel
\RequirePackage[utf8]{inputenc}
\RequirePackage[T1]{fontenc}
\RequirePackage{etoolbox}		% Einfacheres Programmieren in LaTeX
\RequirePackage{calculator}		% Rechnen (auch mit Längen)
\RequirePackage{xfp}			% Genaueres Rechnen (aber nicht mit Längen)
\RequirePackage{pbox}			% \pbox (Parbox flexibler Breite)

\newbool{abbildungsverzeichnis}
\newbool{quellcodeverzeichnis}
\newbool{tabellenverzeichnis}
\newbool{symbolverzeichnis}
\newbool{akronymverzeichnis}
\newbool{abkuerzungsverzeichnis}
\newbool{glossar}

\newbool{verzeichnisseZusammenfassen}
\newbool{verzeichnisseImInhaltsverzeichnis}

\newbool{nichtZitiertInweiterfuehrendeLiteratur}
\newbool{weiterfuehrendeLiteratur}
\newbool{alleAutorenExplizitNennen}

\newbool{verlaengerung}
\newbool{danksagung}
\newbool{zusatzErklaerung}

\newbool{color}
\newbool{linksMarkieren}

\newbool{doppelseitig}

\newlength{\bindekorrektur}
\newlength{\randAussen}
\newcommand{\verzeichnisEintrag}[2]{%
	\ifbool{verzeichnisseImInhaltsverzeichnis}{%
			\clearpage%
		\phantomsection% damit die Seitenzahl auf jeden Fall stimmt
		\nopagebreak%
		\addcontentsline{toc}{chapter}{#1}% 
		\nopagebreak%
	}{%
		\pdfbookmark[0]{#1}{#2}% hier passte die Seitenzahl bisher immer, daher keine \phantomsection nötig
	}%
}

\newcommand{\verzeichnisEintragX}[3]{%
	\ifbool{verzeichnisseImInhaltsverzeichnis}{%
		\clearpage%
		\phantomsection% damit die Seitenzahl auf jeden Fall stimmt
		\nopagebreak%
		\addcontentsline{toc}{chapter}{\texorpdfstring{#1}{#2}}% 
		\nopagebreak%
	}{%
		\pdfbookmark[0]{#2}{#3}% hier passte die Seitenzahl bisher immer, daher keine \phantomsection nötig
	}%
}

\ifbool{doppelseitig}{
	\def\onetwoside{twoside}
}{
	\def\onetwoside{oneside}
}

\documentclass[
	\onetwoside,% twoside: zweiseitig, oneside: einseitig
	titlepage, 	% Titelseite steht für sich allein
	12pt,		% Schriftgröße
	a4paper, 	% A4-Papier
]{report}

\RequirePackage[%
	main=\hauptsprache, %Umsetzung der deutschen Sprache nach neuer Rechtschreibung
	\hauptsprache, % muss hier nochmal erwähnt werden, damit \autoref das auch mitbekommt
	\weitereSprachen %Sprachen, die nur Abschnittsweise (z.B. für Zitate) verwendet werden
	]{babel}

\RequirePackage{translations}

\DeclareTranslation{German}{Bibliography}{Literatur}
\DeclareTranslation{English}{Bibliography}{Bibliography}
\DeclareTranslationFallback{Bibliography}{Bibliography}

\DeclareTranslation{German}{FurtherReading}{Weiterführende Literatur}
\DeclareTranslation{English}{FurtherReading}{Further Reading}
\DeclareTranslationFallback{FurtherReading}{Further Reading}

\DeclareTranslation{German}{abbreviationsname}{Abkürzungen}
\DeclareTranslation{English}{abbreviationsname}{Abbreviations}
\DeclareTranslationFallback{abbreviationsname}{Abbreviations}

\RequirePackage{lmodern} % Schriftart lmodern nutzen
\RequirePackage{microtype} % Besserer Schriftsatz

\RequirePackage[
	autostyle, % Zitierstil an die aktuelle Sprache anpassen
	german=quotes, % wenn Deutsch -> „Anführungszeichen“ benutzen 
]{csquotes} %wird auch von biblatex verwendet

\def\autorenAnzahlImLiteraturverzeichnis{4}
\ifbool{alleAutorenExplizitNennen}{
	\def\autorenAnzahlImLiteraturverzeichnis{99}
}{}

\RequirePackage[backend=bibtex8,  %Daten von Bibtex holen
			style=authoryear, %In der Form [Xil13] statt [1] referenzieren
			sorting=nyt,	      %Name,Year,Title 
			maxbibnames=\autorenAnzahlImLiteraturverzeichnis, %Anzahl maximal/minimal genannter Autoren im Literaturverzeichnis
			minbibnames=3, 
			maxcitenames=2, %Anzahl maximal/minimal genannter Autoren mit \citeauthor{...}
			mincitenames=1,
			maxsortnames=99, % Anzahl der Autoren, die in die Sortierung eingehen
			minsortnames=99,
			maxalphanames=4, % Maximale Anzahl Autoren, bei der jeder Autor mit einem Buchstaben in [ABCD] genannt wird
			minalphanames=3, % Anzahl Autoren, ab der [ABC+] verwendet wird		
   ]{biblatex}

\DeclareNameAlias{default}{family-given}% Format: Nachname, Vorname
\DefineBibliographyStrings{ngerman}{ 
	andothers = {{et\,al\adddot}}, % et al.
}
\DeclareBibliographyCategory{cited} % registrieren, welche Quellen zitiert wurden
\AtEveryCitekey{%
	\addtocategory{cited}{\thefield{entrykey}}%
}

\AtEndDocument{% In manchen Installationen funktionierte das auch in der Präambel, aber so ist es sicherer
	\ifbool{nichtZitiertInweiterfuehrendeLiteratur}{
		\nocite{*} % alle nicht zitierten Quellen mit aufnehmen
	}{
		\explizitesNocite
	}
}

\def\randAussenCm{}
\def\bindekorrekturCm{}
\LENGTHDIVIDE{\randAussen}{1cm}{\randAussenCm} 			% auf cm normieren
\LENGTHDIVIDE{\bindekorrektur}{1cm}{\bindekorrekturCm} 	% auf cm normieren

\newlength{\randRechtsAussen}	% Als Länge anlegen (bessere Wiederverwendbarkeit)
\newlength{\randLinksInnen}

\usepackage[
	layout=a4paper,			% Papierformat
	headheight=1.2cm,		% durch Kopfzeile festgelegt
	top=3.55cm,				% Abstand Text <-> oberer Seitenrand
	left=\randLinksInnen,	% Abstand Text <-> linker Seitenrand
	right=\randRechtsAussen,% Abstand Text <-> rechter Seitenrand
	bottom=3.7cm 			% Abstand Text <-> unterer Seitenrand
]{geometry}

\usepackage[official]{eurosym}

\usepackage{bbding}%Häkchen für abgehakte Dinge

\usepackage[
	\colormodel,% Konvertiere alle Farben in das angegebene Farbmodell
	table, 		% Lade das colortbl package (Farben in Tabellen)
	showerrors,	% Nicht definierte Farbe verwendet -> Fehler erzeugen
        xcdraw
]{xcolor}		% Farben

\definecolor{Buw}			{cmyk}{0.55, 0.00, 1.00, 0.00} % Pantone 376
\definecolor{BuwEtikett}	{cmyk}{0.55, 0.00, 1.00, 0.10} % Pantone 376 + 10K

\definecolor{BuwFak01}		{cmyk}{0.08, 1.00, 0.65, 0.34} % Pantone 201
\definecolor{BuwFak02}		{cmyk}{1.00, 0.03, 0.34, 0.12} % Pantone 321
\definecolor{BuwFak03}		{cmyk}{1.00, 0.57, 0.12, 0.70} % Pantone 540
\definecolor{BuwFak04_oben}	{cmyk}{0.70, 0.30, 0.00, 0.12} % Pantone 646
\definecolor{BuwFak04_unten}{cmyk}{1.00, 0.61, 0.04, 0.26} % Pantone 301
\definecolor{BuwFak05}		{cmyk}{0.23, 0.17, 0.13, 0.46} % Pantone Cool Gray 8
\definecolor{BuwFak06_oben}	{cmyk}{0.37, 1.00, 0.00, 0.26} % Pantone 2425
\definecolor{BuwFak06_unten}{cmyk}{0.37, 1.00, 0.00, 0.47} % Pantone 2425 + 30K
\definecolor{BuwFak07}		{cmyk}{1.00, 0.00, 0.55, 0.40} % Pantone 562
\definecolor{BuwFak08}		{cmyk}{0.00, 0.00, 0.00, 0.00} %
\definecolor{BuwFak09}		{cmyk}{0.00, 0.30, 0.85, 0.00} % Pantone 1235

\definecolor{pureK}			{cmyk}{0.00, 0.00, 0.00, 1.00} % K

\definecolor{AccentStrong}	{cmyk}{0.000,0.420,0.800,0.100} % Orange
\definecolor{AccentWeak}	{cmyk}{0.295,0.142,0.000,0.310} % Bläuliches Grau

\ifbool{color}{
	\colorlet{primaryTop}		{Buw}
	\colorlet{primaryMiddle}	{Buw}
	\colorlet{primaryBottom}	{Buw}
	\colorlet{secondaryTop}		{BuwFak06_oben}
	\colorlet{secondaryMiddle}	{BuwFak06_oben!50!BuwFak06_unten}
	\colorlet{secondaryBottom}	{BuwFak06_unten}
	\colorlet{tertiaryTop}		{black!50}
	\colorlet{tertiaryMiddle}	{black!50}
	\colorlet{tertiaryBottom}	{black!50}
	\colorlet{textBlack}		{black}
}{
	\colorlet{primaryTop}		{pureK}
	\colorlet{primaryMiddle}	{pureK}
	\colorlet{primaryBottom}	{pureK}
	\colorlet{secondaryTop}		{pureK!70}
	\colorlet{secondaryMiddle}	{pureK!70}
	\colorlet{secondaryBottom}	{pureK!70}
	\colorlet{tertiaryTop}		{pureK!50}
	\colorlet{tertiaryMiddle}	{pureK!50}
	\colorlet{tertiaryBottom}	{pureK!50}
	\colorlet{textBlack}		{pureK}
}

\colorlet{primary}			{primaryMiddle}
\colorlet{secondary}		{secondaryMiddle}
\colorlet{tertiary}			{tertiaryMiddle}

\colorlet{pagenumber}		{tertiaryBottom!50!textBlack}
\colorlet{headrule}			{primaryTop}
\colorlet{footnoteRule}		{secondary}
\colorlet{footrule}			{primaryBottom}

\colorlet{captionLabel}		{pureK!80}
\colorlet{subcaptionLabel}	{pureK!70}

\colorlet{captionText}		{pureK!70}
\colorlet{subcaptionText}	{pureK!60}

\colorlet{listingBackground}{pureK!5}

\def\customFootrule{
	\renewcommand{\footrulewidth}{1.5mm}
	\renewcommand{\footrule}{%
		\begingroup
			\color{footrule}
			\hrule height \footrulewidth
			\vspace{0.5em}
		\endgroup
	}
}

\def\customHeadrule{
	\renewcommand{\headrulewidth}{1.5mm}
	\renewcommand{\headrule}{%
		\begingroup
			\color{headrule}
			\hrule height \headrulewidth 
		\endgroup
	}
}

\def\renewMarks{
	\renewcommand{\sectionmark}[1]{%
		\markright{%
			##1%~(\thesection)%
		}%
	}
	\renewcommand{\chaptermark}[1]{%
		\markboth{%
			##1% \chaptername~\thechapter%
		}{%
		}%
	}
}

\def\seitenNummer{%
	\color{pagenumber}\sffamily\bfseries\thepage{}%
}

\usepackage{fancyhdr} % Schönere Kopfzeilen
\def\vertikalZentrieren{%
	\vspace{-\footrulewidth}%
	\vspace{-1.275em}%
}

\ifbool{doppelseitig}{
	\def\rightodd{RO}

	\def\lefteven{LE}
	\def\rightoddlefteven{RO,LE}
	\def\leftoddrighteven{LO,RE}
}{
	\def\rightodd{R}

	\def\lefteven{R}
	\def\rightoddlefteven{R}
	\def\leftoddrighteven{L}
}

\fancypagestyle{thesis-page-regular}{%
	\renewMarks
	\fancyhf{}%clear
	\fancyhead[\rightoddlefteven]{%
		\includegraphics[height=2em]{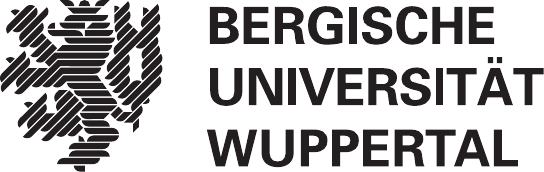}%
	}
	\fancyhead[\leftoddrighteven]{%
		\nouppercase{%
			\sffamily%
			\textbf{\leftmark}%
			\\%
			\rightmark%
		}%
	}
	\fancyfoot[\rightodd]{\vertikalZentrieren\colorbox{white}{\seitenNummer}\hspace{1em}}
	\fancyfoot[\lefteven]{\vertikalZentrieren\hspace{1em}\colorbox{white}{\seitenNummer}}
	
	\customHeadrule
	\customFootrule
}

\fancypagestyle{plain}{%
	\renewMarks
	\fancyhf{} % clear
	\fancyfoot[\rightodd]{\vertikalZentrieren\colorbox{white}{\seitenNummer}\hspace{1em}}
	\fancyfoot[\lefteven]{\vertikalZentrieren\hspace{1em}\colorbox{white}{\seitenNummer}}
	
	\customHeadrule
	\customFootrule
}

\usepackage[%
	nobottomtitles*,	% Überschriften am unteren Seitenrand vermeiden
]{titlesec}
\def\titleformatBase{\sffamily\bfseries}
\def\titleformatChapter{\Huge}
\def\titleformatSection{\LARGE}
\def\titleformatSubsection{\Large}
\def\titleformatSubsubsection{\large}
\def\titleformatParagraph{\normalsize}
\def\titleformatSubparagraph{\normalsize}

\def\titlepartChapter		{\titleformatChapter\ifbool{weAreInTheAppendix}{\Alph{chapter}}{\arabic{chapter}}}
\def\titlepartSection		{\ifnumcomp{\value{secnumdepth}}{>}{0}{\titleformatSection\arabic{section}}{}}
\def\titlepartSubsection	{\ifnumcomp{\value{secnumdepth}}{>}{1}{\titleformatSubsection\arabic{subsection}}{}}
\def\titlepartSubsubsection	{\ifnumcomp{\value{secnumdepth}}{>}{2}{\titleformatSubsubsection\arabic{subsubsection}}{}}
\def\titlepartParagraph		{\ifnumcomp{\value{secnumdepth}}{>}{3}{\titleformatParagraph\arabic{paragraph}}{}}
\def\titlepartSubparagraph	{\ifnumcomp{\value{secnumdepth}}{>}{4}{\titleformatSubparagraph\arabic{subparagraph}}{}}

\def\titleSpaceSubparagraph	{{}}
\def\titleSpaceParagraph	{\hphantom{\titleformatParagraph\ifnumcomp{\value{secnumdepth}}{>}{4}{.}{}\titlepartSubparagraph}\titleSpaceSubparagraph}
\def\titleSpaceSubsubsection{\hphantom{\titleformatSubsubsection\ifnumcomp{\value{secnumdepth}}{>}{3}{.}{}\titlepartParagraph}\titleSpaceParagraph}
\def\titleSpaceSubsection	{\hphantom{\titleformatSubsection\ifnumcomp{\value{secnumdepth}}{>}{2}{.}{}\titlepartSubsubsection}\titleSpaceSubsubsection}
\def\titleSpaceSection		{\hphantom{\titleformatSection\ifnumcomp{\value{secnumdepth}}{>}{1}{.}{}\titlepartSubsection}\titleSpaceSubsection}
\def\titleSpaceChapter		{\hphantom{\titleformatChapter\ifnumcomp{\value{secnumdepth}}{>}{0}{.}{}\titlepartSection}\titleSpaceSection}

\newbool{weAreInTheAppendix}
\setbool{weAreInTheAppendix}{false}
\appto{\appendix}{\setbool{weAreInTheAppendix}{true}}
\def\titleNumberExtraDistance{.5cm}

\titleformat{\chapter} %command
	[hang] %shape
	{\titleformatBase} %format
	{%
		\titlepartChapter\titleSpaceChapter%
	} %label
	{\titleNumberExtraDistance} %sep
	{\titleformatChapter} %before-code
\titleformat{\section}
	[hang]
	{\titleformatBase}
	{%
		\color{textBlack!50}\titlepartChapter.%
		\color{textBlack}\titlepartSection%
		\titleSpaceSection%
	}
	{\titleNumberExtraDistance}
	{\titleformatSection}
\titleformat{\subsection}
	[hang]
	{\titleformatBase}
	{%
		\color{textBlack!33}\titlepartChapter.%
		\color{textBlack!50}\titlepartSection.%
		\color{textBlack}\titlepartSubsection%
		\titleSpaceSubsection%
	}
	{\titleNumberExtraDistance}
	{\titleformatSubsection}
\titleformat{\subsubsection}
	[hang]
	{\titleformatBase}
	{%
		\color{textBlack!20}\titlepartChapter.%
		\color{textBlack!33}\titlepartSection.%
		\color{textBlack!50}\titlepartSubsection.%
		\color{textBlack}\titlepartSubsubsection%
		\titleSpaceSubsubsection%
	}
	{\titleNumberExtraDistance}
	{\titleformatSubsubsection}
\titleformat{\paragraph}
	[runin]
	{\titleformatBase}
	{%
		\color{textBlack!15}\titlepartChapter.%
		\color{textBlack!20}\titlepartSection.%
		\color{textBlack!33}\titlepartSubsection.%
		\color{textBlack!50}\titlepartSubsubsection.%
		\color{textBlack}\titlepartParagraph%
		\titleSpaceParagraph%
	}
	{\titleNumberExtraDistance}
	{\titleformatParagraph}%
\titleformat{\subparagraph}
	[runin]
	{\titleformatBase}
	{%
		\color{textBlack!10}\titlepartChapter.%
		\color{textBlack!15}\titlepartSection.%
		\color{textBlack!20}\titlepartSubsection.%
		\color{textBlack!33}\titlepartSubsubsection.%
		\color{textBlack!50}\titlepartParagraph.%
		\color{textBlack}\titlepartSubparagraph%
		\titleSpaceSubparagraph}
	{\titleNumberExtraDistance}
	{\titleformatSubparagraph}%
\titlespacing*{\chapter}		{0pt}	{-2.1em}{1em}
\titlespacing*{\section}		{0pt}	{1em}	{1em}
\titlespacing*{\subsection}		{0em}	{1em}	{1em}
\titlespacing*{\subsubsection}	{0em}	{1em}	{1em}
\titlespacing*{\paragraph}		{0em}	{1em}	{1em}
\titlespacing*{\subparagraph}	{0em}	{1em}	{1em}

\usepackage{graphicx}				% um Bilder einbinden zu können 
\usepackage{pdfpages}				% Ganzseitig externe PDFs einbinden (z.B. für Aufgabenstellung)

\usepackage{tikz}					% Zeichnen per Code

\usepackage[
]{siunitx}

\usepackage{listings} %Für Quelltexte

\DeclareTranslation{German}{lstlistingname}{Quellcode}
\DeclareTranslation{English}{lstlistingname}{Sourcecode}
\DeclareTranslationFallback{lstlistingname}{\lstlistingname}
\renewcommand{\lstlistingname}{\GetTranslation{lstlistingname}}

\DeclareTranslation{German}{lstlistoflistingname}{Quellcodeverzeichnis}
\DeclareTranslation{English}{lstlistoflistingname}{List of Sourcecodes}
\DeclareTranslationFallback{lstlistoflistingname}{\lstlistlistingname}
\renewcommand{\lstlistlistingname}{\GetTranslation{lstlistoflistingname}}

\ifbool{color}{
	\def\listingBasicstyle		{\color{black}\ttfamily\small}
	\def\listingKeywordstyle	{\color{primary}\bfseries}
	\def\listingStringstyle		{\color{secondary}\slshape}
	\def\listingCommentstyle	{\color{AccentWeak}\fontseries{l}\selectfont}
	\def\listingEmphstyle		{\color{AccentStrong}\bfseries}
	\def\listingRulecolor		{\color{AccentWeak}}
	\def\listingNumbercolor		{\color{AccentWeak!60}}
}{
	\def\listingBasicstyle		{\color{pureK}\ttfamily\small}
	\def\listingKeywordstyle	{\color{pureK}\bfseries}
	\def\listingStringstyle		{\color{pureK}\slshape}
	\def\listingCommentstyle	{\color{pureK}\fontseries{l}\selectfont}
	\def\listingEmphstyle		{\color{pureK}\bfseries\underbar}
	\def\listingRulecolor		{\color{pureK!50}}
	\def\listingNumbercolor		{\color{pureK!26}}
}

\lstdefinelanguage{thesis-latexbeispiel}
	{morekeywords=
		{\\begin, \\end, \\frac, \\pi, \\\\, \\sum, \\varphi, \\pm, \\int, \\limits, \\infty, \\cdot, \\sqrt, \\left, \\right, \\mathrm,
		\\linewidth, \\centering, \\includegraphics, \\caption, \\label, \\lstinline, \\lstinputlisting, \\todo, \\listoftodos, \\url, \\href, \\cite, \\textqoute,
		\\gls, \\glssymbol, \\LaTeX},
		emph={figure, table, tabular, align, lstlisting},
		sensitive=true,
		morecomment=[l]{\%},
		alsoletter={\\},
	}

\usepackage{caption}
\usepackage[labelformat=brace,list=true]{subcaption}
\DeclareCaptionStyle{myCaptionStyle}{%
	format=plain,			% Basis-Format
	labelsep=quad,			% Abstand Typ/Nummer <-> Beschriftung
	margin=0.5em,				% Zusatzabstand Links
	singlelinecheck=false,	%
	labelfont={sf,small,bf,color=captionLabel},% Font Typ/Nummer
	textfont={sf,small, color=captionText},	% Font Beschriftung
}

\DeclareCaptionStyle{mySubCaptionStyle}{%
	format=plain,			% Basis-Format
	labelsep=space,			% Abstand Typ/Nummer <-> Beschriftung
	margin=0.5em,			% Zusatzabstand links
	singlelinecheck=false,	%
	labelfont={sf,small,bf, color=subcaptionLabel},% Font Typ/Nummer
	textfont={sf,small, color=subcaptionText},	% Font Beschriftung
}

\RequirePackage[]{hyperref}

\ifbool{linksMarkieren}{
}{
	\hypersetup{hidelinks} % Link-Markierung verstecken
}

\hypersetup{
	linktoc=all, 						% Welcher Teil im Inhaltsverzeichnis ist klickbar? (none/section/page/all)
    colorlinks=false,					% false: boxed links (wird nicht mitgedruckt); true: colored links (wird auch so gedruckt!)
	linkcolor=primary!50!white,			% color of internal links
    linkbordercolor=primary!20!white,
    citecolor=primary!50!black,			% color of links to bibliography
    citebordercolor=primary!20!gray,
    filecolor=AccentStrong,				% color of file links
    filebordercolor=AccentStrong!20,
    urlcolor=AccentWeak,				% color of external links
    urlbordercolor=AccentWeak!20,
	menucolor=blue,
	menubordercolor=blue,
	runcolor=red,
	runbordercolor=red,
    unicode=true,				% non-Latin characters in Acrobat’s bookmarks
    pdftoolbar=true,			% show Acrobat’s toolbar?
    pdfmenubar=true,			% show Acrobat’s menu?
    pdffitwindow=false,			% window fit to page when opened
    pdfstartview={FitH},		% fits the width of the page to the window
    pdftitle=\thema,			% Titel
    pdfauthor=\autor,			% Autor
    pdfsubject=\artderarbeit,	% subject of the document
    pdfcreator=LaTeX,			% creator of the document
    pdfproducer=,				% producer of the document
    pdfkeywords=\schlagwoerter,	% Schlagworte
    pdfnewwindow=true,			% Links in neuem Fenster öffnen
}

\usepackage[
	toc=false,		% Nicht im Inhaltsverzeichnis anzeigen. Wenn die Verzeichnisse dort erscheinen sollen, machen wir das selber
	translate=babel,% Babel zum Übersetzen verwenden
	acronym,		% Eigenes Verzeichnis für Akronyme	(CRC, BAföG, LASER, ...)
	abbreviations,	% Eigenes Verzeichnis für Abkürzungen (Dr., z.B., bspw., bzw., etc.)
	symbols,		% Eigenes Verzeichnis für Symbole (µ, ...)
        indexonlyfirst, % nur erste orccurence anzeigen
]{glossaries-extra}

\usepackage{glossary-longragged}% Extra Style
\usepackage{glossary-mcols}% Extra Style

\DeclareTranslation{German}{seename}{siehe}
\DeclareTranslation{English}{seename}{see}
\DeclareTranslationFallback{seename}{\seename} 
\renewcommand{\seename}{\expandonce{\GetTranslation{seename}}} %expandonce: es scheint so als würde der Befehl zu früh/spät expandiert - mit expandonce passt es

\makenoidxglossaries %langsames Sortieren, funktioniert aber immer und braucht keine externen Programme
\newabbreviation{ai}{AI}{Artificial Intelligence}
\newabbreviation{api}{API}{Application User Interface}
\newabbreviation{crud}{CRUD}{Create, Read, Update, Delete}
\newabbreviation{cs}{CS}{Computer Science}
\newabbreviation{doi}{DOI}{Digital Object Identifier}
\newabbreviation{era}{ERA}{Excellence in Research for Australia}
\newabbreviation{erm}{ERM}{express-restify-mongoose}
\newabbreviation{lda}{LDA}{Latent Dirichlet Allocation}
\newabbreviation{nlp}{NLP}{Natural Language Processing}
\newabbreviation{rest}{REST}{Representational State Transfer}
\newabbreviation{ui}{UI}{User Interface}

\newglossaryentry{aa}{name={ACL Anthology},description={\url{https://aclanthology.org/}}}
\newglossaryentry{acmdl}{name={ACM Digital Library},description={\url{https://libraries.acm.org/digital-library}}}
\newglossaryentry{arxiv}{name={arXiv},description={\url{https://arxiv.org/}}}
\newglossaryentry{citeseer}{name={CiteSeerX},description={\url{https://citeseerx.ist.psu.edu/}}}
\newglossaryentry{citespace}{name={CiteSpace},description={\url{http://cluster.cis.drexel.edu/~cchen/citespace/}}}
\newglossaryentry{dblp}{name={DBLP},description={\url{https://dblp.org/}}}
\newglossaryentry{drift}{name={DRIFT},description={\url{https://gchhablani-drift-app-t0asgh.streamlitapp.com/}}}
\newglossaryentry{fourtytwo}{name={42Papers},description={\url{https://42papers.com/}}}
\newglossaryentry{grobid}{name={GROBID},description={\url{https://github.com/kermitt2/grobid}}}
\newglossaryentry{gs}{name={Google Scholar},description={\url{https://scholar.google.com/}}}
\newglossaryentry{ieeex}{name={IEEE Xplore},description={\url{https://ieeexplore.ieee.org/}}}
\newglossaryentry{microsoft}{name={Microsoft Academic},description={\url{https://www.microsoft.com/en-us/research/project/academic/}}}
\newglossaryentry{mas}{name={Microsoft Academic Search},description={\url{https://web.archive.org/web/20170105184616/https://academic.microsoft.com/FAQ}}}
\newglossaryentry{mongodb}{name={MongoDB},description={\url{https://www.mongodb.com/}}}
\newglossaryentry{nlp4nlp}{name={NLP4NLP},description={\url{http://www.nlp4nlp.org/}}}
\newglossaryentry{nlpexplorer}{name={NLPExlorer},description={\url{http://nlpexplorer.org}}}
\newglossaryentry{nlpindex}{name={NLP Index},description={\url{https://index.quantumstat.com/}}}
\newglossaryentry{nlpscholar}{name={NLP Scholar},description={\url{http://saifmohammad.com/WebPages/nlpscholar-demo-basic.html}}}
\newglossaryentry{paperswithcode}{name={Papers With Code},description={\url{https://paperswithcode.com/}}}
\newglossaryentry{pubmed}{name={PubMed},description={\url{https://pubmed.ncbi.nlm.nih.gov/}}}
\newglossaryentry{scival}{name={SciVal},description={\url{https://www.scival.com/}}}
\newglossaryentry{scopus}{name={Scopus},description={\url{https://www.scopus.com/}}}
\newglossaryentry{ss}{name={Semantic Scholar},description={\url{https://www.semanticscholar.org/}}}
\newglossaryentry{vosviewer}{name={VOSViewer},description={\url{https://www.vosviewer.com/}}}
\newglossaryentry{wos}{name={Web of Science},description={\url{https://www.webofscience.com/}}}
\newglossaryentry{zetaalpha}{name={Zeta Alpha},description={\url{https://search.zeta-alpha.com/}}}

\newglossaryentry{eacl}{name={EACL},description={\url{https://2023.eacl.org/calls/demos/}}}
\newglossaryentry{lrec}{name={LREC},description={\url{https://lrec2022.lrec-conf.org/en/}}}
\usepackage{blindtext}

\AtBeginDocument{%
	\setlength{\marginparwidth}{10cm}% Dummy-Wert >= 2cm
}
\usepackage[disable,
	\hauptsprache,
	textsize=scriptsize
]{todonotes}

\AtBeginDocument{% korrekte Breite laden
	\setlength{\marginparwidth}{\randAussen-2.5mm}% hieraus ergibt sich die Breite der Todo-Boxen
}

\usepackage[titles]{tocloft}

\usepackage{xpatch}

\makeatletter
\xpatchcmd\l@lstlisting{1.5em}{0em}{}{} % TODO: wirkt nicht sehr robust
\makeatother

\usepackage{tikzducks}

\usepackage{threeparttable}
\usepackage[nameinlink]{cleveref}
\usepackage{enumitem}
\usepackage{wrapfig}
\usepackage{booktabs}
\usepackage{pdflscape}
\usepackage{afterpage}
\usepackage{multirow}
\usepackage{makecell}
\usepackage{xurl}

\newcommand{\projectName}{Computer Science Insights\xspace}
\newcommand{\projectAcronym}{CS-Insights\xspace}
\newcommand{\diii}{D3\xspace}
\newcommand{\diiilong}{DBLP Discovery Dataset\xspace}
\newcommand{\csi}{\projectAcronym}

\newcommand{\nlpScholar}{\gls{nlpscholar}\xspace}
\newcommand{\nlpExplorer}{\gls{nlpexplorer}\xspace}
\newcommand{\googleScholar}{\gls{gs}\xspace}
\newcommand{\semanticScholar}{\gls{ss}\xspace}
\newcommand{\aclAnthology}{\gls{aa}\xspace}
\newcommand{\drift}{\gls{drift}\xspace}
\newcommand{\nlpfournlp}{\gls{nlp4nlp}\xspace}
\newcommand{\wos}{\gls{wos}\xspace}
\newcommand{\scopus}{\gls{scopus}\xspace}
\newcommand{\dblp}{\gls{dblp}\xspace}
\newcommand{\grobid}{\gls{grobid}\xspace}
\newcommand{\arxiv}{\gls{arxiv}\xspace}
\newcommand{\lrec}{\gls{lrec}\xspace}
\newcommand{\eacl}{\gls{eacl}\xspace}
\newcommand{\vosviewer}{\gls{vosviewer}\xspace}
\newcommand{\scival}{\gls{scival}\xspace}
\newcommand{\mas}{\gls{mas}\xspace}
\newcommand{\microsoft}{\gls{microsoft}\xspace}
\newcommand{\ieeex}{\gls{ieeex}\xspace}
\newcommand{\acmdl}{\gls{acmdl}\xspace}
\newcommand{\citeseer}{\gls{citeseer}\xspace}
\newcommand{\zetaalpha}{\gls{zetaalpha}\xspace}
\newcommand{\paperswithcode}{\gls{paperswithcode}\xspace}
\newcommand{\nlpindex}{\gls{nlpindex}\xspace}
\newcommand{\fourtytwo}{\gls{fourtytwo}\xspace}
\newcommand{\mongodb}{\gls{mongodb}\xspace}
\newcommand{\citespace}{\gls{citespace}\xspace}
\newcommand{\pubmed}{\gls{pubmed}\xspace}

\newcommand{\dashboard}{\textit{[Dashboard]}\xspace}

\usepackage{titlesec}
\titleclass{\experiment}{straight}[\chapter]
\newcounter{experiment}
\titleformat{\experiment}{\sffamily\large\bfseries}{}{0em}{Experiment \theexperiment:~}
\titlespacing*{\experiment}{0pt}{3.25ex plus 1ex minus .2ex}{1.5ex plus .2ex}

\titleclass{\showcase}{straight}[\chapter]
\newcounter{showcase}
\titleformat{\showcase}{\sffamily\large\bfseries}{}{0em}{Showcase \theshowcase:~}
\titlespacing*{\showcase}{0pt}{3.25ex plus 1ex minus .2ex}{1.5ex plus .2ex}

\makeatletter
  \def\toclevel@experiment{3}
  \def\l@experiment{\@dottedtocline{3}{3.8em}{3.2em}}
  \def\toclevel@showcase{3}
  \def\l@showcase{\@dottedtocline{3}{3.8em}{3.2em}}
\makeatother
\crefname{experiment}{experiment}{experiments}
\crefname{showcase}{showcase}{showcases}

\crefformat{footnote}{#2\footnotemark[#1]#3}

\newcommand{\mongo}[1]{\lstinline[mathescape]|$\mbox{\textdollar}$#1|}

\newcommand{\printpublication}[1]{\AtNextCite{\defcounter{maxnames}{99}}\fullcite{#1}}

\begin{document} % Hier beginnt das Dokument %%%%%%%%%%%%%%%%%%%%%%%%%%%%%%%%%%%
    \setcounter{secnumdepth}{-2}	
        %% Version 2022-07-08
%% LaTeX-Vorlage für Abschlussarbeiten
%% Erstellt von Nils Potthoff, ab 2020 erneuert und ausgebaut von Simon Lohmann
%% Lehrstuhl Automatisierungstechnik/Informatik Bergische Universität Wuppertal
%%%%%%%%%%%%%%%%%%%%%%%%%%%%%%%%%%%%%%%%%%%%%%%%%%%%%%%%%%%%%%%%%%%%%%%%%%%%%%%%

%%%%%%%%%%%%%%%%%%%%%%%%%%%%%%%%%%%%%%%%%%%%%
%%% ! WARNUNG ! %%%%%%%%%%%%%%%%%%%%%%%%%%%%%
%%%%%%%%%%%%%%%%%%%%%%%%%%%%%%%%%%%%%%%%%%%%%
%%%  Diese Datei bitte nur bearbeiten,    %%%
%%%   wenn du ein LaTeX-Experte bist      %%%
%%%             U N D                     %%%
%%%  die Vorlage unbedingt ändern willst  %%%
%%%%%%%%%%%%%%%%%%%%%%%%%%%%%%%%%%%%%%%%%%%%%
%%%%%%%%%%%%%%%%%%%%%%%%%%%%%%%%%%%%%%%%%%%%%
%
%%%%%%%%%%%%%%%%%%%%%%%%%%%%%%%%%%%%%%%%%%%%%%%%%%%%%%%%%%%%%%%%%%%%%%%%%%%%%%%%
%%% DATEI-INFO %%%%%%%%%%%%%%%%%%%%%%%%%%%%%%%%%%%%%%%%%%%%%%%%%%%%%%%%%%%%%%%%%
%%%%%%%%%%%%%%%%%%%%%%%%%%%%%%%%%%%%%%%%%%%%%%%%%%%%%%%%%%%%%%%%%%%%%%%%%%%%%%%%
%%% Diese Datei enthält die Titelseite und Standardteile am Anfang der Thesis %%
%%%%%%%%%%%%%%%%%%%%%%%%%%%%%%%%%%%%%%%%%%%%%%%%%%%%%%%%%%%%%%%%%%%%%%%%%%%%%%%%
%
\pagenumbering{Roman}
\pagestyle{plain}
%
% Entscheiden, welches der Logopdfs benutzt wird (verschiedene Schwarztöne)
\newcommand{\buwlogopdf}{Medien/Uni_Wuppertal_Logo__cmykblack.pdf}
\ifbool{color}{%
	\renewcommand{\buwlogopdf}{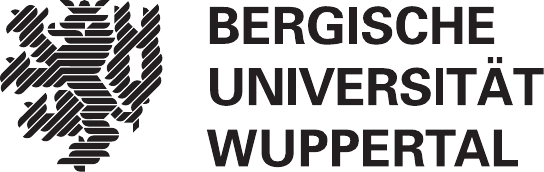}
}{% color disabled
	\ifstrequal{\colormodel}{rgb}{%
		\renewcommand{\buwlogopdf}{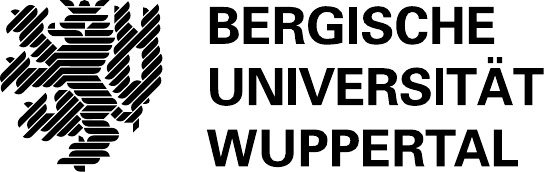}
	}{% cmyk/gray/...
		\renewcommand{\buwlogopdf}{Medien/Uni_Wuppertal_Logo__cmykblack}
	}
}
%
%%%%%%%%%%%%%%%%%%%%%%%%%%%%%%%%%%%%%%%%%%%%%%%%%%%%%%%%%%%%%%%%%%%%%%%%%%%%%%%%
%%% Titelseite %%%%%%%%%%%%%%%%%%%%%%%%%%%%%%%%%%%%%%%%%%%%%%%%%%%%%%%%%%%%%%%%%
%
	\begin{titlepage}
		\sffamily%
		\hspace{-1.5em}%
		\includegraphics[width=4.5cm]{\buwlogopdf}%
		\hfill%
		\begin{minipage}[b]{\linewidth-4.5cm}
			\begin{flushright}
				\normalsize{Fakultät für Elektrotechnik,}%
				\\\vspace{0.2mm}
				\normalsize{Informationstechnik und Medientechnik}%
				\vspace*{0.05mm}
			\end{flushright}
		\end{minipage}
		\vspace{-0.3em}
		\begingroup%
			\color{primary}%
			\rule{\textwidth}{3pt}%
		\endgroup%
		\begin{flushright}
			\lehrstuhl
		\end{flushright}
		
		\vspace{3.2em}
		\begin{center}
			\textbf{\Huge{\artderarbeit}}%
			\vspace{1.4em}
			\\
			\textbf{\large{\thema}}%
		\end{center}

		\vfill

		\begin{center}
			\Large
			\autor%
			\\
			{\Large\matrikelnummer}%
			\bigskip\\%
			\studiengang%
			\\
			{\large\schwerpunkt}%
			\vspace{2em}%
			\hspace{0em}\\%
			\normalsize%
			\ort, den \abgabedatum%
		\end{center}

		\vspace{7.5em}

		\pbox[t][2em]{\linewidth}{Betreuer}%
		\quad%
		\pbox[t][2em]{\linewidth}{\betreuer}%
		\hfill%
		\pbox[t][2em]{\linewidth}{Erstgutachter\\Zweitgutachter}%
		\quad%
		\pbox[t][2em]{\linewidth}{\prueferA\\\prueferB}%
	
	\end{titlepage}
	\ifbool{doppelseitig}{%
		% im doppelseitigen Modus zählt report wie gewünscht
	}{%
		\stepcounter{page}% Seite manuell hochzählen, weil report das im einseitigen Modus nicht macht, im doppelseitigen aber schon (Konsistenz)
	}
	\clearpage{\pagestyle{empty}\cleardoublepage}
%
%
%%%%%%%%%%%%%%%%%%%%%%%%%%%%%%%%%%%%%%%%%%%%%%%%%%%%%%%%%%%%%%%%%%%%%%%%%%%%%%%%
%%% Themenblatt / Aufgabenstellung %%%%%%%%%%%%%%%%%%%%%%%%%%%%%%%%%%%%%%%%%%%%%
%
	% \pdfbookmark[1]{Aufgabenstellung}{aufgabe}%
	% \includepdf[
	% 	pages={1},
	% 	fitpaper=true
	% ]{Medien/Aufgabenstellung.pdf}%
	% \clearpage{\pagestyle{empty}\cleardoublepage}%
%
%
%%%%%%%%%%%%%%%%%%%%%%%%%%%%%%%%%%%%%%%%%%%%%%%%%%%%%%%%%%%%%%%%%%%%%%%%%%%%%%%%
%%% Verlängerung %%%%%%%%%%%%%%%%%%%%%%%%%%%%%%%%%%%%%%%%%%%%%%%%%%%%%%%%%%%%%%%
%
	\ifbool{verlaengerung}{%
		\cleardoublepage%
		\pdfbookmark[1]{Verlängerung}{verlaengerung}%
		\includepdf[pages={1},fitpaper=true]{Medien/Verlaengerung.pdf}%
	  	\clearpage{\pagestyle{empty}\cleardoublepage}%
	}{}%
%
%
%%%%%%%%%%%%%%%%%%%%%%%%%%%%%%%%%%%%%%%%%%%%%%%%%%%%%%%%%%%%%%%%%%%%%%%%%%%%%%%%
%%% Erklärungen %%%%%%%%%%%%%%%%%%%%%%%%%%%%%%%%%%%%%%%%%%%%%%%%%%%%%%%%%%%%%%%%
%
	\pdfbookmark[1]{Erklärungen}{erklaerung}%
	\vspace*{2em}%
	\section*{Eidesstattliche Erklärung}%
	\noindent Hiermit erkläre ich, dass ich die von mir eingereichte Abschlussarbeit (Master-Thesis) selbständig verfasst und keine anderen als die angegebenen Quellen und Hilfsmittel benutzt sowie Stellen der Abschlussarbeit, die anderen Werken dem Wortlaut oder Sinn nach entnommen wurden, in jedem Fall unter Angabe der Quelle als Entlehnung kenntlich gemacht habe.%
	
	\ifbool{zusatzErklaerung}{\vfill}{\vspace{4em}}%
	\begin{tabular}{lc}%
	\ort, den \abgabedatum \hspace*{1cm}& \rule[2px]{5cm}{0.5px}\\%
	                                    &\footnotesize{(Unterschrift)}%
	\end{tabular}
	\ifbool{zusatzErklaerung}{\vfill}{\vspace{5em}}%
	
	\section*{Einverständniserklärung}
	\noindent Hiermit erkläre ich mich damit einverstanden, dass meine Abschlussarbeit (Master-Thesis) wissenschaftlich interessierten Personen oder Institutionen zur Einsichtnahme zur Verfügung gestellt werden kann.
    \\\\
    \noindent Korrektur- oder Bewertungshinweise in meiner Arbeit dürfen nicht zitiert werden.

	\ifbool{zusatzErklaerung}{\vfill}{\vspace{4em}}%
	\begin{tabular}{lc}
	\ort, den \abgabedatum \hspace*{1cm}& \rule[-2px]{5cm}{0.5px} \\ 
	                                       &\footnotesize{(Unterschrift)}
	\end{tabular}
	\ifbool{zusatzErklaerung}{%
		\vfill
		\input{Kapitel/Zusatzerklaerung}
		\vfill
	}{}
	% \clearpage{\thispagestyle{empty}}
%
%
%%%%%%%%%%%%%%%%%%%%%%%%%%%%%%%%%%%%%%%%%%%%%%%%%%%%%%%%%%%%%%%%%%%%%%%%%%%%%%%%
%%% Danksagung %%%%%%%%%%%%%%%%%%%%%%%%%%%%%%%%%%%%%%%%%%%%%%%%%%%%%%%%%%%%%%%%%
	\ifbool{danksagung}{%
		\vspace*{\fill}\vspace{-3em}%
		% \pdfbookmark[1]{Danksagung}{danksagung}%
		\chapter{Acknowledgements}
I would like to express my deepest gratitude to the many people who have supported me throughout my thesis.
Without them, this thesis would not have been possible.

First and foremost, I am incredibly grateful for my supervisors Dr. Terry Ruas and Jan Philip Wahle, who continually provided invaluable guidance throughout the entire time I worked on this thesis.
I deeply appreciate their expertise and constructive feedback, which helped me tackle any issues that arose during the thesis.

I also want to thank Prof. Dr. Bela Gipp for offering to write my thesis at his chair and giving me the opportunity to complete a part of my research at the National Institute of Informatics (NII) in Tokyo.
Thanks should also go to Dr. Norman Meuschke, who helped me organize and review my application for the DAAD scholarship at the NII.

Moreover, I would like to thank my hosting professor at the NII Prof. Dr. Akiko Aizwa for her warm welcome and helpful insights when developing my thesis project, even though I could not visit the NII in person, due to traveling restrictions the COVID-19 pandemic imposed.

I express special thanks to Dr. Saif M. Mohammad for his comments on the project during its development and for essentially shaping how the project turned out.
Many thanks also go to my two fellow students Tom Neuschulten and Alexander von Tottleben for developing the prediction endpoint and the shared efforts to integrate it into the project.

Finally, I want to thank my family and friends for their continued support and encouragement during this very intense academic year.

% daad grant no. 57515252%
		\vfill%
	}{}
 
 	% \clearpage{\thispagestyle{empty}}%
        \vspace*{\fill}\vspace{-3em}%
        % \pdfbookmark[1]{Related Publications}{relatedpublications}%
        \chapter{Related Publications}
    The content of this master's thesis was created as part of an ongoing research project led by my supervisors Dr. Terry Ruas and Jan Philip Wahle.
    Parts of this project I also worked on (and thus parts of my thesis) were already published or are planned to be submitted to computer science conferences, which I list in the following:

    \bigskip\bigskip
    \noindent The \diiilong and its creation (\Crefrange{subs:data_primary}{subs:data_secondary}), and its implementation (\Cref{subs:crawler}).
    \par\bigskip
    \hfill
    \begin{minipage}{\dimexpr\textwidth-1cm}
        \printpublication{wahle_d3_2022}
    \end{minipage}
    
    \bigskip\bigskip
    \noindent The \projectName system (\Cref{subs:interface}), its motivation (\Cref{sec:problem}), and its architecture (\Cref{sec:architecture}).
    \par\bigskip
    \hfill
    \begin{minipage}{\dimexpr\textwidth-1cm}
        \printpublication{ruas_cs-insights_2022}
    \end{minipage}

    \vfill
    \vfill
    In the following the wording ``we'' is used rather than ``I'' as the ongoing research project is a collaborative effort, and I worked closely together with my supervisors.

%
        % \vfill%
%
%
%%%%%%%%%%%%%%%%%%%%%%%%%%%%%%%%%%%%%%%%%%%%%%%%%%%%%%%%%%%%%%%%%%%%%%%%%%%%%%%%
%%% Kurzfassung & Abstract %%%%%%%%%%%%%%%%%%%%%%%%%%%%%%%%%%%%%%%%%%%%%%%%%%%%%
%
	% \clearpage{\pagestyle{empty}\cleardoublepage}
        % \clearpage{\thispagestyle{empty}}
	\vspace*{\fill}\vspace{-3em}
	% \pdfbookmark[1]{Abstract}{abstract}
	\chapter{Abstract}
The number of scientific publications continues to rise exponentially, especially in \gls{cs}.
However, our ability to analyze those publications does not follow the same speed, which prevents us from finding and understanding implicit patterns hidden in their metadata (e.g., venues, document types).
Current solutions are limited by restricting access behind a paywall, offering no features for visual analysis, limiting access to their data, only focusing on niches or sub-fields, and/or not being flexible and modular enough to be transferred to other datasets.

In this thesis, we conduct a scientometric analysis to uncover those implicit patterns hidden in \gls{cs} metadata and to determine the state of \gls{cs} research.
Specifically, we investigate trends of the quantity, impact, and topics for authors, venues, document types (conferences vs. journals), and fields of study (compared to, e.g., medicine).
To achieve this we introduce the \projectName (\csi) system, an interactive web application to analyze \gls{cs} publications through multiple perspectives.
The data underlying this system is the \diiilong (\diii), which contains metadata from 5 million scholarly publications in \gls{cs}.
We create \diii with data from \dblp, the largest open-access bibliography for scientific papers and articles in \gls{cs}, and enrich it with further metadata (e.g., abstracts, citations).
% \dblp also shows that the number of publications per year more than quintupled from 80k records published in 2000 to 426k in 2020.
%source: https://dblp.org/statistics/publicationsperyear.html
\csi offers dedicated dashboards with multiple visualizations for all main features of \diii (e.g., publications, authors, venues, and citations) and multiple filters for more fine-grained analysis.
Both \diii and \csi are open-access, and \csi can be easily adapted to other datasets in the future.

The most interesting findings of our scientometric analysis include that i) there has been a stark increase in publications, authors, and venues in the last two decades, ii) many authors only recently joined the field, iii) the most cited authors and venues focus on computer vision and pattern recognition, while the most productive prefer engineering-related topics, iv) the preference of researchers to publish in conferences over journals dwindles, v) on average, journal articles receive twice as many citations compared to conference papers, but the contrast is much smaller for the most cited conferences and journals, and vi) journals also get more citations in all other investigated fields of study, while only \gls{cs} and engineering publish more in conferences than journals.
	% \vfill
	% \clearpage
        % \clearpage{\thispagestyle{empty}\cleardoublepage}
%
%
%%%%%%%%%%%%%%%%%%%%%%%%%%%%%%%%%%%%%%%%%%%%%%%%%%%%%%%%%%%%%
%  INHALTSVERZEICHNIS
\markboth{\contentsname}{}
\pdfbookmark[0]{\contentsname}{toc}
\begingroup
	\renewcommand{\markboth}[2]{}{}
	\tableofcontents
\endgroup
% \clearpage
%
%
%%WIDMUNG, VORWORT
%
%\nomenclature{$x$}{Description}
%\nomenclature{TEXT}{Abkürz}
%% Ende der Titelei; es folgt der Hauptteil
% \ifbool{doppelseitig}{%
% 	\clearpage%
% 	\hphantom{anker, damit hier auch wirklich eine leere Seite ist}%
% }{}%
% {\pagestyle{empty}\cleardoublepage}% Inhalt soll auf rechter Seite beginnen
\clearpage
\pagenumbering{arabic}%
\pagestyle{thesis-page-regular}%
 % Titelseite etc. (bitte nicht ändern) %%%%%%%%%
    \setcounter{secnumdepth}{\kapitelTiefeNummerierung}
	%%%%%%%%%%%%%%%%%%%%%%%%%%%%%%%%%%%%%%%%%%%%%%%%%%%%%%%%%%%%%%%%%%%%%%%%%%%%
	%%%%%%%%%%%%%  Ab hier ändern und ergänzen  %%%%%%%%%%%%%%%%%%%%%%%%%%%%%%%%
	%%%%%%%%%%%%%  | | | | | | | | | | | | | |  %%%%%%%%%%%%%%%%%%%%%%%%%%%%%%%%   
	%%%%%%%%%%%%%  V V V V V V V V V V V V V V  %%%%%%%%%%%%%%%%%%%%%%%%%%%%%%%%

	% weitere .tex-Dateien werden mit \input eingebunden
	% (z.B. eine für jedes Kapitel)

	% Einführungskapitel
	\chapter{Introduction}\label{chap:intro}
    \Cref{chap:intro} first introduces the problem this thesis tries to solve by explaining the context and motivation (\Cref{sec:problem}).
    We establish the goals of our research through the main research objective and its corresponding tasks and research questions (\Cref{sec:goals}) and list our contributions to \gls{cs} research (\Cref{sec:contribution}).
    Lastly, we outline the remainder of this thesis (\Cref{sec:outline}).

\section{Problem Presentation and Motivation}\label{sec:problem}
    In the last few decades, we have seen an exponential rise in the number of digital scientific publications, while our ability to analyze them does not follow the same speed, preventing us from uncovering implicit patterns among its main features (e.g., authors, venues) \parencite{bornmann_growth_2021}.
    Analyzing these large amounts of publications, and possibly any type of data, is hard, mainly due to its storage and processing challenges.
    There are already existing solutions to mitigate this problem, but all of them show inherent limitations.
    Researchers can use tools or repositories that already implement data storage, crawling, and processing, like \googleScholar\footnote{\url{https://scholar.google.com/}}, \semanticScholar\footnote{\url{https://www.semanticscholar.org/}} or \dblp\footnote{\url{https://dblp.org/}}, to find papers or authors and view their metrics, but these solutions lack details in other areas (e.g., venues) and options for analysis with visual components.
    Other solutions also provide visualizations (e.g., \scopus\footnote{\url{https://www.scopus.com/}}, \wos\footnote{\url{https://www.webofscience.com/}}), but are not open-access and are only available behind paywalls, which is prohibitive to those who would benefit the most from their resources (e.g., institutions in developing countries).
    Therefore, researchers focus on specific research areas, e.g., \nlpScholar \parencite{mohammad_nlp_2020_viz} for \gls{nlp}.
    Areas without such tools rely on data repositories (e.g. \arxiv\footnote{\url{https://arxiv.org/}}) or general tools (e.g., \vosviewer \parencite{van_eck_software_2010}), which also only have a limited set of options for analysis and visualizations.
    
\begin{figure}[!ht]
        \centering
        \includegraphics[width=\textwidth]{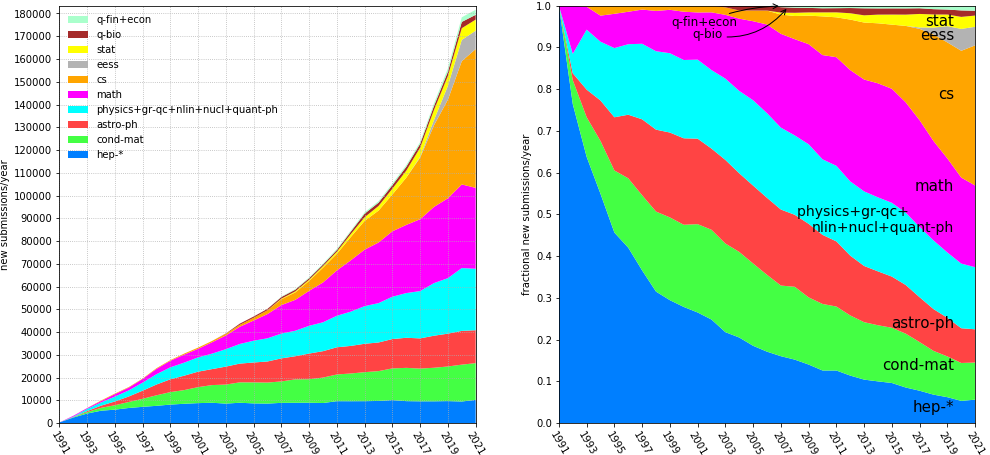}
        \caption{Submission rate statistics of \arxiv 1991-2021; updated 1 January 2021 \parencite{arxiv_arxiv_2022}.}
        \label{fig:arxiv_submissions}
    \end{figure}

    Analyzing the entire research landscape would be prohibitive (\googleScholar alone has more than 389m records \parencite{gusenbauer_google_2019}), so we focus efforts on a specific field of research and conducting a case study on that field, with the goal of our methodology also applying to other areas.
    We decide \gls{cs} is a great candidate for this, for two main reasons.
    First, the presence of \gls{cs} in solving or facilitating other field-related problems is undeniable (e.g., plagiarism detection \parencite{wahle_are_2021} or media bias \parencite{spinde_neural_2021}).
    Advancements in \gls{cs} are also responsible for many benefits, e.g., faster systems, more accurate results, and efficient tools.
    Today there is hardly any area not affected by the vast possibilities of \gls{cs}.
    Consider how difficult it would be to test, develop, and research new vaccines without access to tools of informatics (e.g., public repositories \parencite{kousha_covid-19_2020}, artificial intelligence \parencite{aggarwal_has_2022}).
    Second, \gls{cs} is a massively growing field, especially when compared to other fields, which becomes apparent when we look at the submissions on \arxiv (\Cref{fig:arxiv_submissions}).
    We can not only see that the number of submitted \gls{cs} papers increased over the last 10 years (from less than 10k papers a year in 2011 to 60k in 2021) but also how \gls{cs} is taking up a much larger percentage of the total submissions (from 10\% in 2011 to over a third of the submissions in 2021, the most of all research fields).
    While \arxiv is not a peer-reviewed repository, the same increase in \gls{cs} submissions is also visible for repositories consisting of peer-reviewed publications (e.g., \dblp and \wos, as shown by \textcite{wahle_d3_2022, fiala_computer_2017}, respectively), and sub-fields of \gls{cs} (e.g., \gls{nlp} as shown by \textcite{mohammad_nlp_2020_data}).
    Their studies also reveal that the number of authors has increased significantly over the last few decades.

    % Third, to our knowledge, there is currently no system or tool that is capable of a comprehensive quantitative analysis in \gls{cs}.
    % Previous studies on \gls{cs} use data from repositories like \wos, but conduct their analysis manually, without any pre-existing tool \parencite{coskun_scientometrics-based_2019, fiala_computer_2017}.
    
    As a result, we see \gls{cs} as a promising environment for developing a system to help understand its publications in an automated and democratic way and conducting an analysis with it, that can answer questions like:
    How fast is computer science research growing?
    How many authors are actively publishing in their field?
    What topics are prevalent in specific venues?
    Answering these questions helps other researchers and organizations make more informed decisions in their research and publications.
    Researchers can explore particular topics of interest for individual authors and venues, discover influential publications, or find important venues to inform their own research.
    Conference organizers and research organizations (e.g., ACL\footnote{https://www.aclweb.org/}) can track how their policy changes affect broad publication trends over time or compare the research output of authors and venues.
    For example, with a scientometric analysis one can track citation gaps across authors and venues; and in the future, uncover the influence of big technology companies, highly-funded universities, and governments.
    
    % The issue of losing oversight already became apparent in 2010, when \textcite{newman_information_2013} showed in an interview study, that keeping informed about your field is hard for all research fields, including \gls{cs}.
    % how do we understand CS as a research field -> analyze publications -> how do we analyze publications -> use tools -> tools limited -> provide a methodology that works on the limitations
    	
% \section{Goals}\label{sec:goals}
    % This section first defines the goal of this thesis, i.e., the main research objective (\Cref{subs:objective}) including its corresponding tasks and research questions, and then lists our contributions (\Cref{subs:contribution}) to \gls{cs} research.

\section{Research Objective}\label{sec:goals}
    With our goal of an analysis of the state of computer science research, we define the main research objective as follows:
    \smallskip
    \begin{center}
        \textbf{\textit{Analyze the state of computer science research by inspecting its different core components and uncovering implicit patterns.}}
    \end{center}
    \smallskip
    From the main objective, we distill the following research tasks:
    \begin{enumerate}[label=\textbf{RT\arabic*}, ref={RT\arabic*}]
        \item Review scientometric studies in the area of \gls{cs} and tools/resources for scientometric analyses that already exist.\label{task:review}
        \item Collect, clean, organize, store, and publish data from scientific publications in \gls{cs}.\label{task:data}
        \item Develop a system that uses the collected data to create visualizations to facilitate quantitative analyses.\label{task:system}
        \item Analyze \gls{cs} through its core components (e.g., publications, authors, venues, and document types), evaluate the findings, and compare them to that of previous research.\label{task:analyis}
    \end{enumerate}

    % RQs are still early, so we can reference them further down in the chapter
    \noindent We also derive research questions from our main research objective
    to narrow down which specific questions we have to answer to determine the state of \gls{cs} research.
    Thus, the research questions will guide our analysis in the later parts of this thesis and determine which experiments we will perform during \labelcref{task:analyis}:
    \begin{enumerate}[label=\textbf{RQ\arabic*}, ref={RQ\arabic*}]
        \item How many publications, authors, and venues are in our dataset? How do the numbers change over time? How many authors and venues are currently active?\label{rq:amount}
        \item How are the citations and publications distributed across authors and venues? How do the distributions change over time?\label{rq:distribution}
        \item What are the most prominent authors and venues? Are there preferences for topics? Do the topics change over time?\label{rq:topics}
        \item How do incoming and outgoing citations evolve over time? How do their distributions differ?\label{rq:citations}
        \item How do conferences and journals compare in their number of publications and citations over time? How do the top venues and topics differ? Do top authors prefer conferences or journals?\label{rq:types}
        \item How do the most prominent fields of study differ from \gls{cs} in topics and preference for conferences or journals?\label{rq:fields}
    \end{enumerate}
    
\section{Contributions}\label{sec:contribution}
    Through our goals, this thesis provides multiple contributions to the \gls{cs} research community:
    \begin{itemize}
        \item We propose the \diiilong (\diii), a large and carefully curated dataset of \gls{cs} publications metadata, with metadata of 5m \gls{cs} publications from \dblp, enriched with additional metadata like abstracts and citation counts.
        \item Also, we develop the \projectName (\projectAcronym) system, a modular analysis platform, which allows its users to perform quantitative analyses on the core components of \gls{cs} research.
        \item We use the dataset and system to conduct a scientometric analysis of the publications, authors, venues, document types, fields of study, and their topics and citations, according to our research questions (\labelcref{rq:amount} to \labelcref{rq:fields}).
        \item Both the dataset and system are open-access and open-source to allow everyone, regardless of background, wealth, or institutional affiliation to use and reproduce our research. References for the dataset and system are given at the beginning of \Cref{sec:data_acquisition} and \Cref{subs:interface}, respectively.
    \end{itemize}

\section{Outline}\label{sec:outline}
    \Cref{chap:intro} introduced the current problem in \gls{cs} research and research in general, which this thesis will work on (\Cref{sec:problem}), proposed a solution to address the problem (\Cref{sec:goals}), and listed our specific contributions (\Cref{sec:contribution}).
    
    \Cref{chap:fundamentals} introduces fundamental knowledge the reader needs to understand this thesis, including some technical aspects and concepts (e.g., topic modeling and scientometrics) we use in this thesis.

    \Cref{chap:related} addresses \ref{task:review}, by investigating what scientometric studies other researchers perform and what data sources and tools they use (\Cref{sec:scientometric}).
    We then review those data sources and tools and determine there is currently no free solution that can perform a quantitative analysis on \gls{cs} research and thus show which gap \csi and our research fill (\Cref{sec:resources}).

    \Cref{chap:methodology} presents our methodology, which is composed of  \labelcref{task:data} and \labelcref{task:system}.
    We begin by explaining how we acquire our data from \dblp and how we enrich it with more metadata from full-texts (\Cref{sec:data_acquisition}).
    Next, we cover how we store the data and make it accessible through an \gls{api} (\Cref{sec:data_storage}).
    The rest of the chapter is spent on detailing the \gls{ui} and its features, i.e., dashboards, filters, and visualizations (\Crefrange{subs:design}{subs:interface}).
    We also include some showcases to demonstrate how the system can be used (\Cref{sub:showcases})

    \Cref{chap:implementation} goes over some implementation details that relate to \labelcref{task:data} and \labelcref{task:system}.
    We provide an overview of the \csi system, its components and their architecture (\Cref{sec:architecture}), and some details on our measures for quality assurance (\Cref{sec:qa}).

    \Cref{chap:analysis} addresses \labelcref{task:analyis}
    by conducting an extensive analysis, that follows the research questions given earlier in the chapter.
    We split each research question into multiple smaller experiments, each of which receives a discussion, that evaluates the findings and relates them to other experiments and previous research.
    Before starting with the experiments, we state the general setup of how we conduct them (\Cref{sec:setup}).
    We then group the experiments by the main attribute they investigate (e.g., publications, authors, document types) and for each experiment include a figure or table that contains the results we then discuss (\Crefrange{sec:publications}{sec:fields}).
    In the end, we provide a summary of the chapter that shows we cover each aspect of each research question (\Cref{sec:analysis_summary}).

    \Cref{chap:final} presents the final considerations of this thesis.
    We list our contributions and most interesting findings and conclude \gls{cs} is a very active and growing field, whose characteristics and trends a scientometric analysis with \csi can uncover (\Cref{sec:conclusion}).
    Finally, we present the limitations and future work of this thesis (\Cref{sec:limitations}).

    \Cref{chap:appendix} includes supplementary figures of the \gls{ui} and some of its visualizations, which are referenced throughout this thesis.
    
	\chapter{Fundamentals}\label{chap:fundamentals}
This chapter shortly introduces some basic concepts related to technical aspects (\Cref{sec:technical}), topic modeling (\Cref{sec:topic_modeling}), and scientometrics (\Cref{sec:scientometrics_fund}).

\section{Technical Aspects}\label{sec:technical}
    \begin{wraptable}{r}{0pt}
    \centering
    \footnotesize
    \begin{tabular}{ll}
        \toprule
        \textbf{MySQL} &\textbf{MongoDB} \\
        \midrule
        Database &Database \\
        Table &Collection \\
        Index &Index \\
        Row &Document \\
        Column &Field \\
        Join &Lookup \\
        Primary key &Primary key \\
        Group by &Aggregation \\
        \bottomrule
    \end{tabular}
    \caption{Differences of terms between MySQL and MongoDB; adapted from \textcite{gyorodi_comparative_2015}.}\label{tab:databases}
\end{wraptable}
    The \csi system uses a \mongodb\footnote{https://www.mongodb.com/} database to store its data (\Cref{sec:data_storage}).
    \mongodb is a document-oriented database, and not a relational database (e.g., MySQL).
    The key difference is the way data is stored in a document-oriented database: data is stored in documents, while relational databases use tables with rows and columns.
    Thus, the terminology also changes, which \Cref{tab:databases} highlights.
    In document-oriented databases the schema is not fixed, so adding and removing fields is easier, which allows for quicker iterations during development.
    The documents in \mongodb use JSON's key-value pairs, but add support for more features (e.g., dates) \parencites[3,16-17]{bradshaw_mongodb_2019}{gyorodi_comparative_2015}.

    \glsunset{rest}
    We then provide access to this data through a \gls{rest} \gls{api} on our server (\Cref{subs:api}), i.e., an \gls{api}, which is based on the four principles of \glsfirst{rest}: identification of resources, manipulation of resources through representations, self-descriptive messages, and hypermedia as the engine of application state \parencite[82]{fielding_information_2000}.
    This means REST uses HTTP verbs (e.g., GET, POST, DELETE) to transfer data as representations in a well-defined media type (e.g., JSON, XML).
    The resources are identified through unique resources identifiers (URIs), e.g., \lstinline{shop/products/} to perform operations on all products or \lstinline{shop/products/42} for operations on product 42 \parencite{kopecky_history_2014}.
    \gls{crud} operations \parencite[14]{bradshaw_mongodb_2019} can then be performed on our data through the \gls{rest} \gls{api}.

\section{Topic Modeling}\label{sec:topic_modeling}
    Topic models are statistical models, that cluster a group of words into meaningful ``topics'' from any unstructured text or text corpus (e.g., emails, book chapters, blog posts).
    Each document in a corpus is treated as a ``bag of words'', i.e., the location of words in the document, syntax, and narrative of the document are ignored.
    The models then use these bags to determine the co-occurrence of specific words across the corpus of bags and generate the distribution of words that refers to each topic.
    Topic modeling is an automated approach, so researchers only have to define the number of topics the corpus is supposed to be sorted into \parencite{mohr_introductiontopic_2013}.
    The idea behind this is, e.g., documents about cats often include ``cat'' and ``meow'' and documents about dogs ``dog'' and ``woof'', which show these words occur together in specific documents and relate to the same topic.
    
    \gls{lda} \parencite{blei_latent_2003} is a generative probabilistic model and the most used topic modeling approach.
    It assumes each document contains multiple themes/topics the authors want to discuss.
    The document is then generated by the authors by repeatedly selecting a topic and a word from that topic and placing it in the bag of words of the document until the document is completed.
    Selecting the next topic is based on the distribution of topics across the documents and selecting the next word is based on the distribution of words across the selected topic.
    \gls{lda} then tries to infer the intents of the authors when generating the document by reverse-engineering the two distributions that are used to draw the topics for a document and words from a topic.
    Once both distributions are determined the most probable words for a specific topic can be used by humans to imagine the actual topics, as \gls{lda} cannot generate topic labels \parencite{mohr_introductiontopic_2013}.
    
    The most probable words of specific topics can also include common words that are present across all topics, as they appear in many different scenarios (e.g., ``paper'' or ``present'' for scholarly articles).
    These generic words also appear in the list of most frequent terms of the entire corpus and explain little about the topic's contents or the corpus.
    \textcite{chuang_termite_2012} develop the ranking measure ``saliency'', which is supposed to filter a corpus's list of most frequent words and rank words higher that only appear in a few topics, and words lower that appear in many topics.
    They first define the distinctiveness of a word, which measures how informative the word is for determining the generating topic, e.g., the word ``brain'' would be informative, while ``paper'' would not.
    Saliency is then computed by weighing the distinctiveness of a word against the overall probability of that word in the corpus.
    The list of the most salient words would then rank ``paper'' lower than the list of the most frequent words, making it easier to find differences between topics.

\section{Scientometrics}\label{sec:scientometrics_fund}
    Scientometrics is the study of quantitative aspects of science and technology, i.e., exploration and evaluation of scientific research.
    It covers measuring the quality of research and its impact, tracking and understanding citations, mapping and visualizing scientific fields, and using these measures for policy and management decisions (e.g., by institutions).
    Quantitative measures employed are, e.g., the impact factor for venues (average amount of citations per publication per year of that venue) or the h-index for authors ($h$ papers of the author have at least $h$ citations).
    Bibliometrics is similar to scientometrics and uses statistical methods to analyze publications and books, e.g., citation graphs.
    Both scientometrics and bibliometrics are sub-fields of informetrics, which covers the study of all information as a whole, regardless of form or origin \parencite{mingers_review_2015}.
    An example of a scientometric study in \gls{cs} is \textcite{coskun_scientometrics-based_2019}, which analyzes the trends over time regarding countries, document types, institutions, author collaboration, keywords, and journals.
    \textcite{fiala_computer_2017} conduct a bibliometric study investigating the quantity and the impact of publications according to document types, languages, disciplines, countries, institutions, and publication sources.
    Both use similar approaches (e.g., by examining the distribution of document types or most publishing institutions and countries) even though one is called a scientometric study and one a bibliometric study, which shows the closeness of bibliometrics and scientometrics when analyzing only scientific publications.
    We will further explore the findings of both studies in \Cref{subs:studies_cs}.

	\chapter{Related Work}\label{chap:related}
    This chapter provides an overview of related (scientometric) studies
    and the existing resources that can be used to make these studies easier.
    We start by investigating previous scientometric studies in \gls{cs} and their findings, where we also show what data and tools the authors use for their analyses (\Cref{subs:studies_cs}), and shortly look into unique approaches in \gls{nlp} (\Cref{subs:studies_nlp}).
    Then, we present the data sources those studies use and other available data resources to show their limitations, and which gap \csi fills (\Crefrange{subs:broad_aggregators}{subs:specialized_aggregators}).
    We also show a few general tools that can aid researchers in scientometric studies and highlight their differences from \csi (\Cref{subs:general_tools}).
    Finally, we present selected tools from \gls{nlp}, as some of these tools are very similar to what \csi tries to achieve (\Cref{subs:resources_nlp}).
    
\section{Scientometric Studies}\label{sec:scientometric}
    This section covers previous scientometric studies in \gls{cs} (\Cref{subs:studies_cs}) and shortly explores scientometric studies in \gls{nlp} (\Cref{subs:studies_nlp}).
    We show many researchers rely on paid-access data and most researchers do not use any specific tool to automate their studies.
    
\subsection{Scientometric Studies in Computer Science}\label{subs:studies_cs}
    In this subsection, we review the previous work of other researchers on \gls{cs} research to see which analyses are done, prove useful, and should also be provided in \csi.
    We cover broad studies of \gls{cs}, studies on topics and terms, conferences vs. journals, and lastly some studies on the differences between \gls{cs} and other research fields.

    \subsubsection{Broad Studies on Computer Science Research}
    First, we look into two broad studies on \gls{cs} research that investigate many areas of \gls{cs} research (e.g., publications, authors, venues, and citations).
    \textcite{coskun_scientometrics-based_2019} use data from the \wos core collection to perform a scientometrics-based study of \gls{cs} and Information Science research by looking at two periods (2008-2013 and 2014-2019) with 57,347 and 96,219 documents respectively and comparing the results to discover trends.
    They look into the document types and find that there are slightly fewer conference papers in the second period, but the overall amount of documents increases, as there are more journal articles and documents from other types.
    Conference papers make up most documents in both periods, but the gap between papers and articles closes over time.
    % A list of the top affiliations reveals, there are more universities overall in the second period, but both periods are dominated by US and Chinese universities.
    % Both countries also lead the list of top countries.
    % \textcite{coskun_scientometrics-based_2019} also leverage networks from \vosviewer to analyze the importance of countries using co-authorships and find the USA is the main hub, but China is also becoming a hub.
    % The top research areas that \wos assigns publications show little change in their ranking over time.
    The top journals come mostly from engineering and other technical sub-areas (e.g., from IEEE or IEICE), which also reflects in the top research areas, as they also show a focus on technical and engineering-related issues.
    % similar to the top research areas, which also focus on technical and engineering-related issues.
    Lastly, they use networks from \vosviewer\footnote{\url{https://www.vosviewer.com/}} to investigate the recurrence of keywords, which reveals a shift to current issues, such as privacy, security, IoT, and big data.

    \textcite{fiala_computer_2017} investigate the quantity and impact of 1.9m papers in \gls{cs} based on document type, language, discipline, country, institution, and publication source from 1945 to 2014 available in \wos.
    They investigate the distribution of document types, which shows that proceedings papers make up the biggest part of the collection, but articles have more than 7x the number of citations.
    The number of articles shows a steady rise over time, except for one large drop in 2007 which the authors attribute to papers published in two book series being classified differently from 2007 onward.
    Similarly, the amount of proceedings papers rises over time, except for a drop between 2010-2011, because multiple conferences are not indexed in those years.
    The distribution of document languages shows that 99\% of all documents are in English.
    Considering all seven subject categories in \wos, ``Artificial Intelligence'' has the most papers and citations, while ``Interdisciplinary Applications'' has the most citations per paper.
    In the top 20 sources (i.e., venues) the ``Lecture Notes in Computer Science'' have the most papers, the ``Journal of Computational Physics'' has the most citations, and the ``IEEE Transactions of Information Theory'' has the most citations per paper.
    \textcite{fiala_computer_2017} also compare the top 20 keywords for the entire time frame against those before 1995 and periods of five years after 1995 and find unique keywords in each period except 2005-2009.
    They also investigate the top 20 cited references (\#1 being ``INFORM CONTROL'' from Zadeh, L.A.) and papers (\#1 being ``Fuzzy sets'' from Zadeh, L.A.).
    The distribution of citations shows most citations are two years old, followed by three years, and one year, while 52.2\% of papers remain uncited and less than 1\% get over 100 citations.
    %use a relational database and then query it

    The approach of our analysis in this thesis (\Cref{chap:analysis}) is inspired by that of \textcite{coskun_scientometrics-based_2019, fiala_computer_2017}, as we also conduct a broad study of \gls{cs} research (i.e., looking into publications, authors, venues, and citations).
    While both conduct their analyses manually, we develop and use our system (\csi), which can generate visualizations easily and intuitively.
    \csi can replicate most of their analyses and show the top publications, authors, or venues, the distribution of document types and citations, and how they all change over time.
    Our topic modeling component can also determine the most salient terms for specific periods, venues, or authors.
    We also extend the studies of \textcite{coskun_scientometrics-based_2019} and \textcite{fiala_computer_2017} by using a much larger dataset (\csi has 5m publications and the updated version of \diii 6m\footnote{\url{https://zenodo.org/record/7069915}}), diving deeper into the areas they analyzed (e.g., distribution of citations and papers across authors and venues), and investigating more areas (e.g., the differences between \gls{cs} and other fields of study; see the rest of this section).
    Future researchers can also use \csi to verify and extend our research even further.

    \subsubsection{Other broad Studies}
    Some works allocate a part of their analysis to detail differences between institutions or countries over time or which are the most productive affiliations \parencite{coskun_scientometrics-based_2019, fiala_computer_2017, xia_research_2021}.
    The current version of our dataset does not include any data on the affiliations, so we leave the inclusion and investigation of institutions and countries to future work (\Cref{sec:future_work}).
    For this reason, we also leave out the many studies that focus entirely on analyzing the state and trends over time of \gls{cs} research for specific countries \parencite{uddin_scientometric_2015, supriyadi_bibliometric_2022, faiz_bibliometric_2020} or institutions in general \parencite{zurita_bibliometrics_2020}.
    The country-specific studies focus on the output and performance of publications, authors, and institutions.
    For example, \textcite{uddin_scientometric_2015} compare the performance stats (e.g. publications, citations) of Mexico against the world over time, investigate top countries, institutions, publication sources, and authors, the number of authors per paper, and collaboration patterns of authors, institutions, and countries.
    \textcite{zurita_bibliometrics_2020} only rank the institutions based on citations in seven sub-fields of \gls{cs}.
    This shows their approaches mirror \textcite{coskun_scientometrics-based_2019, fiala_computer_2017} or cannot be replicated with our data, so we do not cover their analyses and results any further.
    The authors of the studies on countries and institutions use different data sources, e.g., \scopus \parencite{supriyadi_bibliometric_2022, faiz_bibliometric_2020}, or \wos \parencite{uddin_scientometric_2015, zurita_bibliometrics_2020}.
    For evaluation \textcite{faiz_bibliometric_2020} uses \scival\footnote{\url{https://www.scival.com/}}, but most researchers use no tool for their evaluation or do not specify it.
    
    Some studies also include analyses with networks on authors, citations, or terms \parencite{coskun_scientometrics-based_2019, uddin_scientometric_2015}.
    While we look into some of these areas, we do not leverage any networks and thus cannot perform any analyses, which require networks or graphs.
    In the future, we intend to also conduct analyses with networks (\Cref{sec:future_work}).

    \subsubsection{Studies on Topics \& Terms}
    Other studies focus more on emerging terms and which areas are researched currently.
    \textcite{tattershall_detecting_2020} apply a stock-market-inspired burst detection algorithm to \dblp data (2.6m documents between 1988 and 2017) to find ``bursty terms'', i.e., the fastest-rising topics in the history of \gls{cs} research.
    They find historic peaks for ``Java'', ``e-commerce'', and ``Smartphone'', and that ``word embeddings'' and ``deep learning'' are still rising.
    Terms like ``neural network'' and ``virtual reality'' have two peaks, while other terms like ``novel'' are linearly increasing.
    Most terms show a life cycle of popularity and their classifier can predict with an accuracy of 80\%, whether a term will rise or fall in popularity.
    \textcite{xia_research_2021} use a different approach and leverage data from \scopus (75m documents from 1996 onward) and its classification system of subject areas.
    They cluster the publications based on direct citation references and then evaluate the prominence of each topic using citations, views, and the impact of recent years.
    The authors investigate the top 20 frontiers in \gls{cs} and find ``Object Detection; CNN; IOU'' to be the most prominent one, followed by ``Bitcoin; Ethereum; Blockchain'', while the most prominent frontier in \gls{nlp} is ``Sentiment Classification; Named Entity Recognition; Entailment'', followed by ``Sentiment Classification; Opinion Mining; Product Review''.
    We do not analyze the evolution of single terms or the most prominent topics as a whole, but using \csi's topic modeling component, we can determine the most salient terms and most prominent topics for venues, authors, and fields of study, and how they change over time.

    \subsubsection{Studies on Conferences vs. Journals}
    Another popular area is the comparison of conferences and journals in \gls{cs} research \parencite{franceschet_role_2010, vrettas_conferences_2015}, which is tied to the characteristics of \gls{cs} research itself.
    Most researchers in \gls{cs} focus their publications on conferences and not journals, unlike other research fields (e.g., medicine), where researchers use journals as the primary way to publish their findings \parencite{vrettas_conferences_2015, vardi_conferences_2009, franceschet_role_2010}.
    \textcite{vrettas_conferences_2015} argue this is the reason why many studies compare conferences and journals in \gls{cs}, as \gls{cs} is an outlier among the research fields in this regard.
    
    \textcite{rahm_citation_2005} analyze the citation frequencies between two conferences and three journals in the database field from \dblp over 10 years (1994-2003), with citation information from \googleScholar.
    They conclude, that the conferences have a higher citation impact than the journals.
    In a later publication, \textcite{rahm_comparing_2008} finds that conferences still have a higher impact than journals, again using select high-quality conferences and journals between 1996 and 2004.
    On the other hand, \textcite{franceschet_role_2010} finds that journals have a higher impact in \gls{cs} than conferences.
    He uses data from \dblp, \googleScholar, and \wos to look into the top authors based on different measures (e.g., number of publications), which shows researchers in \gls{cs} publish more in conferences than journals.
    His study on the most popular topics and nations with the highest scientific impact with separate entries for journals and conferences both show that journals receive significantly more citations and thus have a higher impact.
    \textcite{vrettas_conferences_2015} conclude the differences between the findings of \textcite{rahm_citation_2005,franceschet_role_2010} are due to different data sources, as \textcite{rahm_citation_2005} use \dblp and \textcite{franceschet_role_2010} uses \wos.
    While \textcite{franceschet_role_2010} also uses \googleScholar and \dblp data, his conclusion that journals have a higher impact is based solely on publication and citation information from \wos.
    \textcite{vrettas_conferences_2015} also examine conferences (195,513 papers) and journals (108,600 papers) themselves, by using data from \mas\footnote{The service was retired in 2012.} and aligning it with venue rankings from the Australian government's research assessment \gls{era}.
    They find the difference between citations of journals and conferences in \gls{cs} is marginal.
    Aligning the venues by the \gls{era} ranking, the high-ranked conferences get, on average, more citations than the high-ranked journals.
    Incidentally, this aligns with \textcite{rahm_citation_2005}, who also compare reputable conferences and journals.

    In our research, we are also taking a venue-based approach like \textcite{rahm_citation_2005, vrettas_conferences_2015}, by looking at the BibTeX entries of publications, which determines whether the publication is from a journal (i.e., ``article'') or a conference (i.e., ``inproceedings'').
    We extend their work by covering the number of citations and publications, topics, changes over time, top venues, and preferences of top authors.
    Other researchers use more author-based approaches to analyze publication patterns of authors and affiliations \parencite{kim_author-based_2019, kumari_scientometric_2020}.

    \subsubsection{Studies on Comparisons between \gls{cs} and other Research Fields}
    \textcite{vrettas_conferences_2015} show that there is a prevalence of conferences in \gls{cs}, as 76\% of ranked conferences in the \gls{era} assessment across all fields are from \gls{cs}.
    \textcite{michels_systematic_2014} also show other research fields prefer journals by listing the distributions of publications among journals and conferences for 27 research fields in \wos in 2009.
    Yet, \textcite{subelj_publication_2017} show growth in journal publications in \gls{cs} and attribute this to a rising number of new journals, rather than each journal publishing more.
    This is again different from other research fields, as the rise in journal articles in physics is due to the existing journals publishing more \parencite{subelj_publication_2017}.
    
    We also investigate the differences between \gls{cs} and other research fields (e.g., engineering, medicine), which most other papers only briefly mention if they cover them at all.
    The focus of our analysis is the difference in preferences for journals and conferences regarding citations and publications, and their topics.

\subsection{Scientometric Studies in Natural Language Processing}\label{subs:studies_nlp}
    In \gls{nlp} there are two series of publications on scientometric studies, that are worth noting, as they go into more detail, than \textcite{fiala_computer_2017, coskun_scientometrics-based_2019}.
    
    \textcite{mohammad_nlp_2020_data} uses the \nlpScholar dataset and visualization to examine the state of \gls{nlp} research.
    He discovers an increase in papers and authors in the last two decades and that authors are also publishing more papers yearly.
    The number of workshop and conference papers in the dataset (and thus \gls{nlp}) is also many times larger than that of journals.
    He also finds that the number of publications is higher in alternate years, due to biennial conferences.
    In another paper \textcite{mohammad_examining_2020} investigates the citations of \gls{nlp} literature, where he finds that journals have the highest average and median citation count, even though they make up only 2.5\% of the papers.
    Top-tier conferences (i.e., ACL, EMNLP, NAACL, COLING, EACL) combined ranked second, before the other conferences, workshops, etc.
    The same results are observed when only recent years are considered.
    In his work, he also bins the publications according to citation count and finds 6.4\% do not have any citations and about 56\% have 10 citations or more.
    Lastly, \textcite{mohammad_state_2019} investigates the topics of \gls{nlp} research, by analyzing the top unigrams and bigrams that occur in the titles of the papers and their development over time, number of citations and papers, and average and median citations.
    He finds, that ``language'' is the most occurring unigram and ``machine translation'' the most occurring bigram.
    The diversity of title unigrams was lower in the 1980s compared to recent years and ``neural'' has been the most occurring unigram in titles per year since 2017.
    
    \textcite{mariani_nlp4nlp_2019} perform an extensive analysis of the publications, authors and their collaboration, venues, citations and references, and their trends over time using the \nlpfournlp corpus.
    The authors find the number of papers, authors, and references increased more in the last two decades than before 2000.
    They also list the most productive authors based on their amount of publications and show that publications are higher in alternate years.
    Their analysis shows that recent papers (2015) and old papers (1974 and before) get the least amount of citations on average.
    New papers have not had enough time to accumulate citations yet, and it also becomes apparent there are only very few publications in their data for the 1960s and 1970s, which could explain the low average citations.
    In their second paper using the \nlpfournlp corpus \textcite{mariani_nlp4nlp_2019_2} investigate topics and terms over time and how their occurrences develop.
    \textcite{mariani_nlp4nlp_2019_2} find, e.g., that topics like ``hidden markov models'' and ``speech recognition'' dropped in frequency in the last few years, while ``annotation'' and ``dataset'' were rising.
    ``wordnet'' and ``support vector machine'' were rising for a while, but also dropped in frequency the last few years.
    They also show that single terms like ``bigram'' and ``trigram'' were also less frequent, while ``ngram'' saw a rise and then stagnated in the last few years.
    
    Our case study on \gls{cs} takes multiple aspects from the analysis conducted by \textcite{mohammad_nlp_2020_data, mariani_nlp4nlp_2019} that are not present in the studies on \gls{cs}, e.g., the number of authors over time, or citation binning.
    We use our automated system \csi, similar to \textcite{mohammad_nlp_2020_data}, who uses his \nlpScholar visualization for his analysis.
    \textcite{mariani_nlp4nlp_2019} on the other hand use multiple different tools to process the data and generate the visualizations.
    Finally, the analysis we perform includes the trends of papers, authors, venues, topics, citations, and references in \gls{cs}, but delves deeper into each analysis and adds aspects like discrepancies between conferences and journals, or research fields, which are missing or only shortly mentioned in the scientometric studies on \gls{nlp} mentioned in this subsection.
    
\section{Resources}\label{sec:resources}
    For this thesis we use ``aggregators'' as a broad term to refer to data resources such as digital libraries, repositories, and search engines, that aggregate scientific publications, and index them and their metadata, but might also include additional features to work with the data beyond a search function.
    These features can include, e.g., varying degrees of visualization options or filters to refine the search.
    % I use "aggregator" and not "data source", as "data source" feels like it provides just the data, which is not true for Scopus, etc.
    This section covers broad aggregators, that comprise publications of multiple disciplines (\Cref{subs:broad_aggregators}) and aggregators that are specialized in specific disciplines or fields (\Cref{subs:specialized_aggregators}).
    We also show tools that help to conduct scientometric studies in general (\Cref{subs:general_tools}), and which resources are available in \gls{nlp} (\Cref{subs:resources_nlp}), as some resources are similar to what we try to achieve with \csi.

\subsection{Broad Aggregators}\label{subs:broad_aggregators}
    \googleScholar (estimated to have over 389m records \parencite{gusenbauer_google_2019}) and \semanticScholar (over 206m papers\footnote{\url{https://www.semanticscholar.org/}}) are freely accessible web-based search engines for scholarly literature.
    They include different records, such as peer-reviewed publications and pre-prints.
    The search engines focus on searching and finding publications, authors, and their metrics (e.g., h-index, number of papers, citations), but lack details on venues and publishers.
    Their filter options are also limited, as both do not have a filter for the access type or number of citations, and \googleScholar also cannot filter by authors or venues.
    Additionally, neither \googleScholar nor \semanticScholar offers an interactive platform to browse their databases, preventing users from exploring features not explicitly available on their website (e.g., the fields of study).
    While their web interfaces are freely available, \googleScholar does not provide any \gls{api} or means to download their data.
    Some studies from \Cref{sec:scientometric} use \googleScholar for their analysis, but only to a very limited degree, e.g. to find the h-index of specific authors \parencite{franceschet_role_2010}, find citations for papers of a handful of venues \parencite{rahm_citation_2005}, or get citations for a smaller sub-field in \gls{cs} \parencite{mohammad_nlp_2020_data}.
    % Researchers also found issues with the quality and accuracy of the bibliographic data on \googleScholar \parencite{cavacini_what_2015}.
    \semanticScholar offers their data through an \gls{api} or as a bulk download, but both options to access their data require an access key\footnote{\url{https://www.semanticscholar.org/product/api}}.
    Their \gls{api} also allows 100 requests per 5 minutes without a key, for testing purposes.
    Two datasets are offered by \semanticScholar: S2AG (Semantic Scholar Academic Graph) \parencite{ammar_construction_2018}, which includes all data that makes up the knowledge graph that powers \semanticScholar (over 206m papers); and S2ORC (Semantic Scholar Open Research Corpus) \parencite{lo_s2orc_2020}, which includes a subset of open-access papers from S2AG and their metadata (136m papers), enriched with abstracts and full-texts.
    \googleScholar and \semanticScholar are both limited by not offering any quantitative analysis options and not providing easy access to their data, both of which \csi overcomes.
    While \semanticScholar's S2ORC dataset allows easy access, it is not regularly updated, which limits its use to investigate more current trends.
    Other researchers leverage \googleScholar, but they only use it for its h-index or citation metrics and in much smaller quantities than we require to analyze \gls{cs}.
    
    Some researchers \parencite{vrettas_conferences_2015, bornmann_growth_2021} use Microsoft's academic search engines in their scientometric studies, but Microsoft retired their services \mas in 2012\footnote{\url{https://web.archive.org/web/20170105184616/https://academic.microsoft.com/FAQ}} and its successor \microsoft in 2021\footnote{\url{https://www.microsoft.com/en-us/research/project/academic/}}, so both will not be covered further.

    Two large web-based paid-access platforms are \wos (over 171m records\footnote{\url{https://clarivate.com/webofsciencegroup/solutions/web-of-science/}}) and \scopus (over 87m records\footnote{\url{https://blog.scopus.com/posts/scopus-roadmap-whats-new-in-2022}}).
    They expand on the capabilities of \googleScholar and \semanticScholar (e.g., the search engine and citation index), by each offering more filters (19 in \wos and more than 12 in \scopus) to refine the search (e.g., publisher, field of study, affiliation, keyword).
    The (refined) search results can then be exported and downloaded for further analysis, which other resources (e.g., \googleScholar, \semanticScholar) do not offer.
    For that reason, many scientometric studies use the data from \wos \parencite{coskun_scientometrics-based_2019, fiala_computer_2017, franceschet_role_2010} or \scopus \parencite{xia_research_2021, bornmann_growth_2021}.
    Especially researchers analyzing the research output of affiliations (specific countries or institutions) tend to choose \wos \parencite{uddin_scientometric_2015, zurita_bibliometrics_2020} or \scopus \parencite{faiz_bibliometric_2020, supriyadi_bibliometric_2022}, as other resources might not have data on affiliations (e.g., \dblp).
    Both platforms also allow a basic analysis of the (refined) search results by generating visualizations.
    \wos can group the results by 21 attributes (similar attributes as for the filters, e.g., authors, publication years, document types), and then visualize the results as a treemap, bar chart, and grid with a configurable number of entries.
    \scopus can also group its (refined) search results and visualizes the results either as a line chart (for grouping by year or source), bar chart (author, affiliation, or country/territory), or pie-/ringchart (document type, subject area) while showing a list with the top entities on the left.
    These visualizations allow analyzing distributions, trends, and comparing entities (e.g., authors, venues), but are not used by any studies to our knowledge.
    \wos and \scopus provide many features to their users, but their main limitation is they are paid-access only, which prohibits researchers without the necessary funds (e.g., from developing countries) from accessing the services.
    \csi intends to take their analysis component and make it available to everyone, extending the kind of quantitative analysis that can be done with \wos and \scopus.
    We also conduct a case study with \csi using \gls{cs} to show what it is capable of and that researchers do not have to conduct all their scientometric analysis manually, as most currently do.
    However, we do not offer any ways to search for papers, as there are already multiple open-access solutions for that (e.g., \googleScholar, \semanticScholar, \dblp).

\subsection{Specialized Aggregators}\label{subs:specialized_aggregators}
    The specialized aggregators still have varying degrees of specialization and size.
    We start detailing the larger and broader ones and then move to the smaller and even more specialized ones.

    Two large open-access repositories are \arxiv (over 2.1m scholarly articles\footnote{\url{https://arxiv.org/}}) and \dblp \parencite{ley_trierer_1997} (over 6.3m publications\footnote{\url{https://dblp.org/}}).
    \arxiv stores pre-prints from sciences and some related fields so its contents are not peer-reviewed but it offers multiple ways to download its latest data.
    The Computing and Research Repository (CoRR) \parencite{halpern_corr_2000} is the section of \arxiv that focuses on \gls{cs} and has multiple categories (e.g., Artificial Intelligence, Computation and Language, and Databases).
    \dblp, on the other hand, entirely focuses on \gls{cs} publications, including both peer-reviewed publications and some pre-prints, and their downloadable data gets updated monthly.
    Both \arxiv and \dblp do not offer a citation index or options for analyses with visualizations.
    \citeseer\footnote{\label{foot:citeseer}\url{https://citeseerx.ist.psu.edu/}} (over 10m records\cref{foot:citeseer}) is a digital library, which also focuses on papers in Computer and Information Science.
    It crawls its data from publicly available websites and thus is fully open-access and provides all its data for download.
    Their copyright only covers up to 2019 and an exemplified search for ``machine learning'' only returns papers from 2017 or earlier, so we conclude \citeseer is not further updated.
    No studies we cover use data from \citeseer, but multiple studies use \dblp \parencite{rahm_citation_2005, franceschet_role_2010, tattershall_detecting_2020, kim_author-based_2019} and \arxiv \parencite{sharma_drift_2021}, due to their open-access nature and being up to date.
    More studies use \dblp than \arxiv which we explain with \arxiv consisting solely of pre-prints, while \dblp covers mostly peer-reviewed publications.
    As \dblp, \arxiv, and \citeseer overcome the paid-access issues of \wos and \scopus, and include an easy data download, unlike \googleScholar and \semanticScholar, they are a step in the right direction.
    Unfortunately, they do not provide any visual analysis, but this is the gap \csi fills, as we leverage data from \dblp for our case study (\Cref{subs:data_source}) and thus expand the features of \dblp, to provide a free system to perform a scientometric analysis of \gls{cs}.
    To our knowledge, no one else has created an open-source and open-access (visual) analysis system for \gls{cs} yet.

    Some publishers like IEEE and ACM also have their own platforms, i.e., \ieeex\footnote{\label{note:ieee}\url{https://ieeexplore.ieee.org/}} (over 5.7m items\cref{note:ieee}) and the \acmdl\footnote{\url{https://dl.acm.org/}} (over 550k articles\footnote{\url{https://libraries.acm.org/digital-library}}), for their own publications and those of their partner publishers.
    Both platforms offer a search, citation index, and some filters.
    As both publishers focus on \gls{cs} publications, and in the case of IEEE also engineering, the contents of their platforms reflect that.
    Downloading some of the articles and papers with full-text requires paid-access and they provide little to no options to analyze the search results.
    No study we looked into uses data from \ieeex or the \acmdl, which is not surprising, considering \wos and \scopus offer better features to get data for a scientometric study in general, and repositories like \dblp are fully open-access with a larger number of \gls{cs} publications.
    
    For completion, we also want to mention some of the many small aggregators, that focus on specific areas in \gls{cs} or offer additional features, e.g., linking code or tables.
    \zetaalpha\footnote{\url{https://search.zeta-alpha.com/}} is a discovery and recommendations engine for papers, trends, and code in AI and data science.
    \paperswithcode\footnote{\url{https://paperswithcode.com/})} is a free and open resource of machine learning papers, code, datasets, methods, and evaluation tables.
    \nlpindex\footnote{\url{https://index.quantumstat.com/}} focuses on \gls{nlp} GitHub repositories with papers.
    \fourtytwo\footnote{\url{https://42papers.com/}} aggregates high-quality \gls{cs} and \gls{ai} papers and enabled its users to share them with each other.
    None of the four mentioned aggregators offer any citation counts or analysis, except Zeta Alpha, which maps the search results into a two-dimensional semantic space and links them in a graph using \vosviewer.
    Like \ieeex and the \acmdl, these four small aggregators were also not used in any studies we covered in \Cref{sec:scientometric}, as these are more for niches in \gls{cs} and not \gls{cs} as a whole.
    They do not provide any features that compare to \csi and their data is also not of interest to us, as \dblp provides data more suited for our case study.

    There are also other aggregators for other areas, e.g., \pubmed\footnote{\url{https://pubmed.ncbi.nlm.nih.gov/}} for medicine, but those also do not provide features similar to \csi (i.e., analysis based on visualizations), so these will not be covered, as our case study is on \gls{cs}.

\subsection{General Tools}\label{subs:general_tools}
    Besides aggregators, there are also some tools not specific to any research field, that researchers can use to perform scientometric studies.
    These tools do not focus on a search function, instead, most focus on visualizations and analyses.
    In this subsection, we present a few such tools, their features, and major differences from \csi.

    \scival builds on \scopus's data to visualize research performance for authors, institutions, and countries.
    It allows researchers to benchmark author and institution performance and analyze research trends based on different metrics, including publication and citation metrics from \scopus and additional metrics (e.g., topics, authors, and research areas).
    Both \scopus and \scival belong to Elsevier, and thus \scival is paid-access only like \scopus.
    The main difference between \scival and \csi is, that \scival focuses on determining research performance, while \csi allows for a scientometric analysis of broad trends in \gls{cs} research (e.g., of its publications, authors, and venues).
    % weekly updates
    % evaluate own research activities and of peer

    Some tools use network-based approaches for analysis, which we present two of in this paragraph.
    \citespace\footnote{\url{http://cluster.cis.drexel.edu/~cchen/citespace/}} \parencite{chen_searching_2004} is a free Java application, that visualizes trends and patterns in scientific literature.
    It visualizes the co-citation network of a knowledge domain to make it easy to locate pivoting points, turning points, and cluster centers.
    \textcite{chen_searching_2004} uses his tool to find the two revolutions in the superstring field in theoretical physics.
    He later added features to visualize emerging trends and abrupt changes and uses them to show their effectiveness in mass-extinction research (1981-2003) and terrorism research (1990-2003) \parencite{chen_citespace_2006}.
    \citespace can directly work with data from \wos and includes interfaces to work with data from \pubmed, \arxiv, ADS, and NSF Award Abstracts.
    Similarly, \vosviewer \parencite{van_eck_software_2010} visualizes bibliometric networks.
    It can leverage relations of citations, bibliographic coupling, co-citations, or co-authorships.
    Additionally, it offers text mining capabilities to extract important terms from scientific publications and to visualize them in a co-occurrence network.
    \zetaalpha (mentioned in \Cref{subs:specialized_aggregators}) uses \vosviewer to create their graphs linking publications in a semantic space and \textcite{coskun_scientometrics-based_2019} use it for some of their graph-based analyses.
    % Lastly, the Network Workbench\footnote{\url{https://nwb.cns.iu.edu/}} \parencite{nwb_team_network_2006} also covers network analysis, visualization, and modeling for research in physics, biomedicine, and social sciences.
    % The free program provides multiple algorithms to perform network analysis and includes tools to generate, run, and validate network models, but it does not appear to have been updated for over 10 years.
    Both tools presented in this paragraph focus on analysis with networks, which \csi and our case study on \gls{cs} do not cover.
    \citespace and \vosviewer provide useful insights into research areas or domains that do not have specialized tools (e.g., the tools \gls{nlp} has; see next subsection) available, but they do not help us with a quantitative analysis as \csi does.

\subsection{Resources in Natural Language Processing}\label{subs:resources_nlp}
    In this subsection, we detail resources available for the research field of \gls{nlp}, as some resources strongly correlate to what we want to achieve with \csi.

    The most prominent resource in \gls{nlp} probably is the \aclAnthology\footnote{\url{https://aclanthology.org/}}, which consists of nearly 80k open-access papers from the area of computational linguistics and \gls{nlp}.
    It is used in many other resources and studies, e.g., the \nlpfournlp corpus \parencite{mariani_nlp4nlp_2019}, which includes papers from the \aclAnthology and some other venues with a focus on \gls{nlp} (e.g., ISCA, IEEE, ICASSP, TASLP, LRE); and the \nlpScholar dataset \parencite{mohammad_nlp_2020_data}, which combines the data from the \aclAnthology with citation information from \googleScholar.
    Our analysis also shows over 99.3\% of the papers in the \aclAnthology are also in \dblp \parencite{wahle_d3_2022}, which is the dataset we use for our study.
    We consider the \aclAnthology for \gls{nlp}, what \dblp is for \gls{cs}: the largest available open-access dataset for the respective area.

    Researchers interested in investigating trends in \gls{nlp} can use the interactive visualization \nlpScholar\footnote{\url{http://saifmohammad.com/WebPages/nlpscholar-demo-basic.html}} \parencite{mohammad_nlp_2020_viz}, which is built with Tableau\footnote{\url{https://www.tableau.com/}} and uses the dataset with the same name \parencite{mohammad_nlp_2020_data}.
    The \nlpScholar visualization features a bar chart for papers per year and citations per year, a list with the most cited papers and authors, a boxplot of citations, and a treemap of the venues with the most published papers.
    It offers filters for the year of publication, authors, the number of citations, and paper title unigram or bigram.
    \csi shares certain similarities with \nlpScholar, as it also allows its users to perform a quantitative analysis of its underlying data, which is also our goal, just with another focus on the data.
    Thus, we expand on the capabilities of \nlpScholar and add more filters and aggregation options, while also covering a broader field by investigating \gls{cs}.
    Due to us building a dedicated system and not relying on Tableau, we also have a more scalable solution, that allows processing larger datasets.

    In \nlpExplorer\footnote{\url{http://nlpexplorer.org}} \parencite{parmar_nlpexplorer_2020} users can explore \gls{nlp} papers, venues, authors, and topics with an \gls{lda} \parencite{blei_latent_2003} topic modeling approach.
    The tool curates five topics, each with multiple subcategories.
    These topics and subcategories can be explored by searching and then selecting a specific venue or author.
    \nlpExplorer also shows the paper and citation distribution over years, but selecting a paper or topic only shows its metadata.
    Another tool, called \drift\footnote{\url{https://gchhablani-drift-app-t0asgh.streamlitapp.com/}} \parencite{sharma_drift_2021}, tracks research trends and developments over the years.
    The available analysis methods include keyword extraction, word clouds, predicting trends using productivity, tracking bi-grams, finding the semantic drift of words, tracking trends using similarity, and topic modeling.
    \textcite{sharma_drift_2021} perform a case study on the cs.CL corpus\footnote{\url{https://arxiv.org/list/cs.CL/current}} from \arxiv, which is the subset of \gls{cs} papers that covers Computation and Language (i.e., \gls{nlp}), but users can also upload their own corpus.
    \csi also includes a topic modeling component, but with a different focus than \nlpExplorer, as we generate the topics automatically from the terms used in the titles and abstracts and additionally offer a comparison based on the most used terms.
    Our implementation and visualization also allow for more customizability and exploration than \drift.

        % \setcounter{secnumdepth}{3}
	% weitere (eigene) Kapitel
	\chapter{Methodology}\label{chap:methodology}
    The goal of this thesis is to analyze the state of \gls{cs} research with a scientometric study.
    In the previous chapter (\Cref{chap:related}), we showed that current solutions to analyze the state of \gls{cs} research are limited, so we build our own (\csi) to answer our research questions and achieve our goal.
    This chapter details how we build the \csi system to analyze \gls{cs} research, which takes three steps.
    We first require a large dataset specialized on \gls{cs} publications and their metadata, so we reason the data source we pick (\Cref{subs:data_source}) and explain how we acquire the data (\Crefrange{subs:data_primary}{subs:data_secondary}).
    Second, we need to store the data in a way that makes it easy to extract the information we need again for our study (\Cref{subs:schema}) and then provide ways to manage and interact with the data more efficiently (\Cref{subs:api}), which also makes it available for other researchers.
    Lastly, we create an interactive system that queries the data user-friendly and visualizes it in different plots.
    We explain details about its design (\Cref{subs:design}) and interface (\Cref{subs:interface}) and show some examples of how the system can be used (\Cref{sub:showcases}).
    % An overview of the whole system and all its components from the implementation view is included in the next chapter (\Cref{fig:system_overview}).

\section{Data Acquisition}\label{sec:data_acquisition}
    The first step is to extract a large collection of data which will allow us to answer our research questions.
    For this, we must first decide on a source for our data (\Cref{subs:data_source}).
    We then detail how we extract the data we need and enrich it with more metadata (\Crefrange{subs:data_primary}{subs:data_secondary}).
    This section describes how we create the original version of the \diiilong (\diii) \parencite{wahle_d3_2022}, which we use in this thesis and is available on zenodo\footnote{\url{https://zenodo.org/record/6477785}}.
    A new version is also available\footnote{\url{https://zenodo.org/record/7069915}}, but that is for future work (\Cref{sec:future_work}) as it was not available in time for this thesis.

\subsection{Data Source}\label{subs:data_source}
    We decide to use a preexisting data source (\dblp) over building a new one from scratch as this has multiple benefits, including the source most likely already taking care of issues in matching papers, authors, venues, etc.
    Publications without a \gls{doi} or link might be hard to join because of small differences in the title (e.g., due to hyphens), or there might be multiple versions due to pre-prints.
    The authors might have different spellings (e.g., ``Christopher D. Manning'', ``Christopher Manning'', ``Chris Manning''), or there might be multiple authors with the same name (e.g., ``Yang Liu''\footnote{Search on Google Scholar for ``Yang Liu'': \url{https://scholar.google.com/citations?view_op=search_authors&mauthors=yang+liu}}) \parencite{ley_dblp_2009, ammar_construction_2018}.
    Venues also might be abbreviated (e.g., ``IEEE Transactions on Information Theory'' to ``IEEE Trans. Inf. Theory''), or the venues are just mentioned with their code (e.g., ``EMNLP'' instead of ``Conference on Empirical Methods in Natural Language Processing'').
    The codes can also change, e.g., all codes in the \aclAnthology changed in 2020 \parencite{mohammad_nlp_2020_data}.

    There are multiple reasons for picking \dblp as a data source over the other available aggregators covered in \Cref{sec:resources}.
    All broad aggregators we mentioned in \Cref{subs:broad_aggregators} could not be utilized.
    \googleScholar does not offer easy access to its data (no standardized \gls{api} and rate limitations for webpage crawling), while \wos and \scopus are not open-access.
    Semantic Scholar's contents are also proprietary \parencite{gusenbauer_google_2019}.
    They do offer the S2ORC dataset \parencite{lo_s2orc_2020}, which consists of a subset of their data, but it is not regularly updated and received its last update in 2020\footnote{\url{https://github.com/allenai/s2orc}}.
    Of all the specialized aggregators we covered (\Cref{subs:specialized_aggregators}), we find \dblp the most fitting.
    It is the largest open-access repository of \gls{cs} publications and their metadata, which also gets updated monthly \parencite{wahle_d3_2022}.
    \dblp also takes care of most of the mapping issues mentioned in the last paragraph \parencite{ley_dblp_2002, ley_dblp_2009}.
    Other specialized aggregators all have one or more drawbacks compared to \dblp, because they are smaller (\zetaalpha, \paperswithcode, \nlpindex, \fourtytwo), not peer-reviewed (\arxiv), or not updated anymore (\citeseer).
    \ieeex and the \acmdl are tied to their respective publishers and thus might be focused too much on specific areas and not give enough variety for document types or venues.
    Additionally, some previous scientometric studies already used \dblp data \parencite{rahm_citation_2005, franceschet_role_2010, tattershall_detecting_2020, kim_author-based_2019}, so we feel \dblp is a good choice for our data source.

\subsection{Primary Information from DBLP}\label{subs:data_primary}
    \dblp offers open-access to their data in multiple ways.
    Researchers can use the search, that is available through their website,
    or the search \gls{api} for publications\footnote{\url{https://dblp.org/search/publ/api}}, authors\footnote{\url{https://dblp.org/search/author/api}}, and venues\footnote{\url{https://dblp.org/search/venue/api}}.
    \dblp also provides monthly updated XML dumps of their data\footnote{\url{https://dblp.org/xml/release}}.
    We retrieve the full release of all currently available data, as we are interested in the state of \gls{cs} at a large scale over time, and extract all records from January 1st, 1936 to December 2nd, 2021, which includes \dblp's monthly release from December 1st, 2021.
    In the future, we can also use \dblp's monthly releases to keep \csi up-to-date automatically, i.e., download the latest release each month to add all new entries and update already existing entries \parencite{wahle_d3_2022}.
    An overview of the attributes, including examples of the data we retrieve from \dblp, will be provided in \Cref{tab:schema_new} in \Cref{sec:data_storage}.

    The largest actors in research are the publications, authors, and venues.
    \dblp directly supplies them, so we can extract them using a limited amount of additional work:
    
    \paragraph{Publications}
    Most entries in \dblp are indexed publications with their respective metadata; other examples include webpages and author information.
    \dblp classifies documents according to their BibTeX entry types (e.g., article, inproceedings).
    We transform all records into a standard JSON format based on the document type of the publications and map authors and venues to uniquely identified entities.
    
    \paragraph{Authors}
    \dblp handles multiple authors with the same name using an iterative four-digit counter in their data and when aggregating the data it distinguishes those authors automatically using different heuristics \parencite{ley_dblp_2009}.
    In case authors cannot be clearly distinguished, \dblp uses disambiguation pages\footnote{Disambiguation page for ``Yang Liu'': \url{https://dblp.org/pid/51/3710.html}}.
    Authors with multiple names are mapped in \dblp's author records \parencite{ley_dblp_2009}, which are sparse and rarely contain other informative features besides an URL to the personal webpage of the authors.
    We use the unique ids of the authors to map them to their publications.
    The author's current affiliated institution is not available in its own field and might only be entered in the ``note'' field \parencite{ley_dblp_2009}.
    
    \paragraph{Venues}
    For almost all publications, \dblp provides a venue code, by using the abbreviation of the venue or its acronym (e.g., ``IEEE Trans. Inf. Theory'' instead of ``IEEE Transactions on Information Theory'' or ``EMNLP'' instead of ``Conference on Empirical Methods in Natural Language Processing'').
    We map them to their publications with their unique ids, like the authors.
    In \dblp's data, the venue name is stored in different fields, depending on the document type, i.e., conferences use the field ``booktitle'', while journals use the field ``journal''.
    These are also two different fields in our dataset (\diii), so when extracting information from the dataset, one has to be careful to consider both fields and merge their contents, as both fields are never used in the same record.

    \paragraph{Other Fields}
    \dblp also contains some other fields, which we copy without any modification and directly store in the publication entries.
    The two most notable fields are the \textit{type of paper} (contains information about the BibTeX type of the publications) and the \textit{publishers} (e.g., Springer, IEEE, and ACM; but the data is very scarce, as less than 10\% of publications have publishers annotated).

\subsection{Secondary Information from Full-Texts}\label{subs:data_secondary}
    The full-texts of publications contain valuable information about author affiliations, content, and references currently not present in \dblp or other resources (e.g., \nlpScholar).
    We leverage the different fields \dblp provides for \gls{doi}s and links to the publications (e.g., ``url'' or ``ee'', meaning electronic edition \parencite{ley_dblp_2009}), to crawl the publications and retrieve their corresponding PDF files, which include the full-text.
    We then use \grobid\footnote{\url{https://github.com/kermitt2/grobid}} \parencite{lopez_grobid_2022} to parse the PDFs and extract abstracts, affiliations, and references.
    \grobid stands for \textbf{G}ene\textbf{R}ation \textbf{O}f \textbf{BI}bliographic \textbf{D}ata and is an open-source machine-learning library for extracting, parsing, and converting PDF documents into structured XML documents.
    \Cref{tab:schema_new} in \Cref{sec:data_storage} also includes the attributes we extracted from the full-texts with example values.

    \paragraph{Abstracts}
    For abstract extraction, we use \grobid's CRF Wapiti \parencite{lavergne_practical_2010} engine, which achieves an F1-score (using Levenshtein Matching with a minimum distance of 0.8) of 92.85\% when drawing 1943 \pubmed papers\footnote{\label{foot:grobid_pubmed}\url{https://grobid.readthedocs.io/en/latest/Benchmarking-pmc/}}.
    With this model, we retrieve 3,980,144 abstracts which are 81.33\% of the documents in the dataset.
    \grobid disregards the remaining documents because of poor quality or because there is no accessible document that could be parsed.
    We directly add the extracted information to the records we get from \dblp.

    \paragraph{Affiliations}
    We extract the author names and affiliations with the same engine we use for extracting the abstracts.
    To create author--affiliation pairs, we match author names from extracted affiliations to author names in \dblp using the Levenshtein distance.
    Using name matching to create author--affiliation pairs is also robust in practice, which we demonstrate by performing two small bootstrap and permutation tests \parencite{dror_hitchhikers_2018}.
    In the first test, we randomly draw 20 samples of $n=100$ publications and evaluate how often author names extracted from the PDFs do not match those in \dblp.
    To draw more challenging samples in the second test, we took the first $n=100$ publications from a ranked list in which the average Levenshtein distance between authors' names increased.
    Both tests show less than 5\% of names are mismatched ($p<0.001$).

    While our approach to creating authors--affiliations pairs proves to yield great quality results, \grobid has issues properly extracting and parsing affiliations in the first place.
    There are duplicates, incorrect, incorrectly structured, or missing affiliations.
    Considering the issues \grobid has to parse the affiliations, we decide the incorporation of institutions will be left to future versions of \csi (\Cref{sec:future_work}).
    The information about the countries is also left for future versions, as \dblp does not have that data and we cannot derive the country information from the institutions, due to the issue listed in extracting them.

    \paragraph{Citations}\label{par:citations}
    \googleScholar does not provide large-scale access to their data, i.e., it does not have a standardized \gls{api} and limits access for crawling, so we cannot use \googleScholar to retrieve citations.
    Other services also cannot be used, as they have the same issues (e.g., \semanticScholar), or their data is not open-access (e.g., \wos and \scopus).
    We instead calculate citations within \dblp ourselves, by building a citation graph from the bibliographies of full-texts similar to the \aclAnthology Reference Corpus \parencite{radev_acl_2009}.
    To parse the documents' bibliographies, we use \grobid's BidLSTM-CRF engine, which obtains an F1-score of 87.73\% for the \pubmed samples (using Levenshtein Matching with a minimum distance of 0.8)\cref{foot:grobid_pubmed}.
    We add two fields to each publication entity to create our citation links.
    One for the incoming citations (i.e., for each document that cites the publication) and one for the outgoing citations (i.e., for each reference in the bibliography of the publication).
    From this, we receive two lists of document ids, which we can use to construct a citation graph.

    When measuring the number of citations that come from outside \diii using the \semanticScholar \gls{api} we receive the result that 21.15\% of citations are from papers outside of \diii (i.e., other research fields than \gls{cs}) \parencite{wahle_d3_2022}.
    During that step, we match our data to \semanticScholar using the \dblp-id, which also yields us the entries for the fields of study.

\section{Data Storage \& API}\label{sec:data_storage}
    In the previous section, we acquired the data we need, so the next step is to store it in a way, that makes it easy and efficient to retrieve again.
    To store our data we decide to use \mongodb, which allows great performance and scalability \parencites[6]{bradshaw_mongodb_2019}{gyorodi_comparative_2015}.
    One of our future goals is to make \csi available to work with other datasets, which requires flexibility of the schemas and is one of the benefits of using a non-relational database \parencites[3]{bradshaw_mongodb_2019}{gyorodi_comparative_2015}.
    In this section, we first explain how we design the database schema (\Cref{subs:schema}) and how we manage the data and make it readily available through an \gls{api} (\Cref{subs:api}).

\subsection{Database Schema}\label{subs:schema}
    The current database schema is shown in \Cref{tab:schema_new}.
    It differs from the original schema shown by \textcite{wahle_d3_2022} (\Cref{tab:schema_old}), as we had to make some changes for performance increases and due to some attributes we could not use for any analyses.

    \newcommand{\tabIndent}{\hspace{3mm}}

\begin{table}[p]
\footnotesize
% \tabcolsep=0cm
% \begin{minipage}[t]{.5\linewidth}
\centering
    \begin{tabular}[t]{l r}%{l @{\hspace{-2cm}} r} 
        \toprule
        \textbf{Attribute} &\textbf{Example}\\
        \toprule
        \textbf{publication} & \\
        \tabIndent id                      & 62cc663aeba63d1b526e0689\\
        \tabIndent title                   & NLP Scholar - An Interactive ...\\
        \tabIndent abstractText            & As part of the NLP Scholar ...\\
        \tabIndent yearPublished           & 2020\\
        \tabIndent authors                 & [Saif M. Mohammad]\\
        \tabIndent authorIds               & [62bf2884e9832d137d41fb5e]\\
        \tabIndent venue                   & ACL (demo)\\
        \tabIndent venueId                 & 62bf273022ce6513861ee199\\
        \tabIndent publisher               & ACL\\
        \tabIndent typeOfPaper             & inproceedings\\
        \tabIndent fieldsOfStudy           & [Computer Science]\\
        \tabIndent *inCitations            & []\\
        \tabIndent inCitationsCount        & 9\\
        \tabIndent *outCitations           & []\\
        \tabIndent outCitationsCount       & 34\\
        \tabIndent openAccess              & true\\
        \tabIndent dblpId                  & conf/acl/Mohammad20b\\
        \tabIndent doi                     & https://doi.org/...\\
        \tabIndent pdfUrls                 & []\\
        \tabIndent url                     & db/conf/acl/acl2020-d.html\#Mohammad20b\\
        % \tabIndent createdBy               & 62bf1b1e46d4b21066009fee\\
        % \tabIndent createdAt               & 2022-07-11T18:04:42.639+00:00\\
        \textbf{author} & \\
        \tabIndent id                      & 62bf2884e9832d137d41fb5e\\
        \tabIndent fullname                & Saif M. Mohammad\\
        \tabIndent number                  & 0001\\
        \tabIndent orcid                   & 0000-0003-2716-7516\\
        \tabIndent *timestamp              & -\\
        \tabIndent *email                  & -\\
        \tabIndent *dblpId                 & -\\
        \textbf{venue} & \\
        \tabIndent id                      & 62bf273022ce6513861ee199 \\
        \tabIndent names                   & [ACL (demo)] \\
        \tabIndent *acronyms               & [] \\
        \tabIndent *venueCodes             & [] \\
        \tabIndent *venueDetails           & [] \\
        \tabIndent *dblpId                 & - \\
        *\textbf{affiliation} & \\
        \tabIndent *id             & 4eb3...f094\\
        \tabIndent *name           & National Research Council Canada\\
        \tabIndent *country        & Canada\\
        \tabIndent *city           & Ottawa\\
        \tabIndent *lat            & -\\
        \tabIndent *lng            & -\\
        \tabIndent *dblpId        & -\\
        \bottomrule
    \end{tabular}
    \caption{Database schema currently used in \csi. Unused attributes are marked with an asterisk (*).}
    \label{tab:schema_new}
% }
% \end{minipage}
% \caption{Attributes as proposed by \textcite{wahle_d3_2022} (left) and database schema currently in use (right).} 
% \label{tab:database_schema}
\end{table}
    
    We put the data from the crawler into our database with a Python script.
    The script ignores any attributes in the data that we are not interested in (e.g., pages).
    Some attributes are also unused (i.e., empty fields), as we do not intend to analyze them currently, but maybe in the future (\Cref{sec:future_work}).
    Those attributes are marked with an asterisk (*) in \Cref{tab:schema_new}.
    Citation references (i.e., ids for incoming and outgoing citations) are empty, as those are taking up most of the space, and we are not interested in any network analysis.
    So far, we only use the citation references to extract the citation counts.
    We also leave the affiliations empty, as explained in \Cref{subs:data_secondary}, and many fields for authors or venues, as we do not have that data yet, but might want to investigate them further in the future.
    
    Another change we make is the duplication and denormalization of the author and venue names, so they are also directly available in the publications collection.
    When first testing the system, we quickly realized, that a normalized solution that requires lookups of author and venue names using the unique ids does not work, as it is too slow.
    A denormalized schema helps, as it is quicker to read, but takes longer to write \parencite[211-212]{bradshaw_mongodb_2019}.
    In the normalized schema, we would have to do a \mongo{lookup} operation across the entire dataset, which takes 10-15 minutes from our experience.
    Instead, we denormalize the author and venue names and copy them into the publication collection, so the queries only take 10-15 seconds.
    \csi revolves around analyzing (i.e., reading data) and only rarely writes data (once a month max.), so increased write times are not an issue.
    % this is the way you are supposed to do it: https://www.mongodb.com/developer/products/mongodb/mongodb-schema-design-best-practices/}

    Due to the limitations of the database, we do not perform any further denormalization, even though it could reduce response times even more (i.e., copying the data of all publications into the referenced\footnote{Referenced through the author and venue ids.} entries in the author and venue collection).
    % Even though further denormalization could reduce response times even more (i.e., copying the data of all publications into the referenced\footnote{Referenced through the author and venue ids.} entries in the author and venue collection), we do not perform it due to the limitations of the database.
    \mongodb has a document size limit of 16MB \parencite[207]{bradshaw_mongodb_2019}, which means each publication, venue, and author can only have 16MB of information.
    The largest venue in our dataset (``IEEE Access'') has around 55,000 publications, which means all of its 55,000 publications would need to be saved in the same document with further denormalization.
    16MB is not enough for this, as this would leave less than 300 bytes for each publication including its abstract.
    Implementing this approach would be possible, but would need a lot more schema engineering and the response times at the time were also satisfying, so we saw no need for drastic changes in the schema.
    
\subsection{API: Data Management and Usage}\label{subs:api}
    With our schema ready, the next step is to be able to put the data into the database and efficiently retrieve it again.
    For this, we create the backend of the \csi system, which serves as a \gls{rest} \gls{api} with endpoints, which we can query to manage the data in the database and get the results back.
    Each endpoint serves one function, e.g., read from a collection, update a document in the collection, or aggregate results for analysis.
    
    \subsubsection{Data Management}
    To properly manage our data in the database using the \gls{rest} \gls{api}, the \gls{api} needs to enable the basic \gls{crud} operations to create, read, update, and delete documents in our database.
    We leverage the library \gls{erm}\footnote{\url{https://florianholzapfel.github.io/express-restify-mongoose/}} to automatically generate endpoints for the \gls{crud} operations of all collections (i.e., papers, authors, venues, and affiliations) using our already defined schemas.
    The library also allows customization of the queries to a limited degree, by supporting sorting, skipping, and limiting returned documents, populating documents with documents from other collections using ids, selecting specific attributes, and having some filter capabilities.
    Using a library saves us a lot of work, time, code, and maintenance.

    \subsubsection{User Management}\label{ssubs:user_management}
    Access to \csi requires a user account, which everyone can register without cost, so we can mitigate misuse and better manage our limited server resources.
    To manage the user accounts we add two routes, one to register a new account and one to log in with an existing account.
    All endpoints except those for login and register require authentication with an account, either a normal user account or an administrator account.
    User accounts enable access to the endpoints that aggregate results and are used by the frontend to conduct analyses (We discuss these endpoints in the next paragraph).
    All endpoints for data management discussed in the previous section are only available to administrators, so users cannot modify our data or retrieve parts of the data they are not supposed to retrieve (i.e., abstracts we are not allowed to distribute further due to copyright laws).
    
    \subsubsection{Aggregations for Visualizations}\label{ssubs:aggregations}
    Lastly, we have endpoints that perform aggregations for our visualizations, so the endpoints return exactly what is needed for the visualizations (\Cref{sec:visualizations}).
    These aggregations are performed by directly querying \mongodb.
    We do not use \gls{erm} for this as there were multiple downsides to this approach:
    \begin{itemize}
        \item \gls{erm} does not have the option to aggregate results, i.e., there is no equivalent of the \mongo{group} stage, which we need for most of our queries.
        \item It makes it easier to test the aggregation endpoints and their complex functionality, which make up the largest part of the functionality of the backend.
        \item There are some small issues, e.g., filters not properly working on populated documents.
    \end{itemize}
    Overall \gls{erm} works well for simple tasks in data management but fails for the complex aggregations we require for our visualizations.
    Additionally, we cache the results of any aggregation queries to make repeated queries faster.

\section{Interactive Visualization(s)}\label{sec:visualizations}
    Humans can better understand data if it gets visualized (e.g., in bar charts, line graphs, or scatterplots), than if it is just presented as numbers and text \parencite[552]{shneiderman_designing_2018}.
    We build an interface, which creates interactive visualizations to display our data intuitively, and is integrated into the frontend of our \csi system.
    The first subsection goes over the design decisions of the prototype (\Cref{subs:design}), and the second subsection covers the interface of the \csi system (\Cref{subs:interface}), which is also part of our submission to \arxiv/\eacl \parencite{ruas_cs-insights_2022}.
    Lastly, we showcase how the interface of the frontend can be used with some examples (\Cref{sub:showcases}).

\subsection{Prototype Design}\label{subs:design}
    First, we create a prototype to decide the layout of the interface, its features, and which visualizations we want to include because directly implementing the frontend without a plan would take more time in the end.
    In this subsection, we only cover those basic decisions as many parts of the prototype and finished frontend are identical, and we already explain the finished product in more detail in \Cref{subs:interface}.
    An example of the prototype can be seen in \Cref{fig:prototype}.
    
    We use Figma\footnote{\url{https://www.figma.com/}} to design our prototype and go through four iterations before deciding on the final prototype.
    All four prototypes we create can be navigated, and change their visualizations based on what page is currently selected.
    During the development of the prototypes, we decide our goal is to provide researchers with a system to investigate \gls{cs} research themselves and come up with their own questions they might want to answer.
    We want other researchers to explore what \textit{they} want and not what \textit{we} want, i.e., we do not want to answer our questions from the \lrec paper again and simply reconstruct its plots \parencite{wahle_d3_2022}.
    As a result, we offer multiple dashboards with various visualizations and filter options to give a broad overview of all aspects of \gls{cs}, as shown in the next subsection.
    The selection of plots is inspired by \nlpScholar, as it has proven successful at giving insights into \gls{nlp} and providing a broad overview of the trends of the publications, authors, venues, and citations in \gls{nlp}.
    
\subsection{User Interface}\label{subs:interface}
    \csi offers web-based interactive visualizations to explore \gls{cs} publications through their metadata, such as venues, authors, and abstracts.
    \Cref{fig:frontend} shows an example of the frontend's interface, which we reference throughout this section.
    The interface is composed of three main parts: A. \textit{Dashboards}, B. \textit{Filters}, and C. \textit{Visualizations}.
    \textit{Dashboards} control which \textit{visualizations} are shown and which attribute of the publication metadata is currently visualized.
    \textit{Filters} allow users to select which publications are visualized, by defining criteria the publication metadata has to match.
    For a better understanding of the interface, one can generally make the following analogy: \csi's interface follows a similar structure as a SQL query, in which a dashboard (A) acts as \texttt{GROUP} statement and the filters (B) as \texttt{WHERE} clause.
    % The \textit{Citations} and \textit{LDA Topics} dashboards are an exception to this, as they also aggregate by publications.
    
    A demo for \csi is also available online\footnote{\label{foot:csi_demo}\url{https://cs-insights.uni-goettingen.de/}}.
    To generate the visualizations, the frontend queries the aggregation endpoints in the backend, which requires authentication via a user account, as mentioned in \Cref{ssubs:user_management}.
    \csi is still publicly available, as the account can be created through the interface without cost.
    At the time of writing this thesis, a demo account is available on the main GitHub page\footnote{\url{https://github.com/gipplab/cs-insights-main}}, which is also linked on the homepage of the \gls{ui}\cref{foot:csi_demo}.

    \afterpage{
        \begin{landscape}
            \begin{figure}[p]
                \centering
                \includegraphics[width=\hsize]{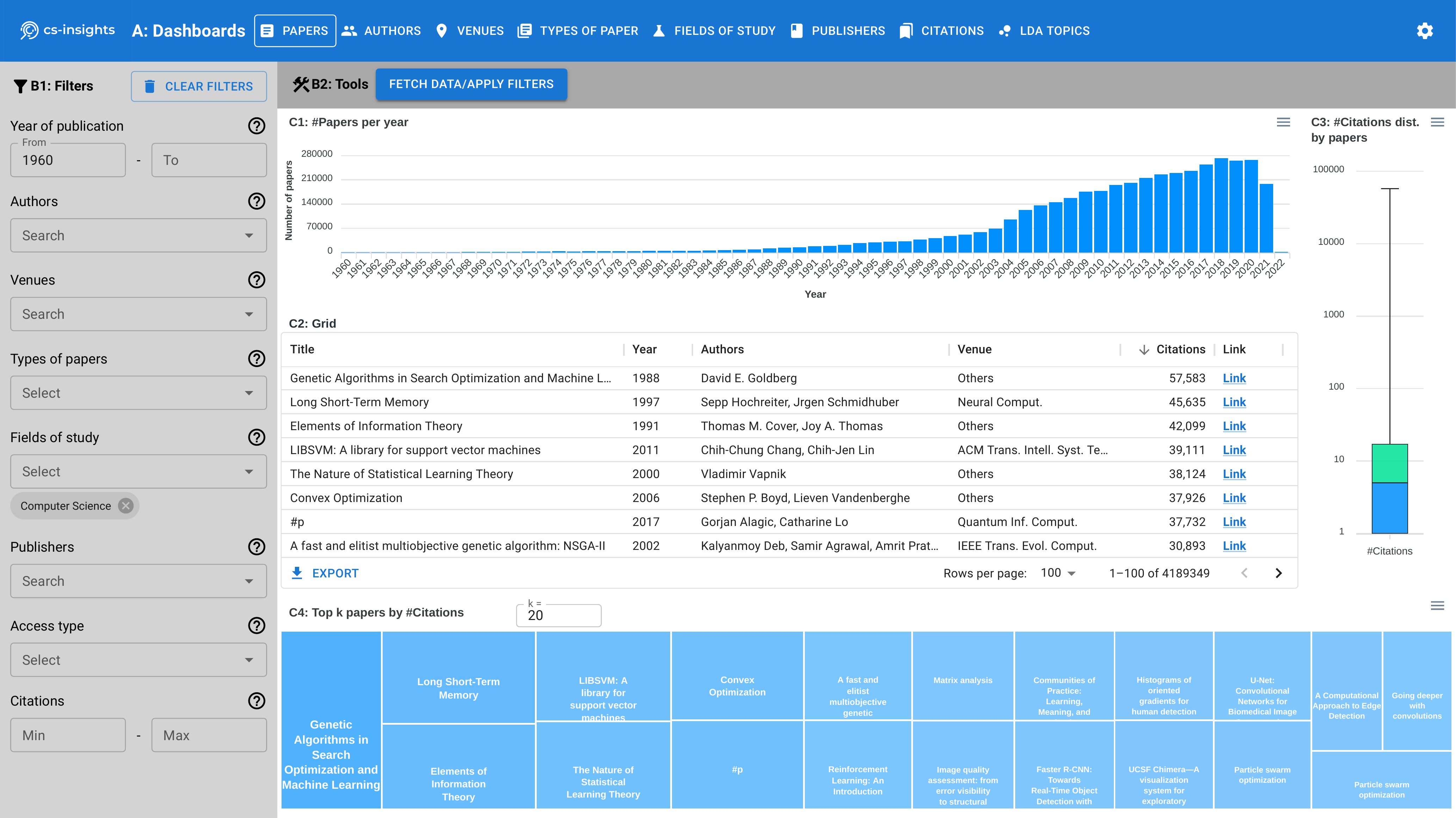}
                \caption{\csi's \gls{ui} with the \textit{Papers} dashboard selected. A. Dashboards, B1. Filters, B2. Tools, C1. \#Papers per year, C2. Grid, C3. \#Citations distribution, and C4. Top $k$ papers by \#Citations.}
                \label{fig:frontend}
            \end{figure}
        \end{landscape}
    }   

    \subsubsection{Dashboards and App Bar}\label{ssubs:dashboards}
    The \textit{dashboards} (A) can be selected from the app bar at the top of the page, which is responsible for the navigation.
    Besides the dashboards, the app bar also shows the \csi logo on the left, which brings the user back to the homepage if clicked on, and a gear on the right to log out again or view the \textit{Account} page.
    There are currently eight different dashboards that users can select from the app bar.
    First is the \textit{Papers} dashboard, which shows information about the publications, as seen in \Cref{fig:frontend}.
    The dashboards for \textit{Authors} (full name), \textit{Venues} (where a paper was published, e.g., ACL, Communications of ACM), \textit{Types of Paper} (according to their BibTeX entries), \textit{Fields of Study} (high-level areas of research, e.g., \gls{cs}, mathematics), and \textit{Publishers} (responsible agency/institution for publishing, but >90\% of publications leave this field blank in \dblp) share the same visualizations with the \textit{Papers} dashboard (\Crefrange{fig:fe_authors}{fig:fe_publishers}).
    However, they aggregate the publications by the respective attribute in the publication metadata, i.e., on the \textit{Venues} dashboard C1 would show the number of venues over time, C2 and C4 the top venues, and C3 the distribution of total citations venues have received.
    These five dashboards include a ``metric switch'' in the \textit{Toolbar} (B2), which switches the metric from \#Citations to \#Papers.
    The \textit{Citations} dashboard consists of a bar chart (C1) and boxplot (C3) each for both the incoming citations (i.e., when a paper gets cited by another publication) and the outgoing citation (i.e., the references in the bibliography of a paper) (\Cref{fig:fe_citations}).
    Lastly, the \textit{LDA Topics} dashboard performs a topic modeling analysis based on \gls{lda} \parencite{blei_latent_2003} with the titles and abstracts of the publications and shows their most frequent and salient \parencite{chuang_termite_2012} terms (\Cref{fig:fe_topics}).
    It uses the topic modeling visualization (C5), which is exclusive to this dashboard and not shown in \Cref{fig:frontend}.
    Both the \textit{Citations} and \textit{LDA Topics} dashboards directly use the metadata of the publications and do not aggregate it beforehand.
    
    \subsubsection{Filters}
    \textit{Filters} (B1) are located in the sidebar on the left and can be configured to select a subset of publications to be visualized.
    Eight different filters can be applied for each available dashboard (A); six for textual values and two for numeric ones.
    Once one or more filter values are set, this modification has to be applied and a new data batch loaded through the ``Fetch Data/Apply Filters'' button (B2).
    The filters can all be cleared again using the ``Clear Filters'' button at the top right corner of the sidebar (B1).

    All textual filters use auto-completion and regular expression; thus, the user is already presented with suggestions while typing, that match the typed string.
    The filters \textit{Authors}, \textit{Venues}, and \textit{Publishers} require the user to stop typing for a predefined amount of time (currently set to one second) before suggestions are loaded, as these filters query the backend for the suggestions.
    For the filters, \textit{Types of papers}, \textit{Field of study}, and \textit{Access type} pre-set values are presented in a drop-down menu, so any suggestions are immediately available, e.g., when clicking on \textit{Types of papers} the suggestions ``Article'', ``Inproceedings'', ``Book'', ``Incollection'', ``Phdthesis'', and ``Mastersthesis'' appear.
    Both numerical filters (i.e., year of publication and citations) function by restricting minimum and maximum values (the filters work inclusive of the selected values).
    To obtain more information about the filters and their match conditions (e.g., case sensitivity), the user can hover over the question mark icon ($?$) to the right of their respective filter heading.
    
    Different filters work together through a logical \texttt{AND}, values on the same textual filter with a logical \texttt{OR}, and values on the same numerical filter with \texttt{AND}.
    For example, if the user decides to search papers from two specific authors in ACL from 2020, the following query would be built: \texttt{author=(Jan Philip Wahle OR Terry Ruas) AND (venue=ACL) AND (yearStart=2020 AND yearEnd=2020)}.

    There is also a hidden feature to show the co-occurrence of authors or fields of study.
    If the user filters by a specific author on the \textit{Authors} dashboard, the grid shows the co-authors of that specific author.
    Similarly, selecting a specific field of study using the filters on the \textit{Fields of study} dashboard shows which other fields of study the selected field of study occurs with the most on the same publication.

    \subsubsection{Visualizations}
    \projectAcronym uses five different visualization elements across its eight dashboards, which \Cref{fig:frontend} (C) shows four of, exemplified for the \textit{Papers} dashboard: \textit{\#Papers per year} (C1), \textit{Paper Details Grid} (C2), \textit{\#Citations distribution} (C3), and \textit{Top $k$ Papers by \#Citations} (C4).
    Our topic modeling visualization (C5) is shown in \Cref{fig:showcase3_topics} in \Cref{show:showcase_topics}.
    The visualization elements C1-C4 can be exported in several formats (e.g., .csv, .svg, .png) using the three bars in the top right of each element, while C5 offers an ``Export'' button to export the entire visualization as an HTML file, which keeps all interactive elements intact.
    All five elements show a loading icon while fetching data from the backend.
    In the following, we use \dashboard as a placeholder for the main dashboard element, i.e., the name of the currently selected dashboard.
    For example ``\#\dashboard per year'' means ``\#papers per year'' on the \textit{Papers} dashboard and ``\#venues per year '' on the \textit{Venues} dashboard.
    
    \paragraph{Bar Chart (C1)}
    The \textbf{\#\dashboard per year} shows the number of unique dashboard main elements per year.
    For example, in the \textit{Venues} dashboard, the user can see the bar chart displaying the number of unique venues where the selected papers were published by year.
    Hovering over a bar reveals the exact number of entries for that year\footnote{We add a label with the exact value for each year for this thesis, which is always shown, to make the analysis in this thesis more comprehensible.}.
    If the number of entries for a given year is $0$, we grey out the year's label to make it easier for the user to distinguish it from a very small number of entries.
    Publications without the year set are aggregated to ``NA'' on the left of the chart, should no filter for the year of publication be selected.
    
    \paragraph{Grid (C2)}
    The \textbf{\dashboard details grid} displays the available details for each dashboard in a table format.
    For example, in the \textit{Papers} dashboard, the first column is the paper's title followed by its year of publication, list of authors, venue, number of citations, and link to the actual paper (if available).
    On the five aggregated dashboards, the grid would change slightly by including the name of the \dashboard (e.g., the name of the author or venue), the first and last year of publication, and the number of papers, citations, and average citations per paper, while also aggregating all \dashboard without the main dashboard element set to ``Others''.
    A link to the search on \dblp is also added on the \textit{Authors} and \textit{Venues} dashboard, and the grid can generally be sorted using any of the column headings.
    As some text values can be too long (e.g., paper titles), we abbreviate them for readability purposes, but hovering over them will still display the entire value.
    
    \paragraph{Boxplot (C3)}
    The \textbf{\#Citations distribution by \dashboard} visualization shows the distribution of total citations for the selected dashboard main element, e.g., for authors, it will show the distribution of total citations authors have received.
    On the five aggregation dashboards (so not on \textit{Papers}), the user can also select the number of papers as an alternative metric using the metric switch (B2).
    Hovering over the boxplot reveals the exact values for the minimum, first quartile, median, third quartile, and maximum.
    All boxplots are log scaled for better usability, as the maximum often is multiple magnitudes larger than the other values.
    
    \paragraph{Treemap (C4)}
    The \textbf{Top $k$ \dashboard by \#Citations} shows the top $k$ elements based on the number of citations.
    As in C3, the \textit{Papers} dashboard uses only the number of citations (\#Citations) as a metric to generate its output, while the five aggregation dashboards also offer the option of showing the top $k$ based on the number of papers (\#Papers) in addition.
    The value of $k$ can be adjusted freely using a text field that reloads the plot automatically.
    When the text in C4's boxes is too large, we collapse them for readability purposes.
    Similar to the other visualizations, one can also hover over the chart and show the entire name and value of the selected field.
    
    \paragraph{Topic Modeling (C5)}
    On the \textit{LDA Topics} dashboard, the user can explore the most frequent and salient \parencite{chuang_termite_2012} terms (stemmed words) of a given collection of documents through our topic modeling visualization, which is adapted from \textcite{sievert_ldavis_2014}.
    \Cref{show:showcase_topics} shows an example of C5 (\Cref{fig:showcase3_topics}) and gives examples of the features of C5 and the \textit{LDA Topics} dashboard.
    The output in this dashboard is divided into two parts: the semantic topic clusters (left) and the list of the most frequent and salient terms (right).
    Both parts are generated based on the text in the titles and abstracts of papers.
    When hovering over a cluster or clicking on it, the 30 most relevant terms of the selected cluster are shown on the right as red bars while continuing to show the overall frequency of those 30 terms in all clusters as blue bars.
    Once a cluster is selected, the user can adjust the relevance metric \parencite{sievert_ldavis_2014} using a slider in the top right.
    When no cluster is selected, the plots consider all titles and abstracts to compose their list of terms.
    The user can also identify clusters associated with a term by hovering over the desired terms directly.
    Overlap of clusters indicates their semantic proximity.

\subsection{Showcases}\label{sub:showcases}
    In this subsection, we show some examples to explain and explore what the \csi system can do, so the reader can better understand how we are doing the experiments for our analysis later in this thesis (\Cref{chap:analysis}).
    During the showcases, we also touch on potential analyses, but we do not discuss the results in detail, as we already cover those in our actual analysis.
        
\begin{figure}[!htb]
        \centering
        \begin{minipage}{.5\textwidth}
            \centering
            \includegraphics[trim={0 1.2cm 0 0}, clip, width=\linewidth]{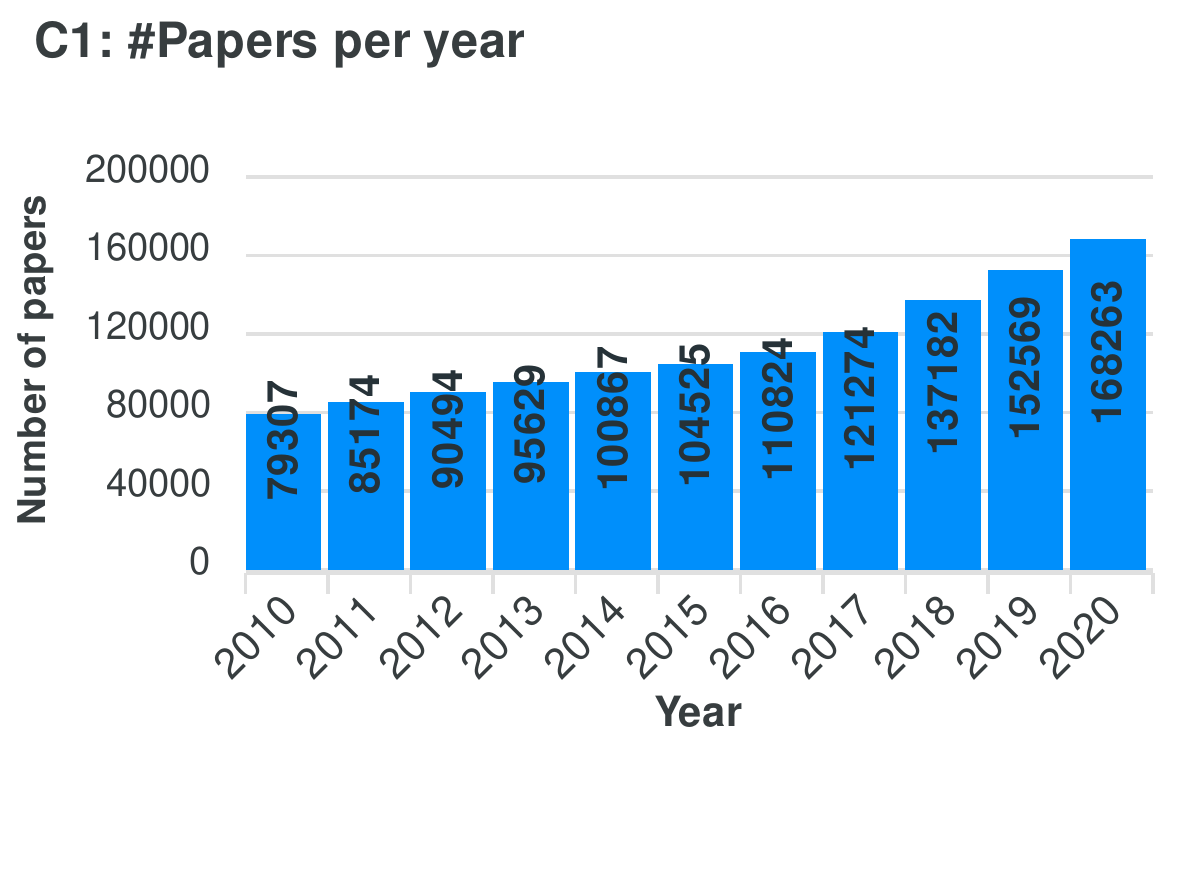}
            \caption{Number of journal articles per year between 2010 and 2020.}
            \label{fig:showcase1_articles}
        \end{minipage}%
        \begin{minipage}{.5\textwidth}
            \centering
            \includegraphics[trim={0 1.2cm 0 0}, clip, width=\linewidth]{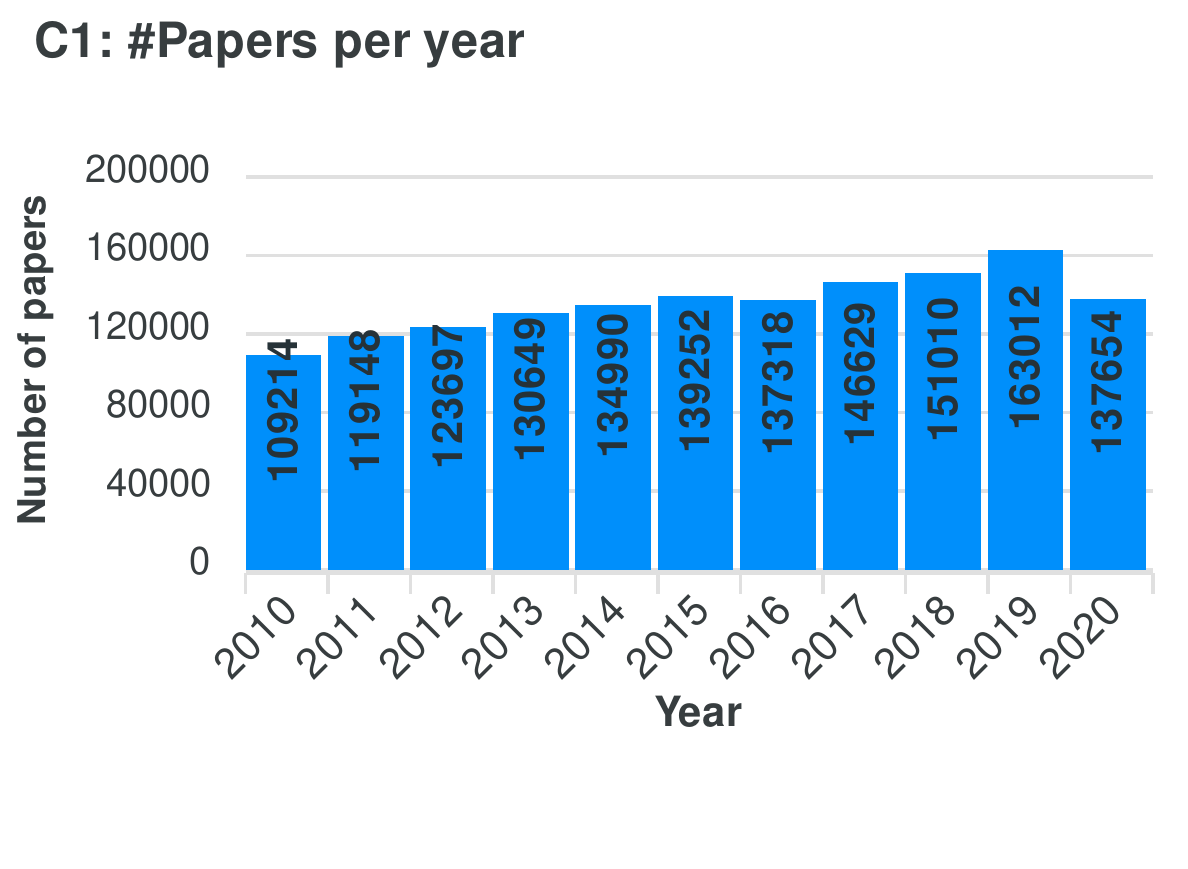}
            \caption{Number of conference papers per year between 2010 and 2020.}
            \label{fig:showcase1_inproceedings}
        \end{minipage}
    \end{figure}

    \showcase{Comparisons with Filters}
    The first showcase exemplifies the functionality of the filters, by comparing the number of conference papers and journal articles over the last few years.
    We use the bar chart (C1) on the dashboard \textit{Papers}, as we are interested in the publications over time and set the \textit{Year of publication} filter to 2010--2020 and the \textit{Type of papers} filter to ``Article'' (\Cref{fig:showcase1_articles}) and ``Inproceedings''  (\Cref{fig:showcase1_inproceedings}).
    In the two charts, we can make multiple observations: i) both journal and conference publications increased over time ii) journal publications increased more and overtook conference publications in 2020 iii) conference papers even had a drop in 2020, which we assume to be related to the COVID-19 pandemic.
    We further investigate the differences between journals and conferences in \Cref{sec:paper_type}, which includes a more detailed look into their number of publications over time.
    
\begin{figure}[!htb]
        \centering
        \includegraphics[width=\textwidth]{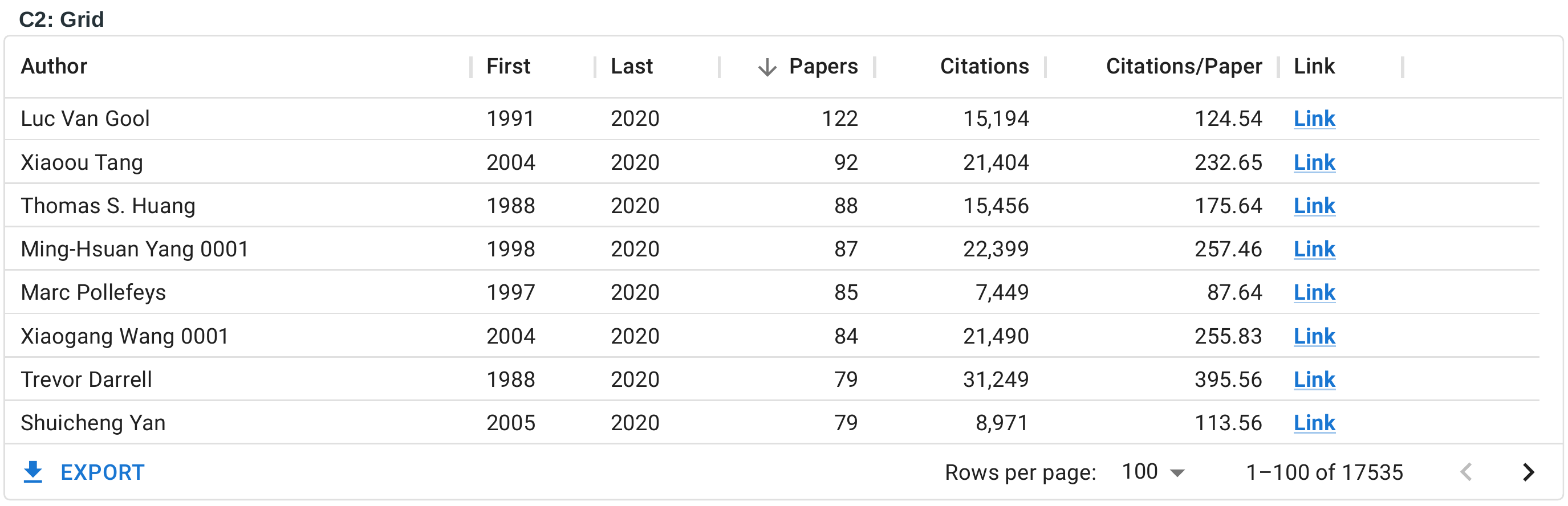}
        \caption{Authors that published the most in the CVPR conference.}
        \label{fig:showcase2_venue}
    \end{figure}

    \showcase{Dashboard--Filter Interaction}
    We use the second showcase to explain the interaction of filters and dashboards and when the user has to use a filter, and when the corresponding dashboard.
    Our first example shows how to find the authors who published the most in CVPR (Computer Vision and Pattern Recognition).
    For this, we select the \textit{Authors} dashboard, ``CVPR'' in the \textit{Venue} filter, and click the heading ``Papers'' in the grid (C2) twice to sort by the papers descending\footnote{An alternative way to sorting the grid is to switch the metric to ``\#Papers'' and then looking at the first entry of the treemap.}.
    The result (\Cref{fig:showcase2_venue}) reveals that ``Luc Van Gool'' published the most papers in ``CVPR'' (122 papers, the first in 1991, the last in 2020).
    
\begin{figure}[!htb]
        \centering
        \includegraphics[width=\textwidth]{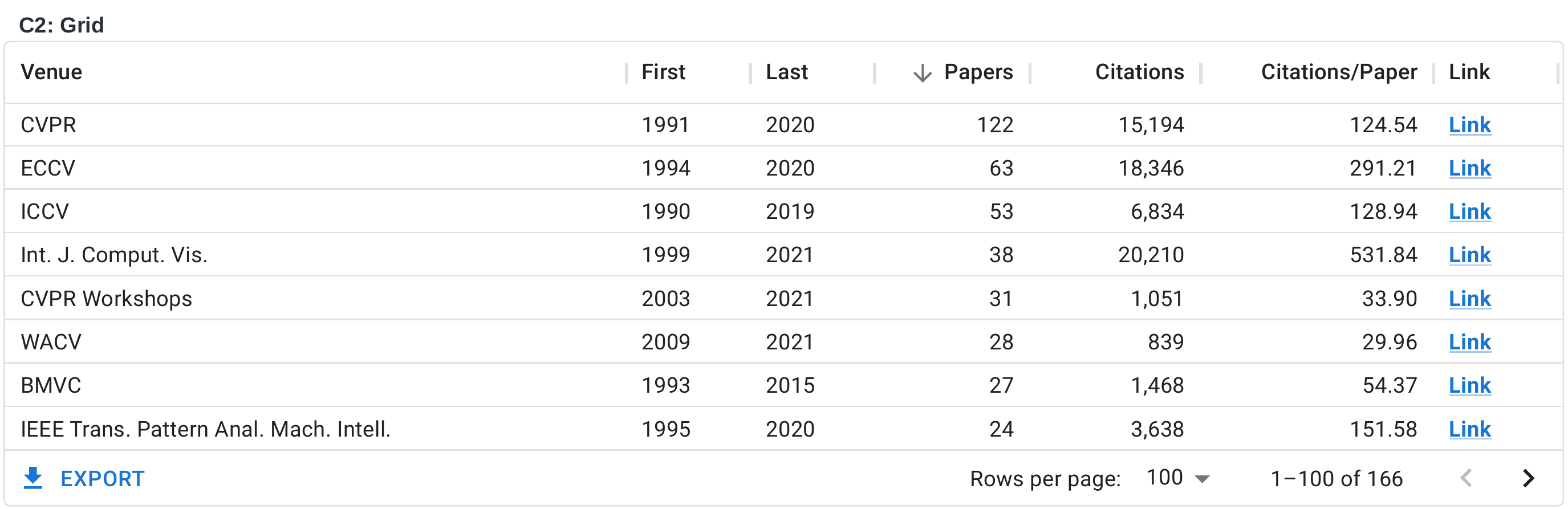}
        \caption{Venues Luc Van Gool published the most in.}
        \label{fig:showcase2_author}
    \end{figure}

    The other way around, if we want to find out which venue a specific author published in the most, we swap what we select as the filter and what as the dashboard.
    We select the \textit{Venues} dashboard, ``Luc Van Gool'' in the \textit{Authors} filter, and again click the ``Papers'' heading twice to sort by paper descending.
    The result (\Cref{fig:showcase2_author}) shows that Luc Van Gool published the most in CVPR (122 papers), with the second-ranked venue being ECCV (63 papers).
    
\begin{figure}[!htb]
        \centering
        \includegraphics[width=\textwidth]{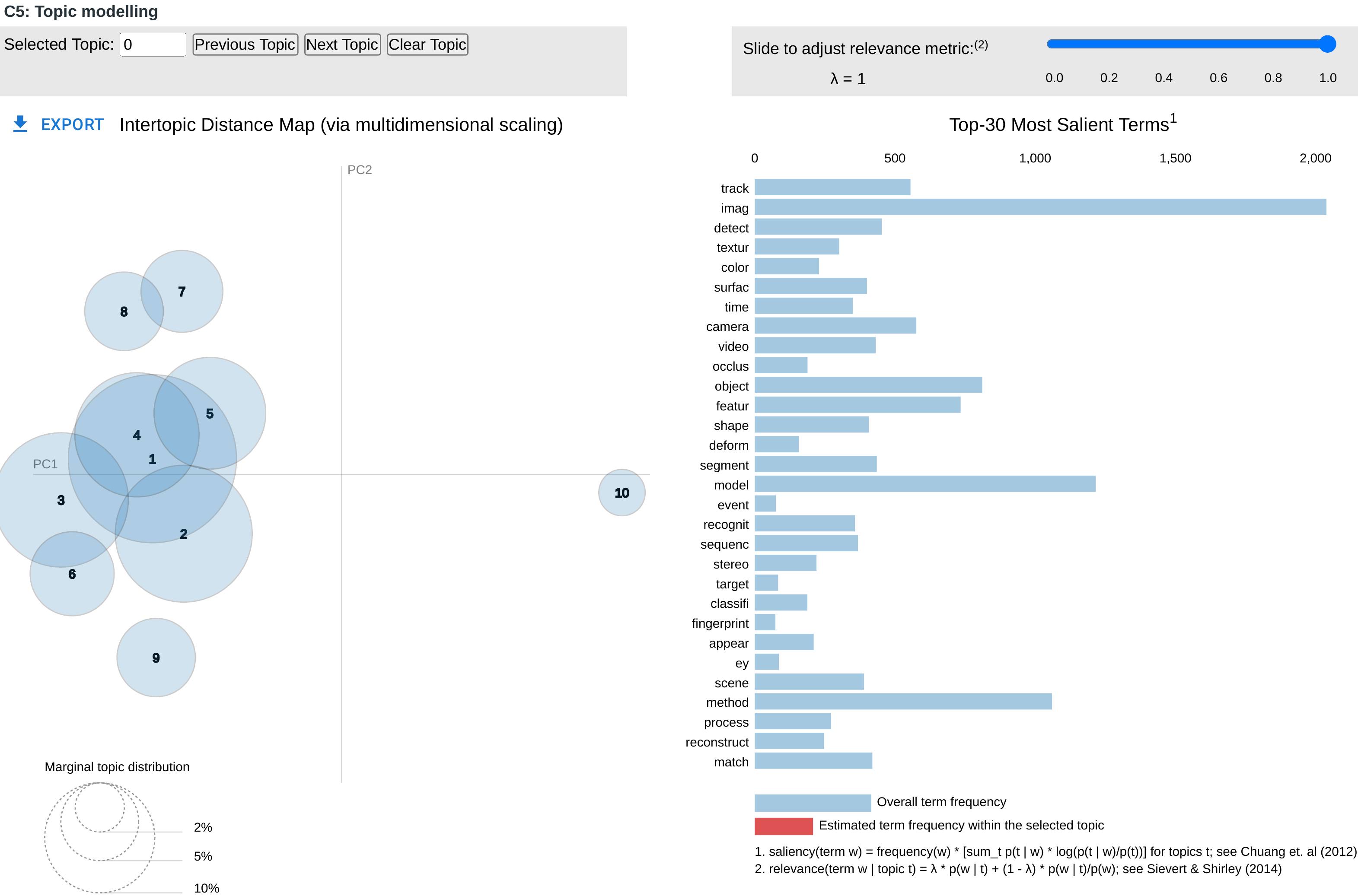}
        \caption{Topic modeling visualization (C5) for the CVPR conference 2000-2005.}
        \label{fig:showcase3_topics}
    \end{figure}

    \showcase{Topic Modeling}\label{show:showcase_topics}
    In our last showcase, we demonstrate the topic modeling component (C5).
    We want to investigate the topics of CVPR between 2000 and 2005 and select the \textit{LDA Topics} dashboard, a fitting model in the toolbar (B2)\footnote{Currently, the user can select any of the available models, as they yield the same results.}, and the filters \textit{Year of publication} 2000--2005 and \textit{Venues} ``CVPR''.
    \Cref{fig:showcase3_topics} shows the results and \Cref{fig:fe_topics} the entire dashboard).
    The terms ``track'', ``imag'', and ``detect'' are the three most salient, while ``imag'', ``model'', and ``method'' are the three most frequent.
    Clicking on topic cluster 1 or hovering over it reveals the adjusted distribution of terms for that topic considering their relevancy, where ``model'', ``imag'', and ``approach'' are the top three (\Cref{fig:showcase3_cluster}).
    Hovering over the term ``track'' reveals, that it appears most in topic cluster 5 (\Cref{fig:showcase3_term}).
    In \Cref{sec:venues} we are further investigating the topics of venues and how they change over time.
    
        \chapter{Implementation}\label{chap:implementation}
    This chapter covers aspects regarding the implementation of \csi.
    We detail the architecture of \csi and its sub-components (\Cref{sec:architecture}) and our measures to improve the quality of our code (\Cref{sec:qa}).

\section{Architecture}\label{sec:architecture}
\begin{figure}[!htb]
        \centering
        \includegraphics[width=\textwidth]{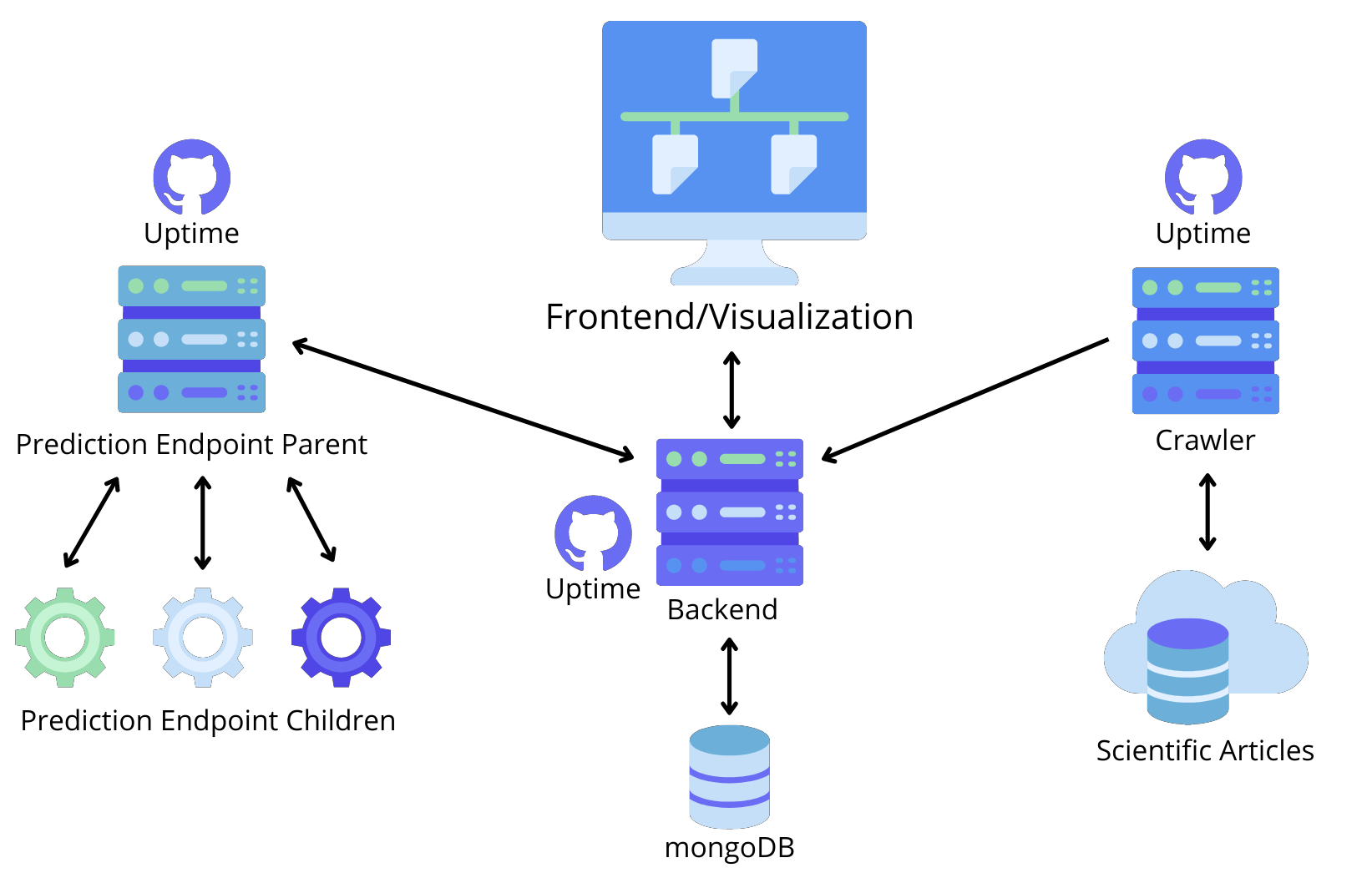}
        \caption{Overview of the \csi system.}
        \label{fig:system_overview}
    \end{figure}

    The \csi system consists of the four sub-components \textit{frontend}, \textit{backend}, \textit{prediction endpoint}, and \textit{crawler}, which we can see in \Cref{fig:system_overview}.
    Our system is available as a free web application without the need for any installation as it runs in any web browser\footnote{\url{https://cs-insights.uni-goettingen.de/}}, providing access to multiple users simultaneously.
    The entire code for all components is available online on GitHub and accessible through the main GitHub repository\footnote{\label{foot:csimain}\url{https://github.com/gipplab/cs-insights-main}}, so researchers can also run the code locally on their machine if they choose to.
    In this case, even though \csi is split into multiple components, other researchers only have to interact with the main component\cref{foot:csimain}, as all sub-components are managed from there automatically.
    To guarantee a flexible and modular setup, every sub-component in \csi runs in its own docker container\footnote{\url{https://www.docker.com}}.
    The architecture discussed in this section is also part of our submission to \arxiv/\eacl \parencite{ruas_cs-insights_2022}.

\subsection{Frontend}\label{subs:frontend}
    The frontend\footnote{\url{https://github.com/gipplab/cs-insights-frontend}} is responsible for presenting the main components of our system (i.e., dashboards, filters, and visualizations), and through the frontend's interface, the user can filter, retrieve, and visualize the metadata of \gls{cs} publications (see \Cref{subs:interface}).
    We use TypeScript\footnote{\url{https://www.typescriptlang.org}; a superset of JavaScript} and as web framework React\footnote{\url{https://reactjs.org/}} because it is open-access and has a large community support\footnote{\url{https://www.statista.com/statistics/1124699/worldwide-developer-survey-most-used-frameworks-web/}}.
    For charts we use ApexCharts\footnote{\url{https://apexcharts.com/react-chart-demos/}} and for other \gls{ui} components, we use Material UI\footnote{\url{https://mui.com/}}.

\subsection{Backend}\label{subs:backend}
    The backend\footnote{\url{https://github.com/gipplab/cs-insights-backend}} serves as \gls{rest} \gls{api} to access, retrieve, aggregate, and analyze data (see \Cref{sec:data_storage}).
    It controls who can access data and how, and performs computationally expensive tasks (e.g., accumulating citations of all authors for each paper available).
    \csi uses \mongodb as a database with mongoose\footnote{\url{https://mongoosejs.com/}} providing the object document mapping.
    We also use TypeScript with Node.js\footnote{\url{https://nodejs.org/en/}} as JavaScript runtime, and Express.js\footnote{\url{https://expressjs.com/}} to handle the HTTP(S) requests.
    The \gls{crud} endpoints of its \gls{api} are auto-generated from the mongoose models with \gls{erm}, while the endpoints the frontend uses are manually written (\Cref{subs:api}).

\subsection{Prediction Endpoint}\label{subs:prediction}
    The prediction endpoint\footnote{\url{https://github.com/gipplab/cs-insights-prediction-endpoint}} is implemented in Python 3, is responsible for the training and prediction of the models in the \textit{LDA Topics} dashboard, and is used to generate the semantic topics and their respective lists of the most frequent and salient terms.
    For topic modeling, we use gensim's\footnote{\url{https://radimrehurek.com/gensim/models/ldamodel.html}} implementation of \gls{lda} \parencite{blei_latent_2003}.
    The visualizations are implemented using pyLDAvis\footnote{\url{https://pyldavis.readthedocs.io/en/latest/readme.html}}, a port of LDAvis \parencite{sievert_ldavis_2014}.
    As the training and inference require processing thousands of documents, we create a dedicated service to maintain these models, distribute them on the available compute infrastructure, assign them to compute jobs, and consolidate all results. 
    The endpoint consists of a manager parent node and a variable amount of child nodes, where the parent node takes requests through a \gls{rest} \gls{api}.
    It also abstracts topic model creation, training, and inference to distribute the computational load across different independent child nodes.
    To preprocess the text (i.e., titles and abstracts) before we create the topics we use \lstinline{preprocess_string()}\footnote{https://radimrehurek.com/gensim/parsing/preprocessing.html} from gensim, which removes HTML tags, punctuation, duplicate whitespaces, numbers, stopwords, and words with less than three characters, and stems the text.
    
\subsection{Crawler}\label{subs:crawler}
    The crawler\footnote{\url{https://github.com/gipplab/cs-insights-crawler}} is also implemented in Python 3 and creates our dataset \diii (see \Cref{sec:data_acquisition}).
    In the future, the crawler can be used to keep \csi and \diii up-to-date with the most recent publications, by running it monthly to add new publications and update existing ones (\Cref{sec:future_work}).
    We use aiohttp\footnote{\url{https://docs.aiohttp.org/en/stable/}} to request the full-texts and \grobid \parencite{lopez_grobid_2022} to extract the metadata from the full-texts, which is a resource-intensive process.
    Therefore, we implement a parallel and asynchronous routine to parse the latest release, retrieve the corresponding full-texts, extract their metadata, and align the information to \dblp.
    We split the dataset into equally sized chunks to work on mutually exclusive parts of the dataset with multiple processes without processing the whole repository at once.
    Then, we launch $n$ processes to retrieve publications, where each process asynchronously requests the PDF link of a paper or, if not present, parses the HTML page of the paper to identify the PDF link and downloads it.
    To restrict requests to the same domain and respect server limits, we use semaphores and wait to respect the retry-after header whenever we receive an HTTP 429 (``Too many requests'') response.
    In parallel to the $n$ retrieval processes, we launch another $n$ processes to work on full-texts of the previously downloaded chunks and extract their metadata.
    % chunk size = 12
    To reduce disk requirements, we delete the full-texts after extraction and only keep their metadata.
    The uncompressed size of \diii is $\approx$18GB in JSON format and $\approx$21.5GB in CSV format \parencite{wahle_d3_2022}.

\section{Quality Assurance}\label{sec:qa}
    To ensure the quality of \csi and its code we employ various measures as \Cref{tab:qa} shows.
    \begin{table}[htb]
    \centering
    \footnotesize
    \begin{tabular}{c|ccccc}
        \toprule
        \textbf{Component} & \textbf{Tests} & \textbf{Linting} & \textbf{Typing} &\textbf{Code Style} \\
        \midrule
        Frontend & - & ESLint\footnote{\url{https://eslint.org/}} & TypeScript & Airbnb\footnote{\url{https://github.com/airbnb/javascript}} \\ 
        Backend & Mocha\footnote{\url{https://mochajs.org/}} & ESLint & TypeScript & Airbnb\\  
        Prediction Endpoint & pytest\footnote{\label{foot:pytest}\url{https://pytest.org/}} & Flake8\footnote{\url{https://flake8.pycqa.org/}} & mypy\footnote{\url{http://mypy-lang.org/}} & Black\footnote{\url{https://github.com/psf/black}} \\
        Crawler & pytest & Flake8 & Pyright\footnote{\url{https://github.com/microsoft/pyright}} & Black \\ 
        \bottomrule
    \end{tabular}
    \caption{Technologies used across the four components of \csi for testing, linting, typing, and code styling.}
    \label{tab:qa}
\end{table}
    We use tests to make sure the logic of our components works correctly and linting for static code analysis to find potential problems (e.g., unused variables, unused imports).
    Checks for typing are added to avoid potential errors from dynamically typed languages, and a common code style is used for better readability.
    Tests, linting, typing, and code style are enforced by checking the code whenever a commit is pushed to one of our GitHub repositories.
    The frontend does not use tests, because normal tests make little sense in our case, where we have a ``dumb'' frontend, that only visualizes data and does not aggregate any data itself.
    \gls{ui} tests were considered but postponed to future work due to the time constraints and little additional value considering the amount of work required.
    
    Documentation for \csi can be found in the README of the main project\footnote{\url{https://github.com/jpwahle/cs-insights/blob/main/README.md}} and more detailed documentation will also be added soon\footnote{\url{https://jpwahle.github.io/cs-insights/}}.
    Automatically generated documentation for the available endpoints of the backend\footnote{\url{https://jpwahle.github.io/cs-insights-backend/}} and prediction endpoint\footnote{\url{https://jpwahle.github.io/cs-insights-prediction-endpoint/}} is already available and will be available soon for the frontend\footnote{\url{https://jpwahle.github.io/cs-insights-frontend/}}.
    
    We also track the uptime of all components from a separate GitHub repository\footnote{\url{https://github.com/gipplab/cs-insights-uptime}} using upptime\footnote{\url{https://github.com/upptime/upptime}}, which automatically creates GitHub issues, when a component goes down and closes the issue when it comes back up again.

	\chapter{Analysis and Discussion}\label{chap:analysis}
    In this chapter, we conduct experiments to answer our research questions (\Cref{sec:goals}) and discuss their results.
    We first explain the setup for our experiments including a data overview (\Cref{sec:setup}), before starting with the experiments for our analysis.
    The experiments are split into multiple sections, each covering one attribute in our data.
    We section the experiments as follows: publications (\Cref{sec:publications}; \Crefrange{ex:publications}{ex:publications_top}), authors (\Cref{sec:authors}; \Crefrange{ex:authors_bar}{ex:authors_topics_time})), venues (\Cref{sec:venues}; \Crefrange{ex:venues_bar}{ex:venues_topics_time}), citations (\Cref{sec:citations}; \Crefrange{ex:citations_bar}{ex:citations_time})), document types (\Cref{sec:paper_type}; \Crefrange{ex:types_distribution}{ex:types_topics_time})), and fields of study (\Cref{sec:fields}; \Crefrange{ex:fields_distribution}{ex:fields_topics})).
    Every experiment references the research question it relates to at the end of its title and includes a discussion of the results.
    Lastly, we close out the chapter with a short summary (\Cref{sec:analysis_summary}).

    \paragraph{Disclaimer}
    Some numbers and results might differ from the paper published in \lrec \parencite{wahle_d3_2022}, as the data we use is from a CSV export of \diii.
    The limitations of our work are further explained in \Cref{sec:limitations}.

\section{Setup}\label{sec:setup}
    This section shortly explains the general setup of the experiments (\Cref{subs:setup_general}) and the setup for specific groups of experiments (\Cref{subs:setup_specific}).
    
\subsection{General Setup}\label{subs:setup_general}
    \begin{wraptable}{r}{0pt}
    \centering
    \footnotesize
    \begin{tabular}{lr}
    \toprule
        \textbf{Attribute} &\textbf{Amount} \\
        \midrule
        Publications &4,893,540 \\
        Abstracts &3,980,144 \\
        Citations &97,053,288 \\
        Authors &2,730,729 \\
        Venues &14,268 \\
        Types of paper &7 \\
        Fields of study &20 \\
        \bottomrule
    \end{tabular}
    \caption{Number of unique entries for each field in \diii.}\label{tab:data_overview}
\end{wraptable}
    We conduct all experiments in this chapter with \csi and its underlying data from \diii, by leveraging its various dashboards and filters in specific ways.
    \Cref{tab:data_overview} shows an overview of the data from \diii, which spans from 1936 to 2022.
    The showcases in \Cref{sub:showcases} already demonstrated some ways \csi can be used to visualize this data and interact with it.
    For our experiments, we use all dashboards and filters except \textit{Publishers} and \textit{Access type}.
    We exclude \textit{Publishers}, because the data is too sparse, and exclude \textit{Access type}, as this is not a focus of our analysis.
    \csi's dashboards also consist of multiple visualization elements, of which we use the bar chart (C1), grid (C2), boxplot (C3), and topic modeling component (C5) in multiple experiments.
    Only the treemap is not used, as the grid provides better formatted and readable results for our analysis and shows exact numbers.
    This results in many possible combinations of specific dashboards, filters, and visualizations, which are too many to detail here.
    
    We extract the results from the experiments with the integrated export functionality for C1-C4 in the format we see fit best for the corresponding experiment (as images or mostly .csv), and for the topic analysis (C5) by copying the list of the most salient terms.
    The results are then either directly included or are preprocessed first, by further aggregating the results to better highlight important aspects and making some results more comprehensible.
    In the analysis in this thesis, we make sure not to go further than two layers deep with our aggregation of results to avoid manual analysis, which would contradict the purpose of \csi.
    For example, directly taking results from a query would be one layer, and aggregating results from multiple (e.g., five) queries into one table would be layer two.

\subsection{Experiment Specific Setup}\label{subs:setup_specific}
    Many experiments share a specific setup, so to avoid repetitiveness during those experiments, we give details in this paragraph.
    % Some readers might not read the entire analysis at once, so we reference this subsection in the corresponding experiments as a reminder.

    \paragraph{Citation and Paper Distributions over distinct Time Periods}\label{par:distribution}
    We investigate the distribution of citations and papers over eight distinct periods for authors (\Cref{ex:authors_citations}), venues (\Cref{ex:venues_citations}), incoming/outgoing citations (\Cref{ex:citations_time}, and conferences/journals (\Cref{ex:types_citations}) with the same setup.
    The first four periods (1960-1999) use a span of 10 years and the other four (2000-2019) use five years.
    This avoids the tables from getting too cluttered, compared to using 12 periods of five years, though we are aware of the issues different lengths of time spans can cause (\Cref{sec:limitations}).
    We leave out data before 1960, because the data is too sparse for meaningful analysis, from 2020, because we want to avoid the influence of the COVID-19 pandemic, whose effect becomes apparent in \Cref{ex:publications,ex:types_bar}, and from 2021 because our data is not complete for that year (\Cref{sec:limitations}).
    The distributions of citations and papers consist of the first quartile, median, third quartile, maximum, and average.
    We leave out the minimums, as for the number of citations they are always 0 and for papers always 1.
    
    \paragraph{Topic Modeling}\label{par:topic}
    We use our topic modeling component to generate 10 topics and find the 30 most salient terms (stemmed words in a ranked list) for subsets of our data and then compare the lists for different subsets.
    For this, we put the 30 most salient terms of each subset into one table column each and then mark in bold which terms are unique to a column and in italics which are common across all columns.
    We use this approach to investigate the trends of the subsets over time, by selecting five distinct periods and generating the most salient terms for each period, but again only use data until 2019 to avoid the influence of the COVID-19 pandemic.
    This approach is applied to the most cited and productive authors (\Cref{ex:authors_topics_time}) and venues (\Cref{ex:venues_topics_time}), and conferences/journals (\Cref{ex:types_topics_time}).
    Alternatively, we compare five distinct entities, i.e., the five most cited or most productive venues (\Cref{ex:venues_topics}) and the five most prominent fields of study (\Cref{ex:fields_topics}).
    The topic modeling experiments do not aggregate any data and directly use the publications.
    Our approach follows that of \textcite{fiala_computer_2017}, who also split their data into distinct periods and highlight terms unique to a period.

\section{Publications}\label{sec:publications}
    This short section focuses on experiments conducted directly on the publications of \csi without any further aggregations and answers a part of \labelcref{rq:amount} by looking into the number of publications per year (\Cref{ex:publications}) and the most cited publications (\Cref{ex:publications_top}).
    
    \experiment{How does the number of publications change over time? (\labelcref{rq:amount})}\label{ex:publications}
    \begin{figure}[!htb]
        \centering
        \includegraphics[trim={0 1cm 0 0}, clip, width=\textwidth]{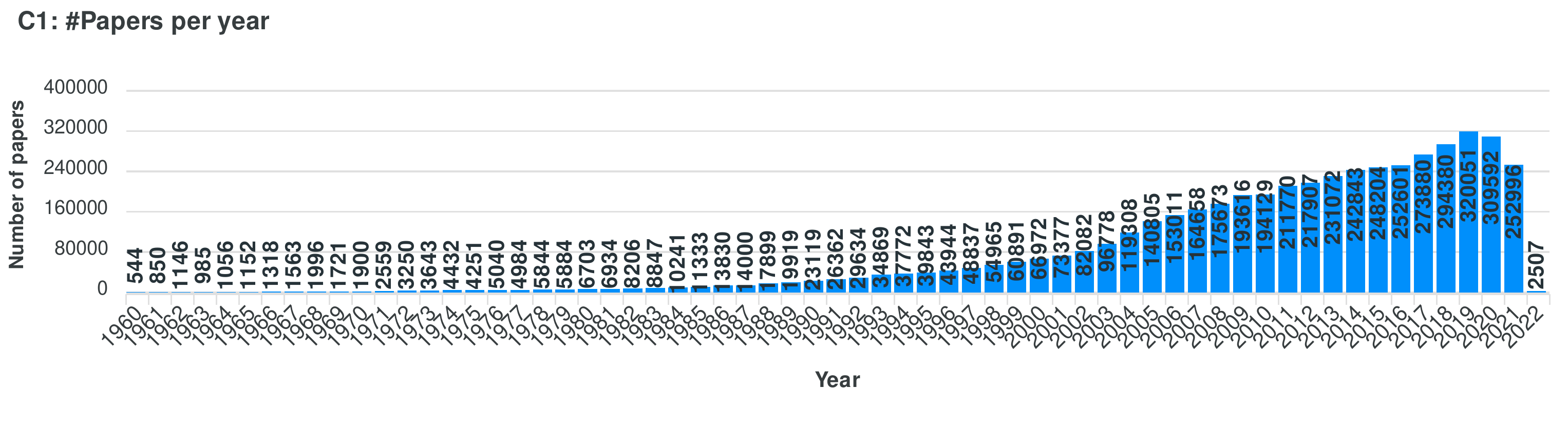}
        \caption{Number of publications per year starting in 1960. See \Cref{fig:publications_bar_app} for the full span.}
        \label{fig:publications_bar}
    \end{figure}
    In total \csi comprises 4,893,540 publications between 1936 and 2022, with 276 publications missing a year of publication.
    The first thing to notice is the exponential increase in publications in the 2000s and a peak around 2019 (\textbf{\Cref{fig:publications_bar}}).
    \textcite{bornmann_growth_2021} also see an exponential increase in scientific publications overall (not just in \gls{cs}) and find a doubling time of 14 years since 1952, while our data shows that in 14 years (2005 to 2019) the number of publications more than doubles.
    A doubling time of 12 years (2007-2019) might be more appropriate for recent years and an even smaller doubling time for earlier years (e.g., in 1990-1997 and 2000-2005 the number of papers also doubled).
    This highlights the boost in \gls{cs} publications compared to other fields of study very well.
    We also observe an increase in publications during the end of the 1980s and 1990s, which we think is caused by the more widespread adoption of personal computers\footnote{\url{https://www.statista.com/statistics/214641/household-adoption-rate-of-computer-in-the-us-since-1997/}} and the internet\footnote{\url{https://www.statista.com/statistics/189349/us-households-home-internet-connection-subscription/}}, respectively.
    Similar increases between the 1980s and 2010s can be observed in \wos \parencite{fiala_computer_2017} and \gls{nlp} \parencite{mohammad_nlp_2020_data, mariani_nlp4nlp_2019}, with \textcite{mohammad_nlp_2020_data} also specifically mentioning the observations.
    The peak around 2019 is not visible in \textcite{fiala_computer_2017, mariani_nlp4nlp_2019, mohammad_nlp_2020_data}, as their data does not cover 2019.
    \textcite{mohammad_nlp_2020_data, mariani_nlp4nlp_2019} also both observe a decrease in papers every second year which they both attribute to biennial conferences, but this trend is not visible in our data.
    We hypothesize the great difference in dataset size (\csi is nearly 100 times larger) and broader coverage of topics (\gls{cs} vs. \gls{nlp}) cause biennial conferences to not affect the overall number of publications or the different biennial conferences compensate for each other.
    The drop in publications in 2020 can be explained by the COVID-19 pandemic.

    \experiment{What are the most cited publications?}\label{ex:publications_top}
    %Please add the following packages if necessary:
%\usepackage{booktabs, multirow} % for borders and merged ranges
%\usepackage{soul}% for underlines
%\usepackage[table]{xcolor} % for cell colors
%\usepackage{changepage,threeparttable} % for wide tables
%If the table is too wide, replace \begin{table}[!htp]...\end{table} with
%\begin{adjustwidth}{-2.5 cm}{-2.5 cm}\centering\begin{threeparttable}[!htb]...\end{threeparttable}\end{adjustwidth}
\begin{table}[!htp]\centering
\scriptsize
\begin{tabular}{rp{4cm}rp{3.5cm}p{2.5cm}r}
\toprule
\textbf{\#} &\textbf{Title} &\textbf{Year} &\textbf{Authors} &\textbf{Venue} &\textbf{\#Citations} \\
\midrule
1& Genetic Algorithms in Search Optimization and Machine Learning &1988 &David E. Goldberg &\textit{Others} &57,583 \\
2& Long Short-Term Memory &1997 &Sepp Hochreiter, Jrgen Schmidhuber &Neural Comput. &45,635 \\
3& Elements of Information Theory &1991 &Thomas M. Cover, Joy A. Thomas &\textit{Others} &42,099 \\
4& LIBSVM: A library for support vector machines &2011 &Chih-Chung Chang, Chih-Jen Lin &ACM Trans. Intell. Syst. Technol. &39,111 \\
5& The Nature of Statistical Learning Theory &2000 &Vladimir Vapnik &\textit{Others} &38,124 \\
6& Convex Optimization &2006 &Stephen P. Boyd, Lieven Vandenberghe &\textit{Others} &37,926 \\
7& \#p &2017 &Gorjan Alagic, Catharine Lo &Quantum Inf. Comput. &37,732 \\
8& A fast and elitist multiobjective genetic algorithm: NSGA-II &2002 &Kalyanmoy Deb, Samir Agrawal, Amrit Pratap, T. Meyarivan &IEEE Trans. Evol. Comput. &30,893 \\
9& Reinforcement Learning: An Introduction &2005 &Richard S. Sutton, Andrew G. Barto &IEEE Trans. Neural Networks &30,815 \\
10& Matrix analysis &1985 &Roger A. Horn, Charles R. Johnson &\textit{Others} &29,323 \\
\bottomrule
\end{tabular}
\caption{Top 10 most cited publications.}\label{tab:publications_top}
\end{table}
    We observe different topics are covered by the most cited publications (\textbf{\Cref{tab:publications_top}}), but machine learning appears to be the most prominent topic with multiple publications.
    Half of the publications in this list have \textit{Others} as the venue, which in this experiment means the publications are books.
    The publication titled ``\#p'' has the same name in \dblp, so this is not an error on our end, though \googleScholar shows it with its full name ``Quantum invariants of 3-manifolds and NP vs \#P''.
    It also only has very few citations on \googleScholar\footnote{\url{https://scholar.google.de/scholar?oi=bibs&hl=de&cluster=13152820713440837432}} and \semanticScholar\footnote{\url{https://api.semanticscholar.org/CorpusID:233455290}} compared to 37,732 citations in \csi.
    We believe this is due to an error in the matching of the citations (\Cref{par:citations}), because of the broken and very short title.
    Our top 10 is entirely different from the top 20 of \textcite{fiala_computer_2017}, which might be because of the different underlying datasets (\dblp vs. \wos).
    Only our \#8 shows up as \#17 on their list.
    There are no overlaps with the top 20 from \textcite{mariani_nlp4nlp_2019} and the top 15 from \textcite{mohammad_nlp_2020_data}, which we explain by them only using publications from \gls{nlp}.
    We also note some author names are missing special characters (\Cref{sec:limitations}).

\section{Authors}\label{sec:authors}
    This section covers experiments that aggregate the data in \csi by the authors and then analyzes the authors to answer parts of \labelcref{rq:amount}, \labelcref{rq:distribution}, and \labelcref{rq:topics}.
    We start with general trends for the number of authors and their number of publications and citations  (\Crefrange{ex:authors_bar}{ex:authors_citations}), before covering the most cited and most productive (i.e., most published) authors, and what topics/venues they publish in (\Crefrange{ex:authors_top}{ex:authors_topics_time}).
    
    \experiment{How many authors publish in \csi per year? (\labelcref{rq:amount})}\label{ex:authors_bar}
    \begin{figure}[!htb]
        \centering
        \includegraphics[trim={0 1cm 0 0}, clip, width=\textwidth]{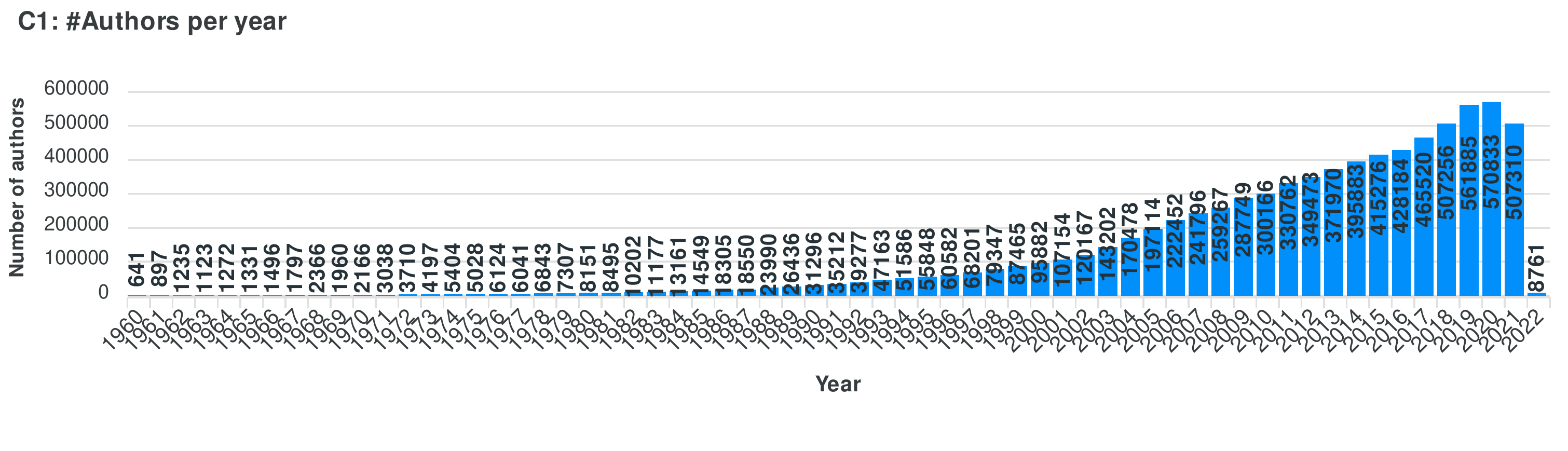}
        \caption{Number of unique authors per year starting in 1960. See \Cref{fig:authors_bar_app} for the full span.}
        \label{fig:authors_bar}
    \end{figure}
    In total there are 2,730,729 authors in \csi, while there are 54,604 publications without any authors listed.
    Similar to the number of publications (\Cref{ex:publications}) there is a continuous growth in the number of authors (\textbf{\Cref{fig:authors_bar}}), even in 2020, though the increase was smaller between 2019 and 2020 than between 2018 and 2019.
    The patterns from \Cref{ex:publications} are also reflected; there has been a major increase since the 2000s, which began as a spurt in the late 1980s, and a peak around 2020 (for publications 2019).
    \textcite{mohammad_nlp_2020_data, mariani_nlp4nlp_2019} observe similar trends in \gls{nlp} and \textcite{mohammad_nlp_2020_data} concludes that attracting more researchers every year means the research field is in good health.
    
    \experiment{How many authors published in the last $x$ years? How many authors are new to \gls{cs}? (\labelcref{rq:amount})}\label{ex:authors_active}
    \begin{table}[!htb]
\centering
\footnotesize
\begin{tabular}{l|rrrrr|r}
\toprule
\textbf{Time span} & \textbf{1 year} & \textbf{2 years} & \textbf{3 years} & \textbf{4 years} & \textbf{5 years} & \textbf{Total} \\
\midrule
Years &2020 &2019-2020 &2018-2020 &2017-2020 &2016-2020 &1936-2020 \\
\#Authors &570,833 &875,697 &1,081,137 &1,243,146 &1,378,348 &2,579,224 \\
\%Authors &22.13\% &33.95\% &41.92\% &48.20\% &53.44\% &100.00\% \\
\#Authors new &183,559 &370,122 &534,718 &683,923 &817,879 & \\
\%Authors new &7.12\% &14.35\% &20.73\% &26.52\% &31.71\% & \\
\midrule
Years (inv.) &1936-2019 &1936-2018 &1936-2017 &1936-2016 &1936-2015 & \\
\#Authors &2,395,665 &2,209,102 &2,044,506 &1,895,301 &1,761,345 & \\
\bottomrule
\end{tabular}
\caption{Number of authors who were active (at least one publication) in the last $x$ years compared to the total amount of authors. Also includes the number of new authors and their percentage considering the total number of authors (calculated using the inverted time span).}\label{tab:authors_active}
\end{table}
    We include the number of authors that were active before the last $x$ years at the bottom (inverted time span), to determine the number of new authors by calculating the difference between authors in the inverted time span and total authors (\textbf{\Cref{tab:authors_active}}).
    
    In 2020 over one-fifth of all authors published at least one paper, over a third in the last two years, and over half of all researchers in \csi in the last five years.
    Over 30\% of all authors in \csi only joined \gls{cs} in the last five years, which also infers that more than half of the authors that published between 2016 and 2020 were new authors.
    From this, it again becomes apparent that the field is massively growing but also very active and healthy.
    Similar trends for more and more new authors joining can also be observed in \gls{nlp} \parencite{mariani_nlp4nlp_2019}.
    \textcite{wahle_d3_2022} give some more insights into the activity of authors in \diii, which we cannot reproduce with \csi's \gls{ui} at the current time.
    
    \experiment{How are the citations and papers distributed across authors? How do the distributions change over time? (\labelcref{rq:distribution})}\label{ex:authors_citations}
    \begin{table}[!htp]
\centering
\footnotesize
\begin{tabular}{l|rrrr|rrrr}\toprule
\multirow{2}{*}{\textbf{Time span}} &\multicolumn{4}{c|}{\textbf{\#Citations}} &\multicolumn{4}{c}{\textbf{\#Papers}} \\
&Q1 &Med. &Q3 &Max. &Q1 &Med. &Q3 &Max. \\\midrule
1960-1969 &0 &3 &19 &20,532 &1 &1 &2 &39 \\
1970-1979 &0 &2 &16 &19,871 &1 &1 &2 &78 \\
1980-1989 &0 &3 &24 &57,729 &1 &1 &2 &171 \\
1990-1999 &0 &7 &44 &48,209 &1 &1 &3 &256 \\
\midrule
2000-2004 &1 &12 &61 &52,250 &1 &1 &3 &214 \\
2005-2009 &3 &17 &70 &53,063 &1 &1 &3 &360 \\
2010-2014 &3 &14 &55 &41,367 &1 &1 &3 &399 \\
2015-2019 &2 &8 &30 &89,647 &1 &1 &3 &586 \\
\midrule
1960-2019 &2 &11 &51 &133,020 &1 &2 &4 &1,332 \\
\bottomrule
\end{tabular}
\caption{Distribution of the number of total citations and papers over authors per time period showing the first quartile, median, third quartile, and maximum. The upper block covers 10 years per time period and the lower block five years.}\label{tab:authors_citations}
\end{table}
    It is not possible to compute the average number of authors per paper (or papers per author) purely from \csi's \gls{ui}, as it only supplies the unique number of authors per year and not the total number of authors per year, so authors who publish multiple papers in a year would only count once.

    The overall citations in the investigated period show a median of 11 citations with the first quartile being at 2 citations and the third quartile being at 51 citations, meaning half of all authors received between 2 and 51 citations (\textbf{\Cref{tab:authors_citations}}).
    This is a difference from \gls{nlp} \parencite{mariani_nlp4nlp_2019}, where 42\% of authors have no citations, while our first quartile (25\%) already has 2 citations.
    They conclude the high percentage of authors without citations is due to many citations coming from neighboring domains not covered in their dataset.
    As we include the entirety of \gls{cs} and not only a sub-field, this issue is smaller in our work.
    For the citations, we also see a peak in 2005-2009, which slowly falls off for the earlier years and more quickly for the more recent years.
    While this trend might appear interesting, we later see in \Cref{ex:citations_bar} that it mirrors the overall trend for citations quite well.
    The maximum for citations fluctuates but also shows a general trend upward, meaning singular recent papers get cited more, most likely due to the increase in publications (\Cref{ex:publications}) and researchers not looking so far back for references \parencite{fiala_computer_2017}.

    For the number of papers, the distribution nearly stays the same but shows a slight increase since 1990.
    A general increase is visible in the maximum, which shows singular authors push to always publish more papers in the same span.
    The low median makes sense as most authors in our dataset (45.85\%) only published one paper \parencite{wahle_d3_2022}.
    This is even more extreme in \gls{nlp}, where \textcite{mohammad_nlp_2020_data} shows 57.9\% of all authors only published one paper in \gls{nlp}.
    Considering the large number of authors in recent years (\Cref{ex:authors_bar}), and many of them being new contributors (\Cref{ex:authors_active}), it is possible they have only published one paper so far, or quickly dropped out of the field again.
    Determining the exact reason requires more research in the future, e.g., by investigating the number of papers authors publish, similar to \textcite{wahle_d3_2022}, who show only very few authors stay active in \gls{cs} for a long time.

    \experiment{Who are the most cited and most productive authors? (\labelcref{rq:topics})}\label{ex:authors_top}
    %Please add the following packages if necessary:
%\usepackage{booktabs, multirow} % for borders and merged ranges
%\usepackage{soul}% for underlines
%\usepackage[table]{xcolor} % for cell colors
%\usepackage{changepage,threeparttable} % for wide tables
%If the table is too wide, replace \begin{table}[!htp]...\end{table} with
%\begin{adjustwidth}{-2.5 cm}{-2.5 cm}\centering\begin{threeparttable}[!htb]...\end{threeparttable}\end{adjustwidth}
\begin{table}[!htb]\centering
\footnotesize
\begin{tabular}{rlrrrr}\toprule
\textbf{\#} &\textbf{Author (\#Citations)} &\textbf{First} &\textbf{\#Papers} &\textbf{\#Citations} &\textbf{Avg. Cit.} \\\midrule
1 &\textit{Others} &1936 &54,604 &676,548 &12.39 \\
2 &Ross B. Girshick &2004 &69 &146,867 &2,128.51 \\
3 &Anil K. Jain 0001 &1974 &662 &123,682 &186.83 \\
4 &Kaiming He &2009 &66 &114,330 &1,732.27 \\
5 &Jitendra Malik &1987 &231 &109,821 &475.42 \\
6 &Andrew Zisserman &1985 &454 &105,025 &231.33 \\
7 &Li Fei-Fei 0001 &2003 &194 &102,735 &529.56 \\
8 &Luc Van Gool &1984 &784 &96,530 &123.13 \\
9 &Jiawei Han 0001 &1985 &874 &95,371 &109.12 \\
10 &Trevor Darrell &1987 &288 &91,451 &317.54 \\
\midrule
&Average &1991 &402 &109,535 &648.19 \\
\midrule
\textbf{\#} &\textbf{Author (\#Papers)} &\textbf{First} &\textbf{\#Papers} &\textbf{\#Citations} &\textbf{Avg. Cit.} \\
\midrule
1 &\textit{Others} &1936 &54,604 &676,548 &12.39 \\
2 &H. Vincent Poor &1977 &1,649 &74,467 &45.16 \\
3 &Mohamed-Slim Alouini &1997 &1,445 &39,300 &27.20 \\
4 &Lajos Hanzo &1989 &1,382 &35,887 &25.97 \\
5 &*Wei Wang &1986 &1,334 &22,805 &17.10 \\
6 &Philip S. Yu &1980 &1,288 &73,436 &57.02 \\
7 &*Lei Zhang &1992 &1,269 &12,299 &9.69 \\
8 &*Yu Zhang &1991 &1,261 &11,380 &9.02 \\
9 &Victor C. M. Leung &1982 &1,260 &30,704 &24.37 \\
10 &*Yang Liu &2001 &1,247 &8,778 &7.04 \\
\midrule
&Average &1988 &1,348 &34,340 &24.73 \\
\bottomrule
\end{tabular}
\caption{Top 10 authors based on the number of citations received (top) and publications (bottom). The average is computed excluding \textit{Others}. Asterisks (*) denote entries, which refer to disambiguation pages in \dblp and not singular authors.}\label{tab:authors_top}
\end{table}
    It is interesting to see that the most productive and most cited authors have no overlap (\textbf{\Cref{tab:authors_top}}).
    We also note not only are the total citations higher for the most cited authors but also the average citations.
    Currently, it seems to be a quantity vs. quality matter, but subsequent experiments (\Cref{ex:authors_topics,ex:authors_topics_time}) show the topics the authors mainly cover are different (and thus also the venues they publish in), which could cause this.
    \textcite{franceschet_role_2010} also looks into the top 10 authors based on publications by using \dblp data from 2010.
    In his work, Philip S. Yu was \#1 with 547 publications and can still be seen in our list at \#6, now with 1288 publications.
    The other authors who were in the top 10 in 2010 are not in the top 10 anymore but are now somewhere in the top 100, e.g., Elisa Bertino was \#3 with 494 publications and is now \#31 with 966 publications.
    As with the most cited publications (\Cref{ex:publications_top}), there is no overlap to \gls{nlp}-specific authors, both for the most cited and most productive authors \parencite{mariani_nlp4nlp_2019}.
    This makes sense once we look into the next experiments, which investigate the topics of the most cited and most productive authors and show \gls{nlp} is not among the most researched topics.
    % no overlap except Yang Liu, which we assume to be a disambiguation entry, as they matched names
    
    \experiment{What are the preferred venues/topics of the most cited and most productive authors? (\labelcref{rq:topics})}\label{ex:authors_topics}
    \begin{table}[p]\centering
\scriptsize
\begin{tabular}{l|ll}\toprule
\textbf{Author (\#Citations)} &\textbf{Venue (\#Citations)} &\textbf{Venue (\#Publications)} \\
\midrule
\multirow{5}{*}{Ross B. Girshick} &CVPR (53,294) &CVPR (30) \\
&IT. Pattern Anal. Mach. Intell. (53,179) &ICCV (12) \\
&ICCV (22,197) &IT. Pattern Anal. Mach. Intell. (9) \\
&ACM Multimedia (13,371) &ECCV (8) \\
&ECCV (4,566) &ACM Multimedia (2) \\
\midrule
\multirow{5}{*}{Anil K. Jain 0001} &IT. Pattern Anal. Mach. Intell. (35,962) &IT. Pattern Anal. Mach. Intell. (104) \\
&ACM Comput. Surv. (13,647) &ICPR (75) \\
&\textit{Others} (10,794) &Pattern Recognit. (49) \\
&ICPR (7,387) &ICB (41) \\
&IEEE Trans. Inf. Forensics Secur. (5,375) &IEEE Trans. Inf. Forensics Secur. (35) \\
\midrule
\multirow{5}{*}{Kaiming He} &IT. Pattern Anal. Mach. Intell. (53,089) &CVPR (29) \\
&CVPR (22,051) &ICCV (13) \\
&ICCV (21,211) &IT. Pattern Anal. Mach. Intell. (11) \\
&ECCV (16,580) &ECCV (10) \\
&ECCV (13) (1,309) &ACM Trans. Graph. (1) \\
\midrule
\multirow{5}{*}{Jitendra Malik} &CVPR (46249) &CVPR (65) \\
&IT. Pattern Anal. Mach. Intell. (32,564) &ICCV (41) \\
&ICCV (12,451) &ECCV (29) \\
&ECCV (8,048) &IT. Pattern Anal. Mach. Intell. (24) \\
&SIGGRAPH (3,876) &Int. J. Comput. Vis. (12) \\
\midrule
\multirow{5}{*}{Andrew Zisserman} &CVPR (25,570) &CVPR (70) \\
&Int. J. Comput. Vis. (21,130) &BMVC (53) \\
&ICCV (16,604) &ECCV (45) \\
&BMVC (9,016) &ICCV (41) \\
&ECCV (8,992) &Int. J. Comput. Vis. (31) \\
\midrule
\textbf{Author (\#Papers)} &\textbf{Venue (\#Citations)} &\textbf{Venue (\#Publications)} \\\midrule
\multirow{5}{*}{H. Vincent Poor} &IEEE Trans. Inf. Theory (10,369) &IEEE Trans. Inf. Theory (134) \\
&IEEE Trans. Signal Process. (7,207) &ISIT (132) \\
&IEEE Trans. Commun. (7,005) &IEEE Trans. Wirel. Commun. (123) \\
&IEEE Trans. Wirel. Commun. (6,723) &IEEE Trans. Commun. (109) \\
&IEEE Signal Process. Mag. (5,384) &IEEE Trans. Signal Process. (99) \\
\midrule
\multirow{5}{*}{Mohamed-Slim Alouini} &IEEE Trans. Commun. (8,637) &IEEE Trans. Wirel. Commun. (164) \\
&IEEE Trans. Wirel. Commun. (6,993) &IEEE Trans. Commun. (138) \\
&Others (4,468) &ICC (105) \\
&IEEE Trans. Veh. Technol. (2,241) &GLOBECOM (84) \\
&ICC (2,022) &IEEE Trans. Veh. Technol. (83) \\
\midrule
\multirow{5}{*}{Lajos Hanzo} &IEEE Trans. Veh. Technol. (5,462) &IEEE Trans. Wirel. Commun. (164) \\
&IEEE Commun. Surv. Tutorials (4,796) &IEEE Trans. Commun. (138) \\
&IEEE Trans. Commun. (3,665) &ICC (105) \\
&Proc. IEEE (3,193) &GLOBECOM (84) \\
&IEEE Trans. Wirel. Commun. (2,928) &IEEE Trans. Veh. Technol. (83) \\
\midrule
\multirow{5}{*}{Philip S. Yu} &KDD (9,702) &IEEE Trans. Knowl. Data Eng. (94) \\
&IEEE Trans. Knowl. Data Eng. (8,442) &ICDM (77) \\
&Knowl. Inf. Syst. (5,494) &SDM (74) \\
&SIGMOD Conference (5,260) &KDD (73) \\
&ICDM (4,437) &CIKM (71) \\
\midrule
\multirow{5}{*}{Victor C. M. Leung} &IEEE Commun. Mag. (3,498) &IEEE Trans. Veh. Technol. (89) \\
&IEEE Trans. Veh. Technol. (2,801) &ICC (88) \\
&IEEE J. Sel. Areas Commun. (2,087) &GLOBECOM (73) \\
&IEEE Wirel. Commun. (1,830) &IEEE Trans. Wirel. Commun. (65) \\
&IEEE Trans. Wirel. Commun. (1,814) &WCNC (46) \\
\bottomrule
\end{tabular}
\caption{The top 5 venues for each of the top 5 most cited authors (top) and most productive authors (bottom) they got most cited in (left) and most published in (right).
``IEEE Trans. Pattern Anal. Mach. Intell.'' is abbreviated with ``IT. Pattern Anal. Mach. Intell.''}\label{tab:authors_venues}
\end{table}
    To determine the topics we look into the venues of the top 5 authors (ignoring \textit{Others} and disambiguation entries) by citations and publications (\textbf{\Cref{tab:authors_venues}}).
    We are aware that the top 5 authors might not resemble every author in \csi, but we believe they are a good approximation, and some valuable insights can be gained.
    
    In the top and bottom halves themselves, there is not much difference between the venues the authors get most cited in and most published in, but there are obvious differences between the top and bottom halves.
    The top 5 venues for the most cited authors show a clear trend toward the topics of computer vision and pattern recognition, both for the venues these five authors got most cited in and most published in.
    Most venues are conferences, with the most reoccurring being CVPR (Computer Vision and Pattern Recognition), ICCV (International Conference on Computer Vision), and ECCV (European Conference on Computer Vision), but the journal IEEE Trans. Pattern Anal. Mach. Intell. also appears just as often as CVPR.
    On the other hand, the top 5 venues of the most productive authors appear to be more on the engineering side of \gls{cs} and focus on signal processing, communication, and information theory.
    Most of the venues in this list are also IEEE journals, with a few conferences in between.
    The topics of the most cited and most productive authors also explain why there is no overlap with the top authors from \gls{nlp} (\Cref{ex:authors_top}); the most covered topics in \csi do not include \gls{nlp}.
    
    \experiment{Do the topics of the most cited and productive authors change over time? (\labelcref{rq:topics})}\label{ex:authors_topics_time}
    \afterpage{
\begin{landscape}
\begin{table}[p]
\scriptsize
\begin{center}
\begin{tabular}{lllll|l|l|lllll}
\multicolumn{ 6}{c|}{\textbf{Top 5 most cited authors}} & \multicolumn{ 6}{c}{\textbf{Top 5 most productive authors}} \\ 
1974-1999 & 2000-2004 & 2005-2009 & 2010-2014 & 2015-2019 & 1974-2019 & 1977-2019 & 1977-1999 & 2000-2004 & 2005-2009 & 2010-2014 & 2015-2019 \\ 
\midrule
cluster & cluster & cluster & latent & \textit{fingerprint} & imag & network & estim & data & \textbf{schedul} & \textit{network} & \textbf{paper} \\ 
\textit{object} & face & face & \textit{fingerprint} & face & fingerprint & channel & \textbf{signal} & \textbf{cluster} & scheme & \textbf{code} & \textbf{simul} \\ 
\textbf{invari} & \textit{recognit} & \textit{fingerprint} & face & \textit{match} & textur & problem & \textbf{handoff} & algorithm & relai & \textbf{graph} & result \\ 
textur & \textbf{featur} & biometr & \textit{imag} & \textit{object} & model & differ & \textbf{time} & queri & power & \textit{channel} & \textit{channel} \\ 
\textit{imag} & biometr & minutia & \textit{recognit} & latent & object & paper & \textbf{detect} & \textbf{index} & \textbf{sep} & mobil & problem \\ 
\textbf{distanc} & \textit{fingerprint} & \textit{imag} & \textit{object} & segment & face & code & \textit{network} & \textbf{pattern} & user & \textbf{social} & estim \\ 
\textbf{curv} & algorithm & algorithm & biometr & \textbf{human} & pose & data & \textbf{parallel} & \textbf{mine} & \textit{channel} & \textbf{featur} & \textbf{present} \\ 
\textbf{project} & model & data & minutia & cluster & cluster & present & user & base & averag & \textbf{decod} & power \\ 
\textbf{structur} & data & \textit{object} & segment & \textit{imag} & motion & relai & cach & adapt & adapt & \textbf{spectrum} & \textit{network} \\ 
\textit{match} & document & \textbf{classif} & model & \textbf{task} & track & consid & algorithm & divers & combin & \textbf{sens} & \textbf{exist} \\ 
\textit{recognit} & \textbf{person} & \textbf{deform} & user & pose & algorithm & model & \textbf{buffer} & \textbf{cdma} & divers & \textbf{predict} & \textbf{numer} \\ 
\textbf{form} & \textbf{select} & segment & \textit{match} & \textbf{network} & descriptor & result & mobil & \textbf{equal} & \textbf{select} & \textbf{complex} & \textbf{compar} \\ 
\textbf{camera} & view & secur & templat & \textbf{text} & human & propos & \textbf{disk} & \textbf{system} & \textbf{propos} & \textbf{detector} & \textbf{noma} \\ 
\textit{fingerprint} & \textit{match} & shape & \textbf{qualiti} & \textbf{instanc} & learn & fade & \textbf{comput} & cach & data & error & \textbf{express} \\ 
segment & \textbf{facial} & \textit{recognit} & retriev & shape & biometr & scheme & \textbf{multius} & \textit{channel} & \textbf{transmit} & data & relai \\ 
\textbf{descriptor} & \textbf{writer} & \textbf{express} & detect & \textbf{box} & kernel & energi & \textbf{join} & \textbf{burst} & perform & \textbf{privaci} & \textbf{solut} \\ 
pose & system & model & \textbf{composit} & \textit{recognit} & segment & express & queri & mobil & \textbf{spectral} & \textbf{cooper} & scheme \\ 
\textbf{vision} & \textit{object} & system & \textbf{ag} & \textbf{spoof} & data & close & \textbf{quantiz} & \textbf{structur} & \textbf{effici} & \textbf{link} & \textbf{analyz} \\ 
estim & partit & templat & shape & \textbf{reader} & method & simul & \textbf{block} & \textit{network} & distribut & \textbf{capac} & \textbf{demonstr} \\ 
view & templat & partit & secur & visual & plane & error & problem & perform & \textbf{transmiss} & \textbf{label} & \textbf{devic} \\ 
plane & user & \textbf{region} & \textbf{sketch} & detect & retriev & investig & \textbf{filter} & object & \textbf{fade} & \textbf{cognit} & \textbf{optic} \\ 
document & textur & pose & \textbf{orient} & identif & classif & user & \textbf{execut} & combin & \textbf{rate} & \textbf{aid} & \textbf{end} \\ 
\textbf{determin} & shape & retriev & \textbf{alter} & \textbf{depth} & match & select & \textit{channel} & \textbf{wireless} & base & relai & \textbf{number} \\ 
algorithm & \textbf{ensembl} & \textbf{scene} & \textbf{dataset} & \textbf{map} & featur & develop & scheme & \textbf{probabl} & \textbf{antenna} & user & mobil \\ 
model & identif & \textbf{fusion} & \textbf{ridg} & \textbf{target} & train & numer & \textbf{processor} & scheme & \textbf{optim} & base & \textbf{term} \\ 
data & pattern & \textit{match} & \textbf{method} & estim & correspond & decod & data & \textbf{segment} & \textbf{switch} & video & \textbf{achiev} \\ 
\textbf{frame} & \textbf{line} & \textbf{supervis} & learn & \textbf{transfer} & perform & order & web & averag & result & \textbf{applic} & video \\ 
pattern & learn & learn & \textbf{class} & \textbf{embed} & video & time & \textbf{node} & \textbf{gener} & \textbf{threshold} & power & \textbf{function} \\ 
\textbf{motion} & plane & \textbf{reconstruct} & \textbf{field} & \textbf{train} & bodi & power & \textbf{codec} & \textbf{interfer} & error & \textbf{radio} & \textbf{appli} \\ 
\textbf{integr} & \textit{imag} & visual & \textbf{databas} & \textbf{cnn} & geometri & graph & \textbf{server} & web & \textit{network} & object & distribut \\ 
\end{tabular}
\end{center}
\caption{Top 30 most salient terms for the top 5 most cited and most productive authors in different time periods.
Terms that only appear in one time period are \textbf{bold} and terms that appear in all five time periods are in \textit{italics}.}
\label{tab:authors_topics_time}
\end{table}
\end{landscape}
}
    In \textbf{\Cref{tab:authors_topics_time}} the first period includes all publications before 2000 to reduce issues with data sparsity in the earlier years, similar to \textcite{fiala_computer_2017}, and starts with the year of the first publication, i.e., 1974 and 1977.

    We observe the findings from the 30 most salient words align with the general topics of the venues the authors publish and get cited in the most (\Cref{ex:authors_topics}), and thus barely overlap.
    For the most cited authors, the three most salient terms over the whole period ``imag'', ``fingerprint'', and ``textur'' match the computer vision and pattern recognition focus of their preferred venues, which ``object'', ``motion'', and ``track'' also support.
    Similarly, the terms ``object'', ``imag'', ``match'', ``recognit'', and ``fingerprint'' appear in all five periods, again indicating a strong tie to the focus of the venues.
    In 2000-2004 terms like ``face'', ``biometr'', and ``shape'' first appear, indicating a greater focus on biometrics.
    The most salient term ``cluster'' (1974-2009) is replaced by ``latent'' in 2010-2014, which we hypothesize to be related to clustering algorithms with the term ``latent'' (e.g., Latent Dirichlet Allocation (\gls{lda})) gaining more popularity.
    We also note the first appearance of ``network'', ``train'', and ``cnn'' in 2015-2019, which shows the rise of approaches leveraging neural networks.
    \textcite{fiala_computer_2017} show the popularity of the term ``neural network'' already before 1999 and a decline in popularity for ``neural network'' after 2005.
    On the other hand, \textcite{tattershall_detecting_2020} find an ongoing increase for ``neural network'' and ``Convolutional neural (CNN)'', and \textcite{xia_research_2021} even show ``Object Detection; CNN; IOU'' is the \#1 research frontier in \gls{cs}, which matches more with our results.
    We assume the differences are due to the different approaches and underlying data.

    For the most productive authors, the two most salient terms over the whole period ``network'' and ``channel'', and the terms ``code'', ``data'' and ``relai'' are related to the signal processing focus of their venues.
    There are also some general terms (e.g., ``problem'', ``present'', and ``consid'') with high saliency, which is not the case for the terms of the most cited authors.
    The terms ``network'' and ``channel'' are also the two only terms, that appear in every period, but we believe the term ``network'' in this context is not related to neural networks and instead, physical networks considering the associated venues and other terms.
    We also see generally more unique terms per period compared to the most cited authors, e.g., ``codec'' before 2000, ``wireless'' in 2000-2004, and ``privaci'' in 2010-2014.
    The shift of topics to issues such as privacy, security, IoT, and big data, that \textcite{coskun_scientometrics-based_2019} find is thus only somewhat visible for our most cited and most productive authors.
    
\section{Venues}\label{sec:venues}
    This section covers experiments that aggregate the data in \csi by the venues and then analyzes the venues to answer the remaining parts of \labelcref{rq:amount}, \labelcref{rq:distribution}, and \labelcref{rq:topics}.
    We start with general trends for the number of venues and their number of publications and citations  (\Crefrange{ex:venues_bar}{ex:venues_citations}), before covering the most cited and most productive venues, and what topics they cover (\Crefrange{ex:venues_top}{ex:venues_topics_time}).

    \experiment{How many venues publish in \csi per year? (\labelcref{rq:amount})}\label{ex:venues_bar}
    \begin{figure}[htbp]
        \centering
        \includegraphics[trim={0 1cm 0 0}, clip, width=\textwidth]{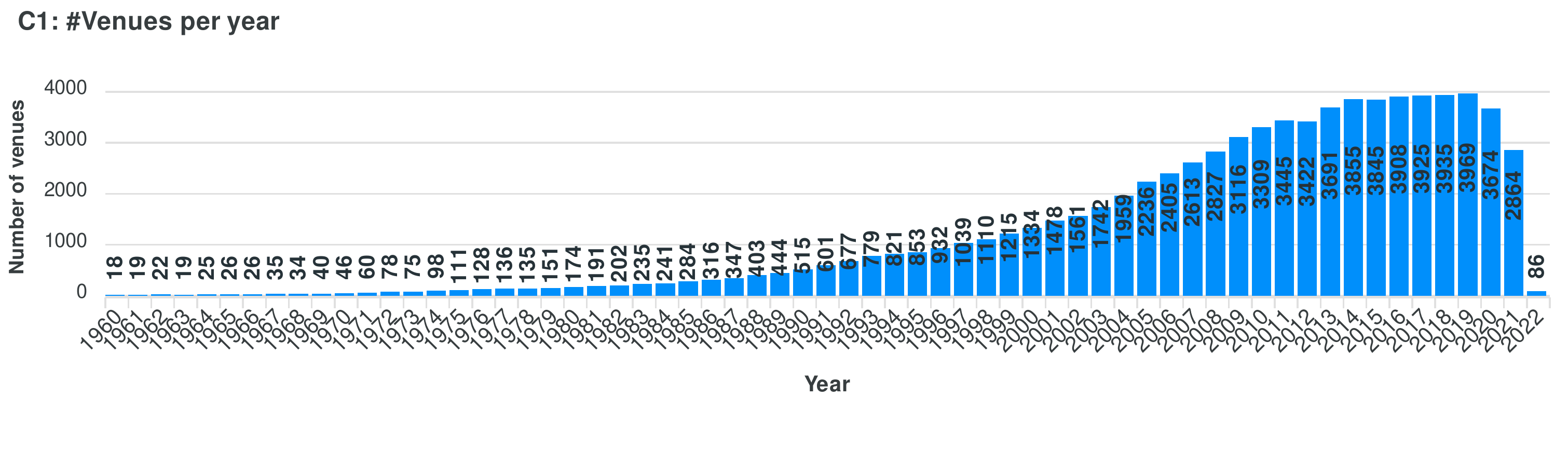}
        \caption{Number of unique venues per year starting in 1960. See \Cref{fig:venues_bar_app} for the full span.}
        \label{fig:venues_bar}
    \end{figure}
    In total there are 14,268 venues in \csi, while there are 14,212 publications without any venue listed.
    12,683 publications without a venue (nearly 90\%) are books, and another 1,131 are Ph.D. and master's theses.
    For the number of venues we see an increase during the late 1980s and a large increase since the 2000s (\textbf{\Cref{fig:venues_bar}}), similar to the publications (\Cref{ex:publications}) and authors (\Cref{ex:authors_bar}).
    The number of venues seems to reach a plateau in 2014, unlike publications and authors, which show an even bigger increase since 2017.
    In 2020, the number of venues went down, which was likely caused by the COVID-19 pandemic and in-person events being canceled.
    This is also the reason why the number of publications goes down in 2020, as seen in \Cref{ex:publications}.
    
    \experiment{How many unique venues published in the last $x$ years? How many venues are new? (\labelcref{rq:amount})}\label{ex:venues_active}
    \begin{table}[!htb]\centering
\footnotesize
\begin{tabular}{l|rrrrr|r}\toprule
\textbf{Time span} &\textbf{1 year} &\textbf{2 years} &\textbf{3 years} &\textbf{4 years} &\textbf{5 years} &\textbf{Total} \\
\midrule
Years &2020 &2019-2020 &2018-2020 &2017-2020 &2016-2020 &1936-2020 \\
\#Venues &3,674 &4,670 &5,338 &5,890 &6,456 &14,044 \\
\%Venues &26.16\% &33.25\% &38.01\% &41.94\% &45.97\% &100.00\% \\
\#Venues new &448 &963 &1,488 &2,009 &2,604 & \\
\%Venues new &3.19\% &6.86\% &10.60\% &14.31\% &18.54\% & \\
\midrule
Years (inv.) &1936-2019 &1936-2018 &1936-2017 &1936-2016 &1936-2015 & \\
\#Venues &13,596 &13,081 &12,556 &12,035 &11,440 & \\
\bottomrule
\end{tabular}
\caption{Number of venues that were active (at least one publication) in the last $x$ years compared to the total amount of venues. Also includes the number of new venues and their percentage considering the total number of venues (calculated using the inverted time span).}\label{tab:venues_active}
\end{table}
    We show the number of venues that published in the last few years and how many are new, i.e., published for the first time (\textbf{\Cref{tab:venues_active}}).
    Our goal is to investigate if new venues replace old venues or if the number of unique venues per year is actually stagnating in recent years (\Cref{ex:venues_bar}).
    We include the number of venues that were active before the last $x$ years at the bottom (inverted time span), to determine the number of new venues by calculating the difference between venues in the inverted time span and total venues.
    
    In the last five years, only 46\% of all venues were active, while 40\% of those were new venues.
    This is surprising, as we expected the bulk of venues to be recurring over the years, with some new ones every year.
    We think this has multiple reasons.
    First, up to 2020, there are 1,774 venues only for books, which only have one publication assigned nearly all the time.
    Second, some venues are listed on \dblp and in \diii with an increasing counter added to their name (e.g., ``iConference (1)'', ``TOOLS (48)'', ``HCI (42)''), which causes a few more new venues and fewer older ones (see \Cref{sec:limitations} for all limitations of our work).
    We try to remedy this by removing the counter with a script to fix the data from \diii, but five venues with a total of around 100 occurrences are still in \csi, so the total number of venues should be around 100 less.
    However, this does not explain the effect seen in this experiment, as the 1,774 book venues and 100 duplicate venues are spread over 1936-2020 and thus not enough to have an effect this large.
    We conclude the experiment shows an actual trend, that is not just caused by issues in the data.
    There are new venues in the last few years even though \Cref{ex:venues_bar} shows a plateau in the number of unique venues per year, which implies the new venues compensate for an apparent loss of older venues.
    The number of new venues is smaller compared to the number of new authors (\Cref{ex:authors_active}) in the last few years, which means the number of authors is faster growing than the number of venues.
    As mentioned in \Cref{ex:publications}, in \gls{nlp} the number of publications goes down every second year due to biennial conferences \parencite{mohammad_nlp_2020_data, mariani_nlp4nlp_2019}, but we can still not replicate the same trend in \csi.

    \experiment{How are the citations and papers distributed across venues? How do the distributions change over time? (\labelcref{rq:distribution})}\label{ex:venues_citations}
    \begin{table}[!htp]\centering
\footnotesize
\begin{tabular}{l|rrrrr|rrrrr}\toprule
\multirow{2}{*}{\textbf{Time span}} &\multicolumn{5}{c|}{\textbf{\#Citations}} &\multicolumn{5}{c}{\textbf{\#Papers}} \\
&Q1 &Med. &Q3 &Max. &Avg. &Q1 &Med. &Q3 &Max. &Avg. \\\midrule
1960-1969 &0 &111 &1,699 &60,481 &4,950.42 &3 &41 &231 &2,179 &171.26 \\
1970-1979 &0 &75 &851 &157,404 &4,128.27 &15 &36 &138 &2,133 &140.22 \\
1980-1989 &0 &117 &1,125 &184,718 &4,269.73 &15 &35 &136 &5,276 &137.43 \\
1990-1999 &1 &296 &2,319 &318,304 &5,023.76 &18 &44 &156 &8,821 &160.09 \\
\midrule
2000-2004 &47 &711 &3,200 &237,082 &4,947.81 &19 &59 &155 &5,716 &145.40 \\
2005-2009 &92 &553 &2,819 &479,768 &4,725.48 &16 &51 &169 &6,881 &167.19 \\
2010-2014 &91 &419 &1,797 &492,469 &3,549.21 &15 &54 &166 &7,724 &170.56 \\
2015-2019 &49 &243 &1,094 &722,350 &2,523.54 &17 &66 &195 &25,371 &211.79 \\
\midrule
1960-2019 &37 &239 &1,494 &1,609,420 &7,029.40 &14 &33 &193 &42,071 &318.30 \\
\bottomrule
\end{tabular}
\caption{Distribution of the number of total citations and papers across venues per time period showing the first quartile, median, third quartile, maximum, and average. The upper block covers 10 years per time period and the lower block five years.}\label{tab:venues_citations}
\end{table}
    
    We see both the distribution of citations and papers have a higher average than the respective median for every period (\textbf{\Cref{tab:venues_citations}}), which implies that most citations and papers fall to a few venues and are not spread evenly.
    For the citations, the first quartile peaks in 2005-2009, which is also the period with the most citations (\Cref{ex:citations_bar}) and lines up with the authors (\Cref{ex:authors_citations}), while the median and third quartile peak in 2000-2004.
    We explain the difference in 2000-2004 with the distribution across venues being the most even (highest median over average ratio of all periods), and also 2005-2009 showing the highest increase in venues per year (3,000 unique venues in 2000-2004 to nearly 5,000 in 2005-2009), which causes the citations to split up more.
    The large drop-off in citations for the first quartile before 2000 also matches the citations of the authors (\Cref{ex:authors_citations}).
    If we consider the total period, half of all venues receive between 37 and 1494 citations in total.
    The maximum steadily increases for both citations and papers, showing some venues are publishing more papers every period and some also receiving more citations.
    This again matches the authors, who show a slight trend upward of the maximum of citations and a steady increase in publications (\Cref{ex:authors_citations}).
    The number of papers reveals that most venues publish only slightly more on average over the decades and most venues do not show the same increase as the venues that are responsible for the maximums.
    Overall we see a correlation between overall citations, authors, and venues and a small correlation between the number of papers for authors and venues.
    
    \experiment{What are the most cited and most productive venues? (\labelcref{rq:topics})}\label{ex:venues_top}
    \begin{table}[!htb]\centering
\footnotesize
\begin{tabular}{rlrrrrrr}\toprule
\textbf{\#} &\textbf{Venue (\#Citations)} &\textbf{First} &\textbf{\#Papers} &\textbf{\#Citations} &\textbf{Avg. Cit.} \\\midrule
1 &CVPR &1988 &12,757 &1,621,492 &127.11 \\
2 &\textit{Others} &1938 &14,212 &1,505,675 &105.94 \\
3 &NeuroImage &1996 &16,947 &1,377,202 &81.27 \\
4 &IT. Pattern Anal. Mach. Intell. &1975 &6,559 &1,337,060 &203.85 \\
5 &IEEE Trans. Inf. Theory &1963 &16,325 &1,147,862 &70.31 \\
6 &Commun. ACM &1958 &12,742 &948,274 &74.42 \\
7 &IEEE Trans. Ind. Electron. &1990 &12,777 &802,571 &62.81 \\
8 &ICRA &1984 &25,017 &790,170 &31.59 \\
9 &IEEE Trans. Signal Process. &1990 &13,328 &762,802 &57.23 \\
10 &IEEE Trans. Autom. Control. &1991 &10,762 &717,910 &66.71 \\
\midrule
&Average &1982 &14,143 &1,101,102 &88.12 \\
\midrule
\textbf{\#} &\textbf{Venue (\#Papers)} &\textbf{First} &\textbf{\#Papers} &\textbf{\#Citations} &\textbf{Avg. Cit.} \\\midrule
1 &IEEE Access &2013 &54,961 &452,363 &8.23 \\
2 &ICASSP &1975 &45,660 &714,945 &15.66 \\
3 &Sensors &2009 &36,718 &526,251 &14.33 \\
4 &IGARSS &2002 &29,421 &119,000 &4.04 \\
5 &ICRA &1984 &25,017 &790,170 &31.59 \\
6 &ISCAS &1993 &23,549 &174,126 &7.39 \\
7 &ICIP &1993 &22,714 &330,302 &14.54 \\
8 &ICC &1984 &22,296 &303,031 &13.59 \\
9 &Appl. Math. Comput. &1998 &19,983 &336,211 &16.82 \\
10 &IROS &1988 &19,561 &422,430 &21.60 \\
\midrule
&Average &1994 &29,988 &416,883 &14.78 \\
\bottomrule
\end{tabular}
\caption{Top 10 venues based on the number of citations received (top) and publications (bottom). The average is computed excluding \textit{Others}. ``IEEE Trans. Pattern Anal. Mach. Intell.'' is abbreviated with ``IT. Pattern Anal. Mach. Intell.''}\label{tab:venues_top}
\end{table}
    Both top 10s have barely any overlap, except for the ICRA (IEEE Robotics and Automation Society) conference, which also has the highest average citations of the top 10 most productive venues, but the lowest of the top 10 most cited venues (\textbf{\Cref{tab:venues_top}}).
    Some venues do appear in the other's top 20 though, e.g., NeuroImage at \#14 and IEEE Trans. Inf. Theory at \#16 in the top 20 venues with the most publications.
    In the top 20 of the most cited venues ICASSP (IEEE International Conference on Acoustics, Speech and Signal Processing) appears at \#11, and Sensors at \#19.
    The most cited venues include many venues we already saw as venues the most productive and most cited authors publish in (\Cref{ex:authors_topics}), e.g., CVPR, IEEE Trans. Pattern Anal. Mach. Intell., and IEEE Trans. Inf. Theory.
    This implies the topics of the most cited venues also align with computer vision, pattern recognition, signal processing, and communication.
    We investigate the topics further in \Cref{ex:venues_topics}.
    In a later experiment, we see engineering has a high preference for conferences, which explains why many of the most productive venues in this experiment are IEEE conferences with a focus on topics from the field of engineering (\Cref{ex:fields_types}).
    
    The most cited venues are dominated by journals, but the first place is taken by the CVPR conference.
    On the contrary, the most productive venues are mostly conferences, while the first and third places are taken by the open-access journals IEEE Access and Sensors, respectively.
    In general, the average citations are also lower for most productive venues compared to the most cited venues, and IEEE Trans. Pattern Anal. Mach. Intell. has the highest average citations (203.85) of all covered venues.
    When we compare the list to the most productive venues of \textcite{fiala_computer_2017} we recognize some venues, e.g., IEEE Transactions on Information Theory, and Communications of the ACM.
    Most venues in their list appear to be journals, which is different from our list, where it is mostly conferences.
    \textcite{coskun_scientometrics-based_2019} only investigate the top 10 journals for two periods (2008-2013 and 2014-2019).
    We again see a little overlap, e.g., IEEE Access is \#1 in the second period, and IEEE Transactions On Information Theory also appears.
    Their most productive journals are dominated by IEEE and IEICE journals, which we do not see in our most productive venues, but instead, we see five IEEE journals in our 10 most cited venues.
    The \#1 of both works has fewer papers than our \#10, and they use a different and smaller dataset (from \wos) which can explain the differences.
    
    An experiment for the future would be to look into other measures of quality (e.g., impact factor) for these most prominent venues, considering the most cited venues are, on average, 12 years older and thus more established and possibly more prestigious.
    Another interesting experiment for future research is to look into open-access publications only, as both lists change drastically, with IEEE Access and Sensors leading the most productive venues and them being \#3 and \#1 respectively for the most cited venues, where they do not appear in the top 10 or 20 before at all.

    \experiment{What are the most popular topics of the most cited and most productive venues? How do the venues' topics differ from each other? (\labelcref{rq:topics})}\label{ex:venues_topics}
    \afterpage{
\begin{landscape}
\begin{table}[p]
\scriptsize
\begin{center}
\begin{tabular}{lllll|lllll}
\multicolumn{ 5}{c|}{\textbf{Top 5 most cited venues}} & \multicolumn{ 5}{c}{\textbf{Top 5 most productive venues}} \\ 
CVPR & NeuroImage & IT. PA. M. Int. & IT. Inf. Theory & Commun. ACM & IEEE Access & ICASSP & Sensors & IGARSS & ICRA \\ 
\midrule
propos & \textbf{respons} & propos & \textbf{code} & algorithm & \textit{propos} & \textit{propos} & \textit{propos} & \textbf{sar} & \textit{propos} \\ 
result & \textbf{connect} & paper & \textbf{channel} & \textbf{program} & result & \textbf{speech} & sensor & \textit{imag} & control \\ 
object & \textbf{ag} & result & network & \textbf{time} & control & \textit{imag} & network & \textit{propos} & \textit{present} \\ 
network & data & \textbf{face} & \textbf{sequenc} & \textbf{softwar} & network & paper & measur & \textbf{surfac} & \textbf{object} \\ 
present & method & model & \textbf{decod} & develop & \textit{imag} & model & paper & \textbf{area} & result \\ 
imag & function & learn & present & problem & data & result & \textit{present} & method & \textbf{manipul} \\ 
method & imag & \textbf{recognit} & \textbf{capac} & \textbf{new} & \textit{present} & \textit{present} & provid & \textbf{radar} & learn \\ 
state & \textbf{suggest} & \textbf{cluster} & paper & provid & paper & network & \textit{imag} & model & \textbf{robot} \\ 
motion & network & \textbf{label} & estim & \textbf{languag} & learn & estim & data & \textbf{classif} & paper \\ 
\textbf{art} & \textbf{activ} & object & \textbf{signal} & \textbf{need} & detect & \textbf{filter} & method & \textbf{soil} & estim \\ 
\textbf{camera} & process & train & algorithm & \textbf{inform} & model & \textbf{recognit} & featur & \textbf{land} & \textbf{task} \\ 
problem & model & work & \textbf{nois} & \textbf{technolog} & \textbf{power} & algorithm & work & \textit{present} & \textit{imag} \\ 
\textbf{dataset} & \textbf{left} & network & problem & \textbf{system} & \textbf{differ} & \textbf{nois} & \textbf{posit} & \textbf{chang} & \textbf{plan} \\ 
learn & \textbf{associ} & imag & \textbf{set} & \textbf{includ} & featur & \textbf{rate} & achiev & featur & design \\ 
\textbf{featur} & \textbf{cortex} & develop & \textbf{studi} & \textbf{number} & method & signal & detect & data & \textbf{grasp} \\ 
\textbf{challeng} & \textbf{visual} & achiev & \textbf{obtain} & point & algorithm & achiev & \textbf{monitor} & perform & provid \\ 
\textbf{gener} & result & surfac & \textbf{sourc} & \textbf{acm} & provid & problem & energi & \textbf{resolut} & problem \\ 
\textbf{experi} & \textbf{right} & \textbf{track} & function & \textbf{commun} & obtain & \textbf{train} & result & measur & sensor \\ 
segment & \textbf{particip} & import & \textbf{construct} & model & \textbf{comput} & \textbf{code} & node & \textbf{forest} & \textbf{map} \\ 
\textbf{us} & \textbf{diffus} & point & \textbf{perform} & \textbf{follow} & \textbf{predict} & provid & obtain & \textbf{sens} & time \\ 
\textbf{deep} & provid & motion & \textbf{sub} & \textbf{design} & achiev & \textbf{speaker} & \textbf{requir} & \textbf{cover} & \textbf{motion} \\ 
surfac & \textbf{brain} & segment & \textbf{user} & \textbf{servic} & energi & channel & \textbf{optic} & \textbf{retriev} & perform \\ 
\textbf{demonstr} & \textbf{area} & requir & given & \textbf{internet} & design & learn & \textbf{sensit} & \textbf{remot} & \textbf{actuat} \\ 
\textbf{detect} & \textbf{show} & \textbf{match} & \textbf{rate} & data & time & featur & estim & work & data \\ 
data & \textbf{matter} & \textbf{deriv} & process & paper & \textbf{commun} & \textbf{demonstr} & learn & \textbf{spectral} & \textbf{dynam} \\ 
train & state & term & achiev & \textbf{discuss} & signal & \textbf{frequenc} & temperatur & \textbf{inform} & \textbf{mechan} \\ 
requir & \textbf{eeg} & view & \textbf{scheme} & import & channel & \textbf{consid} & signal & \textbf{moistur} & \textbf{forc} \\ 
point & \textbf{stimuli} & problem & \textbf{receiv} & given & \textbf{user} & work & accuraci & time & \textbf{real} \\ 
view & \textbf{stimul} & \textbf{recent} & \textbf{distort} & \textbf{possibl} & accuraci & \textbf{transform} & \textbf{wireless} & temperatur & \textbf{camera} \\ 
work & \textbf{region} & estim & term & \textbf{describ} & node & \textbf{experiment} & \textbf{develop} & \textbf{polarimetr} & \textbf{path} \\ 
\end{tabular}
\end{center}
\caption{Top 30 most salient terms for the top 5 most cited and most productive venues. Terms that only appear in one venue are \textbf{bold} and terms that appear in all five venues are in \textit{italics}. ``IEEE Trans. Pattern Anal. Mach. Intell.'' is abbreviated with ``IT. PA. M. Int.'' and ``IEEE Trans. Inf. Theory'' with ``IT. Inf. Theory''}\label{tab:venues_topics}
\end{table}
\end{landscape}
}
    While there are no terms that appear in all five of the most cited venues, we can use the unique terms (bold) to see how the venues differentiate from each other and what makes them unique (\textbf{\Cref{tab:venues_topics}}).
    For CVPR it is ``camera'', ``deep'', ``detect'' (computer vision), for NeuroImage ``cortex'', ``brain'', and ``stimuli'' (brain imagery), and for IEEE Trans. Pattern Anal. Mach. Intell. ``face'', ``cluster'', and ``track'' (facial recognition).
    IEEE Trans. Inf. Theory is more related to signal processing and communication, which is visible due to ``code'', ``channel'', ``decode'', and ``signal'', which makes it also overlap with the top 5 most productive authors, who focus on the same area.
    The Commun. ACM is a broader venue about current trends in \gls{cs}, which can be somewhat seen through its unique terms ``program'', ``softwar'', ``inform'', ``technolog'', and ``internet''.
    We see that for three of the five venues, the focus on computer vision and pattern recognition becomes visible and their specialization in that field.
    Other topics are present in the other two venues (i.e., information theory and current trends respectively), so there is not just one clear direction for the topics.

    The top 5 most productive venues have three words that overlap, ``propos'', ``imag'', and ``present''.
    IEEE Access is a multidisciplinary open-access journal, which also shows in the data as it has only a few unspecific unique terms, but the other four venues all have unique terms, that describe the topic of the respective venue well.
    ICASSP (International Conference on Acoustics, Speech and Signal Processing) has ``speech'', ``filter'', and ``noise'', Sensors has ``optic'', ``sensit'', and ``wireless'', IGARSS (International Geoscience and Remote Sensing Symposium) has ``surface'', ``area'', and ``radar'', and lastly ICRA (International Conference on Robotics and Automation) has ``object'', ``robot'', and ``motion''.
    We again see the specialization of each venue, and additionally a general focus on engineering topics, or more specifically, topics that heavily use sensors in four of the five venues, the exception being IEEE Access.
    
    \experiment{Do the topics of the most prominent venues change over time? (\labelcref{rq:topics})}\label{ex:venues_topics_time}
    \afterpage{
\begin{landscape}
\begin{table}[p]
\scriptsize
\begin{center}
\begin{tabular}{lllll|l|l|lllll}
\multicolumn{ 6}{c|}{\textbf{Top 5 most cited venues}} & \multicolumn{ 6}{c}{\textbf{Top 5 most productive venues}} \\ 
1958-1999 & 2000-2004 & 2005-2009 & 2010-2014 & 2015-2019 & 1958-2019 & 1975-2019 & 1975-1999 & 2000-2004 & 2005-2009 & 2010-2014 & 2015-2019 \\ 
\midrule
code & code & brain & \textit{imag} & propos & code & paper & \textit{robot} & \textit{propos} & paper & \textit{propos} & \textit{propos} \\ 
algorithm & activ & activ & brain & \textbf{art} & algorithm & present & \textit{present} & paper & \textit{propos} & paper & \textit{imag} \\ 
present & \textbf{spl} & \textit{imag} & propos & brain & present & speech & filter & \textit{present} & \textit{robot} & \textit{model} & paper \\ 
paper & \textbf{sub} & paper & network & \textbf{dataset} & paper & result & \textit{control} & \textit{robot} & \textit{imag} & \textit{robot} & network \\ 
\textit{imag} & \textit{imag} & object & activ & compar & imag & signal & estim & data & \textit{present} & result & result \\ 
develop & area & present & \textbf{cortic} & network & develop & estim & \textit{propos} & \textit{model} & sensor & \textit{present} & \textbf{demonstr} \\ 
\textbf{program} & \textit{model} & result & data & \textit{imag} & program & propos & \textit{speech} & \textit{imag} & data & \textit{imag} & method \\ 
\textit{model} & investig & subject & present & method & model & network & \textbf{adapt} & result & compar & achiev & \textit{robot} \\ 
error & compar & code & area & paper & error & time & algorithm & \textbf{develop} & \textit{control} & \textit{control} & \textbf{optim} \\ 
\textbf{estim} & respons & algorithm & subject & \textbf{function} & estim & problem & \textbf{manipul} & \textit{speech} & sar & \textit{speech} & \textit{control} \\ 
user & task & experi & task & studi & user & achiev & \textit{model} & \textit{control} & \textit{speech} & estim & \textbf{learn} \\ 
object & develop & channel & respons & task & object & algorithm & \textit{signal} & sar & filter & problem & \textit{present} \\ 
\textbf{possibl} & studi & \textit{model} & method & present & possibl & featur & \textbf{recognit} & compar & featur & nois & featur \\ 
\textbf{softwar} & user & cortex & paper & \textbf{learn} & softwar & model & \textbf{code} & \textit{provid} & \textbf{us} & data & data \\ 
\textbf{languag} & problem & studi & \textit{model} & \textbf{train} & languag & provid & object & obtain & work & work & algorithm \\ 
\textbf{inform} & bound & increas & \textbf{demonstr} & \textbf{end} & inform & vector & task & radar & network & algorithm & \textit{provid} \\ 
\textbf{author} & data & task & cortex & differ & author & control & problem & network & \textit{model} & \textit{signal} & problem \\ 
result & work & specif & state & object & result & recognit & \textbf{motion} & \textbf{area} & algorithm & \textit{provid} & \textit{speech} \\ 
work & cortex & develop & develop & \textbf{visual} & work & imag & \textbf{rate} & channel & method & featur & estim \\ 
\textbf{nois} & \textbf{sup} & obtain & \textbf{show} & includ & nois & nois & \textbf{describ} & achiev & \textbf{number} & \textbf{error} & \textbf{train} \\ 
\textbf{probabl} & \textbf{motion} & compar & investig & data & probabl & error & \textbf{process} & filter & \textbf{surfac} & filter & \textit{model} \\ 
\textbf{rate} & experi & differ & \textbf{connect} & work & rate & method & \textit{imag} & differ & \textbf{resolut} & \textbf{simul} & obtain \\ 
\textbf{sourc} & \textbf{signific} & \textbf{us} & studi & featur & sourc & demonstr & \textit{provid} & object & achiev & sensor & \textbf{term} \\ 
\textbf{signal} & error & problem & obtain & state & signal & channel & \textbf{forc} & \textit{signal} & \textit{provid} & channel & work \\ 
\textbf{term} & channel & area & \textbf{provid} & \textbf{domain} & term & filter & sensor & \textbf{land} & \textit{signal} & \textbf{shown} & \textbf{time} \\ 
problem & \textbf{relat} & \textbf{matter} & increas & \textbf{control} & problem & word & nois & \textbf{remot} & \textbf{requir} & method & \textbf{perform} \\ 
bound & subject & featur & channel & code & bound & train & \textbf{design} & featur & problem & \textbf{speaker} & comput \\ 
\textbf{number} & specif & \textbf{effect} & \textbf{patient} & area & number & qualiti & obtain & \textbf{sens} & radar & comput & \textbf{state} \\ 
includ & \textbf{chang} & \textbf{fmri} & \textbf{memori} & \textit{model} & includ & requir & \textbf{transform} & \textbf{band} & nois & task & \textbf{commun} \\ 
\textbf{requir} & differ & \textbf{region} & \textbf{segment} & activ & requir & data & \textbf{joint} & \textbf{satellit} & differ & \textbf{oper} & \textit{signal} \\ 
\end{tabular}
\end{center}
\caption{Top 30 most salient terms for the top 5 most cited and most productive venues in different time periods. Terms that only appear in one time period are \textbf{bold} and terms that appear in all five time periods are in \textit{italics}.}\label{tab:venues_topics_time}
\end{table}
\end{landscape}
}
    In \textbf{\Cref{tab:venues_topics_time}} the first period includes all publications before 2000 to reduce issues with data sparsity in the earlier years, similar to \textcite{fiala_computer_2017}, and starts with the year of the first publication, i.e., 1958 and 1975.
    We group the venues to try and find trends over time that overlap between the most successful venues, e.g., usage of specific technologies, even if the specific venues might have a different focus.
    The idea is venue specific terms should be ranked lower, and (common) terms that exist across all venues are ranked higher.

    For the most cited venues, the terms ``model'' and ``imag'' appear in each period, which is probably due to three of the five venues focusing on computer vision and pattern recognition.
    The term ``brain'' is ranked high starting in 2005, which we attribute to NeuroImage being exclusively about brain imagery and first starting publication in 1996.
    However, in 2015-2019 we also see ``dataset'', ``learn'', and ``train'' appearing for the first time, and ``network'' appearing again, which indicates a shift to methods leveraging neural networks across venues, similar to the most cited authors in that period (\Cref{ex:authors_topics_time}).
    In that experiment, we already related the findings from other researchers regarding neural networks, so we refrain from repeating them.
    The most productive venues share multiple terms across all periods.
    Some are general terms (``present'', ``propos'', ``provid''), and some were venue specific in \Cref{ex:venues_topics} (``speech'', ``robot''), similar to ``brain'' for the most cited venues, which leaves ``signal'', ``imag'', ``control'', and ``model''.
    These last four terms indicate a general focus more toward engineering for the most productive venues, which matches the findings of the last experiment (\Cref{ex:venues_topics}).
    Again, 2015-2019 indicates a shift toward approaches using neural networks because ``learn'', ``train'', and ``optim'' are new terms, and ``network'' also appears again on a high rank.
    
    We conclude that this experiment (i.e., grouping venues for overarching topics) only had partly the effect we had hoped for.
    Generic words are now more present (e.g., ``present'', ``paper'', and ``propos''), which we hoped the saliency measure would prevent (as it takes distinctiveness into account), but also specific technologies are not ranked higher in most cases, or there were simply no overlaps across venues as we hoped for.
    Especially the two full periods show no clear direction compared to earlier experiments (\Cref{ex:authors_topics_time,ex:venues_topics}).
    Additionally, we see terms that formerly only related to one specific venue are still being ranked very high, e.g., ``brain'', ``robot'', and ``speech''.
    It is possible these terms also have some importance in the other venues and thus are ranked high up, but we believe it is more likely, that the measure of saliency causes these issues.
    Saliency measures the distinctiveness of words considering all topics, and in this case, the topic model correctly sorts all publications with, e.g., ``brain'' into one topic which gives it a high saliency, so the problem appears to be that the topics model can perfectly sort five venues into 10 different topics.
    We hope to overcome these issues and eliminate the venue-specific terms in a better way, by grouping more venues together, which we will do when we investigate the topics of the different document types (\Cref{ex:types_topics_time}).
    Another possibility that \csi supports is to analyze each venue separately, but our goal is to analyze the state of \gls{cs}, so we have to move this to possible future work (\Cref{sec:future_work}).

\section{Citations}\label{sec:citations}
    This section exclusively deals with experiments conducted directly on the citations of publications in \csi without any further aggregations to answer \labelcref{rq:citations}.
    We cover both incoming and outgoing citations, including their numbers per year (\Cref{ex:citations_bar}) and their distributions (\Crefrange{ex:citations_distribution}{ex:citations_time}).
    
    \experiment{How do incoming and outgoing citations compare over time? (\labelcref{rq:citations})}\label{ex:citations_bar}
    \begin{figure}[htbp]
        \centering
        \includegraphics[trim={0 1cm 0 0}, clip, width=\textwidth]{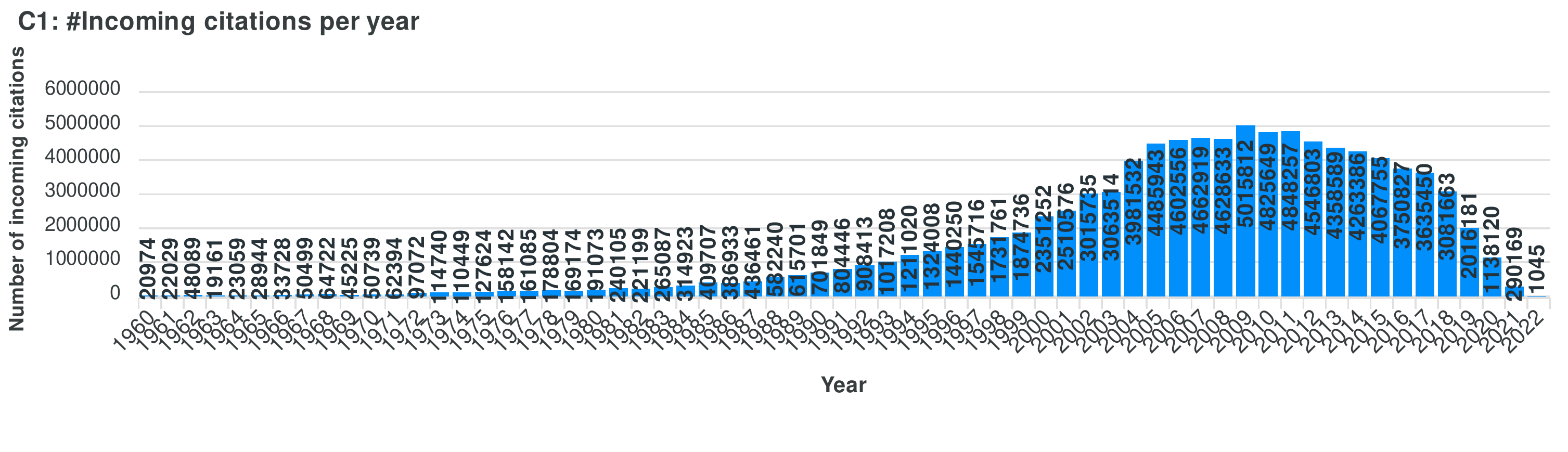}
        \includegraphics[trim={0 1cm 0 0}, clip, width=\textwidth]{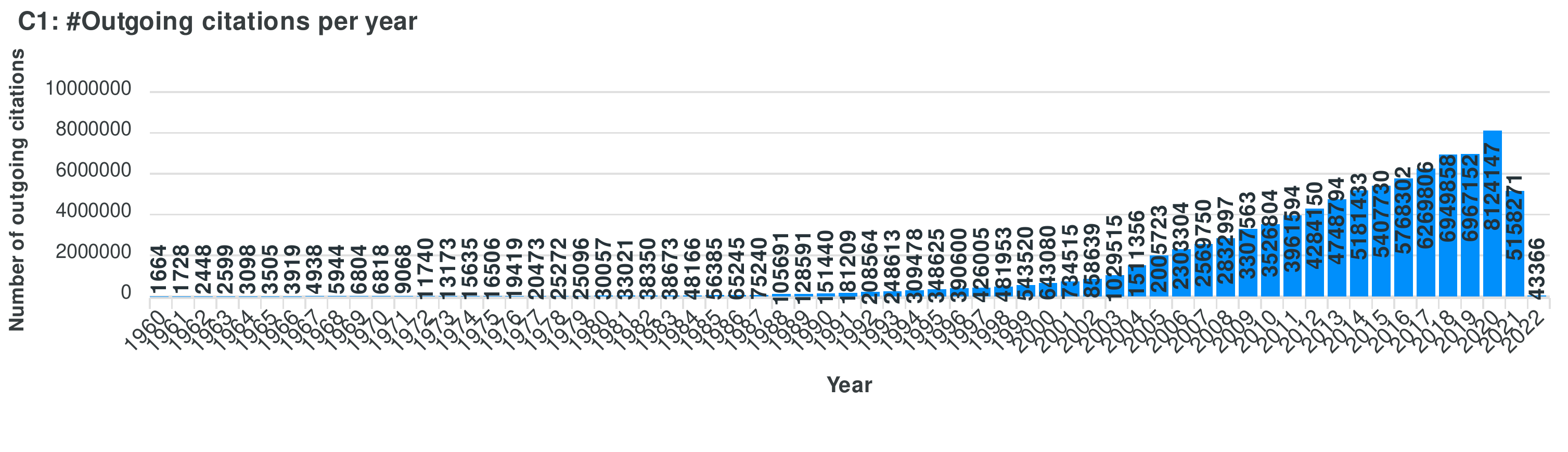}
        \caption{Incoming citations per year (top), that publications from that year received in their lifetime, and outgoing citations (references) per year (bottom), that publications from that year listed in their bibliography; starting in 1960. See \Cref{fig:citations_in_bar_app,fig:citations_out_bar_app} for the full span.}
        \label{fig:citations_bar}
    \end{figure}
    In total \csi contains 97,053,288 incoming citations and 88,302,512 outgoing citations.
    The incoming citations that publications receive have a peak in 2009 and consistently fall off in earlier and later years (\textbf{\Cref{fig:citations_bar}}), which we explain with earlier years having fewer publications and thus fewer citations overall, and the publications in later years not being old enough to aggregate enough citations yet.
    This is supported by \textcite{fiala_computer_2017} who show new citations become fewer every year after publication, which explains why older publications have fewer citations.
    We also explain this with researchers focusing more on current work when citing other publications.
    The drop-off for newer publications comes from them just not having accumulated so many citations yet, as \textcite{fiala_computer_2017} also show that many citations still come in years after publication.
    We believe this causes a certain point (i.e., the peak around 2009), where the aforementioned effects related to the drop-offs balance each other out.
    We already observed a similar trend when investigating the authors (\Cref{ex:authors_citations}) and explain this with the general trend for authors simply following the general trend of the overall citations.
    The trend of citations for venues appears different, as some peaks are during earlier periods, which we already covered in \Cref{ex:venues_citations}.
    \textcite{mohammad_nlp_2020_data} sees a similar curve for incoming citations, even though 2009 is not the peak, as the curve is more susceptible to irregularities caused by singular highly cited publications, due to the smaller dataset size and fewer publications per year.
    We can thus infer, that the peak of received (i.e., incoming) citations always is a few years back in time, even though we cite recent papers the most \parencite{fiala_computer_2017} and produce more citations every year (see next paragraph).
    
    For the outgoing citations (references) we observe a consistent increase, that follows similar trends as the number of publications per year, with the first larger increase in the late 1980s and a second large increase since the 2000s (\Cref{ex:publications}).
    The consistent increase in outgoing citations shows us we are citing more other publications with each passing year.
    Whether this is solely due to the increase in publications (\Cref{ex:publications}) or if we are also citing more publications per paper is investigated in \Cref{ex:citations_time}.
    
    In \Cref{subs:data_secondary} we explained the number of citations is derived from linking citations between publications in our dataset, so each incoming citation should have one matching outgoing citation, and thus, the total number of incoming and outgoing citations should be equal.
    We believe the total numbers do not match in this experiment, as the numbers were calculated and saved individually for each publication before the export of \diii, which possibly lost some publications (\Cref{sec:limitations}).
    This issue should be fixed in the new version of \diii which was not yet available when conducting the experiments.
    
    \experiment{How are the publications distributed based on the number of incoming citations? (\labelcref{rq:citations})}\label{ex:citations_distribution}
    \begin{table}[!htp]\centering
\footnotesize
\begin{tabular}{l|rrrrr|r}\toprule
\textbf{\#Citations} &\textbf{0} &\textbf{1-9} &\textbf{10-99} &\textbf{100-999} &\textbf{1000+} &\textbf{Total} \\\midrule
\#Publications &1,441,029 &1,909,897 &1,372,511 &164,601 &5,502 &4,893,540 \\
\%Publications &29.45\% &39.03\% &28.05\% &3.36\% &0.11\% &100.00\% \\
\bottomrule
\end{tabular}
\caption{Incoming citations sorted into citation bins.}\label{tab:citations_distribution}
\end{table}
    We choose the same binning sizes as \textcite{mohammad_nlp_2020_data} and as a result, sort the 128 publications with 10,000 citations or more into our largest bin (1000+) (\textbf{\Cref{tab:citations_distribution}}).
    
    Our citations are heavily skewed to the bins with fewer citations.
    29\% of all publications in \csi never receive any citations, 39\% receive 1-9, and only close to a third of all publications receive 10 or more citations.
    \textcite{mariani_nlp4nlp_2019} also see 44\% of publications are never cited in \gls{nlp} and \textcite{fiala_computer_2017} see 52\% are never cited in \gls{cs} using \wos data.
    Only \textcite{mohammad_examining_2020} has just 6\% of publications with 0 citations, but 48\% with 10-99 citations when investigating \gls{nlp} publications.
    This is quite unusual, as \textcite{fiala_computer_2017} state, that it is a well-known fact in scientometrics, that most papers remain without citations.
    We believe this large difference in the distribution of citations is due to different datasets (\dblp vs. \aclAnthology), and different ways of obtaining the citation counts (matching citations in the corpus itself vs. \googleScholar).

    \experiment{How do the distributions of incoming and outgoing citations change over time? (\labelcref{rq:citations})}\label{ex:citations_time}
    \begin{table}[!htp]\centering
\footnotesize
\begin{tabular}{l|rrrrr|rrrrr}\toprule
\multirow{2}{*}{\textbf{Time span}} &\multicolumn{5}{c|}{\textbf{Incoming Citations}} &\multicolumn{5}{c}{\textbf{Outgoing Citations}} \\
&Q1 &Med. &Q3 &Max. &Avg. &Q1 &Med. &Q3 &Max. &Avg. \\
\midrule
1960-1969 &0 &1 &10 &10,747 &28.91 &0 &0 &4 &509 &2.97 \\
1970-1979 &0 &1 &9 &18,702 &29.44 &0 &0 &6 &376 &3.91 \\
1980-1989 &0 &1 &12 &57,583 &31.07 &0 &0 &8 &885 &5.25 \\
1990-1999 &0 &2 &17 &45,635 &31.38 &0 &4 &13 &2,665 &8.22 \\
\midrule
2000-2004 &0 &5 &22 &38,124 &34.03 &0 &8 &16 &1,348 &10.89 \\
2005-2009 &1 &7 &23 &37,926 &28.26 &5 &12 &22 &1,313 &15.73 \\
2010-2014 &1 &6 &18 &39,111 &20.81 &8 &16 &27 &4,627 &19.77 \\
2015-2019 &1 &3 &10 &37,732 &11.92 &8 &18 &31 &1,292 &22.58 \\
\midrule
1960-2019 &0 &4 &16 &57,583 &22.08 &4 &13 &24 &4,627 &17.33 \\
\bottomrule
\end{tabular}
\caption{Distribution of the number of total incoming and outgoing citations per time period showing the first quartile, median, third quartile, maximum, and average.}\label{tab:citations_citations}
\end{table}
    We observe an increase in the median and third quartile of incoming citations until 2005-2009, after which they fall off again (\textbf{\Cref{tab:citations_citations}}), which matches the overall trends of incoming citations per year, as that period also has the most citations overall (\Cref{ex:citations_bar}).
    The average number of incoming citations slightly increases between 1960 and 2004, and peaks in 2000-2004, before dropping off, due to newer papers being less cited and the increase in papers since the 2000s (\Cref{ex:publications}), which drags the average down.
    This could also explain why the peak of the average is in 2000-2004 and not 2005-2009.
    \textcite{mariani_nlp4nlp_2019} see a similar distribution of incoming citations, except they observe a spike in the 1970s and a larger fall-off toward the 1960s, which we explain with only a small number of \gls{nlp} papers existing during that time.
    The maximum for both incoming and outgoing citations is lower in the first periods and fluctuates afterward.
    Outgoing citations seem to increase with every period, with the respect to the first quartile, median, third quartile, and average which means not only are there more outgoing citations in total (\Cref{ex:citations_bar}) but also on average, each publication cites more other publications.
    The same trend of increasing outgoing citations per paper is observed by \textcite{mariani_nlp4nlp_2019} in \gls{nlp}.

\section{Document Types}\label{sec:paper_type}
    This section covers experiments that answer \labelcref{rq:types} by investigating the document types and differences between conferences and journals in \csi.
    We start with an overview of the distribution of document types (\Cref{ex:types_distribution}), by aggregating \csi's data by document type.
    The experiments thereafter use no aggregation and instead leverage the \textit{Types of papers} filter to find differences between conferences and journals regarding their trends over time (\Crefrange{ex:types_bar}{ex:types_venues2}), and the most cited and productive authors, most cited venues, and topics (\Crefrange{ex:types_authors}{ex:types_topics_time}).
    We refer to the document types in \csi's \gls{ui} with ``types of paper'', and in this section refer to publications of conferences with ``papers'' and publications in journals with ``articles''.

    \experiment{How are the publications distributed across the document types? (\labelcref{rq:types})}\label{ex:types_distribution}
    \begin{table}[!htp]\centering
\footnotesize
\begin{tabular}{lrrrr}\toprule
\textbf{Document Type} &\textbf{First} &\textbf{\#Publications} &\textbf{\#Citations} &\textbf{Avg. Citations} \\
\midrule
inproceedings &1951 &2,574,226 (52.60\%) &35,420,550 (36.50\%) &13.76 \\
article &1936 &2,210,231 (45.17\%) &58,782,837 (60.57\%) &26.60 \\
incollection &1941 &64,936 (1.33\%) &910,658 (0.94\%) &14.02 \\
proceedings &1951 &28,408 (0.58\%) &334,822 (0.34\%) &11.79 \\
book &1949 &14,563 (0.30\%) &1,602,926 (1.65\%) &110.07 \\
phdthesis &1938 &1,171 (0.02\%) &1,495 (0.00\%) &1.28 \\
mastersthesis &1984 &5 (0.00\%) &0 (0.00\%) &0.00 \\
\bottomrule
\end{tabular}
\caption{Distribution of publications and citations across document types.}\label{tab:types_distribution}
\end{table}
    The document types are based on the BibTeX types of publications (\textbf{\Cref{tab:types_distribution}}).
    
    We see just over half of all publications \csi are conference papers (``inproceedings''), 45\% are journal articles, and together they make up around 98\% of our dataset.
    Thus, \csi is more evenly distributed concerning conference papers and journal articles compared to \textcite{fiala_computer_2017} with 56.1\% vs. 34.8\% (an additional 8.7\% are classified as ``Article; Paper''), and \textcite{coskun_scientometrics-based_2019} with 59.75\% vs. 38.18\% (second period only, the first one is even less evenly distributed), both in favor of conference papers.
    Even though there are fewer journal articles, they make up 60\% of the total citations and receive on average twice as many citations as conference papers, while books receive on average by far the most citations (110), but they also only make up 0.3\% of \csi.
    \textcite{fiala_computer_2017} also show a disbalance in citations, as journals contribute 75.6\% of all citations and conference papers only 10.7\% (an additional 10.9\% are for the type ``Article; Paper'').
    \textcite{franceschet_role_2010} also finds in general journal articles receive more citations, while there are more publications in conferences, which \textcite{mohammad_examining_2020} also finds for \gls{nlp}, as only 2.5\% of the publications in \nlpScholar are journals, but journals have the highest median and average citations of all document types.
    On the other hand, \textcite{vrettas_conferences_2015} find that disregarding the quality of the venue, there are few differences between conferences and journals regarding citations.
    We assume the differences are due to the different data sources (\dblp vs. \mas), which is the same reason \textcite{vrettas_conferences_2015} themselves state regarding contradicting results of previous research in this area.
    One might argue the reason for the lower citations of conference papers in \csi might be the way workshop papers are handled, as \textcite{mohammad_nlp_2020_data} shows they make up the bulk of the papers and are classified as conference papers in \csi.
    However, \textcite{mohammad_examining_2020} shows that workshop papers still have a higher citation average and median compared to non-top-tier conferences.
    This means workshop papers are not dragging down the citations for conferences in \gls{nlp} and rather low-quality conferences, which we assume is also the case for \csi.

    \experiment{How do the number of publications per year differ for conferences and journals? (\labelcref{rq:types})}\label{ex:types_bar}
    \begin{figure}[htbp]
        \centering
        \includegraphics[trim={0 1cm 0 0}, clip, width=\textwidth]{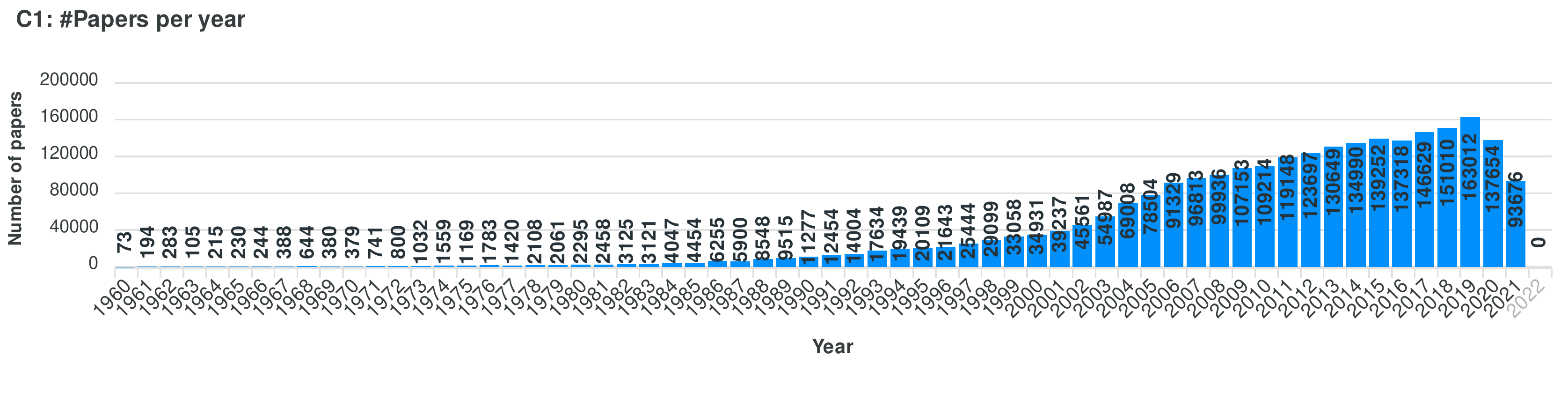}
        \includegraphics[trim={0 1cm 0 0}, clip, width=\textwidth]{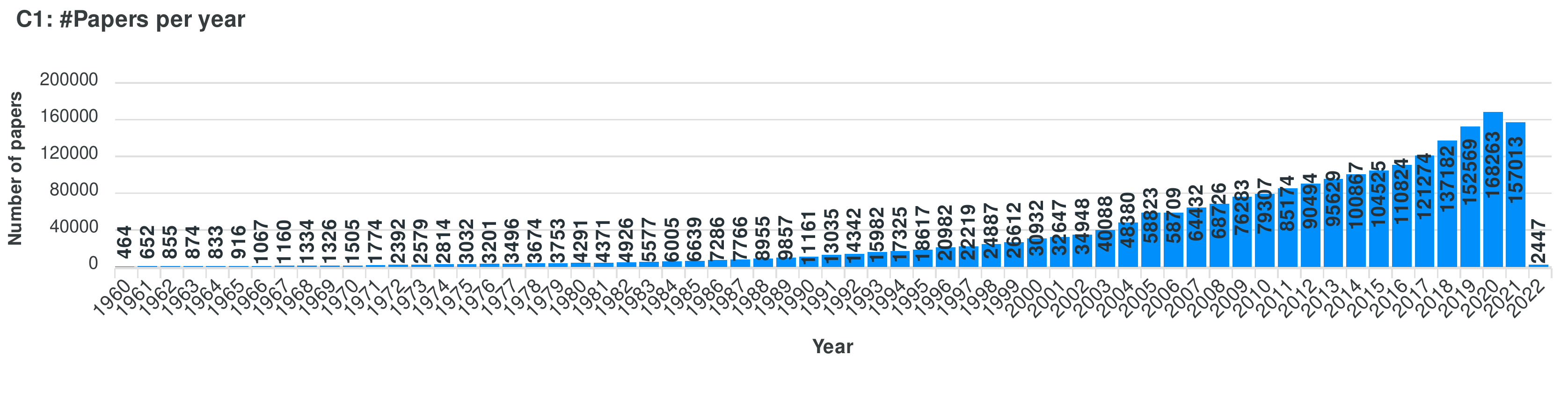}
        \caption{Number of publications per year in conferences (top) and journals (bottom). See \Cref{fig:types_inproceedings_bar_app,fig:types_articles_bar_app} for the full span.}
        \label{fig:types_bar}
    \end{figure}
    Journals appear in \csi since the start of our dataset in 1936, while the first conference was in 1951 (AIEE-IRE Computer Conference) (\textbf{\Cref{fig:types_bar}}).
    Both journals and conferences have published at least one document each year since then, but journals published more each year until 1993, which was the first year conferences published consistently more than journals (17,634 papers vs. 15,982 articles).
    This trend continues until 2020 when journals published 168,263 articles, but conferences only 137,654 papers.
    2020 was also the first year the number of conference papers per year went down compared to the previous year, which is likely linked to the COVID-19 pandemic and in-person events being canceled.
    Journals, on the other hand, see another increase of 15,000 publications compared to 2019, and their number of publications appears unaffected by the pandemic based on the graph.
    We can now link this insight to \Cref{ex:publications}, where we saw a drop in the overall number of publications in 2020, and know this drop is solely caused by conference papers and the increase in journal articles could not compensate.
    
    In general, conferences saw the first spurt in publications in the late 1980s, a second one in the early-mid 2000s, and another one around 2017.
    The publications per year in journals grow at a more steady rate, but a small spurt during the early-mid 2000s and a larger one around 2017 are also visible.
    Similar trends and bursts are visible for authors (\Cref{ex:authors_bar}) and publications (\Cref{ex:publications}), which show there are general trends visible for both conferences and journals (except for the effect of COVID-19), which makes sense, as more authors mean more publications overall and thus also more publications for conferences and journals.
    In a later experiment, we will investigate how the number of unique conferences and journals affects the number of publications (\Cref{ex:types_venues2}).
    \textcite{fiala_computer_2017} also investigate the number of articles and papers per year in \wos and find similar spikes in the early 2000s, but their data shows a drop for both conferences and journals soon after.
    For journals, this is related to the ``Lecture Notes in Computer Science'' and ``Lecture Notes in Artificial Intelligence'' being classified differently starting in 2007 and conferences being indexed less from 2008 onward.
    Conferences are also indexed less before the 1990s in \wos, as there are nearly zero indexed publications before 1985, while our data shows a more natural increase in conference publications over the decades.
    \textcite{coskun_scientometrics-based_2019} see a similar shift of researchers publishing more in journals, as the disbalance of document types (conferences vs. journals) evened out when comparing 2008-2013 (73\% vs. 26\%) to 2014-2019 (60\% vs. 38\%).

    \textcite{vardi_conferences_2009} attributes the preference of researchers in \gls{cs} for conferences to a best practices memo from the Computing Research Association in 1999, but we show that conference papers overtook journal articles already in 1993 and that the increase in conference papers started in the late 1980s.
    Similarly, the increase in conference publications in the early 2000s should also not be linked to the memo, as journal papers also saw an increase during the early 2000s.
    We believe it is possible the memo was merely coincidental or a result of the already existing trend.

    \experiment{How do the distributions of citations differ for conferences and journals? How do they change over time? (\labelcref{rq:types})}\label{ex:types_citations}
    \begin{table}[!htp]\centering
\footnotesize
\begin{tabular}{l|rrrrr|rrrrr}\toprule
\multirow{2}{*}{\textbf{Time span}} &\multicolumn{5}{c|}{\textbf{Conferences \#Citations}} &\multicolumn{5}{c}{\textbf{Journals \#Citations}} \\
&Q1 &Med. &Q3 &Max. &Avg. &Q1 &Med. &Q3 &Max. &Avg. \\
\midrule
1960-1969 &0 &2 &8 &4,207 &16.80 &0 &1 &11 &10,474 &31.75 \\
1970-1979 &0 &1 &5 &6,582 &14.57 &0 &1 &13 &18,702 &34.89 \\
1980-1989 &0 &1 &10 &13,895 &20.39 &0 &0 &15 &27,195 &35.26 \\
1990-1999 &0 &3 &15 &26,281 &23.71 &0 &1 &21 &45,635 &38.16 \\
\midrule
2000-2004 &0 &5 &18 &25,843 &25.61 &0 &4 &30 &30,893 &44.10 \\
2005-2009 &1 &5 &15 &28,322 &18.09 &2 &13 &40 &30,815 &43.40 \\
2010-2014 &1 &4 &11 &16,712 &13.28 &3 &11 &31 &39,111 &31.28 \\
2015-2019 &0 &2 &6 &28,076 &8.14 &1 &6 &17 &37,732 &16.51 \\
\midrule
1960-2019 &0 &3 &11 &28,322 &14.99 &0 &7 &25 &45,635 &30.62 \\
\bottomrule
\end{tabular}
\caption{Distribution of the total number of citations for conferences and journals per time period showing the first quartile, median, third quartile, maximum, and average.}\label{tab:types_citations}
\end{table}
    Generally, both conferences and journals follow similar trends, with the difference of journals having more citations (\textbf{\Cref{tab:types_citations}}), which shows the overall trend of journals getting more citations (\Cref{ex:types_distribution}) is not new and already existed for a few decades.
    The average peaks in 2000-2004 and the median in 2005-2009 for both conferences and journals, and both fall off before and after, which we also saw for the general trend of citations in \Cref{ex:citations_time}.
    Again, the same reasons for the trends of the overall citations from \Cref{ex:citations_time} apply here, so we do not explain the trends again.
    
    \experiment{How does the number of venues influence the number of publications and citations? (\labelcref{rq:types})}\label{ex:types_venues2}
    \begin{table}[!htb]\centering
\footnotesize
\begin{tabular}{l|rrrrr}\toprule
\multirow{2}{*}{\textbf{Time span}} &\multicolumn{5}{c}{\textbf{Conferences \#Venues}} \\%\cmidrule{2-6}
&\#Venues &\#Papers &\#Citations &Avg. Papers &Avg. Citations \\
\midrule
1960-1969 &18 &2,756 &46,289 &153.11 &2,571.61 \\
1970-1979 &151 &13,052 &190,218 &86.44 &1,259.72 \\
1980-1989 &503 &49,718 &1,013,942 &98.84 &2,015.79 \\
1990-1999 &1,571 &204,161 &4,840,908 &129.96 &3,081.42 \\
\midrule
2000-2004 &1,850 &243,724 &6,242,358 &131.74 &3,374.25 \\
2005-2009 &2,826 &473,735 &8,571,443 &167.63 &3,033.07 \\
2010-2014 &3,768 &617,698 &8,204,522 &163.93 &2,177.42 \\
2015-2019 &4,204 &737,221 &6,001,438 &175.36 &1,427.55 \\
\midrule
1960-2019 &8,637 &2,342,065 &35,111,118 &271.17 &4,065.20 \\
\midrule

\multirow{2}{*}{\textbf{Time span}} &\multicolumn{5}{c}{\textbf{Journals \#Venues}} \\
&\#Venues &\#Articles &\#Citations &Avg. Articles &Avg. Citations \\
\midrule
1960-1969 &45 &9,481 &301,017 &210.69 &6,689.27 \\
1970-1979 &121 &28,220 &984,559 &233.22 &8,136.85 \\
1980-1989 &263 &65,673 &2,315,765 &249.71 &8,805.19 \\
1990-1999 &607 &185,162 &7,066,360 &305.04 &11,641.45 \\
\midrule
2000-2004 &807 &186,995 &8,245,893 &231.72 &10,217.96 \\
2005-2009 &1,178 &326,973 &14,190,957 &277.57 &12,046.65 \\
2010-2014 &1,428 &451,471 &14,122,256 &316.16 &9,889.54 \\
2015-2019 &1,514 &626,374 &10,342,030 &413.72 &6,830.93 \\
\midrule
1960-2019 &1,775 &1,880,349 &57,568,837 &1,059.35 &32,433.15 \\
\bottomrule
\end{tabular}
\caption{Number of publications and citations in relation to the number of venues for conferences (top) and journals (bottom).}\label{tab:types_venues2}
\end{table}
    Our goal is to investigate if the rise in publications is caused by venues publishing more or new venues emerging, which \textcite{subelj_publication_2017} also investigate for journals in \gls{cs} and physics, but with a different approach.

    We observe, there has been a constant and noticeable increase in both conferences and journals (\textbf{\Cref{tab:types_venues2}}).
    There were less than 100 venues total in 1960-1969 and over 4000 conferences and 1500 journals in 2015-2019.
    At the same time, the average number of publications per venue per period has roughly doubled (ignoring the outlier for conferences in 1960-1969), even though the length of the period has halved.
    We consider redoing this experiment in the future with a constant length of periods (\Cref{sec:future_work}) to be able to compare the numbers from earlier periods to the later ones with different lengths more easily.
    
    Each journal receives on average more citations in the same period compared to conferences, and again, journals also receive more citations in total than conferences.
    Yet, journals publish less in total (during 1990-2019, as discussed in \Cref{ex:types_bar}), but due to there being far fewer journals, they publish more than conferences on average.
    We think this is due to the nature of journals, which publish multiple issues yearly, while most conferences are only held once a year or less.
    Additionally, 85\% of all journals published at least one issue in 2015-2019, while only half of all conferences were still held in 2015-2019, which we attribute to journals staying the same, while conferences are more fast-paced.

    \textcite{subelj_publication_2017} attribute the rise in \gls{cs} journal articles to new journals, and not journals publishing more, which might seem contradictory to our findings, but they argue based on a comparison to journals in physics.
    They find the number of \gls{cs} journals increases significantly (from $\approx$50 in 1975 to almost 450 in 2010), but in physics, the number of journals barely increases (from 100 to 150 in 35 years).
    The increase in the number of journals is certainly partly responsible for the rise in publications, but we would add to \textcite{subelj_publication_2017}'s findings, that journals still publish twice as much on average, or maybe even four times as much when we consider the difference in length of periods.
    We look into the research fields in \Cref{sec:fields}, but we cannot compare the number of venues over time for different research fields, as our dataset has a focus on \gls{cs} and is missing dedicated venues in physics to make this comparison properly, but we might look into this in the future (\Cref{sec:future_work}).

    \experiment{How are the citations and publications of the most prominent authors split between conferences and journals? (\labelcref{rq:types})}\label{ex:types_authors}
    \begin{table}[!htb]\centering
\footnotesize
\begin{tabular}{rlrrrr}\toprule
\textbf{\#} &\textbf{Author (\#Citations)} &\textbf{\#Pub. (C)} &\textbf{\#Pub. (J)} &\textbf{\#Cit. (C)} &\textbf{\#Cit. (J)} \\
\midrule
1 &\textit{Others} &3,263 (6\%) &10,019 (18\%) &3,370 (0\%) &28,177 (4\%) \\
2 &Ross B. Girshick &56 (81\%) &13 (19\%) &93,437 (64\%) &53,430 (36\%) \\
3 &Anil K. Jain 0001 &369 (56\%) &278 (42\%) &32,696 (26\%) &77,851 (63\%) \\
4 &Kaiming He &53 (80\%) &13 (20\%) &61,151 (53\%) &53,179 (47\%) \\
5 &Jitendra Malik &181 (78\%) &48 (21\%) &75,449 (69\%) &34,238 (31\%) \\
6 &Andrew Zisserman &350 (77\%) &98 (22\%) &73,043 (70\%) &29,794 (28\%) \\
7 &Li Fei-Fei 0001 &161 (83\%) &25 (13\%) &69,435 (68\%) &33,182 (32\%) \\
8 &Luc Van Gool &622 (79\%) &151 (19\%) &56,774 (59\%) &39,570 (41\%) \\
9 &Jiawei Han 0001 &615 (70\%) &188 (22\%) &59,869 (63\%) &19,585 (21\%) \\
10 &Trevor Darrell &243 (84\%) &43 (15\%) &66,277 (72\%) &24,333 (27\%) \\
\midrule
&Average &294 (73\%) &95 (24\%) &65,348 (60\%) &40,574 (37\%) \\
\midrule
\textbf{\#} &\textbf{Author (\#Publications)} &\textbf{\#Pub. (C)} &\textbf{\#Pub. (J)} &\textbf{\#Cit. (C)} &\textbf{\#Cit. (J)} \\
\midrule
1 &\textit{Others} &3,263 (6\%) &10,019 (18\%) &3,370 (0\%) &28,177 (4\%) \\
2 &H. Vincent Poor &698 (42\%) &949 (58\%) &9,390 (13\%) &61,109 (82\%) \\
3 &Mohamed-Slim Alouini &643 (44\%) &799 (55\%) &7,206 (18\%) &27,625 (70\%) \\
4 &Lajos Hanzo &429 (31\%) &949 (69\%) &3,382 (9\%) &31,376 (87\%) \\
5 &*Wei Wang &710 (53\%) &624 (47\%) &3,575 (16\%) &19,230 (84\%) \\
6 &Philip S. Yu &850 (66\%) &406 (32\%) &45,676 (62\%) &26,798 (36\%) \\
7 &*Lei Zhang &749 (59\%) &518 (41\%) &6,222 (51\%) &6,077 (49\%) \\
8 &*Yu Zhang &687 (54\%) &574 (46\%) &4,722 (41\%) &6,658 (59\%) \\
9 &Victor C. M. Leung &571 (45\%) &685 (54\%) &6,111 (20\%) &24,548 (80\%) \\
10 &*Yang Liu &717 (57\%) &528 (42\%) &6,222 (71\%) &5,596 (64\%) \\
\midrule
&Average &673 (50\%) &670 (50\%) &10,278 (30\%) &23,224 (68\%) \\
\bottomrule
\end{tabular}
\caption{Top 10 authors based on the number of citations (top) and publications (bottom) with their publications and citations split by conferences (C) and journals (J). The average is computed excluding \textit{Others}. Asterisks (*) denote entries, which refer to disambiguation pages in \dblp and not singular authors.}\label{tab:types_authors}
\end{table}
    We split the number of publications and citations of the most prominent authors by conferences and journals (\textbf{\Cref{tab:types_authors}}), which is inspired by the approach of \textcite{franceschet_role_2010}.
    Some numbers do not add up to 100\% due to publications with different document types (e.g., books), which are not covered in this table.
    
    When we investigated the topics of the most prominent authors through their venues (\Cref{ex:authors_topics}), we got a hint that the top 5 most cited authors prefer conferences, while the most productive authors prefer journals.
    This trend is visible again, as all of the most cited authors publish more in conference proceedings.
    At the same time, eight out of nine authors also receive more citations in conferences, the exception being Anil K. Jain 0001.
    The publications of the most productive authors are evenly split between conferences and journals, but more than two-thirds of their citations come from journals.
    \textcite{franceschet_role_2010} finds the top 10 most productive authors publish more in conferences (63\% to 34\%) while using the same approach and data source (i.e., \dblp), but over 12 years this appears to have evened out.
    He also shows that the most prestigious authors publish more in conferences, both based on the h-index (59\% to 40\%), and for the winners of the ACM A.M. Turing Award (65\% to 33\%).
    This matches our finding that the most popular authors based on citations also publish more in conference proceedings.
    
    \experiment{Do the most cited conferences and journals show different trends considering the average citations compared to the average mass? (\labelcref{rq:types})}\label{ex:types_venues}
    \begin{table}[!htb]\centering
\footnotesize
\begin{tabular}{rlrrrrr}\toprule
\textbf{\#} &\textbf{Conference Name} &\textbf{First} &\textbf{\#Papers} &\textbf{\#Citations} &\textbf{Avg. Cit.} \\
\midrule
1 &CVPR &1988 &12,757 &1,621,492 &127.11 \\
2 &ICRA &1984 &24,997 &790,107 &31.61 \\
3 &ICASSP &1975 &45,655 &714,945 &15.66 \\
4 &ICCV &1988 &5,198 &601,139 &115.65 \\
5 &CHI &1982 &8,864 &551,752 &62.25 \\
6 &INFOCOM &1983 &8,785 &503,263 &57.29 \\
7 &ECCV &1990 &3,857 &490,664 &127.21 \\
8 &IROS &1988 &19,560 &422,428 &21.60 \\
9 &KDD &1999 &4,367 &386,964 &88.61 \\
10 &STOC &1969 &3,662 &353,241 &96.46 \\
\midrule
&Average &1985 &13,770 &643,600 &74.34 \\
\midrule
\textbf{\#} &\textbf{Journal Name} &\textbf{First} &\textbf{\#Articles} &\textbf{\#Citations} &\textbf{Avg. Cit.} \\
\midrule
1 &NeuroImage &1996 &16,947 &1,377,202 &81.27 \\
2 &IEEE Trans. Pattern Anal. Mach. Intell. &1975 &6,559 &1,337,060 &203.85 \\
3 &IEEE Trans. Inf. Theory &1963 &16,325 &1,147,862 &70.31 \\
4 &Commun. ACM &1958 &12,742 &948,274 &74.42 \\
5 &IEEE Trans. Ind. Electron. &1990 &12,777 &802,571 &62.81 \\
6 &IEEE Trans. Signal Process. &1990 &13,328 &762,802 &57.23 \\
7 &IEEE Trans. Autom. Control. &1991 &10,762 &717,910 &66.71 \\
8 &IEEE Trans. Image Process. &1991 &8,627 &702,460 &81.43 \\
9 &IEEE Trans. Commun. &1972 &15,395 &660,511 &42.90 \\
10 &IEEE Trans. Geosci. Remote. Sens. &1987 &11,297 &619,250 &54.82 \\
\midrule
&Average &1981 &12,476 &907,590 &79.58 \\
\bottomrule
\end{tabular}
\caption{Top 10 most cited conferences (top) and journals (bottom).}\label{tab:types_venues}
\end{table}
    We try to approximate the most prestigious venues by taking the most cited ones and then looking at the average citations per publication (\textbf{\Cref{tab:types_venues}}).
    Directly taking the venues with the highest average citations would yield venues with mostly less than 100 publications.
    
    The top 10 most cited journals (most of which are from IEEE) have in total more citations than the top 10 most cited conferences, which we already expected, as the list of the top 10 most cited venues leans heavily toward journals (\Cref{ex:venues_top}).
    Yet, the CVPR conference is the most cited venue overall and has the third-highest average citations, with IEEE Trans. Pattern Anal. Mach. Intell. having the highest average citations (203.85).
    Generally, conferences take most of the spots going by highest average citations (\#2-\#6; ECCV, CVPR, ICCV, STOC, KDD), but they also take the last spots (\#18-\#20; ICRA, IROS, ICASSP), which shows a greater fluctuation in average citations compared to journals.
    Considering the top ranks are occupied by IEEE Trans. Pattern Anal. Mach. Intell., ECCV, CVPR, and ICCV we see a heavy focus on computer vision and pattern recognition again, similar to the most cited authors (\Cref{ex:authors_topics}) and most cited venues (\Cref{ex:venues_top}).
    We conclude that on average the top journals are still a bit more prestigious regarding citations compared to conferences, but the gap compared to the average mass, where journals have twice as many citations (\Cref{ex:types_citations}), is a lot smaller.
    A few highly cited conferences are also able to rank higher based on average citations compared to highly-cited journals.
    
    Other works show elite conferences are getting more citations than elite journals, based on different measures for quality.
    \textcite{rahm_citation_2005} use select high-quality venues from their research field (databases), compare their citations, and find that conferences have a higher citation impact than journals.
    \textcite{vrettas_conferences_2015} also compare conference and journal citations and find little difference overall, but when the quality of the venues is considered (using a ranking from the \gls{era} assessment), high-quality conferences have higher average citation rates than high-quality journals, and low-quality journals get more citations than low-quality conferences.
    In \gls{nlp} \textcite{mohammad_examining_2020}, on the other hand, finds that journal publications have higher average and median citations than both top-tier and non-top-tier conferences.
    We also find a higher fluctuation of average citations in the top 10 most cited conferences, which reflects the fluctuation of conference quality \textcite{vrettas_conferences_2015} find.
    Higher citation rates for elite conferences are only visible to a limited degree in our research, which we attribute to us only using an approximation of the highest-quality conferences and not a more refined solution to measure the quality of venues.

    Lastly, the average number of publications is a bit higher for conferences but also fluctuates more compared to journals.
    Thus, this experiment also reflects the findings from the most cited and most productive venues (\Cref{ex:venues_top}), where the most productive venues are mostly conferences and the most cited venues are mostly journals.

    \experiment{How do the topics of conferences and journals change over time? How do the topics differ? (\labelcref{rq:types})}\label{ex:types_topics_time}
    \newcommand{\greyed}[1]{\color{darkgray} #1}
\afterpage{
\begin{landscape}
\begin{table}[p]
\begin{center}
\scriptsize
\begin{tabular}{lllll|lllll}
\toprule
\multicolumn{5}{c|}{\textbf{Conferences}} & \multicolumn{5}{c}{\textbf{Journals}} \\
1999 & 2004 & 2009 & 2014 & 2019 & 1999 & 2004 & 2009 & 2014 & 2019 \\
\midrule
\greyed{\textit{paper}} & \greyed{\textit{propos}} & \greyed{\textit{paper}} & \greyed{\textit{propos}} & \greyed{\textit{propos}} & \greyed{\textit{propos}} & \greyed{\textit{paper}} & \greyed{\textit{paper}} & \greyed{\textit{propos}} & \greyed{\textit{propos}} \\
\greyed{\textit{propos}} & \greyed{\textit{paper}} & \greyed{\textit{present}} & \greyed{\textit{network}} & \greyed{\textit{algorithm}} & \greyed{\textit{present}} & \greyed{\textit{problem}} & \greyed{\textit{propos}} & \greyed{\textit{paper}} & \greyed{\textit{paper}} \\
data & \greyed{\textit{present}} & \greyed{\textit{propos}} & \greyed{\textit{paper}} & \greyed{\textit{paper}} & \greyed{\textit{control}} & \greyed{\textit{propos}} & \greyed{\textit{algorithm}} & \greyed{\textit{model}} & \greyed{\textit{network}} \\
\greyed{\textit{network}} & \greyed{\textit{network}} & \greyed{\textit{network}} & \greyed{\textit{present}} & result & \greyed{\textit{paper}} & \greyed{\textit{provid}} & \greyed{\textit{network}} & \greyed{\textit{algorithm}} & \textit{result} \\
\greyed{\textit{imag}} & \greyed{\textit{algorithm}} & \textit{time} & \greyed{\textit{problem}} & \textit{time} & \greyed{\textit{algorithm}} & \greyed{\textit{network}} & \greyed{\textit{present}} & \greyed{\textit{network}} & learn \\
\greyed{\textit{problem}} & \greyed{\textit{provid}} & result & \textit{time} & \greyed{\textit{problem}} & \greyed{\textit{problem}} & \greyed{\textit{algorithm}} & \greyed{\textit{control}} & \greyed{\textit{present}} & time \\
\greyed{\textit{provid}} & \textit{time} & \greyed{\textit{method}} & data & \textbf{learn} & \greyed{\textit{provid}} & \textit{result} & \textit{result} & \greyed{\textit{provid}} & \greyed{\textit{provid}} \\
\greyed{\textit{present}} & \greyed{\textit{imag}} & \greyed{\textit{model}} & \greyed{\textit{provid}} & \greyed{\textit{network}} & \textit{result} & \greyed{\textit{present}} & \greyed{\textit{problem}} & \greyed{\textit{problem}} & \greyed{\textit{present}} \\
object & result & \greyed{\textit{imag}} & \greyed{\textit{model}} & \greyed{\textit{imag}} & \greyed{\textit{network}} & \textit{data} & \textit{data} & \greyed{\textit{method}} & \greyed{\textit{imag}} \\
user & \greyed{\textit{problem}} & \greyed{\textit{provid}} & \greyed{\textit{imag}} & \greyed{\textit{present}} & us & time & \greyed{\textit{model}} & \greyed{\textit{imag}} & \greyed{\textit{control}} \\
\greyed{\textit{algorithm}} & \greyed{\textit{method}} & process & user & differ & \greyed{\textit{model}} & \greyed{\textit{imag}} & order & number & \greyed{\textit{model}} \\
set & work & user & featur & \textbf{experi} & number & \greyed{\textit{control}} & \greyed{\textit{method}} & develop & \textit{data} \\
\greyed{\textit{method}} & \textbf{mobil} & \greyed{\textit{algorithm}} & \greyed{\textit{method}} & featur & perform & set & \greyed{\textit{provid}} & time & featur \\
\textit{develop} & requir & \greyed{\textit{problem}} & \textbf{implement} & \greyed{\textit{control}} & inform & inform & time & \greyed{\textit{control}} & \greyed{\textit{problem}} \\
\greyed{\textit{control}} & perform & \greyed{\textit{control}} & \greyed{\textit{control}} & data & includ & \greyed{\textit{method}} & develop & optim & \greyed{\textit{method}} \\
test & real & perform & differ & \textbf{state} & \textbf{bound} & perform & function & \textit{result} & studi \\
\textit{time} & user & set & process & \greyed{\textit{provid}} & obtain & \greyed{\textit{model}} & inform & featur & \greyed{\textit{algorithm}} \\
allow & \greyed{\textit{control}} & work & result & \greyed{\textit{model}} & set & \textbf{consid} & research & \textit{data} & optim \\
design & \textit{develop} & featur & power & real & \greyed{\textit{imag}} & \textbf{system} & \textbf{work} & \textbf{design} & test \\
\greyed{\textit{model}} & allow & requir & optim & detect & \textit{data} & solut & \greyed{\textit{imag}} & requir & develop \\
\textbf{inform} & \textbf{program} & optim & \greyed{\textit{algorithm}} & test & \textbf{given} & \textbf{compar} & number & obtain & \textbf{accuraci} \\
\textbf{robot} & \textbf{scheme} & \textit{develop} & servic & \textbf{comput} & \textbf{exampl} & function & solut & order & object \\
requir & servic & test & detect & optim & graph & test & obtain & demonstr & function \\
\textbf{environ} & node & power & \textit{develop} & compar & code & includ & graph & perform & us \\
compar & \greyed{\textit{model}} & servic & \textbf{demonstr} & \greyed{\textit{method}} & \greyed{\textit{method}} & number & servic & set & \textbf{task} \\
\textbf{obtain} & introduc & node & studi & \textit{develop} & \textbf{describ} & \textbf{signal} & us & studi & \textbf{estim} \\
introduc & \textbf{reduc} & \textbf{achiev} & \textbf{signal} & \textbf{consid} & object & \textbf{exist} & code & servic & demonstr \\
\textbf{distribut} & \textbf{protocol} & design & \textbf{increas} & \textbf{train} & requir & \textbf{structur} & featur & research & \textbf{increas} \\
servic & object & differ & \textbf{research} & power & \textbf{differ} & \textbf{paramet} & learn & function & obtain \\
\textbf{import} & \textbf{estim} & \textbf{interact} & \textbf{larg} & studi & \textbf{technolog} & develop & includ & includ & \textbf{condit} \\
\bottomrule
\end{tabular}
\end{center}
\caption{Top 30 most salient terms for conferences and journals per year. Terms that only appear in one year are \textbf{bold} and terms that appear in all five years are in \textit{italics}. Terms that appear in all 10 years are additionally greyed out.}\label{tab:types_topics}
\end{table}   
\end{landscape}
}
    Compared to previous experiments (\Cref{ex:authors_topics_time,ex:venues_topics_time}), we only use single years (\textbf{\Cref{tab:types_topics}}), as full five-year periods are currently not possible due to technical limitations (e.g., memory size).
    
    The words that are common across all columns (e.g., ``paper'') are high up again, and there are more than in previous experiments, which we expected after the results from grouping the most cited and most productive venues (\Cref{ex:venues_topics_time}).
    There are now even more titles and abstracts used to generate these lists and saliency already failed to rank these terms lower when investigating the trends in venues over time (\Cref{ex:venues_topics_time}).
    Many of the terms that appear in all five years and occur in both conferences and journals either tell us nothing about the topics (i.e., ``paper'', ``propos'', ``problem'', ``provid'', ``method'', and ``present'') or only very little (i.e., ``network'', ``imag'', ``algorithm'', ``model'', and ``control'').
    The same is true for the terms ``develop'' and ``time'', which are present in all five years for conferences, and ``data'' and ``result'' for journals.
    We can see ``imag'' which relates to computer vision again and ``network'' and ``model'' might relate to neural networks, but in this case, we have to consider there are more engineering-related venues in our dataset, as we saw in \Cref{ex:authors_topics,ex:venues_top}, so neural networks might only make up a part of it and physical networks the other.
    However, in 2019 ``learn'' and ``train'' appear for the first time for conferences together with the repeating terms ``featur'' and ``optim'', which does show a rise in approaches leveraging neural networks and also matches the late appearance for the top venues (\Cref{ex:venues_topics_time}) and most cited authors (\Cref{ex:authors_topics_time}), where we also related this finding to previous research.
    The term ``learn'' also appears in 2019 for journals together with ``feature'', ``optim'', and ``object'' which can also indicate usage of neural networks, but none of those terms appear for the first time in 2019.
    Also, the term ``mobil'' only appears in 2004 for conferences and ``signal'' for journals in 2004, which could indicate more research related to wireless communication, e.g. mobile phones and their technologies.
    
    Overall, there are only a few differences between journals and conferences visible in \Cref{tab:types_topics}, which shows the topics are mostly the same.
    We also do not see many trends over time.
    Considering \textcite{fiala_computer_2017} can leverage the approach of highlighting terms to find trends, we attribute our issues to the high number of common terms and using saliency instead of frequency, similar to the issues already mentioned when investigating the trends of the top venues over time (\Cref{ex:venues_topics_time}).
    In our experiments, we use only 10 topics due to technical limitations (e.g., available memory size), but \textcite{anderson_towards_2012} use 73 topics just to cluster publication in \gls{nlp}, and in \textcite{xia_research_2021}'s research \gls{cs} covers 15,460 topics, which makes up 16\% of their total topics across all research fields.
    As saliency is calculated from the distinctiveness of terms across all topics we believe the saliency measure would work better with more topics, but this would exceed our current technical capabilities.
    In the future, we might conduct these term-related experiments again and then include the necessary changes (i.e., filtering common words beforehand and using a different measure) (\Cref{sec:future_work}).
    Then we might also be able to see some of the trends \textcite{coskun_scientometrics-based_2019} see (e.g., a shift to topics such as privacy, security, IoT, and big data), or the important current topics \textcite{tattershall_detecting_2020,xia_research_2021} uncover.
    
\section{Fields of Study}\label{sec:fields}
    This section answers the last research question (\labelcref{rq:fields}), which focuses on the fields of study (e.g. \gls{cs}, medicine).
    We conduct the experiments by aggregating \csi's data by the fields of study and then investigating their distribution (\Cref{ex:fields_distribution}) and the differences between \gls{cs} and other fields of study regarding preferences for conferences/journals, and topics (\Crefrange{ex:fields_types}{ex:fields_topics}).
    \csi and \diii are built with data from \dblp, which focuses on \gls{cs} publications, but \diii also includes data on the fields of study, which makes this analysis possible.
    As \Cref{ex:fields_distribution} shows, most publications in \csi are from \gls{cs}.
    Even though other fields are also included (each publication can have multiple fields of study), we have to be aware that the publications likely still have strong ties to \gls{cs} or they would not be included in \dblp.

    \experiment{How are the publications distributed across the fields of study? (\labelcref{rq:fields})}\label{ex:fields_distribution}
    \begin{table}[!htp]\centering
\footnotesize
\begin{tabular}{lrrrr}\toprule
\textbf{Field of Study} &\textbf{First} &\textbf{\#Papers} &\textbf{\#Citations} &\textbf{Avg. Citations} \\\midrule
Computer Science &1936 &4,192,059 (86\%) &97,037,289 (100\%) &23.15 \\
\textit{Others} &1936 &698,734 (14\%) &781 (0\%) &0.00 \\
Mathematics &1936 &696,143 (14\%) &21,778,856 (22\%) &31.29 \\
Engineering &1936 &324,195 (7\%) &7,554,187 (8\%) &23.30 \\
Medicine &1936 &267,808 (5\%) &10,179,977 (10\%) &38.01 \\
Psychology &1948 &80,578 (2\%) &3,200,102 (3\%) &39.71 \\
Physics &1946 &67,195 (1\%) &1,400,731 (1\%) &20.85 \\
Business &1953 &57,282 (1\%) &1,497,035 (2\%) &26.13 \\
Materials Science &1951 &50,049 (1\%) &619,066 (1\%) &12.37 \\
Biology &1961 &30,751 (1\%) &985,557 (1\%) &32.05 \\
Economics &1953 &30,084 (1\%) &1,063,618 (1\%) &35.35 \\
Sociology &1955 &27,719 (1\%) &762,886 (1\%) &27.52 \\
Environmental Science &1955 &26,356 (1\%) &410,730 (0\%) &15.58 \\
Chemistry &1952 &17,180 (0\%) &547,508 (1\%) &31.87 \\
Geology &1957 &16,928 (0\%) &304,352 (0\%) &17.98 \\
Geography &1953 &16,010 (0\%) &407,273 (0\%) &25.44 \\
Political Science &1941 &15,987 (0\%) &245,115 (0\%) &15.33 \\
Philosophy &1936 &5,718 (0\%) &61,511 (0\%) &10.76 \\
Art &1956 &4,334 (0\%) &18,521 (0\%) &4.27 \\
History &1939 &2,760 (0\%) &49,628 (0\%) &17.98 \\
\bottomrule
\end{tabular}
\caption{Distribution of publications across fields of study. One publication can have multiple fields of study, so the numbers exceed 100\% when added.}\label{tab:fields_distribution}
\end{table}
    As expected, most publications are from \gls{cs} (86\%), while nearly 700,000 publications (14\%) do not have any field of study assigned (\textbf{\Cref{tab:fields_distribution}}).
    This leaves only 2,747 publications, that have a field of study assigned but are not from \gls{cs}, so when we analyze other fields of study besides \gls{cs} in the next experiments, we should keep in mind those publications likely all additionally have \gls{cs} as a field of study.
    Mathematics, Engineering, and Medicine then take up rank \#3-\#5, and Psychology has the highest average citations per publication (39.71).
    Interestingly, nearly all citations are also for publications from \gls{cs}, and publications without fields of study only have a total of 781 citations.
    We assume the publications without fields of study cause general issues in the matching processes for both the fields of study and the citations.
    Possible reasons could be no or only a bad quality full-text, but these issues should be fixed with the new version of the dataset (\Cref{sec:limitations}).
    
    \experiment{How is the split between conference papers and journal articles for different fields of study? (\labelcref{rq:fields})}\label{ex:fields_types}
    \begin{table}[!htp]\centering
\scriptsize
\begin{tabular}{lrrrrr}\toprule
\textbf{Field of Study} &\textbf{\#Papers (C)} &\textbf{\#Articles (J)} &\textbf{\#Citations (C)} &\textbf{\#Citations (J)} \\\midrule
Computer Science &2,318,335 (55\%) &1,786,215 (43\%) &35,420,219 (37\%) &58,767,760 (61\%) \\
\textit{Others} &255,474 (37\%) &421,687 (60\%) &118 (15\%) &73 (9\%) \\
Mathematics &246,873 (35\%) &444,253 (64\%) &4,942,536 (23\%) &16,251,144 (75\%) \\
Engineering &197,828 (61\%) &121,161 (37\%) &2,645,846 (35\%) &4,725,451 (63\%) \\
Medicine &51,424 (19\%) &215,234 (80\%) &492,404 (5\%) &9,637,120 (95\%) \\
Psychology &31,924 (40\%) &47,179 (59\%) &379,265 (12\%) &2,725,165 (85\%) \\
Physics &26,168 (39\%) &42,704 (64\%) &236,146 (17\%) &1,110,054 (79\%) \\
Business &9,725 (17\%) &29,307 (51\%) &142,053 (9\%) &1,236,968 (83\%) \\
Materials Science &22,922 (46\%) &26,818 (54\%) &121,386 (20\%) &493,004 (80\%) \\
Biology &9,163 (30\%) &21,202 (69\%) &77,018 (8\%) &887,691 (90\%) \\
Economics &6,140 (20\%) &23,595 (78\%) &76,043 (7\%) &972,169 (91\%) \\
Sociology &9,725 (35\%) &17,179 (62\%) &142,053 (19\%) &586,270 (77\%) \\
Environmental Science &12,242 (46\%) &13,904 (53\%) &48,661 (12\%) &360,013 (88\%) \\
Chemistry &2,911 (17\%) &14,077 (82\%) &19,666 (4\%) &525,262 (96\%) \\
Geology &5,665 (33\%) &11,113 (66\%) &30,379 (10\%) &272,702 (90\%) \\
Geography &6,783 (42\%) &8,272 (52\%) &85,174 (21\%) &299,778 (74\%) \\
Political Science &6,140 (38\%) &8,762 (55\%) &52,171 (21\%) &141,718 (58\%) \\
Philosophy &1,233 (22\%) &4,054 (71\%) &3,230 (5\%) &47,468 (77\%) \\
Art &2,364 (55\%) &1,740 (40\%) &7,564 (41\%) &9,211 (50\%) \\
History &1,005 (36\%) &1,424 (52\%) &31,992 (64\%) &13,606 (27\%) \\
\bottomrule
\end{tabular}
\caption{Split of publications and citations between conferences (C) and journals (J) per field of research.}\label{tab:fields_types}
\end{table}
    We observe nearly every field publishes more in journals than conferences, except \gls{cs} and Engineering (\textbf{\Cref{tab:fields_types}}).
    Engineering has an even larger percentage of publications in conferences (61\% to 37\%) than \gls{cs} (55\% to 43\%).
    \textcite{michels_systematic_2014} find that \gls{cs} has the highest preference for conferences of all research fields ($\approx$77\% to 23\%), followed by Engineering (e.g., Electrical Engineering: $\approx$72\% to 28\%), based on \wos data from 2009.
    In previous experiments, we already showed researchers in \gls{cs} slowly shift their publications more toward journals over time (\Cref{ex:types_bar,ex:types_venues}), which can explain the shift from the numbers \textcite{michels_systematic_2014} report ($\approx$77\% to 23\%) to our numbers (55\% to 43\%).
    As we do not investigate Engineering, we cannot say if there was a shift toward conferences or if our data on Engineering is insufficient and skewed too much by \gls{cs}.
    We assume it is the latter as \textcite{vrettas_conferences_2015} also find Engineering has more than 10 times the number of journals compared to conferences, even though it is placed second after \gls{cs} in terms of the total number of conferences.
    Investigating this further would require a dataset more focused on Engineering, which we might look into in the future.
    This preference of Engineering toward conferences also helps to explain why many of the most productive venues were IEEE conferences with a focus on engineering-related tasks (\Cref{ex:venues_top}).
    For the other fields of study, we can still see the preference for journals, which \textcite{michels_systematic_2014} also report.
    They also find all fields of study (including \gls{cs} and Engineering) either have the same citation rates or higher ones for journals compared to conferences, which is also visible in \Cref{tab:fields_types}.
    Only \textit{Others} and History have a higher percentage of citations in conferences in our case, but we explain this with the small number of citations for \textit{Others} and the small number of publications in History being affected too easily by fluctuations.

    \experiment{Are the topics of the top research fields different from \gls{cs}? Does \gls{cs} influence other research fields? (\labelcref{rq:fields})}\label{ex:fields_topics}
    \afterpage{
\begin{landscape}
\begin{table}[p]
\begin{center}
\scriptsize
\begin{tabular}{lllll|lllll}
\toprule
\multicolumn{5}{c|}{\textbf{1999}} & \multicolumn{5}{c}{\textbf{2019}} \\ 
\textbf{\gls{cs}} & \textbf{Mathematics} & \textbf{Engineering} & \textbf{Medicine} & \textbf{Psychology} & \textbf{\gls{cs}} & \textbf{Mathematics} & \textbf{Engineering} & \textbf{Medicine} & \textbf{Psychology} \\ 
\midrule
paper & paper & control & \textbf{patient} & \textbf{internet} & result & \textit{propos} & \textit{propos} & imag & \textbf{student} \\ 
propos & present & paper & imag & learn & network & \textit{paper} & network & \textit{propos} & \textit{learn} \\ 
problem & problem & propos & fuzzi & \textbf{group} & \textit{propos} & \textbf{graph} & provid & \textit{paper} & data \\ 
imag & \textbf{graph} & imag & \textbf{sequenc} & health & \textit{learn} & problem & imag & present & \textbf{game} \\ 
present & imag & present & \textbf{medic} & \textbf{technolog} & \textit{paper} & method & \textbf{neural} & \textbf{mri} & \textit{paper} \\ 
provid & code & robot & health & \textbf{face} & provid & \textbf{consid} & \textit{paper} & \textbf{edr} & \textbf{task} \\ 
algorithm & \textbf{given} & \textbf{power} & algorithm & \textbf{knowledg} & imag & present & present & signal & show \\ 
object & control & provid & \textbf{movement} & task & present & \textbf{comput} & research & \textit{learn} & \textbf{aim} \\ 
network & system & \textbf{circuit} & care & \textbf{team} & \textbf{featur} & control & method & model & includ \\ 
result & obtain & describ & inform & \textbf{agent} & data & algorithm & includ & \textbf{shore} & \textbf{conduct} \\ 
\textbf{data} & algorithm & \textbf{demonstr} & virtual & softwar & algorithm & network & develop & \textbf{respir} & \textbf{interact} \\ 
design & motion & result & \textbf{cluster} & care & time & \textit{learn} & deep & network & experi \\ 
inform & \textbf{filter} & \textbf{voltag} & \textbf{clinic} & virtual & method & model & \textit{learn} & result & \textbf{virtual} \\ 
time & \textbf{set} & algorithm & \textbf{detect} & commun & develop & \textbf{system} & optim & \textbf{measur} & \textbf{signific} \\ 
control & result & signal & \textbf{rule} & \textbf{emot} & \textbf{studi} & provid & signal & method & \textbf{educ} \\ 
\textbf{test} & signal & \textbf{color} & code & \textbf{copi} & problem & \textbf{set} & \textbf{vehicl} & deep & \textbf{design} \\ 
code & \textbf{estim} & \textbf{manipul} & \textbf{model} & inform & \textbf{us} & data & \textbf{convolut} & \textbf{shell} & \textbf{social} \\ 
develop & time & consid & learn & \textbf{social} & experi & \textbf{obtain} & train & \textbf{fetal} & \textbf{emot} \\ 
\textbf{requir} & object & motion & \textbf{relat} & \textbf{cours} & \textbf{reduc} & \textbf{equat} & \textbf{increas} & power & \textbf{robot} \\ 
describ & \textbf{function} & softwar & servic & \textbf{support} & \textbf{commun} & \textbf{bound} & \textbf{accuraci} & data & perform \\ 
\textbf{differ} & \textbf{exampl} & \textbf{engin} & commun & \textbf{journal} & \textbf{user} & \textbf{state} & result & \textbf{patient} & \textbf{languag} \\ 
servic & network & develop & \textbf{standard} & research & compar & \textbf{code} & control & \textbf{apnea} & \textbf{find} \\ 
\textbf{allow} & \textbf{us} & \textbf{fault} & develop & \textbf{comput} & \textbf{detect} & \textbf{numer} & power & provid & result \\ 
\textbf{featur} & \textbf{bound} & \textbf{current} & \textbf{studi} & \textbf{visual} & optim & \textbf{solut} & \textbf{segment} & optim & \textbf{suggest} \\ 
obtain & fuzzi & research & task & \textbf{teach} & \textbf{process} & \textbf{introduc} & compar & \textbf{sensor} & \textbf{discuss} \\ 
consid & \textbf{known} & \textbf{track} & \textbf{neuron} & \textbf{work} & \textbf{differ} & \textbf{distribut} & \textbf{end} & show & \textbf{subject} \\ 
robot & \textbf{appli} & design & system & \textbf{older} & research & time & achiev & estim & \textbf{brain} \\ 
\textbf{discuss} & \textbf{nois} & spl & object & network & train & estim & \textbf{energi} & achiev & \textit{propos} \\ 
\textbf{includ} & spl & \textbf{forc} & \textbf{measur} & \textbf{user} & perform & \textbf{given} & \textbf{better} & \textbf{demonstr} & \textbf{function} \\ 
\textbf{implement} & \textbf{sub} & \textbf{manag} & motion & \textbf{eb} & \textbf{term} & \textbf{number} & \textbf{speech} & \textbf{predict} & research \\
\bottomrule
\end{tabular}
\end{center}
\caption{Top 30 most salient terms for the top 5 fields of study in 1999 and 2019. Terms that only appear in one field of study are \textbf{bold} and terms that appear in all five fields of study are in \textit{italics}.}\label{tab:fields_topics}
\end{table}   
\end{landscape}
}
    We only analyze the top fields of study based on the number of publications excluding \textit{Others} and any field of study with only 1\% of papers or less, which leaves us with the top 5 fields of study (\textbf{\Cref{tab:fields_topics}}).
    
    There clearly are some differences between \gls{cs} and other fields of study even though we are using a dataset with a \gls{cs} focus.
    In 1999 the unique terms show that Mathematics deals with graph theory (``graph'' and ``function'') and Engineering with electrical engineering (``power'', ``circuit'', and ``voltag''), while we see more medical terms in Medicine (``patient'', ``medic'', and ``neuron''), and Psychology deals with more social aspects (``group'', ``team'', ``emot'').
    \gls{cs} only has a few unique terms in 1999 (``data'', ``test'', and ``implement''), because many are shared across other fields of study (``imag'', ``network'', and ``object'').
    We explain this with our data source \dblp having a focus on \gls{cs} and thus the publications of the other fields of study automatically being related to \gls{cs} in some way, or they would not be in \dblp.
    In 2019 one could argue for \gls{cs} the unique terms (``featur'', ``commun'', and ``detect'') together with some not unique terms (``train'', ``learn'') indicate a trend toward neural networks again, which we already saw in previous experiments (\Cref{ex:authors_topics_time,ex:venues_topics_time,ex:types_topics_time}).
    The term ``learn'' is present in each field in 2019, but the other fields also show unique terms from their research areas again.
    Mathematics again covers graph theory (``graph'') in 2019, but also other fields (``bound'', ``equat'', ``numer'', ``distribut'').
    Engineering deals with ``vehicl'' and ``energi'', but also seems to leverage neural networks more (``convolut'', ``neural''), which is highlighted even more by the not unique terms (``deep'', ``learn''), possibly in the context of autonomous driving.
    In Medicine, we again see medical terms (``mri'', ``respir'', ``apnea''), but also a trend to neural networks with the not unique terms (``network'', ``deep'', ``learn''), while Psychology sticks with social aspects (``game'', ``social'', ``emot'').
    We conclude there are certainly distinct topics in other fields of study besides \gls{cs}, even though the underlying data of \csi focuses on \gls{cs}.
    The trend toward adopting approaches leveraging neural networks is also visible outside of \gls{cs}.
    This is also covered in other literature, e.g., for medicine \textcite{aggarwal_has_2022} show the importance of artificial intelligence, machine learning, and deep learning and their gains for healthcare in face of the COVID-19 pandemic.
    % No full analysis because of technical limitations (JS engine only allows 512MB in string length, we have to send it to the server). Math with over 600k papers already breaks that. Even if longer strings would be possible, we would still not have enough RAM for \gls{cs}.

\section{Summary}\label{sec:analysis_summary}
    In this chapter, we conducted \arabic{experiment} experiments to answer our research questions (\Cref{sec:goals}).
    Here, we shortly show that all research questions were answered and reference the corresponding experiments.
    Our most interesting findings are listed in the conclusion in the next chapter (\Cref{sec:conclusion}).

    \paragraph{\labelcref{rq:amount}} How many publications, authors, and venues are in our dataset? How do the numbers change over time? How many authors and venues are currently active?
    
    The dataset overview showed the number of publications, authors, venues, and more (\Cref{tab:data_overview}).
    We investigated the changes in the numbers per year for publications (\Cref{ex:publications}), authors (\Cref{ex:authors_bar}), and venues (\Cref{ex:venues_bar}).
    The activity for the last five years was also shown for authors (\Cref{ex:authors_active}) and venues (\Cref{ex:venues_active}).

    \paragraph{\labelcref{rq:distribution}} How are the citations and publications distributed across authors and venues? How do the distributions change over time?
    
    We covered the trends in the distribution of the number of citations and papers over multiple periods for authors (\Cref{ex:authors_citations}) and venues (\Cref{ex:venues_citations}).

    \paragraph{\labelcref{rq:topics}} What are the most prominent authors and venues? Are there preferences for topics? Do the topics change over time?
    
    We showed the most cited and most productive authors (\Cref{ex:authors_top}), their preferences for venues/topics (\Cref{ex:authors_topics}), and how the topics changed over time (\Cref{ex:authors_topics_time}).
    Similarly, we covered the most cited and publishing venues (\Cref{ex:venues_top}), their preferences for topics (\Cref{ex:venues_topics}), and how the topics changed over time (\Cref{ex:venues_topics_time}).

    \paragraph{\labelcref{rq:citations}} How do incoming and outgoing citations evolve over time? How do their distributions differ?
        
    We investigated the changes in the number of incoming and outgoing citations per year (\Cref{ex:citations_bar}), the distribution of incoming citations based on citation bins (\Cref{ex:citations_distribution}), and the trends in the distribution of incoming and outgoing citations over multiple periods (\Cref{ex:citations_time}).

    \paragraph{\labelcref{rq:types}} How do conferences and journals compare in their number of publications and citations over time? How do the top venues and topics differ? Do top authors prefer conferences or journals?
    
    After showing the distribution of document types (\Cref{ex:types_distribution}), we covered the development of conferences and journals over time regarding their number of publications per year (\Cref{ex:types_bar}), their distribution of citations (\Cref{ex:types_citations}), and topics (\Cref{ex:types_topics_time}).
    The most cited conferences and journals (\Cref{ex:types_venues}), and the preferences of top authors for conferences and journals (\Cref{ex:types_authors}) were also examined.

    \paragraph{\labelcref{rq:fields}} How do the most prominent fields of study differ from \gls{cs} in topics and preference for conferences or journals?
    
    Lastly, we investigated the distribution of the fields of study (\Cref{ex:fields_distribution}) and how other fields (e.g., medicine) differed from \gls{cs} regarding their preference for conferences or journals (\Cref{ex:fields_types}) and their topics (\Cref{ex:fields_topics}).

	% Schlusskapitel
	\chapter{Final Considerations}\label{chap:final}
This chapter presents the final considerations of this thesis.
We start with the conclusion of our experiments from the previous chapter and a summary of our contributions (\Cref{sec:conclusion}) and close this thesis with the current limitations and future work of our research (\Cref{sec:limitations}).

\section{Conclusion}\label{sec:conclusion}
    Our goal was to analyze the state of \gls{cs} research, covering authors, venues, document types, fields of study, and their publications, citations, and topics to uncover implicit patterns in \gls{cs} literature.
    To achieve this, we introduced \projectName (\csi), an interactive, responsive open-source browser-based visualization system to facilitate the exploration of \gls{cs} publications, and the \diiilong (\diii) that contains metadata associated with 6m \gls{cs} papers in its newest version\footnote{\url{https://zenodo.org/record/7069915}}.
    The \csi system crawls and processes publications from \dblp and enriches them with additional metadata (e.g., citations and abstracts) from their full-texts to create \diii.
    \csi is also built in a modular architecture to facilitate the maintenance and incorporation of more efficient components in the future.
    Both \csi\footnote{\url{https://github.com/gipplab/cs-insights}} and \diii\footnote{\url{https://github.com/jpwahle/lrec22-d3-dataset}} are fully open-access and freely available online in their respective GitHub repositories.
    To reproduce the results from our work the original version of \diii is also available online\footnote{\url{https://zenodo.org/record/6477785}}.
    
    We then used \csi to conduct a case study on \gls{cs} and demonstrate its capabilities.
    Some of the most interesting and relevant findings are listed below.
    \begin{itemize}
        \item \gls{cs} attracts increasingly more new authors (30\% joined in the last five years and those 30\% make up half of all authors who published in the last 5 years), who also publish more papers per year. At the same time, the number of venues and their publications per year also increases.
        \item Incoming (received) citations peak in 2009, and fall off before and after, while on average each paper cites more other papers with each passing year. Yet, 29\% of all publications have no citations, and only a third get 10 or more citations.
        \item The most cited authors do their research in computer vision and pattern recognition and prefer to publish in conferences, while the most productive authors cover signal processing and communication, and prefer to publish in journals. Similarly, the most cited venues are journals and focus on computer vision/pattern recognition or signal processing/communication, while the most productive venues are mostly conferences and more focused on engineering topics regarding sensors.
        \item In total, most publications in \gls{cs} are from conferences (53\%). Due to the COVID-19 pandemic, the number of conference papers dropped in 2020, also affecting the overall number of publications in 2020, but the number of journal articles appeared unaffected. This made journals again overtake conferences in the number of publications per year, a first since 1992. A shift back to journal articles was visible before 2020, as the gap between journal articles and conference papers per year was already getting smaller in recent years.
        \item Journal articles get on average twice as many citations as conference papers, a trend that has been visible for decades. The citation gap between the most cited journals and conferences is much smaller, but still favors journals.
        Some highly cited conferences reach on average more citations per publication than highly cited journals, but the average citations of highly cited conferences also fluctuate more than for highly cited journals.
        \item \gls{cs} and engineering are the only fields favoring conferences over journals considering the number of publications, but all investigated fields of study get more citations in journals than conferences.
        \item Overall, an increase in the popularity of approaches leveraging neural networks was visible in 2015-2019 across conferences, journals, and top fields of study, authors, and venues.
    \end{itemize}
    We conclude that \gls{cs} appears to be a strongly growing and very active field, and with a scientometric analysis supported by \csi, we can show its characteristics, trends, and implicit patterns through its core components and attributes (i.e., publications, authors, venues, citations, topics, document types, and differences to other fields of study).
    %(e.g., shifts in citations and publications for authors or venues, and preferences for conferences/journals).

    The same methodology we used to analyze \gls{cs} can also be applied to other areas, by using the \csi system to conduct the same experiments on different datasets (see future work; \Cref{sec:future_work}) from other research fields (e.g., medicine) or sub-fields (e.g., \gls{nlp}).
    % It would then be possible to compare results for different fields solely based on the data itself, and not different approaches.
    To the best of our knowledge, no other researchers have conducted a study into \gls{cs} as extensive as we did in this thesis.
    While many authors already looked into certain aspects of \gls{cs}, it was always only partially, with always different datasets, and different approaches.
    Some authors then get contradicting results and it becomes hard to determine the exact reason, e.g., when investigating if conferences or journals get more citations \parencite{rahm_citation_2005,franceschet_role_2010,vrettas_conferences_2015}.
    Having a common and flexible system that can efficiently perform the same analysis on different datasets allows for better comparability of scientometric research in the future.
    Comparing datasets more easily might also enable researchers to make more informed decisions about which dataset they want to use for their research.

\section{Limitations \& Future Work}\label{sec:limitations}\label{sec:future_work}
    In this section, we cover limitations and future work regarding the data, backend structure, features of the frontend, and analysis, some of which were already mentioned in \Cref{chap:methodology,chap:analysis}.
    We show most of the current limitations are already planned to be fixed in future work.
    %, but some limitations we cannot fix, e.g., we cannot make some abstracts available in our dataset due to copyright reasons.
    More details on future versions, features, and improvements of the \csi system are also available in a roadmap on GitHub\footnote{\url{https://github.com/users/jpwahle/projects/1}}.

\subsubsection{Data}
    Even though \dblp is the largest repository of \gls{cs} publications, with an extensive list of features at its disposal, it does not contain all publications about \gls{cs} (e.g., journal and conference volumes without openly available metadata cannot be automatically indexed by \dblp\footnote{\url{https://dblp.org/faq/5210229.html}}).
    We are now in the process of switching the data source to \semanticScholar, as we can access their data, we are allowed to use it, and it gives us more flexibility in the future.
    The crawler was already changed to get the data from \semanticScholar and not \dblp\footnote{We already used the new crawler to get the \dblp data from \semanticScholar in our newest version of \diii available on zenodo: \url{https://zenodo.org/record/7069915}}, but the structure in the backend is not yet adjusted accordingly.
    Using data from \semanticScholar allows us to extract the same publications we already covered with \dblp, but also cover other research fields (e.g., physics) or datasets (e.g., \pubmed, \aclAnthology) integrated into \semanticScholar in a unified way.
    This way we will be able to compare research fields and datasets more easily in the future, as explained in \Cref{sec:conclusion}.
    An additional benefit of using \semanticScholar data is that it fixes the current issues from the broken export of \diii, e.g.,  publications that got missing, duplicate publications, author names missing special characters (e.g., umlauts; see \Cref{ex:publications_top}) or not being disambiguated (e.g., Saif M. Mohammad, Saif Mohammad), and venues with an increasing counter in \dblp (e.g., ``HCI (42)''; see \Cref{ex:venues_active}).
    We can then also use \semanticScholar's citation counts, which removes the necessity to use \grobid for the extraction of bibliographies.
    This should fix the citation counts for publications without research fields (\Cref{ex:fields_distribution}) and the incoming and outgoing citation counts not matching (\Cref{ex:citations_bar}).
    The new dataset also adds newer publications, as our current dataset from \dblp was crawled on 2 December 2021 and thus is not complete for 2021.
    Some limitations remain with \semanticScholar data, e.g., the missing affiliations (institutions and countries) and publishers, or not being able to make some abstracts available in \diii due to copyright reasons.
    However, \semanticScholar contains other measures for authors (e.g., influential citations and h-index) and we also thought about adding other quality measures for venues in the future (e.g., impact factor or grade) from other sources.

    We also plan some data-related features besides switching to \semanticScholar, e.g., adding an update functionality that automatically updates our data by adding new publications and updating old ones.
    Another feature would be an import functionality for other data sources where we cannot provide the data ourselves due to restricted access (e.g., \wos, \scopus).
    Should the users have access to those services (e.g., through their institution) and be able to download the data, they could then import it into their local version of \csi to analyze it.

\subsubsection{Backend Structure}
    Some queries for the authors take up to a minute on a good machine, but a machine with less memory and a worse hard drive might need several minutes.
    This severely limits people without access to a better (and more expensive) machine and requires more server resources to host our demo.
    As we are already changing the backend schema for the \semanticScholar data, we are also looking into increasing the performance of the backend.
    One idea is to change the schema and copy all data to each collection to pre-aggregate it by authors, venues, etc.
    %, but this would require more schema engineering to overcome issues with the document size limit of \mongodb 
    (\Cref{subs:schema}).
    Another idea is to use a different database, e.g., PostgreSQL\footnote{\url{https://www.postgresql.org/}} or Neo4j\footnote{\url{https://neo4j.com/}} (a graph database).
    % In \Cref{subs:schema} we already explained that we left further changes to the schema for future work.

\subsubsection{Frontend Features}
    We also intend to expand the frontend functionalities in the future by improving the current dashboards and visualizations and building new ones based on the things we learned from the analysis.
    The goal is to automate more analyses but still allow a great variety of analyses and visualizations, so researchers can continue to explore the field of \gls{cs} with \csi and \diii.
    For example, we plan to add line charts for the average authors per paper per year and/or papers per author per year, as we currently cannot compute either from the currently available information (\csi only shows the unique number of authors per year; see \Cref{ex:authors_citations}).
    Line charts to visualize the average number of publications or citations over time are also possible.
    % e.g. log switch for bar chart
    New features apart from the visualizations are also planned (e.g., a ``NOT'' option for filters or better account management), or new features leveraging new models in the prediction endpoint (e.g., a frequency measure for terms; see \Cref{ex:venues_topics_time,ex:types_topics_time}).
    \csi does not have a search function for publications or authors and we do not plan to add one, as both \dblp and \semanticScholar already provide this functionality.
    
    We also consider some larger features, which would take a considerable amount of time to implement and which we present two of in this paragraph.
    The first idea is to make comparisons easier by selecting two sets of filters and showing the information side by side, e.g., with two lines or bars in the same visualization.
    Another idea includes leveraging our data for analyses with networks and graphs.
    We already have data on incoming and outgoing citations in \diii through our crawler (\Cref{subs:data_secondary}), which \semanticScholar will continue to provide after the switch, so citation networks would be possible.
    One potential approach to implementing this would be to integrate \vosviewer just like \zetaalpha does for the terms of publications.
    Creating networks for co-authorship or terms from titles or abstracts would then also be a possibility.

\subsubsection{Analysis}
    Generally, the analysis in this thesis (\Cref{chap:analysis}) is limited by the features of \csi's \gls{ui}, as we do not conduct any analysis directly on the data and only through the \gls{ui} \csi provides.
    During the analysis we also found drawbacks of our approaches and some other interesting aspects to look into in the future, which we cover in the following.
   
    We conducted multiple experiments using data from 1960-2019, and used two different lengths for the time periods (i.e., 1960-1999 used 10 years and 2000-2019 five years).
    Originally, this worked well to uncover trends in the number of citations and papers over time (\Cref{ex:citations_time,ex:types_citations,ex:venues_citations,ex:authors_citations}) and the distribution appeared unaffected (i.e., first quartile, median, third quartile, average).
    The maximum was only introduced later on and showed the first issues, as it sometimes sank when crossing from 1990-1999 to 2000-2004.
    \Cref{ex:types_venues2} then showed a noticeable effect, which also skewed the numbers for the average publications per venue but, unfortunately, that was the last experiment we conducted using this approach.
    In retrospect, it might have been better to stick to one length for the periods, but start in 1980 to keep it at eight periods and not blow up the tables too much compared to 12 five-year periods starting in 1960 would result in.
    
    Another issue we ran into was the saliency measure used in our topic modeling visualization.
    For some experiments, the measure worked well and we got good results (\Cref{ex:authors_topics_time,ex:venues_topics,ex:fields_topics}), but others caused issues (\Cref{ex:venues_topics_time,ex:types_topics_time}).
    Saliency measures the distinctiveness of terms across all topic clusters and boosts the terms that are exclusive to specific topic clusters.
    \Cref{ex:types_venues} grouped five venues, but the topic model could perfectly sort those five venues into 10 topics, which made venue-specific terms (e.g. ``brain'') nearly exclusive to one topic and thus saliency boosted the terms.
    Generic terms (e.g., ``paper'') also still appeared in high ranks, which saliency should prevent in theory.
    In \Cref{ex:types_topics_time}, similar issues became apparent, as most terms were generic terms, and the saliency measure failed to rank them lower.
    This might be due to trying to fit all publications of \gls{cs} from various sub-fields into 10 topics, while other researchers use more topics, which was not possible for us due to technical limitations.
    Considering the drawbacks of these two experiments and the resources required to overcome the technical limitations to make saliency more viable, a normal frequency measure with a filter for generic terms might prove more useful in the future.
    
    We also found some interesting aspects while conducting our experiments we did not further explore in this thesis.
    When investigating the most cited and most productive venues (\Cref{ex:venues_top}) it appeared as if older venues are more respected and get more citations.
    It might be interesting to see how the citations differ between more established venues and newer ones, and how long it might take newer once to get more established and also receive more citations.
    In the same experiment, we found large differences once we only considered open-access publications.
    Investigating the difference between venues that make their publications open-access, and venues that lock them behind a paywall might show differences in their impact, considering authors can more easily verify the contents of open-access papers for references, but paid-access papers might be published in more renowned venues.
    Lastly, a comparison between singular venues (\Cref{ex:venues_topics_time}) and their authors, topics, citations, etc. might generally yield interesting results.

	%%%%%%%%%%%%%%%%%%%%%%%%%%%%%%%%%%%%%%%%%%%%%%%%%%%%%%%%%%%%%%%%%%%%%%%%%%%%
	\appendix % Ende des Inhalts, hier beginnt der Anhang %%%%%%%%%%%%%%%%%%%%%%
        \chapter{Appendix}\label{chap:appendix}
\section{Additional Tables/Figures}

% \newcommand{\tabIndent}{\hspace{3mm}}

% \begin{table}[p]
\begin{table}[htb]
\footnotesize
\centering
% \subfloat[caption1]{
% \parbox[t][][t]{.50\linewidth}{
% \begin{minipage}[t]{.5\linewidth}
% \centering
    % \begin{tabular}[t]{l @{\hspace{-1.4cm}} r} 
    \begin{tabular}{l r} 
        \toprule
        \textbf{Attribute} &\textbf{Example}\\
        \toprule
        \textbf{publication} & \\
        \tabIndent id                      & conf/acl/Mohammad20b\\
        \tabIndent modified date           & 2021-09-12\\
        \tabIndent title                   & NLP Scholar - An Interactive ...\\
        \tabIndent pages                   & 232-255\\
        \tabIndent year                    & 2020\\
        \tabIndent type                    & Conference and Workshop Papers\\
        \tabIndent access                  & open\\
        \tabIndent links                   & [https://doi.org/...]\\
        \tabIndent doi                     & 10.18653/v1/2020.acl-demos.27\\
        \tabIndent publisher               & ACL\\
        \textbf{author} & \\
        \tabIndent id                      & 58/380\\
        \tabIndent fullname                & Saif M. Mohammad\\
        \tabIndent webpage                 & http://saifmohammad.com/\\
        \textbf{venue} & \\
        \tabIndent names                   & [International Conference on Lang...] \\
        \tabIndent acronyms                & [LREC] \\
        \tabIndent type                    & Conference or Workshop \\
        \tabIndent id                     & conf/lrec \\
        \midrule
        \textbf{affiliation}                & \\
        \tabIndent id             & 4eb3...f094\\
        \tabIndent name           & National Research Council Canada\\
        \tabIndent country        & Canada\\
        \tabIndent city           & Ottawa\\
        \tabIndent postcode       & K1A 0R6\\
        \tabIndent addressline    & 1200 Montreal Road, Bldg. M-58\\
        \textbf{publication} & \\
        \tabIndent outgoing citations          & \\
        \tabIndent\tabIndent ids            & [7615..., 76af...]\\
        \tabIndent\tabIndent count          & 2\\
        \tabIndent incoming citations           & \\
        \tabIndent\tabIndent ids            & [7ca5..., 7d0e...] \\
        \tabIndent\tabIndent count          & 11\\
        \tabIndent keywords                    & [Scientometrics, Citations, ...]\\
        \tabIndent ocr title                       & NLP Scholar: An Interactive ...\\
        \tabIndent ocr abstract                    & As part of the NLP Scholar ...\\
        \bottomrule
    \end{tabular}
    \caption{\diii attributes as proposed in \textcite{wahle_d3_2022} (top half: data from \dblp, bottom half: data extracted from full-texts).}
    \label{tab:schema_old}
\vspace{-2cm}
\end{table}
% \vspace{-5cm}

% \afterpage{
\begin{landscape}
%Prototype
    \begin{figure}
        \centering
        \includegraphics[width=\hsize]{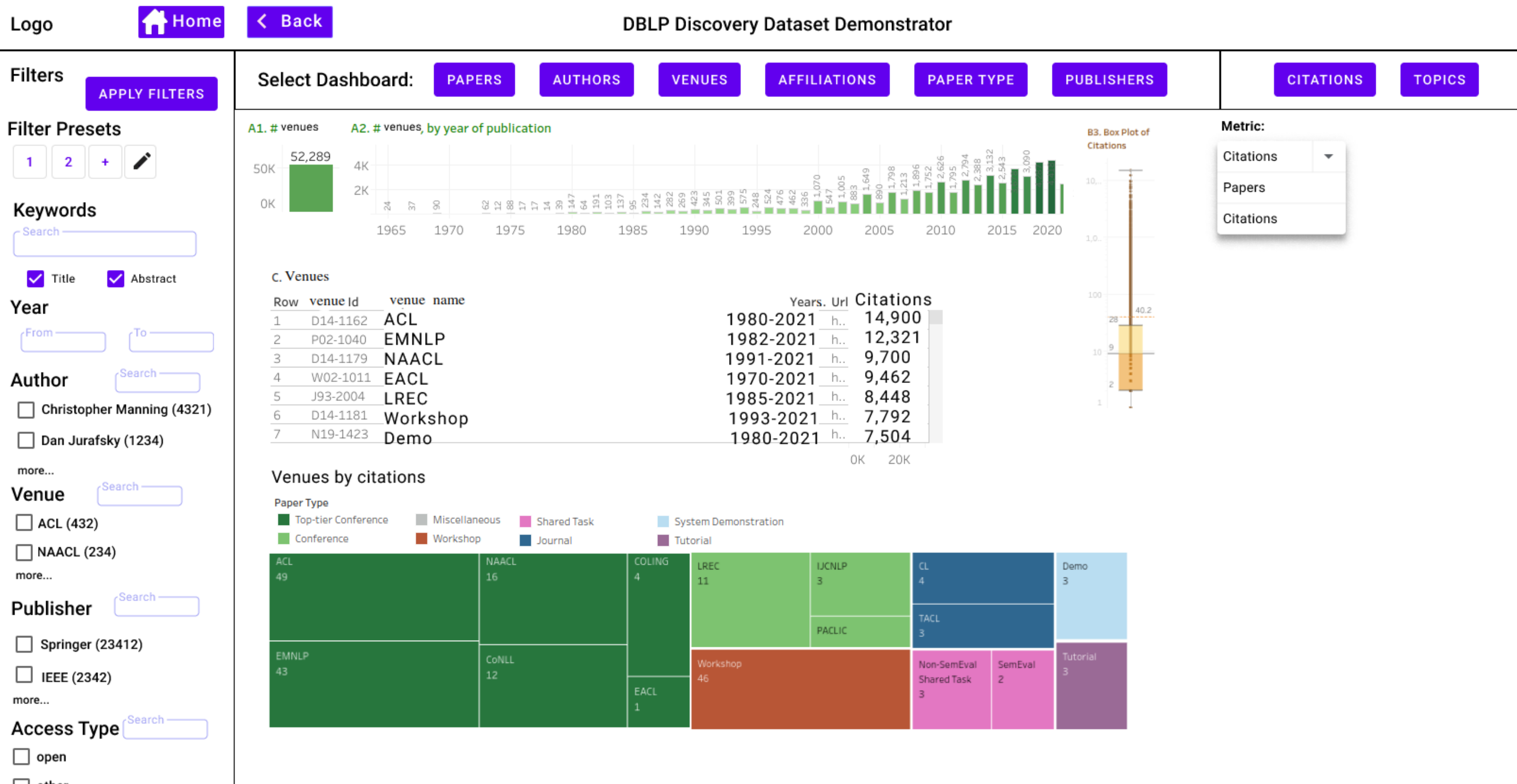}
        \caption{Final prototype of the \projectAcronym frontend (current selection: venues dashboard). The graphs shown are taken from \nlpScholar \parencite{mohammad_nlp_2020_viz} with the author's permission and altered for the prototype.}
        \label{fig:prototype}
    \end{figure}

%Other Dashboards
    \begin{figure}
        \centering
        \includegraphics[width=\hsize]{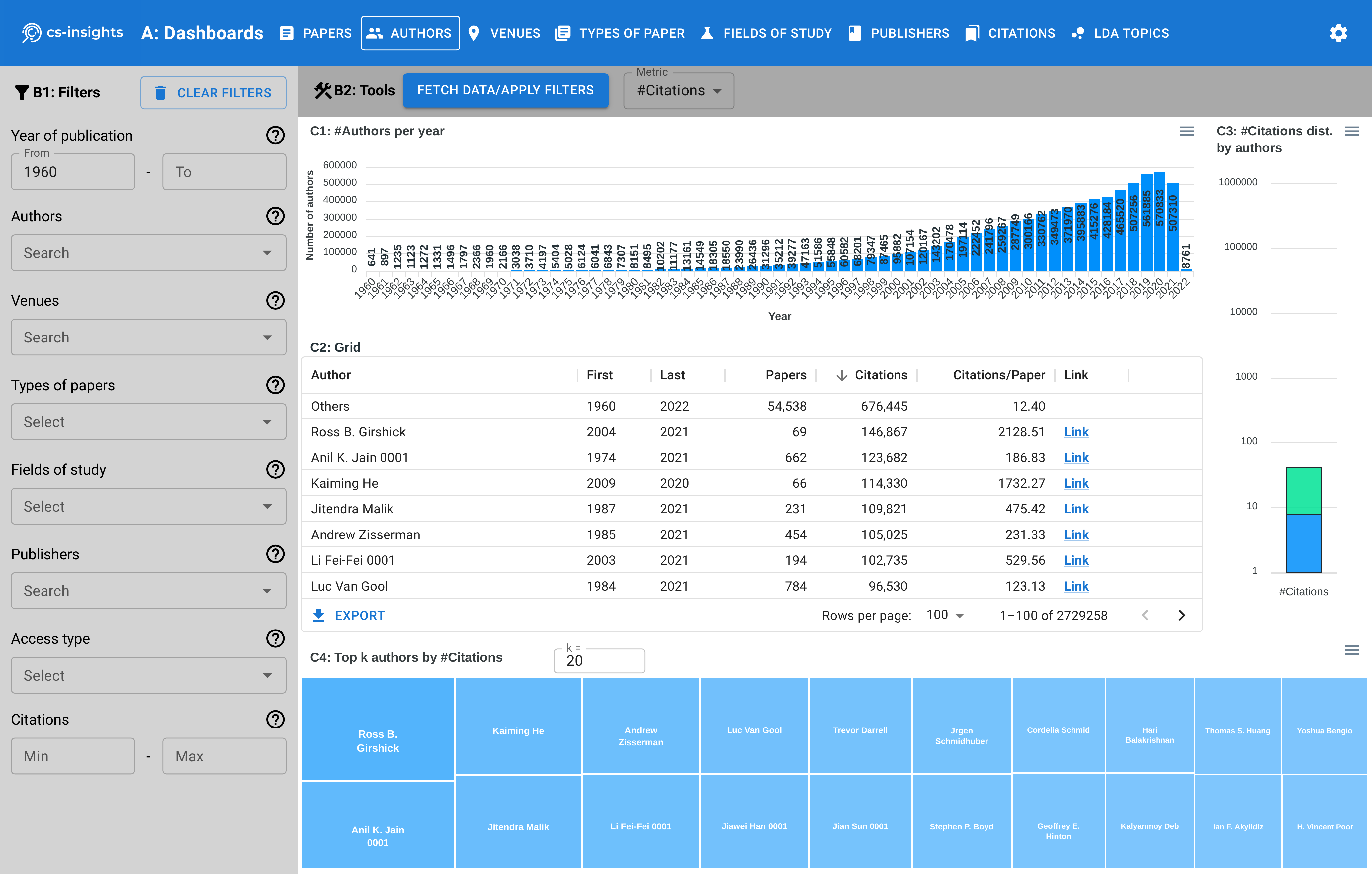}
        \caption{Authors dashboard in \csi.}
        \label{fig:fe_authors}
    \end{figure}

    \begin{figure}
        \centering
        \includegraphics[width=\hsize]{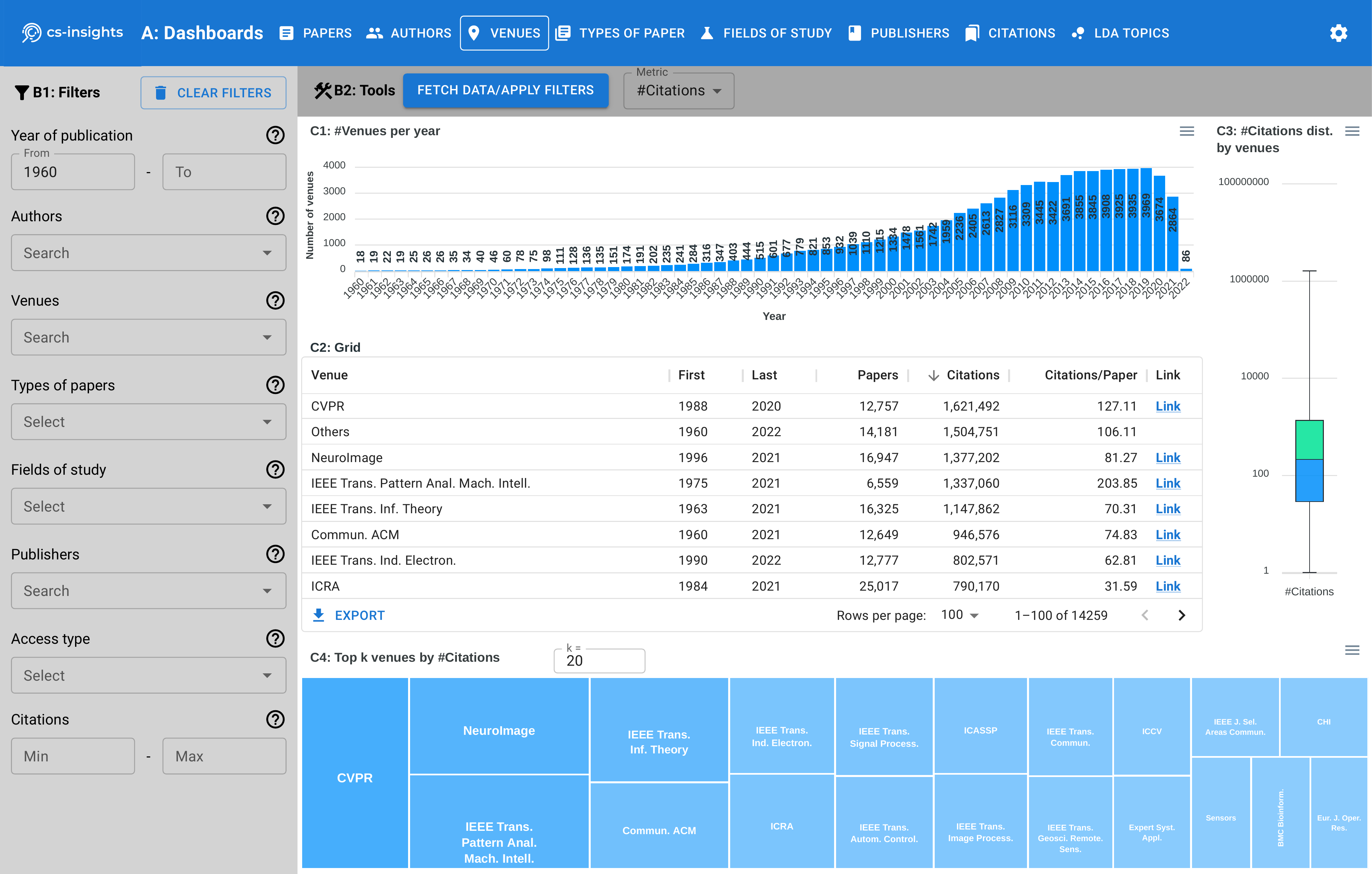}
        \caption{Venues dashboard in \csi.}
        \label{fig:fe_venues}
    \end{figure}

    \begin{figure}
        \centering
        \includegraphics[width=\hsize]{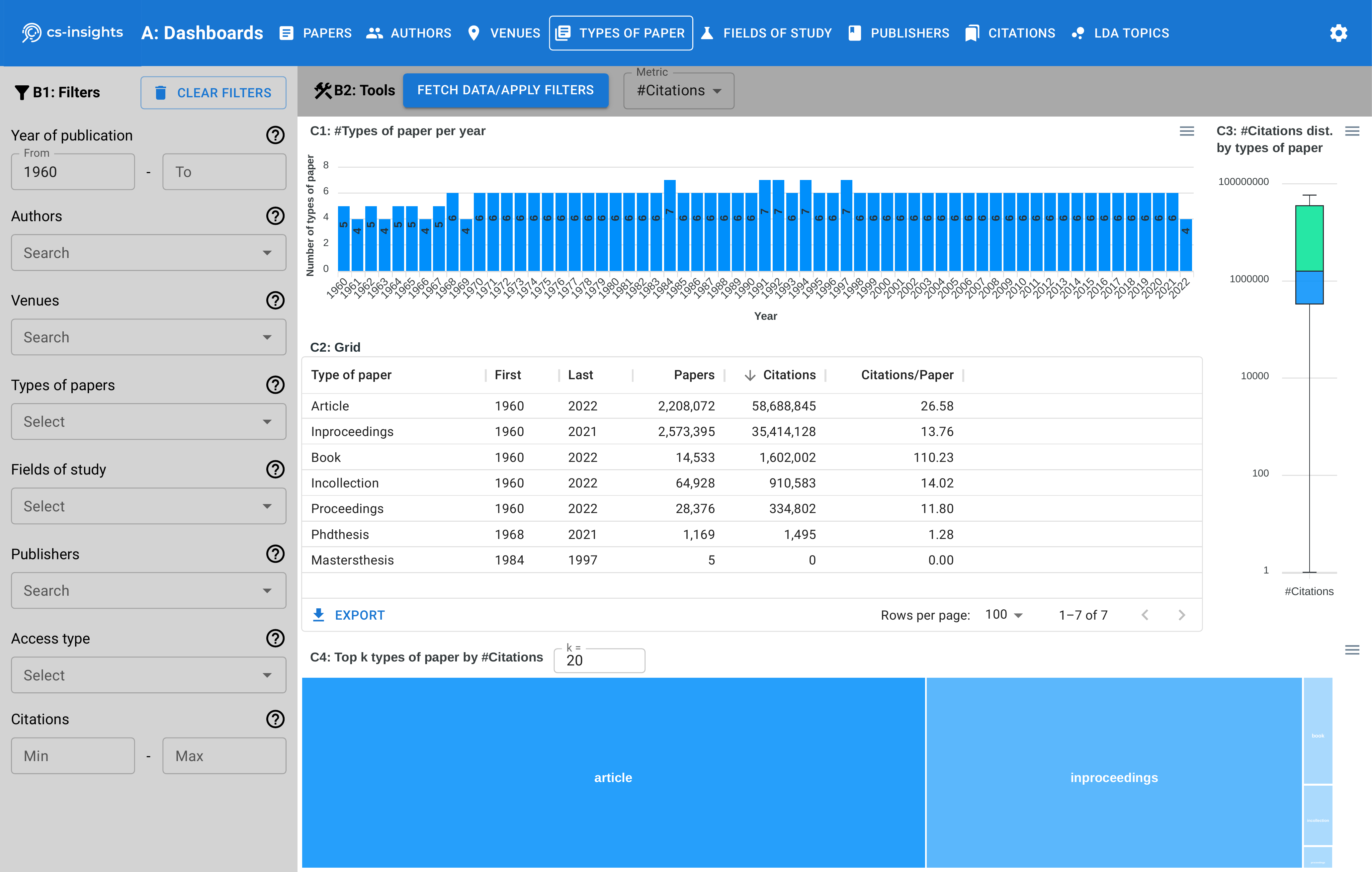}
        \caption{Types of Paper (document types) dashboard in \csi.}
        \label{fig:fe_types}
    \end{figure}

    \begin{figure}
        \centering
        \includegraphics[width=\hsize]{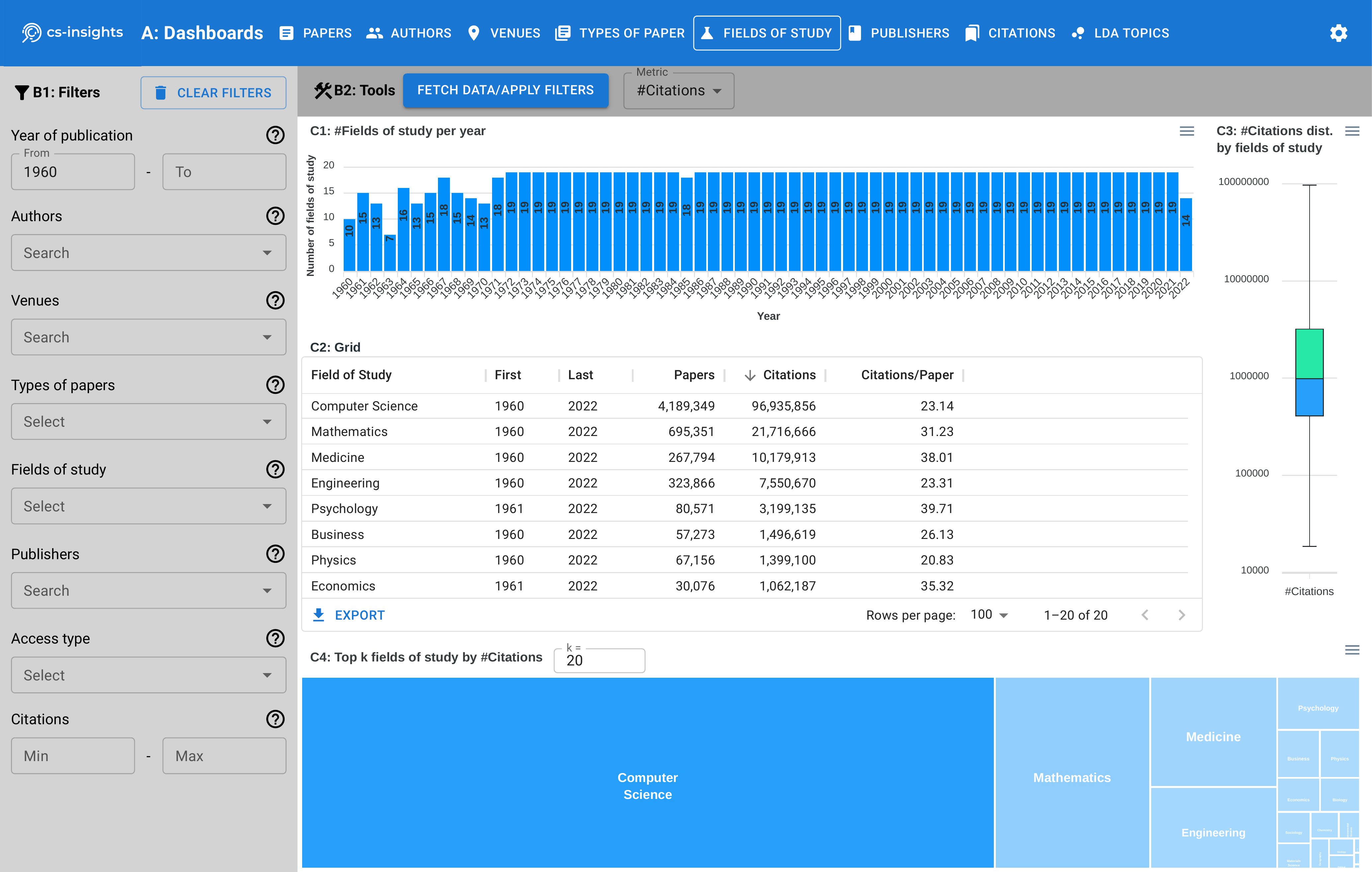}
        \caption{Fields of Study dashboard in \csi.}
        \label{fig:fe_fields}
    \end{figure}

    \begin{figure}
        \centering
        \includegraphics[width=\hsize]{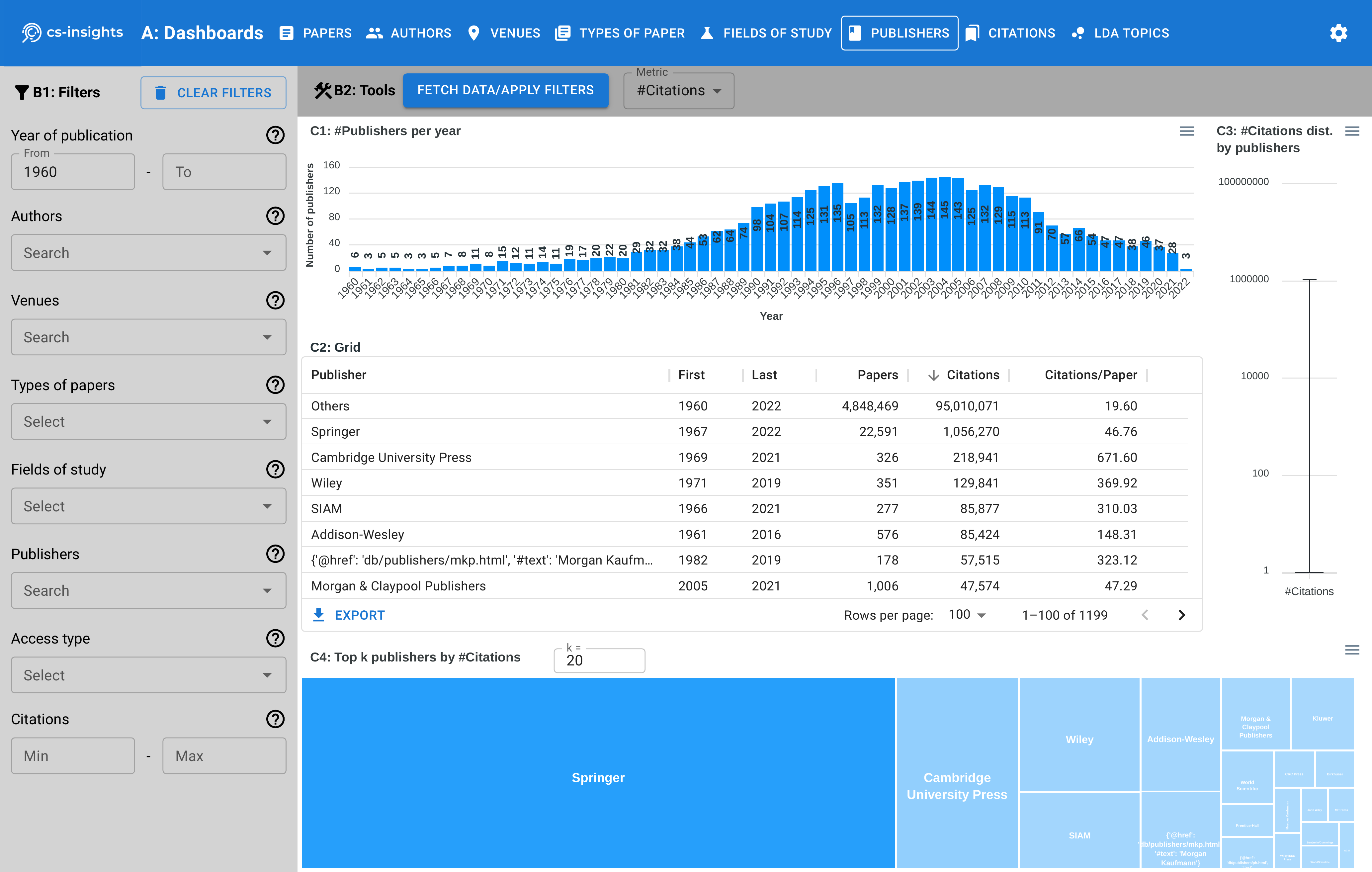}
        \caption{Publishers dashboard in \csi. The grid shows most publications have no publisher and fall under \textit{Others}.}
        \label{fig:fe_publishers}
    \end{figure}

    \begin{figure}
        \centering
        \includegraphics[width=\hsize]{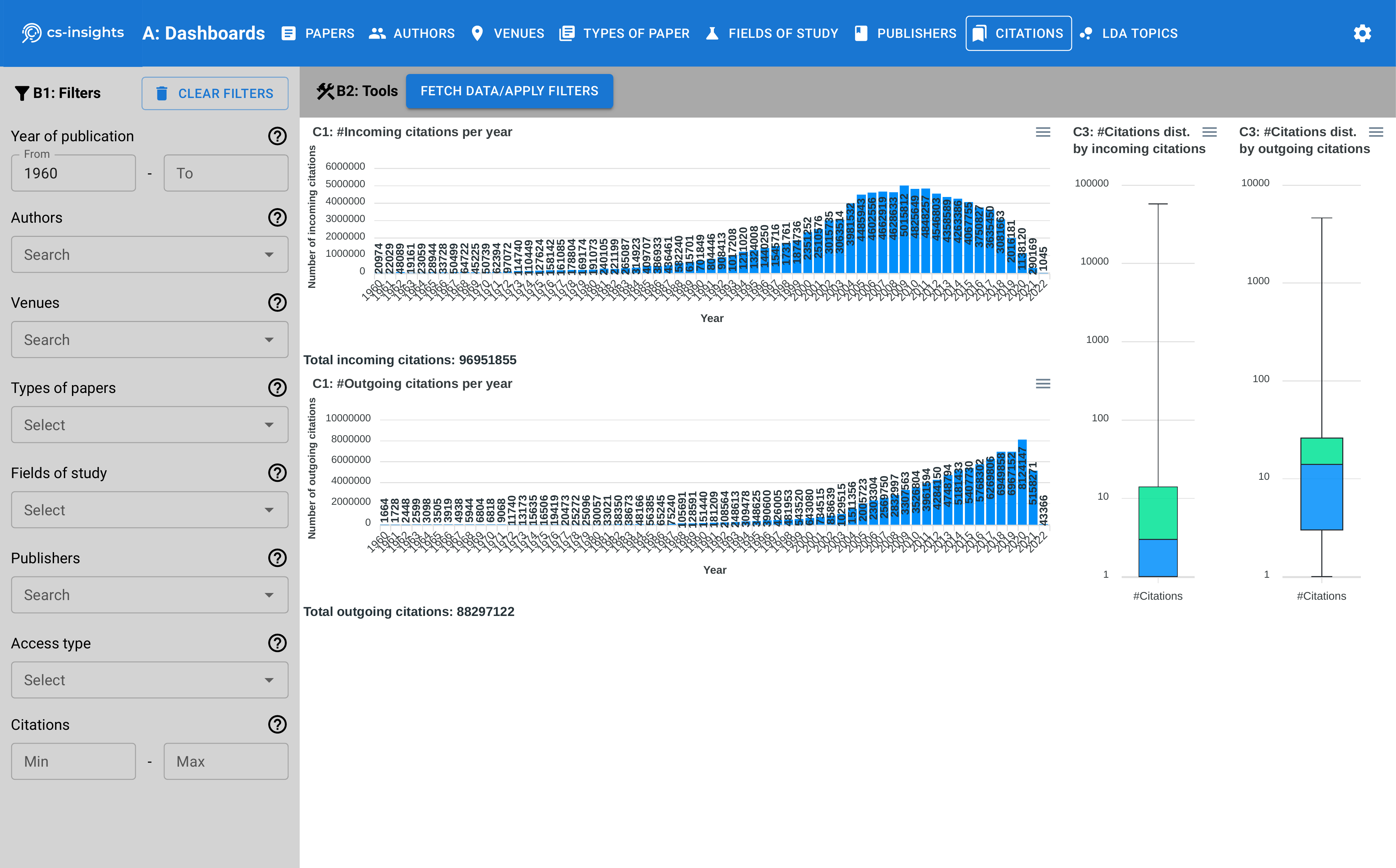}
        \caption{Citations dashboard in \csi.}
        \label{fig:fe_citations}
    \end{figure}

    \begin{figure}
        \centering
        \includegraphics[width=\hsize]{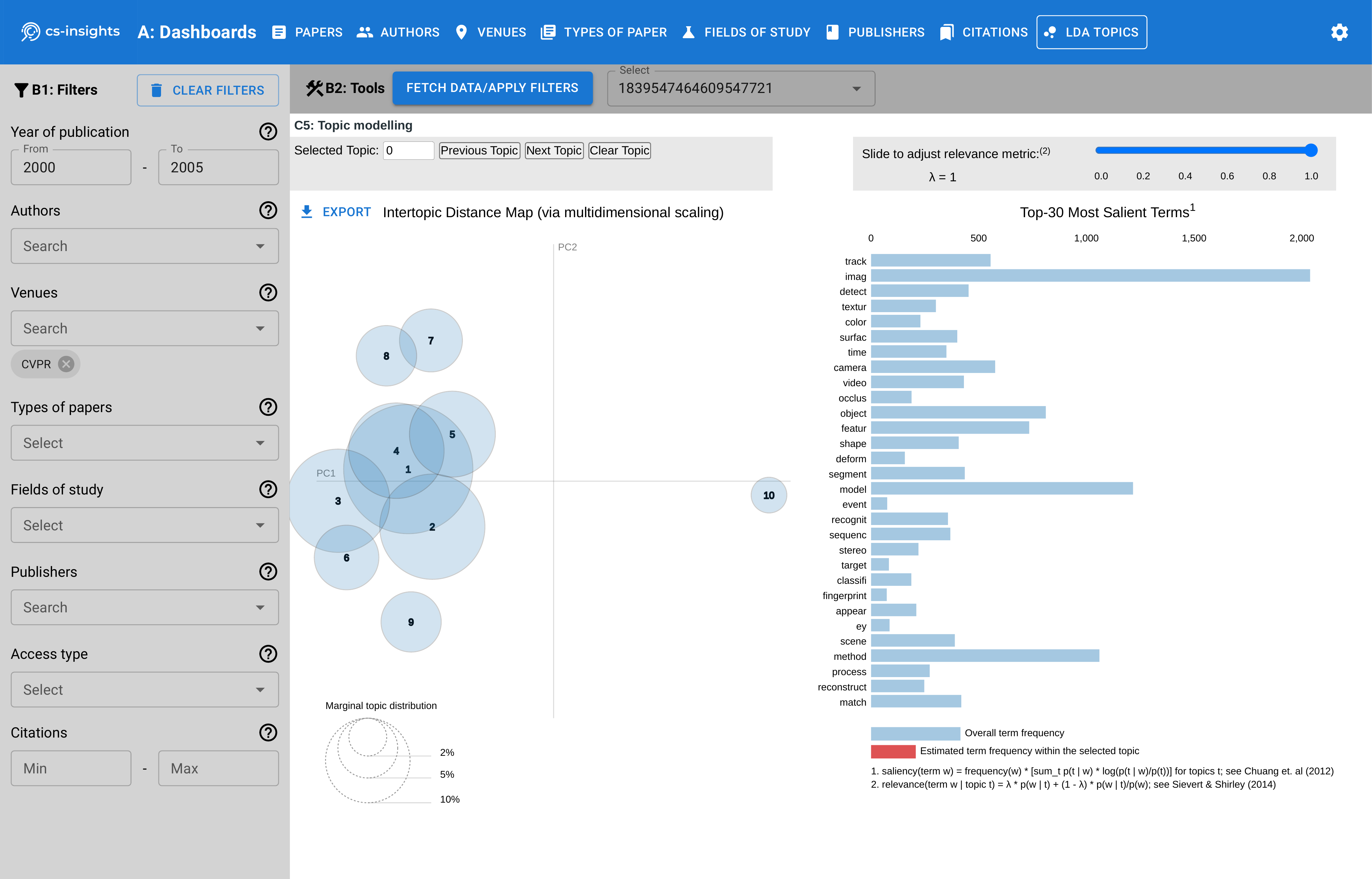}
        \caption{Topics dashboard in \csi showing a visualization of the topics for CVPR (2000-2005).}
        \label{fig:fe_topics}
    \end{figure}

% Topic modeling
    \begin{figure}
        \centering
        \includegraphics[width=\hsize]{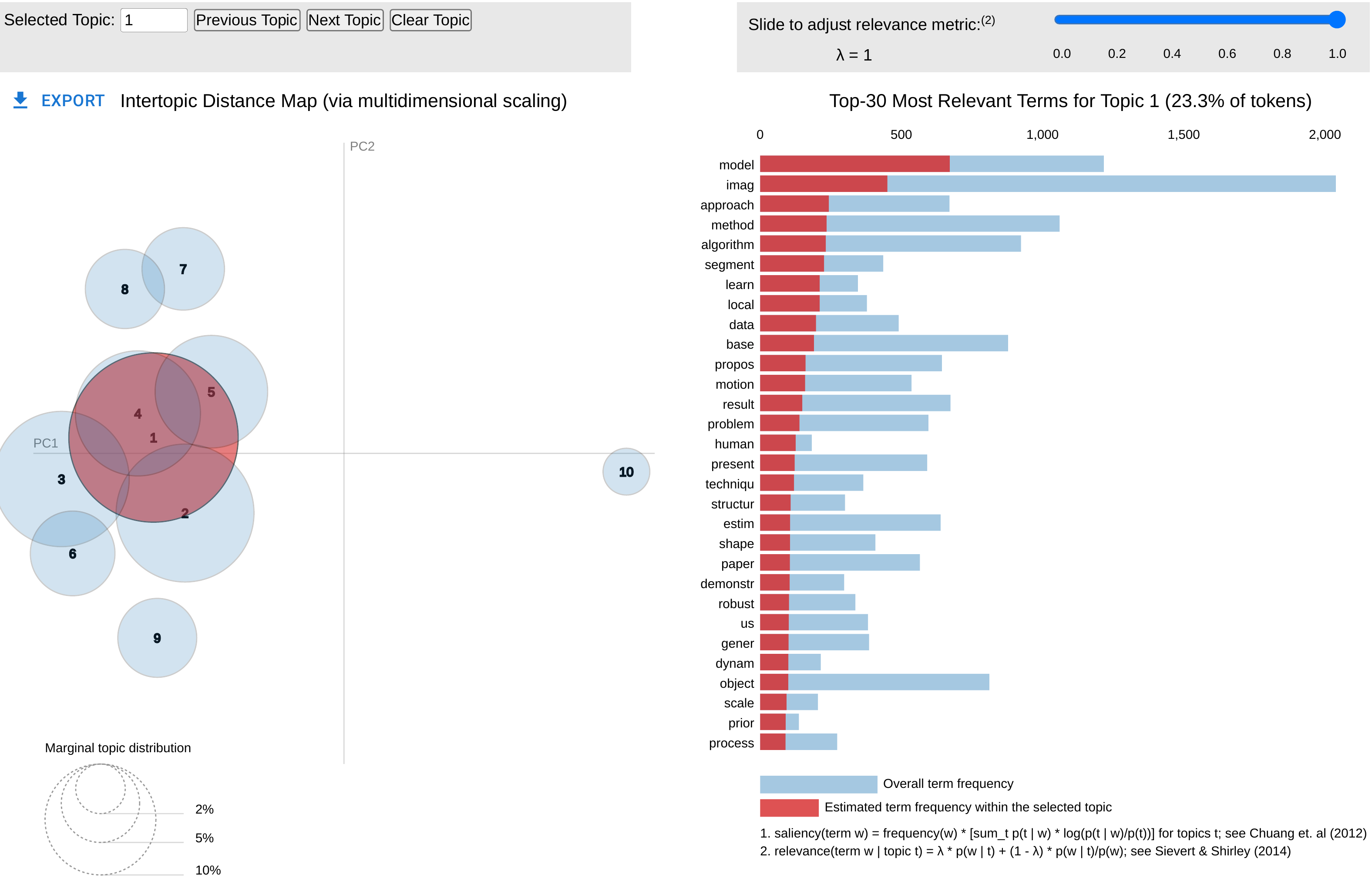}
        \caption{Visualization of the topics for CVPR (2000-2005) with cluster 1 selected.}
        \label{fig:showcase3_cluster}
    \end{figure}

    \begin{figure}
        \centering
        \includegraphics[width=\hsize]{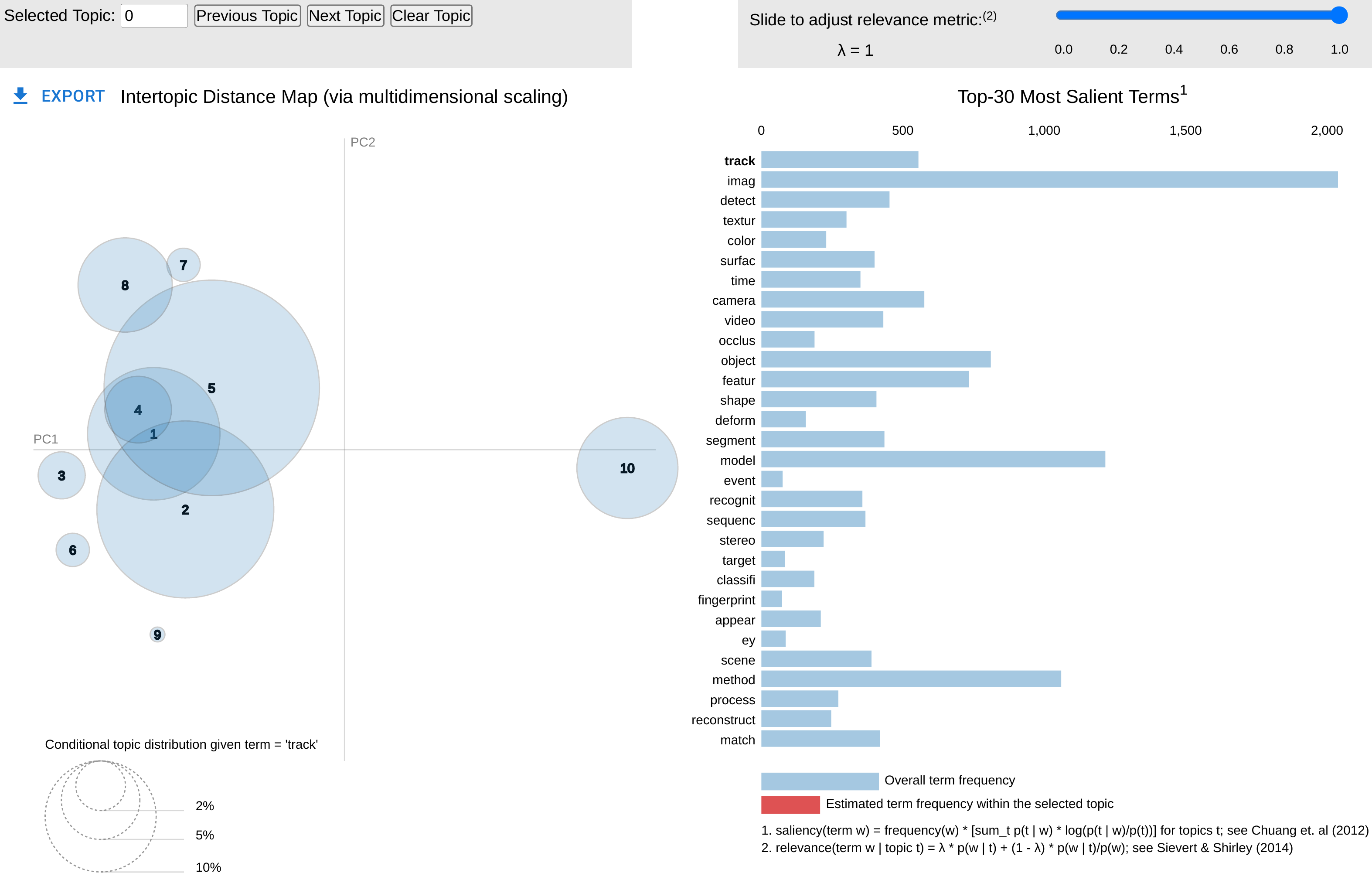}
        \caption{Visualization of the topics for CVPR (2000-2005) with the term ``track'' selected.}
        \label{fig:showcase3_term}
    \end{figure}

% Long Figures from Analysis
    \begin{figure}
        \centering
        \includegraphics[trim={0 1.2cm 0 0}, clip, width=\hsize]{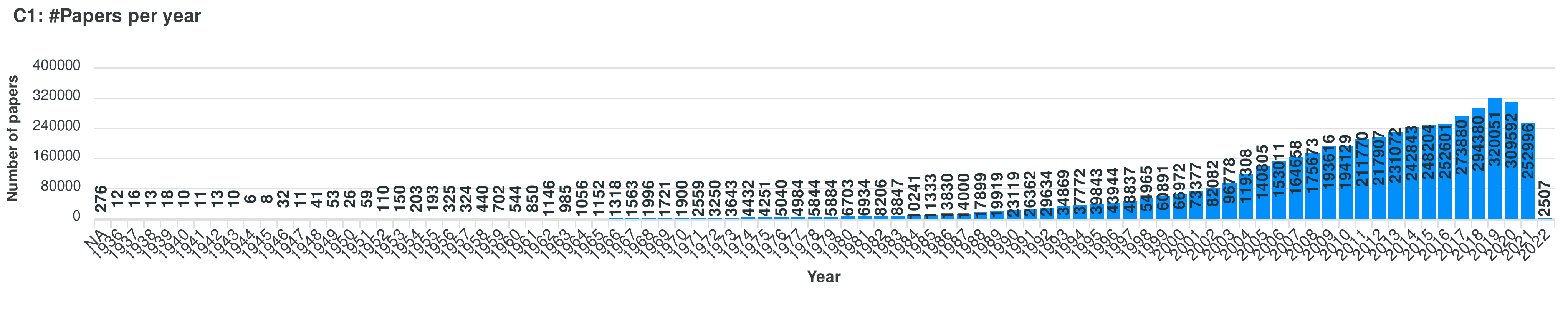}
        \caption{Number of publications per year.}
        \label{fig:publications_bar_app}
    \end{figure}

    \begin{figure}
        \centering
        \includegraphics[trim={0 1.2cm 0 0}, clip, width=\hsize]{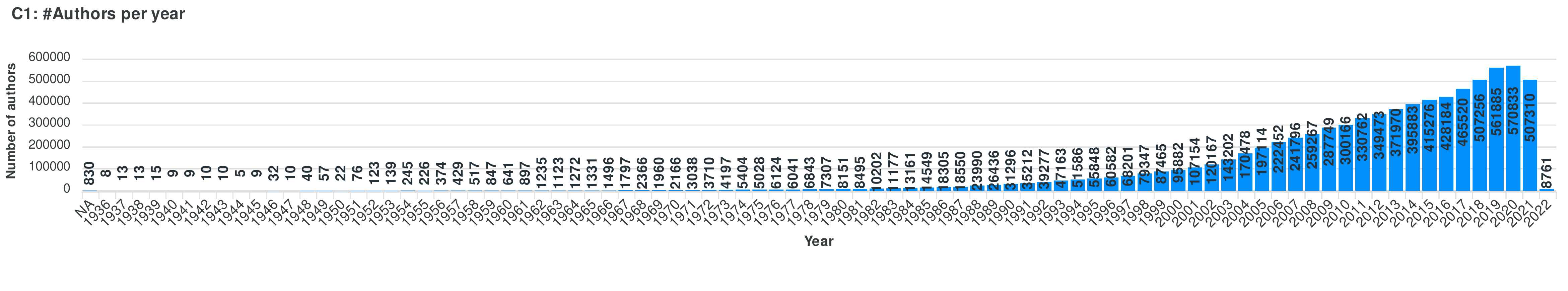}
        \caption{Number of unique authors per year.}
        \label{fig:authors_bar_app}
    \end{figure}

    \begin{figure}
        \centering
        \includegraphics[trim={0 1.2cm 0 0}, clip, width=\hsize]{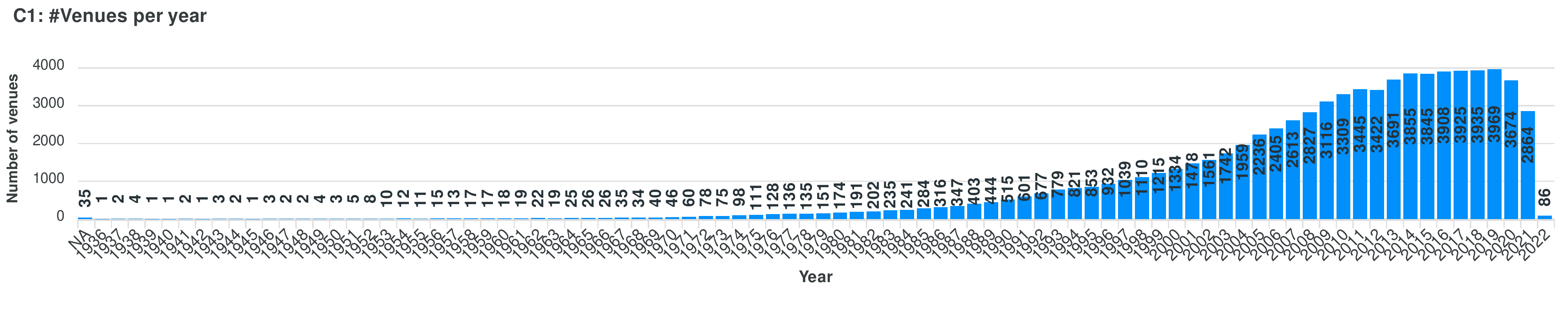}
        \caption{Number of unique venues per year.}
        \label{fig:venues_bar_app}
    \end{figure}

    \begin{figure}
        \centering
        \includegraphics[trim={0 1.2cm 0 0}, clip, width=\hsize]{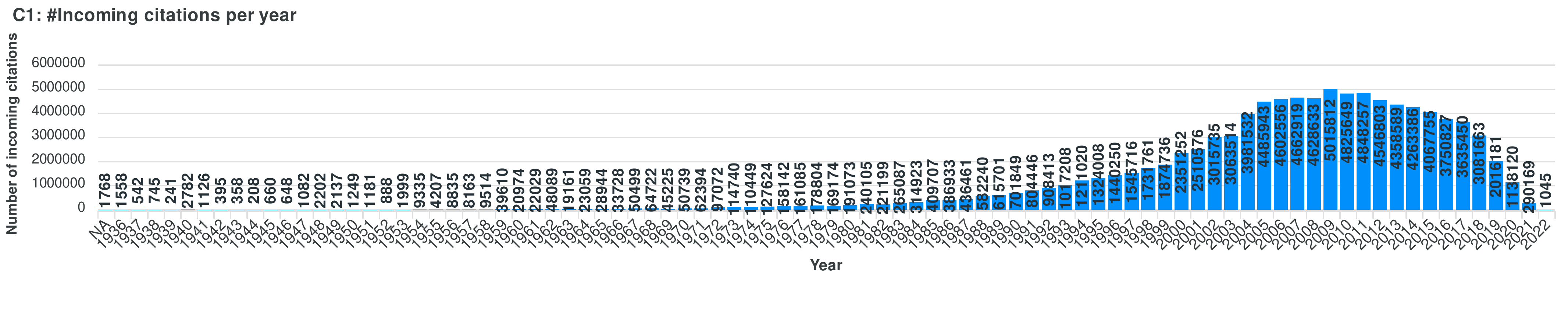}
        \caption{Number of incoming (received) citations per year.}
        \label{fig:citations_in_bar_app}
    \end{figure}

    \begin{figure}
        \centering
        \includegraphics[trim={0 1.2cm 0 0}, clip, width=\hsize]{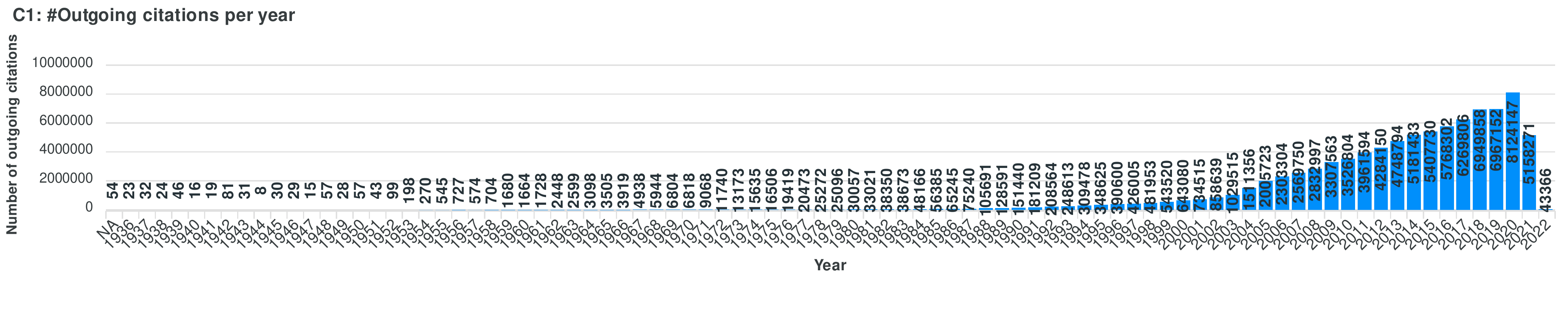}
        \caption{Number of outgoing citations (references) per year.}
        \label{fig:citations_out_bar_app}
    \end{figure}

    \begin{figure}
        \centering
        \includegraphics[trim={0 1.2cm 0 0}, clip, width=\hsize]{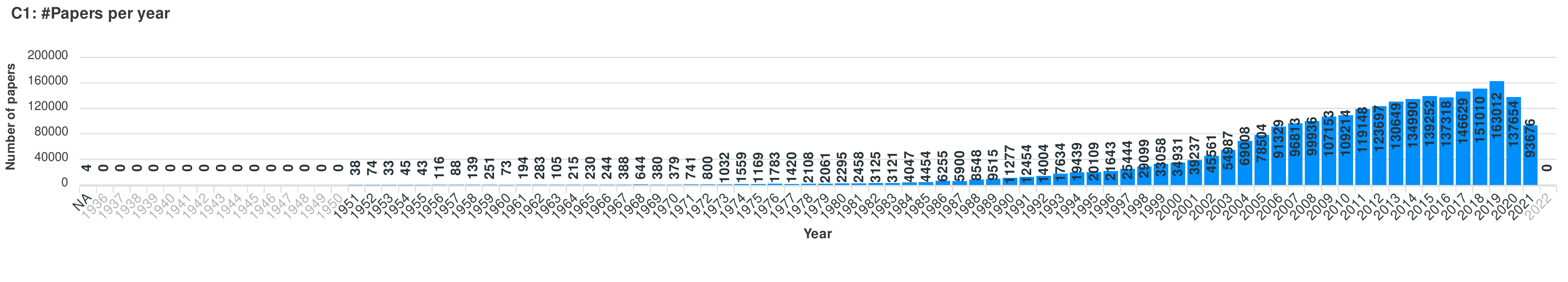}
        \caption{Number of conference papers per year.}
        \label{fig:types_inproceedings_bar_app}
    \end{figure}

    \begin{figure}
        \centering
        \includegraphics[trim={0 1.2cm 0 0}, clip, width=\hsize]{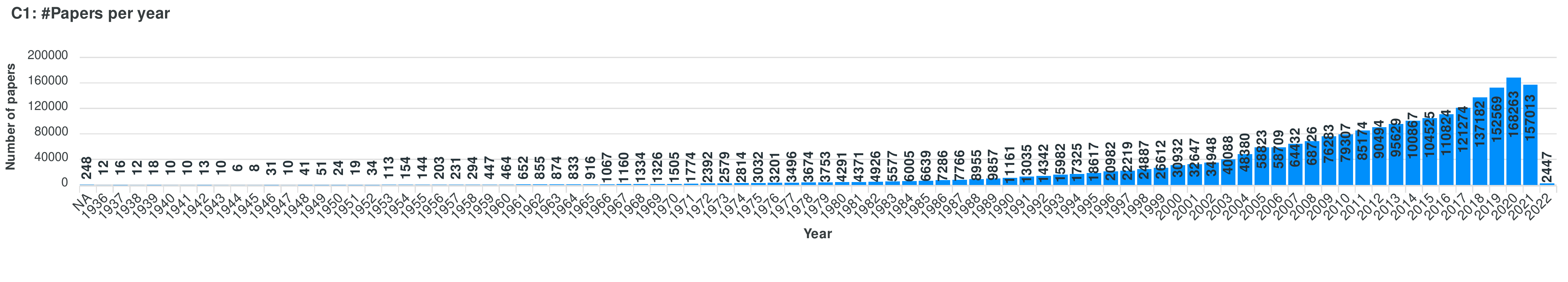}
        \caption{Number of journal articles per year.}
        \label{fig:types_articles_bar_app}
    \end{figure}
\end{landscape}
% }
	%%%%%%%%%%%%%%%%%%%%%%%%%%%%%%%%%%%%%%%%%%%%%%%%%%%%%%%%%%%%%%%%%%%%%%%%%%%%
	%% Version 2022-07-08
%% LaTeX-Vorlage für Abschlussarbeiten
%% Erstellt von Nils Potthoff, ab 2020 erneuert und ausgebaut von Simon Lohmann
%% Lehrstuhl Automatisierungstechnik/Informatik Bergische Universität Wuppertal
%%%%%%%%%%%%%%%%%%%%%%%%%%%%%%%%%%%%%%%%%%%%%%%%%%%%%%%%%%%%%%%%%%%%%%%%%%%%%%%%

%%%%%%%%%%%%%%%%%%%%%%%%%%%%%%%%%%%%%%%%%%%%%
%%% ! WARNUNG ! %%%%%%%%%%%%%%%%%%%%%%%%%%%%%
%%%%%%%%%%%%%%%%%%%%%%%%%%%%%%%%%%%%%%%%%%%%%
%%%  Diese Datei bitte nur bearbeiten,    %%%
%%%   wenn du ein LaTeX-Experte bist      %%%
%%%             U N D                     %%%
%%%  die Vorlage unbedingt ändern willst  %%%
%%%%%%%%%%%%%%%%%%%%%%%%%%%%%%%%%%%%%%%%%%%%%
%%%%%%%%%%%%%%%%%%%%%%%%%%%%%%%%%%%%%%%%%%%%%
%
%%%%%%%%%%%%%%%%%%%%%%%%%%%%%%%%%%%%%%%%%%%%%%%%%%%%%%%%%%%%%%%%%%%%%%%%%%%%%%%%
%%% DATEI-INFO %%%%%%%%%%%%%%%%%%%%%%%%%%%%%%%%%%%%%%%%%%%%%%%%%%%%%%%%%%%%%%%%%
%%%%%%%%%%%%%%%%%%%%%%%%%%%%%%%%%%%%%%%%%%%%%%%%%%%%%%%%%%%%%%%%%%%%%%%%%%%%%%%%
%%% Diese Datei generiert diverse Verzeichnisse %%%%%%%%%%%%%%%%%%%%%%%%%%%%%%%%
%%%%%%%%%%%%%%%%%%%%%%%%%%%%%%%%%%%%%%%%%%%%%%%%%%%%%%%%%%%%%%%%%%%%%%%%%%%%%%%%
%
%SONSTIGE VERZEICHNISSE
\clearpage{\pagestyle{empty}\cleardoublepage}%
\begingroup
	% Lokaler Override -> Verzeichnisse tragen sich gerne mal sowohl als aktuelles Kapitel als auch aktuelles Unterkapitel ein - dass steht in der Kopfzeile dann doppelt da und sieht hässlich aus!
	\let\oldmarkboth\markboth
	\renewcommand{\markboth}[2]{
		\oldmarkboth{#1}{}
	}

	\ifbool{verzeichnisseZusammenfassen}{% Seitenumbruch durch Abstand ersetzen, wenn gewünscht
		\def\clearpage{\vspace{2em}}%
	}{}%
	\ifbool{abbildungsverzeichnis}{%
		\clearpage%
		\verzeichnisEintrag{\listfigurename}{lof}
		\setlength{\cftfigindent}{0em}% Verzeichniseinträge ohne extra Indent darstellen
		\listoffigures%
	}{}%
	\ifbool{quellcodeverzeichnis}{%
		\clearpage%
		\verzeichnisEintrag{\lstlistlistingname}{listings}
		% indent wird in Präambel entfernt
		\lstlistoflistings%
	}{}%
\clearpage
	\ifbool{tabellenverzeichnis}{%
		\clearpage%
		\verzeichnisEintrag{\listtablename}{lot}
		\setlength{\cfttabindent}{0em}% Verzeichniseinträge ohne extra Indent darstellen
		\listoftables%
	}{}%
%
	
%		\clearpage%
%		%Redefinition des Stils von "siehe {anderer Begriff}":
%		\renewcommand\glsseeformat[3][\seename]{%
%			\\*% non breaking new line
%			\emph{#1} \glsseelist{#2}%
%		}
%
		% Anwendbare Styles (es gibt noch viel mehr): 
		% mcolist 			: mehrspaltig
		% mcolindexgroup 	: spalten+Anfangsbuchstaben über Gruppen
		% altlist			: 		
		% long				:
		%
		% nogroupskip deaktiviert den Abstand zwischen Gruppen (CLK und CRC gehören zu einer Gruppe, weil sie beide mit C beginnen)
	\ifbool{symbolverzeichnis}{%
		\vspace{-1em}
		\setlength{\glsdescwidth}{.75\linewidth}
		% TODO: Verzeichnisbreite dynamisch anpassen, z.B. mit https://tex.stackexchange.com/questions/174652/glossaries-package-how-to-format-the-positions-of-the-columns-and-width-of-the
		\verzeichnisEintrag{\glssymbolsgroupname}{listofsymbols}
		\printnoidxglossary[type=symbols,		style=altlongragged4col,nogroupskip]
		\vspace{-1em}
	}{}%
 % \renewcommand*\glspostdescription{\dotfill}
 % \setlength{\glsdescwidth}{3in}
 % \setlength{\glspagelistwidth}{3in}
 % \newglossarystyle{clong}{%
 % \renewenvironment{theglossary}%
 %     {\begin{longtable}{p{.3\linewidth}p{\glsdescwidth}}}%
 %     {\end{longtable}}%
 %  \renewcommand*{\glossaryheader}{}%
 %  \renewcommand*{\glsgroupheading}[1]{}%
 %  \renewcommand*{\glossaryentryfield}[5]{%
 %    \glstarget{##1}{##2} & ##3\glspostdescription\space ##5\\}%
 %  \renewcommand*{\glossarysubentryfield}[6]{%
 %     & \glstarget{##2}{\strut}##4\glspostdescription\space ##6\\}%
     
  %\renewcommand*{\glsgroupskip}{ & \\}%
% }
% \renewcommand{\glossarypreamble}{\glsfindwidesttoplevelname[\acronymtype] \setlength{\parskip}{0pt}}

% \renewcommand{\glossarypreamble}{\glsfindwidesttoplevelname[abbreviations] \setlength{\parskip}{0pt}} % <--------- THAT IS THE KEY, NOW USING alttree style.

% \renewcommand*{\glspostdescription}{\medskip}

% \newglossarystyle{owngloss}{%
%     \setglossarystyle{treegroup}%
%     \renewcommand*{\glossentry}[2]{%
%         \glsentryitem{##1}\textbf{\glstarget{##1}{\glossentryname{##1}}}%
%         \\ \glossentrydesc{##1} \\ \par
%     }%
% }

	\ifbool{abkuerzungsverzeichnis}{%
%		\setlength{\glsdescwidth}{.85\linewidth}
		% TODO: Verzeichnisbreite dynamisch anpassen, z.B. mit https://tex.stackexchange.com/questions/174652/glossaries-package-how-to-format-the-positions-of-the-columns-and-width-of-the
            \glsfindwidesttoplevelname[abbreviations] 
            \renewcommand*\glspostdescription{\dotfill\bigskip}
		\SaveTranslation{\abbreviationsnamesaved}{abbreviationsname}
		\verzeichnisEintragX{\GetTranslation{abbreviationsname}}{\abbreviationsnamesaved}{listofabbreviations}
		\printnoidxglossary[type=abbreviations,	style=alttree, nogroupskip, title={\GetTranslation{abbreviationsname}}]
		\vspace{-1em}
	}{}%
	\ifbool{akronymverzeichnis}{%
		\verzeichnisEintrag{\acronymname}{listofacronyms}
		\printnoidxglossary[type=acronym, style=mcolindex,nogroupskip]
		\vspace{-1em}
	}{}%
	\ifbool{glossar}{%
            \renewcommand*\glspostdescription{\dotfill}
		\verzeichnisEintrag{\glossaryname}{glossary}
		\printnoidxglossary[type=main, style=altlist,nogroupskip]
%		\par
	}{}%
	%	
	%%Literaturverzeichnis
	\ifbool{weiterfuehrendeLiteratur}{%
		\verzeichnisEintrag{\GetTranslation{Bibliography}}{literature}
		\printbibliography[title={\GetTranslation{Bibliography}},category=cited]%
		
		\SaveTranslation{\furtherreadingsaved}{FurtherReading}% Umweg, da ggf. Umlaute enthalten sind und das sonst direkt nicht klappt
		\verzeichnisEintragX{\GetTranslation{FurtherReading}}{\furtherreadingsaved}{furtherreading}
		\printbibliography[title={\GetTranslation{FurtherReading}},notcategory=cited]%
	}{%
		\verzeichnisEintrag{\GetTranslation{Bibliography}}{literature}
		\printbibliography[title={\GetTranslation{Bibliography}}]%

@article{aggarwal_has_2022,
  title = {Has the {{Future Started}}? {{The Current Growth}} of {{Artificial Intelligence}}, {{Machine Learning}}, and {{Deep Learning}}},
  shorttitle = {Has the {{Future Started}}?},
  author = {Aggarwal, Karan and Mijwil, Maad M. and Sonia and Al-Mistarehi, Abdel-Hameed and Alomari, Safwan and Gök, Murat and Alaabdin, Anas M. Zein and Abdulrhman, Safaa H.},
  date = {2022-01-30},
  journaltitle = {Iraqi Journal For Computer Science and Mathematics},
  volume = {3},
  number = {1},
  pages = {115--123},
  issn = {2788-7421},
  doi = {10.52866/ijcsm.2022.01.01.013},
  url = {https://journal.esj.edu.iq/index.php/IJCM/article/view/100},
  urldate = {2022-09-27},
  issue = {1},
  langid = {english}
}

@inproceedings{ammar_construction_2018,
  title = {Construction of the {{Literature Graph}} in {{Semantic Scholar}}},
  booktitle = {Proceedings of the 2018 {{Conference}} of the {{North American Chapter}} of the {{Association}} for {{Computational Linguistics}}: {{Human Language Technologies}}, {{Volume}} 3 ({{Industry Papers}})},
  author = {Ammar, Waleed and Groeneveld, Dirk and Bhagavatula, Chandra and Beltagy, Iz and Crawford, Miles and Downey, Doug and Dunkelberger, Jason and Elgohary, Ahmed and Feldman, Sergey and Ha, Vu and Kinney, Rodney and Kohlmeier, Sebastian and Lo, Kyle and Murray, Tyler and Ooi, Hsu-Han and Peters, Matthew and Power, Joanna and Skjonsberg, Sam and Wang, Lucy and Wilhelm, Chris and Yuan, Zheng and van Zuylen, Madeleine and Etzioni, Oren},
  options = {useprefix=true},
  date = {2018-06},
  pages = {84--91},
  publisher = {{Association for Computational Linguistics}},
  location = {{New Orleans - Louisiana}},
  doi = {10.18653/v1/N18-3011},
  url = {https://aclanthology.org/N18-3011},
  urldate = {2022-09-06},
  eventtitle = {{{NAACL-HLT}} 2018}
}

@inproceedings{anderson_towards_2012,
  title = {Towards a {{Computational History}} of the {{ACL}}: 1980-2008},
  shorttitle = {Towards a {{Computational History}} of the {{ACL}}},
  booktitle = {Proceedings of the {{ACL-2012 Special Workshop}} on {{Rediscovering}} 50 {{Years}} of {{Discoveries}}},
  author = {Anderson, Ashton and Jurafsky, Dan and McFarland, Daniel A.},
  date = {2012-07},
  pages = {13--21},
  publisher = {{Association for Computational Linguistics}},
  location = {{Jeju Island, Korea}},
  url = {https://aclanthology.org/W12-3202},
  urldate = {2022-10-10}
}

@article{anderson_towards_nodate,
  title = {Towards a {{Computational History}} of the {{ACL}}: 1980–2008},
  author = {Anderson, Ashton and McFarland, Dan and Jurafsky, Dan},
  pages = {9},
  langid = {english}
}

@online{arxiv_arxiv_2022,
  title = {{{arXiv}} Submission Rate Statistics | {{arXiv}} E-Print Repository},
  author = {{arXiv}},
  date = {2022-01-03},
  url = {https://arxiv.org/help/stats/2021_by_area},
  urldate = {2022-09-04}
}

@article{blei_latent_2003,
  title = {Latent {{Dirichlet Allocation}}},
  author = {Blei, David M. and Ng, Andrew Y. and Jordan, Michael I.},
  date = {2003},
  journaltitle = {Journal of Machine Learning Research},
  volume = {3},
  pages = {993--1022},
  issn = {ISSN 1533-7928},
  url = {https://jmlr.org/papers/v3/blei03a.html},
  urldate = {2022-08-26},
  issue = {Jan}
}

@article{bornmann_growth_2021,
  title = {Growth Rates of Modern Science: A Latent Piecewise Growth Curve Approach to Model Publication Numbers from Established and New Literature Databases},
  shorttitle = {Growth Rates of Modern Science},
  author = {Bornmann, Lutz and Haunschild, Robin and Mutz, Rüdiger},
  date = {2021-10-07},
  journaltitle = {Humanities and Social Sciences Communications},
  shortjournal = {Humanit Soc Sci Commun},
  volume = {8},
  number = {1},
  pages = {1--15},
  publisher = {{Palgrave}},
  issn = {2662-9992},
  doi = {10.1057/s41599-021-00903-w},
  url = {https://www.nature.com/articles/s41599-021-00903-w},
  urldate = {2022-09-04},
  issue = {1},
  langid = {english}
}

@book{bradshaw_mongodb_2019,
  title = {{{MongoDB}}: {{The Definitive Guide}}: {{Powerful}} and {{Scalable Data Storage}}},
  shorttitle = {{{MongoDB}}},
  author = {Bradshaw, Shannon and Brazil, Eoin and Chodorow, Kristina},
  date = {2019-12-09},
  eprint = {pIrCDwAAQBAJ},
  eprinttype = {googlebooks},
  publisher = {{"O'Reilly Media, Inc."}},
  isbn = {978-1-4919-5443-0},
  langid = {english},
  pagetotal = {514}
}

@article{burnham_scopus_2006,
  title = {Scopus Database: A Review},
  shorttitle = {Scopus Database},
  author = {Burnham, Judy F.},
  date = {2006-03-08},
  journaltitle = {Biomedical Digital Libraries},
  shortjournal = {Biomedical Digital Libraries},
  volume = {3},
  number = {1},
  pages = {1},
  issn = {1742-5581},
  doi = {10.1186/1742-5581-3-1},
  url = {https://doi.org/10.1186/1742-5581-3-1},
  urldate = {2022-10-12}
}

@article{cavacini_what_2015,
  title = {What Is the Best Database for Computer Science Journal Articles?},
  author = {Cavacini, Antonio},
  date = {2015-03-01},
  journaltitle = {Scientometrics},
  shortjournal = {Scientometrics},
  volume = {102},
  number = {3},
  pages = {2059--2071},
  issn = {1588-2861},
  doi = {10.1007/s11192-014-1506-1},
  url = {https://doi.org/10.1007/s11192-014-1506-1},
  urldate = {2022-08-28},
  langid = {english}
}

@article{chen_citespace_2006,
  title = {{{CiteSpace II}}: {{Detecting}} and Visualizing Emerging Trends and Transient Patterns in Scientific Literature},
  shorttitle = {{{CiteSpace II}}},
  author = {Chen, Chaomei},
  date = {2006-02-01},
  journaltitle = {Journal of the American Society for Information Science and Technology},
  shortjournal = {J. Am. Soc. Inf. Sci.},
  volume = {57},
  number = {3},
  pages = {359--377},
  issn = {15322882, 15322890},
  doi = {10.1002/asi.20317},
  url = {https://onlinelibrary.wiley.com/doi/10.1002/asi.20317},
  urldate = {2022-09-09},
  langid = {english}
}

@article{chen_searching_2004,
  title = {Searching for Intellectual Turning Points: {{Progressive}} Knowledge Domain Visualization},
  shorttitle = {Searching for Intellectual Turning Points},
  author = {Chen, Chaomei},
  date = {2004-04-06},
  journaltitle = {Proceedings of the National Academy of Sciences},
  volume = {101},
  pages = {5303--5310},
  publisher = {{Proceedings of the National Academy of Sciences}},
  doi = {10.1073/pnas.0307513100},
  url = {https://www.pnas.org/doi/10.1073/pnas.0307513100},
  urldate = {2022-09-09},
  issue = {suppl\_1}
}

@article{chen_structure_2010,
  title = {The {{Structure}} and {{Dynamics}} of {{Co-Citation Clusters}}: {{A Multiple-Perspective Co-Citation Analysis}}},
  shorttitle = {The {{Structure}} and {{Dynamics}} of {{Co-Citation Clusters}}},
  author = {Chen, Chaomei and Ibekwe-SanJuan, Fidelia and Hou, Jianhua},
  date = {2010-03-18},
  journaltitle = {Journal of the American Society for Information Science and Technology},
  shortjournal = {J. Am. Soc. Inf. Sci.},
  volume = {61},
  number = {7},
  eprint = {1002.1985},
  eprinttype = {arxiv},
  primaryclass = {cs},
  pages = {1386--1409},
  issn = {15322882, 15322890},
  doi = {10.1002/asi.21309},
  url = {http://arxiv.org/abs/1002.1985},
  urldate = {2022-09-09},
  archiveprefix = {arXiv}
}

@inproceedings{chuang_termite_2012,
  title = {Termite: Visualization Techniques for Assessing Textual Topic Models},
  shorttitle = {Termite},
  booktitle = {International {{Working Conference}} on {{Advanced Visual Interfaces}}, {{AVI}} 2012, {{Capri Island}}, {{Naples}}, {{Italy}}, {{May}} 22-25, 2012, {{Proceedings}}},
  author = {Chuang, Jason and Manning, Christopher D. and Heer, Jeffrey},
  editor = {Tortora, Genny and Levialdi, Stefano and Tucci, Maurizio},
  date = {2012},
  pages = {74--77},
  publisher = {{ACM}},
  doi = {10.1145/2254556.2254572}
}

@inproceedings{coskun_scientometrics-based_2019,
  title = {{{SCIENTOMETRICS-BASED STUDY OF COMPUTER SCIENCE AND INFORMATION SYSTEMS RESEARCH COMMUNITY MACRO LEVEL PROFILES}}},
  booktitle = {12th {{IADIS International Conference Information Systems}} 2019},
  author = {Coşkun, Erman and Özdağoğlu, Güzin and Damar, Muhammet and Çallı, Büşra Alma},
  date = {2019-04-11},
  pages = {180--188},
  publisher = {{IADIS Press}},
  doi = {10.33965/is2019_201905L023},
  url = {http://www.iadisportal.org/digital-library/scientometrics-based-study-of-computer-science-and-information-systems-research-community-macro-level-profiles},
  urldate = {2022-08-26},
  eventtitle = {12th {{IADIS International Conference Information Systems}} 2019},
  isbn = {978-989-8533-87-6}
}

@inproceedings{dror_hitchhikers_2018,
  title = {The {{Hitchhiker}}'s {{Guide}} to {{Testing Statistical Significance}} in {{Natural Language Processing}}},
  booktitle = {Proceedings of the 56th {{Annual Meeting}} of the {{Association}} for {{Computational Linguistics}} ({{Volume}} 1: {{Long Papers}})},
  author = {Dror, Rotem and Baumer, Gili and Shlomov, Segev and Reichart, Roi},
  date = {2018-07},
  pages = {1383--1392},
  publisher = {{Association for Computational Linguistics}},
  location = {{Melbourne, Australia}},
  doi = {10.18653/v1/P18-1128},
  url = {https://aclanthology.org/P18-1128},
  urldate = {2022-09-13},
  eventtitle = {{{ACL}} 2018}
}

@article{faiz_bibliometric_2020,
  title = {Bibliometric {{Analysis}} of {{Computer Science Literature}} of {{Pakistan}}},
  author = {Faiz, Adnan},
  date = {2020},
  volume = {8},
  number = {2},
  pages = {21},
  langid = {english}
}

@inproceedings{fatima_google_2020,
  title = {Google {{Scholar}} vs. {{Dblp}} vs. {{Microsoft Academic Search}}: {{An Indexing Comparison}} for {{Software Engineering Literature}}},
  shorttitle = {Google {{Scholar}} vs. {{Dblp}} vs. {{Microsoft Academic Search}}},
  booktitle = {2020 {{IEEE}} 44th {{Annual Computers}}, {{Software}}, and {{Applications Conference}} ({{COMPSAC}})},
  author = {Fatima, Rubia and Yasin, Affan and Liu, Lin and Wang, Jianmin},
  date = {2020-07},
  pages = {1097--1098},
  issn = {0730-3157},
  doi = {10.1109/COMPSAC48688.2020.0-122},
  eventtitle = {2020 {{IEEE}} 44th {{Annual Computers}}, {{Software}}, and {{Applications Conference}} ({{COMPSAC}})}
}

@article{fiala_computer_2017,
  title = {Computer {{Science Papers}} in {{Web}} of {{Science}}: {{A Bibliometric Analysis}}},
  shorttitle = {Computer {{Science Papers}} in {{Web}} of {{Science}}},
  author = {Fiala, Dalibor and Tutoky, Gabriel},
  date = {2017-09-29},
  journaltitle = {Publications},
  shortjournal = {Publications},
  volume = {5},
  doi = {10.3390/publications5040023}
}

@thesis{fielding_information_2000,
  title = {Architectural Styles and the Design of Network-based Software Architectures},
  author = {Fielding, Roy Thomas},
  date = {2000},
  langid = {english}
}

@article{franceschet_collaboration_2011,
  title = {Collaboration in Computer Science: {{A}} Network Science Approach},
  shorttitle = {Collaboration in Computer Science},
  author = {Franceschet, Massimo},
  date = {2011-10-01},
  journaltitle = {Journal of the American Society for Information Science and Technology},
  shortjournal = {J. Am. Soc. Inf. Sci. Technol.},
  volume = {62},
  number = {10},
  pages = {1992--2012},
  issn = {1532-2882},
  doi = {10.1002/asi.21614},
  url = {https://doi.org/10.1002/asi.21614},
  urldate = {2022-08-28}
}

@article{franceschet_comparison_2010-1,
  title = {A Comparison of Bibliometric Indicators for Computer Science Scholars and Journals on {{Web}} of {{Science}} and {{Google Scholar}}},
  author = {Franceschet, Massimo},
  date = {2010-04-01},
  journaltitle = {Scientometrics},
  shortjournal = {Scientometrics},
  volume = {83},
  number = {1},
  pages = {243--258},
  issn = {1588-2861},
  doi = {10.1007/s11192-009-0021-2},
  url = {https://doi.org/10.1007/s11192-009-0021-2},
  urldate = {2022-08-31},
  langid = {english}
}

@article{franceschet_role_2010,
  title = {The Role of Conference Publications in {{CS}}},
  author = {Franceschet, Massimo},
  date = {2010-12-01},
  journaltitle = {Communications of the ACM},
  shortjournal = {Commun. ACM},
  volume = {53},
  number = {12},
  pages = {129--132},
  issn = {0001-0782},
  doi = {10.1145/1859204.1859234},
  url = {https://doi.org/10.1145/1859204.1859234},
  urldate = {2022-08-28}
}

@article{freyne_relative_2010,
  title = {Relative Status of Journal and Conference Publications in Computer Science},
  author = {Freyne, Jill and Coyle, Lorcan and Smyth, Barry and Cunningham, Padraig},
  date = {2010-11-01},
  journaltitle = {Communications of the ACM},
  shortjournal = {Commun. ACM},
  volume = {53},
  number = {11},
  pages = {124--132},
  issn = {0001-0782},
  doi = {10.1145/1839676.1839701},
  url = {https://doi.org/10.1145/1839676.1839701},
  urldate = {2022-09-14}
}

@article{gusenbauer_google_2019,
  title = {Google {{Scholar}} to Overshadow Them All? {{Comparing}} the Sizes of 12 Academic Search Engines and Bibliographic Databases},
  shorttitle = {Google {{Scholar}} to Overshadow Them All?},
  author = {Gusenbauer, Michael},
  date = {2019-01-01},
  journaltitle = {Scientometrics},
  shortjournal = {Scientometrics},
  volume = {118},
  number = {1},
  pages = {177--214},
  issn = {1588-2861},
  doi = {10.1007/s11192-018-2958-5},
  url = {https://doi.org/10.1007/s11192-018-2958-5},
  urldate = {2022-09-12},
  langid = {english}
}

@inproceedings{gyorodi_comparative_2015,
  title = {A Comparative Study: {{MongoDB}} vs. {{MySQL}}},
  shorttitle = {A Comparative Study},
  booktitle = {2015 13th {{International Conference}} on {{Engineering}} of {{Modern Electric Systems}} ({{EMES}})},
  author = {Győrödi, Cornelia and Győrödi, Robert and Pecherle, George and Olah, Andrada},
  date = {2015-06},
  pages = {1--6},
  doi = {10.1109/EMES.2015.7158433},
  eventtitle = {2015 13th {{International Conference}} on {{Engineering}} of {{Modern Electric Systems}} ({{EMES}})}
}

@article{halpern_corr_2000,
  title = {{{CoRR}}: A Computing Research Repository},
  shorttitle = {{{CoRR}}},
  author = {Halpern, Joseph Y.},
  date = {2000-05-01},
  journaltitle = {ACM Journal of Computer Documentation},
  shortjournal = {ACM J. Comput. Doc.},
  volume = {24},
  number = {2},
  pages = {41--48},
  issn = {1527-6805},
  doi = {10.1145/337271.337274},
  url = {https://doi.org/10.1145/337271.337274},
  urldate = {2022-09-07}
}

@article{ivancheva_scientometrics_2008,
  title = {Scientometrics {{Today}}: {{A Methodological Overview}}},
  shorttitle = {Scientometrics {{Today}}},
  author = {Ivancheva, Ludmila},
  date = {2008-12},
  journaltitle = {Collnet Journal of Scientometrics and Information Management},
  shortjournal = {Collnet Journal of Scientometrics and Information Management},
  volume = {2},
  number = {2},
  pages = {47--56},
  issn = {0973-7766, 2168-930X},
  doi = {10.1080/09737766.2008.10700853},
  url = {http://www.tandfonline.com/doi/abs/10.1080/09737766.2008.10700853},
  urldate = {2022-10-18},
  langid = {english}
}

@article{jelodar_latent_2019,
  title = {Latent {{Dirichlet}} Allocation ({{LDA}}) and Topic Modeling: Models, Applications, a Survey},
  shorttitle = {Latent {{Dirichlet}} Allocation ({{LDA}}) and Topic Modeling},
  author = {Jelodar, Hamed and Wang, Yongli and Yuan, Chi and Feng, Xia and Jiang, Xiahui and Li, Yanchao and Zhao, Liang},
  date = {2019-06-01},
  journaltitle = {Multimedia Tools and Applications},
  shortjournal = {Multimed Tools Appl},
  volume = {78},
  number = {11},
  pages = {15169--15211},
  issn = {1573-7721},
  doi = {10.1007/s11042-018-6894-4},
  url = {https://doi.org/10.1007/s11042-018-6894-4},
  urldate = {2022-10-17},
  langid = {english}
}

@article{johnson_stm_2018,
  title = {The {{STM Report}}, 5th Edition: {{An}} Overview of Scientific and Scholarly Publishing},
  shorttitle = {The {{STM Report}}, 5th Edition},
  author = {Johnson, Rob and Watkinson, Anthony and Mabe, Michael},
  date = {2018-10-15},
  publisher = {{STM: International Association of Scientific, Technical and Medical Publishers}},
  url = {https://policycommons.net/artifacts/1575771/2018_10_04_stm_report_2018/2265545/},
  urldate = {2022-09-04},
  langid = {english}
}

@article{katuk_scientometric_2020,
  title = {A {{SCIENTOMETRIC ANALYSIS OF THE EMERGING TOPICS IN GENERAL COMPUTER SCIENCE}}},
  author = {Katuk, Norliza and Ku-Mahamud, Ku Ruhana and Zakaria, Nur Haryani and Jabbar, Ayad Mohammed},
  date = {2020-08-20},
  journaltitle = {Journal of Information and Communication Technology},
  volume = {19},
  number = {4},
  pages = {583--622},
  issn = {2180-3862},
  doi = {10.32890/jict2020.19.4.6},
  url = {https://e-journal.uum.edu.my/index.php/jict/article/view/jict2020.19.4.6},
  urldate = {2022-08-31},
  issue = {4},
  langid = {english}
}

@article{kim_author-based_2019,
  title = {Author-Based Analysis of Conference versus Journal Publication in Computer Science},
  author = {Kim, Jinseok},
  date = {2019},
  journaltitle = {Journal of the Association for Information Science and Technology},
  volume = {70},
  number = {1},
  pages = {71--82},
  issn = {2330-1643},
  doi = {10.1002/asi.24079},
  url = {https://onlinelibrary.wiley.com/doi/abs/10.1002/asi.24079},
  urldate = {2022-08-28},
  langid = {english}
}

@article{kopecky_history_2014,
  title = {A History and Future of {{Web APIs}}},
  author = {Kopecký, Jacek and Fremantle, Paul and Boakes, Rich},
  date = {2014-01-28},
  journaltitle = {it - Information Technology},
  shortjournal = {it - Information Technology},
  volume = {56},
  doi = {10.1515/itit-2013-1035}
}

@article{kousha_covid-19_2020,
  title = {{{COVID-19}} Publications: {{Database}} Coverage, Citations, Readers, Tweets, News, {{Facebook}} Walls, {{Reddit}} Posts},
  shorttitle = {{{COVID-19}} Publications},
  author = {Kousha, Kayvan and Thelwall, Mike},
  date = {2020-08-01},
  journaltitle = {Quantitative Science Studies},
  shortjournal = {Quantitative Science Studies},
  volume = {1},
  number = {3},
  pages = {1068--1091},
  issn = {2641-3337},
  doi = {10.1162/qss_a_00066},
  url = {https://doi.org/10.1162/qss_a_00066},
  urldate = {2022-09-27}
}

@article{kumar_scientometrics_2005,
  title = {Scientometrics of Computer Science Research in {{India}} and {{China}}},
  author = {Kumar, Suresh and Garg, Kailash},
  date = {2005-08-01},
  journaltitle = {Scientometrics},
  shortjournal = {Scientometrics},
  volume = {64},
  pages = {121--132},
  doi = {10.1007/s11192-005-0244-9}
}

@article{kumari_scientometric_2020,
  title = {Scientometric {{Analysis}} of {{Computer Science Publications}} in {{Journals}} and {{Conferences}} with {{Publication Patterns}}},
  author = {Kumari, Priti and Kumar, Rajeev},
  date = {2020},
  journaltitle = {Journal of Scientometric Research},
  volume = {9},
  number = {1},
  pages = {9},
  langid = {english}
}

@inproceedings{lavergne_practical_2010,
  title = {Practical {{Very Large Scale CRFs}}},
  booktitle = {Proceedings of the 48th {{Annual Meeting}} of the {{Association}} for {{Computational Linguistics}}},
  author = {Lavergne, Thomas and Cappé, Olivier and Yvon, François},
  date = {2010-07},
  pages = {504--513},
  publisher = {{Association for Computational Linguistics}},
  location = {{Uppsala, Sweden}},
  url = {https://aclanthology.org/P10-1052},
  urldate = {2022-09-13},
  eventtitle = {{{ACL}} 2010}
}

@inproceedings{ley_dblp_2002,
  title = {The {{DBLP Computer Science Bibliography}}: {{Evolution}}, {{Research Issues}}, {{Perspectives}}},
  shorttitle = {The {{DBLP Computer Science Bibliography}}},
  booktitle = {String {{Processing}} and {{Information Retrieval}}},
  author = {Ley, Michael},
  editor = {Laender, Alberto H. F. and Oliveira, Arlindo L.},
  date = {2002},
  series = {Lecture {{Notes}} in {{Computer Science}}},
  pages = {1--10},
  publisher = {{Springer}},
  location = {{Berlin, Heidelberg}},
  doi = {10.1007/3-540-45735-6_1},
  isbn = {978-3-540-45735-0},
  langid = {english}
}

@article{ley_dblp_2009,
  title = {{{DBLP}}: Some Lessons Learned},
  shorttitle = {{{DBLP}}},
  author = {Ley, Michael},
  date = {2009-08-01},
  journaltitle = {Proceedings of the VLDB Endowment},
  shortjournal = {Proc. VLDB Endow.},
  volume = {2},
  number = {2},
  pages = {1493--1500},
  issn = {2150-8097},
  doi = {10.14778/1687553.1687577},
  url = {https://doi.org/10.14778/1687553.1687577},
  urldate = {2022-08-28}
}

@inproceedings{ley_trierer_1997,
  title = {Die Trierer Informatik-Bibliographie DBLP},
  booktitle = {Informatik ’97 Informatik als Innovationsmotor},
  author = {Ley, Michael},
  editor = {Jarke, Matthias and Pasedach, Klaus and Pohl, Klaus},
  date = {1997},
  series = {Informatik aktuell},
  pages = {257--266},
  publisher = {{Springer}},
  location = {{Berlin, Heidelberg}},
  doi = {10.1007/978-3-642-60831-5_34},
  isbn = {978-3-642-60831-5},
  langid = {ngerman}
}

@inproceedings{lo_s2orc_2020,
  title = {{{S2ORC}}: {{The Semantic Scholar Open Research Corpus}}},
  shorttitle = {{{S2ORC}}},
  booktitle = {Proceedings of the 58th {{Annual Meeting}} of the {{Association}} for {{Computational Linguistics}}},
  author = {Lo, Kyle and Wang, Lucy Lu and Neumann, Mark and Kinney, Rodney and Weld, Daniel},
  date = {2020-07},
  pages = {4969--4983},
  publisher = {{Association for Computational Linguistics}},
  location = {{Online}},
  doi = {10.18653/v1/2020.acl-main.447},
  url = {https://aclanthology.org/2020.acl-main.447},
  urldate = {2022-09-06},
  eventtitle = {{{ACL}} 2020}
}

@software{lopez_grobid_2022,
  title = {{{GROBID}}},
  author = {Lopez, Patrice},
  date = {2022-08-26},
  origdate = {2012-09-13},
  url = {https://github.com/kermitt2/grobid},
  urldate = {2022-08-26}
}

@article{mariani_nlp4nlp_2019,
  title = {The {{NLP4NLP Corpus}} ({{I}}): 50 {{Years}} of {{Publication}}, {{Collaboration}} and {{Citation}} in {{Speech}} and {{Language Processing}}},
  shorttitle = {The {{NLP4NLP Corpus}} ({{I}})},
  author = {Mariani, Joseph and Francopoulo, Gil and Paroubek, Patrick},
  date = {2019-02-07},
  journaltitle = {Frontiers in Research Metrics and Analytics},
  shortjournal = {Front. Res. Metr. Anal.},
  volume = {3},
  pages = {36},
  issn = {2504-0537},
  doi = {10.3389/frma.2018.00036},
  url = {https://www.frontiersin.org/article/10.3389/frma.2018.00036/full},
  urldate = {2022-09-20}
}

@article{mariani_nlp4nlp_2019_2,
  title = {The {{NLP4NLP Corpus}} ({{II}}): 50 {{Years}} of {{Research}} in {{Speech}} and {{Language Processing}}},
  shorttitle = {The {{NLP4NLP Corpus}} ({{II}})},
  author = {Mariani, Joseph and Francopoulo, Gil and Paroubek, Patrick and Vernier, Frédéric},
  date = {2019-02-07},
  journaltitle = {Frontiers in Research Metrics and Analytics},
  shortjournal = {Front. Res. Metr. Anal.},
  volume = {3},
  pages = {37},
  issn = {2504-0537},
  doi = {10.3389/frma.2018.00037},
  url = {https://www.frontiersin.org/article/10.3389/frma.2018.00037/full},
  urldate = {2022-09-20}
}

@inproceedings{mcburney_what_2002,
  title = {What Is Bibliometrics and Why Should You Care?},
  author = {McBurney, Melissa and Novak, Pamela},
  date = {2002-02-01},
  pages = {108--114},
  doi = {10.1109/IPCC.2002.1049094},
  eventtitle = {{{IEEE International Professional Communication Conference}}},
  isbn = {978-0-7803-7591-8}
}

@article{michels_systematic_2014,
  title = {Systematic Analysis of Coverage and Usage of Conference Proceedings in Web of Science},
  author = {Michels, Carolin and Fu, Jun-Ying},
  date = {2014-08-01},
  journaltitle = {Scientometrics},
  shortjournal = {Scientometrics},
  volume = {100},
  number = {2},
  pages = {307--327},
  issn = {1588-2861},
  doi = {10.1007/s11192-014-1309-4},
  url = {https://doi.org/10.1007/s11192-014-1309-4},
  urldate = {2022-09-14},
  langid = {english}
}

@article{mingers_review_2015,
  title = {A Review of Theory and Practice in Scientometrics},
  author = {Mingers, John and Leydesdorff, Loet},
  date = {2015-10-01},
  journaltitle = {European Journal of Operational Research},
  shortjournal = {European Journal of Operational Research},
  volume = {246},
  number = {1},
  pages = {1--19},
  issn = {0377-2217},
  doi = {10.1016/j.ejor.2015.04.002},
  url = {https://www.sciencedirect.com/science/article/pii/S037722171500274X},
  urldate = {2022-10-18},
  langid = {english}
}

@inproceedings{mohammad_examining_2020,
  title = {Examining {{Citations}} of {{Natural Language Processing Literature}}},
  booktitle = {Proceedings of the 58th {{Annual Meeting}} of the {{Association}} for {{Computational Linguistics}}},
  author = {Mohammad, Saif M.},
  date = {2020},
  pages = {5199--5209},
  publisher = {{Association for Computational Linguistics}},
  location = {{Online}},
  doi = {10.18653/v1/2020.acl-main.464},
  url = {https://www.aclweb.org/anthology/2020.acl-main.464},
  urldate = {2022-08-26},
  eventtitle = {Proceedings of the 58th {{Annual Meeting}} of the {{Association}} for {{Computational Linguistics}}},
  langid = {english}
}

@inproceedings{mohammad_gender_2020,
  title = {Gender {{Gap}} in {{Natural Language Processing Research}}: {{Disparities}} in {{Authorship}} and {{Citations}}},
  shorttitle = {Gender {{Gap}} in {{Natural Language Processing Research}}},
  booktitle = {Proceedings of the 58th {{Annual Meeting}} of the {{Association}} for {{Computational Linguistics}}},
  author = {Mohammad, Saif M.},
  date = {2020-07},
  pages = {7860--7870},
  publisher = {{Association for Computational Linguistics}},
  location = {{Online}},
  doi = {10.18653/v1/2020.acl-main.702},
  url = {https://aclanthology.org/2020.acl-main.702},
  urldate = {2022-08-26},
  eventtitle = {{{ACL}} 2020}
}

@inproceedings{mohammad_nlp_2020_data,
  title = {{{NLP Scholar}}: {{A Dataset}} for {{Examining}} the {{State}} of {{NLP Research}}},
  shorttitle = {{{NLP Scholar}}},
  booktitle = {Proceedings of the 12th {{Language Resources}} and {{Evaluation Conference}}},
  author = {Mohammad, Saif M.},
  date = {2020-05},
  pages = {868--877},
  publisher = {{European Language Resources Association}},
  location = {{Marseille, France}},
  url = {https://aclanthology.org/2020.lrec-1.109},
  urldate = {2022-08-26},
  eventtitle = {{{LREC}} 2020},
  isbn = {979-10-95546-34-4},
  langid = {english}
}

@inproceedings{mohammad_nlp_2020_viz,
  title = {{{NLP Scholar}}: {{An Interactive Visual Explorer}} for {{Natural Language Processing Literature}}},
  shorttitle = {{{NLP Scholar}}},
  booktitle = {Proceedings of the 58th {{Annual Meeting}} of the {{Association}} for {{Computational Linguistics}}: {{System Demonstrations}}},
  author = {Mohammad, Saif M.},
  date = {2020},
  pages = {232--255},
  publisher = {{Association for Computational Linguistics}},
  location = {{Online}},
  doi = {10.18653/v1/2020.acl-demos.27},
  url = {https://www.aclweb.org/anthology/2020.acl-demos.27},
  urldate = {2022-08-26},
  eventtitle = {Proceedings of the 58th {{Annual Meeting}} of the {{Association}} for {{Computational Linguistics}}: {{System Demonstrations}}},
  langid = {english}
}

@misc{mohammad_state_2019,
  title = {The {{State}} of {{NLP Literature}}: {{A Diachronic Analysis}} of the {{ACL Anthology}}},
  shorttitle = {The {{State}} of {{NLP Literature}}},
  author = {Mohammad, Saif M.},
  date = {2019-11-08},
  number = {arXiv:1911.03562},
  publisher = {{arXiv}},
  doi = {10.48550/arXiv.1911.03562},
  url = {http://arxiv.org/abs/1911.03562},
  urldate = {2022-08-26}
}

@article{mohr_introductiontopic_2013,
  title = {Introduction—{{Topic}} Models: {{What}} They Are and Why They Matter},
  shorttitle = {Introduction—{{Topic}} Models},
  author = {Mohr, John W. and Bogdanov, Petko},
  date = {2013-12-01},
  journaltitle = {Poetics},
  shortjournal = {Poetics},
  series = {Topic {{Models}} and the {{Cultural Sciences}}},
  volume = {41},
  number = {6},
  pages = {545--569},
  issn = {0304-422X},
  doi = {10.1016/j.poetic.2013.10.001},
  url = {https://www.sciencedirect.com/science/article/pii/S0304422X13000685},
  urldate = {2022-10-17},
  langid = {english}
}

@report{nalimov_measurement_1971,
  title = {Measurement of {{Science}}. {{Study}} of the {{Development}} of {{Science}} as an {{Information Process}}},
  author = {Nalimov, V. V. and Mulchenko, Z. M.},
  date = {1971-10},
  institution = {{National Technical Information Service, Springfield, Virginia 22151 (AD 735 634, MF-\$0}},
  langid = {english}
}

@article{newman_information_2013,
  title = {Information Workflow of Academic Researchers in the Evolving Information Environment: An Interview Study},
  shorttitle = {Information Workflow of Academic Researchers in the Evolving Information Environment},
  author = {Newman, Michael L. and Sack, John},
  date = {2013},
  journaltitle = {Learned Publishing},
  volume = {26},
  number = {2},
  pages = {123--131},
  issn = {1741-4857},
  doi = {10.1087/20130208},
  url = {https://onlinelibrary.wiley.com/doi/abs/10.1087/20130208},
  urldate = {2022-09-04},
  langid = {english}
}

@report{noauthor_measurement_nodate,
  title = {Measurement of {{Science}}. {{Study}} of the {{Development}} of {{Science}} as an {{Information Process}},},
  url = {https://apps.dtic.mil/sti/citations/AD0735634},
  urldate = {2022-10-18},
  langid = {english}
}

@online{nwb_team_network_2006,
  title = {Network {{Workbench Tool}}},
  author = {{NWB Team}},
  date = {2006},
  url = {https://nwb.cns.iu.edu/},
  urldate = {2022-09-10}
}

@report{park_software_1992,
  title = {Software {{Size Measurement}}: {{A Framework}} for {{Counting Source Statements}}},
  shorttitle = {Software {{Size Measurement}}},
  author = {Park, Robert E.},
  date = {1992}
}

@inproceedings{parmar_nlpexplorer_2020,
  title = {{{NLPExplorer}}: {{Exploring}} the {{Universe}} of {{NLP Papers}}},
  shorttitle = {{{NLPExplorer}}},
  booktitle = {Advances in {{Information Retrieval}}},
  author = {Parmar, Monarch and Jain, Naman and Jain, Pranjali and Jayakrishna Sahit, P. and Pachpande, Soham and Singh, Shruti and Singh, Mayank},
  editor = {Jose, Joemon M. and Yilmaz, Emine and Magalhães, João and Castells, Pablo and Ferro, Nicola and Silva, Mário J. and Martins, Flávio},
  date = {2020},
  series = {Lecture {{Notes}} in {{Computer Science}}},
  pages = {476--480},
  publisher = {{Springer International Publishing}},
  location = {{Cham}},
  doi = {10.1007/978-3-030-45442-5_61},
  isbn = {978-3-030-45442-5},
  langid = {english}
}

@inproceedings{radev_acl_2009,
  title = {The {{ACL Anthology Network}}},
  booktitle = {Proceedings of the 2009 {{Workshop}} on {{Text}} and {{Citation Analysis}} for {{Scholarly Digital Libraries}} ({{NLPIR4DL}})},
  author = {Radev, Dragomir R. and Muthukrishnan, Pradeep and Qazvinian, Vahed},
  date = {2009-08},
  pages = {54--61},
  publisher = {{Association for Computational Linguistics}},
  location = {{Suntec City, Singapore}},
  url = {https://aclanthology.org/W09-3607},
  urldate = {2022-09-13}
}

@article{rahm_citation_2005,
  title = {Citation Analysis of Database Publications},
  author = {Rahm, Erhard and Thor, Andreas},
  date = {2005-12-01},
  journaltitle = {ACM SIGMOD Record},
  shortjournal = {SIGMOD Rec.},
  volume = {34},
  number = {4},
  pages = {48--53},
  issn = {0163-5808},
  doi = {10.1145/1107499.1107505},
  url = {https://doi.org/10.1145/1107499.1107505},
  urldate = {2022-08-28}
}

@article{rahm_comparing_2008,
  title = {Comparing the {{Scientific Impact}} of {{Conference}} and {{Journal Publications}} in {{Computer Science}}},
  author = {Rahm, Erhard},
  date = {2008-10-02},
  journaltitle = {Information Services and Use},
  shortjournal = {Information Services and Use},
  volume = {28},
  doi = {10.3233/ISU-2008-0562}
}

@inproceedings{rehurek_software_2010,
  title = {Software Framework for Topic Modelling with Large Corpora},
  booktitle = {Proceedings of the {{LREC}} 2010 Workshop on New Challenges for {{NLP}} Frameworks},
  author = {Řehůřek, Radim and Sojka, Petr},
  date = {2010-05-22},
  pages = {45--50},
  publisher = {{ELRA}},
  location = {{Valletta, Malta}},
  langid = {english}
}

@misc{ruas_cs-insights_2022,
  title = {{{CS-Insights}}: {{A System}} for {{Analyzing Computer Science Research}}},
  shorttitle = {{{CS-Insights}}},
  author = {Ruas, Terry and Wahle, Jan Philip and Küll, Lennart and Mohammad, Saif M. and Gipp, Bela},
  date = {2022-10-13},
  number = {arXiv:2210.06878},
  publisher = {{arXiv}},
  doi = {10.48550/arXiv.2210.06878},
  url = {http://arxiv.org/abs/2210.06878},
  urldate = {2022-10-18},
  addendum = {Submission planned for EACL'23 (System Demonstrations)}
}

@article{saheb_mapping_2021,
  title = {Mapping Research Strands of Ethics of Artificial Intelligence in Healthcare: {{A}} Bibliometric and Content Analysis},
  shorttitle = {Mapping Research Strands of Ethics of Artificial Intelligence in Healthcare},
  author = {Saheb, Tahereh and Saheb, Tayebeh and Carpenter, David O.},
  date = {2021-08-01},
  journaltitle = {Computers in Biology and Medicine},
  shortjournal = {Computers in Biology and Medicine},
  volume = {135},
  pages = {104660},
  issn = {0010-4825},
  doi = {10.1016/j.compbiomed.2021.104660},
  url = {https://www.sciencedirect.com/science/article/pii/S0010482521004546},
  urldate = {2022-09-10},
  langid = {english}
}

@article{sengupta_bibliometrics_1992,
  title = {Bibliometrics, {{Informetrics}}, {{Scientometrics}} and {{Librametrics}}: {{An Overview}}},
  shorttitle = {Bibliometrics, {{Informetrics}}, {{Scientometrics}} and {{Librametrics}}},
  author = {Sengupta, I. N.},
  date = {1992-01-01},
  volume = {42},
  number = {2},
  pages = {75--98},
  publisher = {{De Gruyter Saur}},
  issn = {1865-8423},
  doi = {10.1515/libr.1992.42.2.75},
  url = {https://www.degruyter.com/document/doi/10.1515/libr.1992.42.2.75/html},
  urldate = {2022-10-18},
  langid = {english}
}

@inproceedings{sharma_drift_2021,
  title = {{{DRIFT}}: {{A Toolkit}} for {{Diachronic Analysis}} of {{Scientific Literature}}},
  shorttitle = {{{DRIFT}}},
  booktitle = {Proceedings of the 2021 {{Conference}} on {{Empirical Methods}} in {{Natural Language Processing}}: {{System Demonstrations}}},
  author = {Sharma, Abheesht and Chhablani, Gunjan and Pandey, Harshit and Patil, Rajaswa},
  date = {2021-11},
  pages = {361--371},
  publisher = {{Association for Computational Linguistics}},
  location = {{Online and Punta Cana, Dominican Republic}},
  doi = {10.18653/v1/2021.emnlp-demo.40},
  url = {https://aclanthology.org/2021.emnlp-demo.40},
  urldate = {2022-08-26}
}

@book{shneiderman_designing_2018,
  title = {Designing the User Interface: Strategies for Effective Human-Computer Interaction},
  shorttitle = {Designing the User Interface},
  author = {Shneiderman, Ben and Plaisant, Catherine and Cohen, Maxine and Jacobs, Steven and Elmqvist, Niklas},
  date = {2018},
  edition = {Sixth edition, global edition},
  publisher = {{Pearson}},
  location = {{Boston Columbus Indianapolis New York San Francisco Hoboken Amsterdam Cape Town Dubai London Madrid Milan Munich Paris Montréal Toronto Delhi Mexico City Sao Paulo Sydney Hong Kong Seoul Singapore Taipei Tokyo}},
  isbn = {978-1-292-15391-9},
  langid = {english},
  pagetotal = {622}
}

@inproceedings{sievert_ldavis_2014,
  title = {{{LDAvis}}: {{A}} Method for Visualizing and Interpreting Topics},
  shorttitle = {{{LDAvis}}},
  booktitle = {Proceedings of the {{Workshop}} on {{Interactive Language Learning}}, {{Visualization}}, and {{Interfaces}}},
  author = {Sievert, Carson and Shirley, Kenneth},
  date = {2014-06},
  pages = {63--70},
  publisher = {{Association for Computational Linguistics}},
  location = {{Baltimore, Maryland, USA}},
  doi = {10.3115/v1/W14-3110},
  url = {https://aclanthology.org/W14-3110},
  urldate = {2022-08-26}
}

@inproceedings{spinde_neural_2021,
  title = {Neural {{Media Bias Detection Using Distant Supervision With BABE}} - {{Bias Annotations By Experts}}},
  booktitle = {Findings of the {{Association}} for {{Computational Linguistics}}: {{EMNLP}} 2021},
  author = {Spinde, Timo and Plank, Manuel and Krieger, Jan-David and Ruas, Terry and Gipp, Bela and Aizawa, Akiko},
  date = {2021-11},
  pages = {1166--1177},
  publisher = {{Association for Computational Linguistics}},
  location = {{Punta Cana, Dominican Republic}},
  doi = {10.18653/v1/2021.findings-emnlp.101},
  url = {https://aclanthology.org/2021.findings-emnlp.101},
  urldate = {2022-09-30},
  eventtitle = {{{EMNLP-Findings}} 2021}
}

@article{subelj_publication_2017,
  title = {Publication {{Boost}} in {{Web}} of {{Science Journals}} and {{Its Effect}} on {{Citation Distributions}}},
  author = {Šubelj, Lovro and Fiala, Dalibor},
  date = {2017-04-01},
  journaltitle = {Journal of the Association for Information Science and Technology},
  shortjournal = {Journal of the Association for Information Science and Technology},
  volume = {68},
  pages = {1018--1023},
  doi = {10.1002/asi.23718}
}

@article{supriyadi_bibliometric_2022,
  title = {A {{Bibliometric Analysis}}: {{Computer Science Research From Indonesia}}},
  shorttitle = {A {{Bibliometric Analysis}}},
  author = {Supriyadi, Edi},
  date = {2022-08-29},
  journaltitle = {TIERS Information Technology Journal},
  volume = {3},
  number = {1},
  pages = {28--34},
  issn = {2723-4541},
  doi = {10.38043/tiers.v3i1.3706},
  url = {https://journal.undiknas.ac.id/index.php/tiers/article/view/3706},
  urldate = {2022-09-11},
  issue = {1},
  langid = {english}
}

@article{tattershall_detecting_2020,
  title = {Detecting Bursty Terms in Computer Science Research},
  author = {Tattershall, E. and Nenadic, G. and Stevens, R. D.},
  date = {2020-01},
  journaltitle = {Scientometrics},
  shortjournal = {Scientometrics},
  volume = {122},
  number = {1},
  pages = {681--699},
  issn = {0138-9130, 1588-2861},
  doi = {10.1007/s11192-019-03307-5},
  url = {http://link.springer.com/10.1007/s11192-019-03307-5},
  urldate = {2022-08-27},
  langid = {english}
}

@article{uddin_scientometric_2015,
  title = {Scientometric Mapping of Computer Science Research in {{Mexico}}},
  author = {Uddin, Ashraf and Singh, Vivek Kumar and Pinto, David and Olmos, Ivan},
  date = {2015-10-01},
  journaltitle = {Scientometrics},
  shortjournal = {Scientometrics},
  volume = {105},
  number = {1},
  pages = {97--114},
  issn = {0138-9130},
  doi = {10.1007/s11192-015-1654-y},
  url = {https://doi.org/10.1007/s11192-015-1654-y},
  urldate = {2022-08-27}
}

@article{van_eck_software_2010,
  title = {Software Survey: {{VOSviewer}}, a Computer Program for Bibliometric Mapping},
  shorttitle = {Software Survey},
  author = {van Eck, Nees Jan and Waltman, Ludo},
  options = {useprefix=true},
  date = {2010-08-01},
  journaltitle = {Scientometrics},
  shortjournal = {Scientometrics},
  volume = {84},
  number = {2},
  pages = {523--538},
  issn = {1588-2861},
  doi = {10.1007/s11192-009-0146-3},
  url = {https://doi.org/10.1007/s11192-009-0146-3},
  urldate = {2022-09-10},
  langid = {english}
}

@article{vardi_conferences_2009,
  title = {Conferences vs. {{Journals}} in {{Computing Research}}},
  author = {Vardi, Moshe},
  date = {2009-05-01},
  journaltitle = {Commun. ACM},
  shortjournal = {Commun. ACM},
  volume = {52},
  pages = {5},
  doi = {10.1145/1506409.1506410}
}

@article{vrettas_conferences_2015,
  title = {Conferences versus Journals in Computer Science: {{Conferences}} vs. {{Journals}} in {{Computer Science}}},
  shorttitle = {Conferences versus Journals in Computer Science},
  author = {Vrettas, George and Sanderson, Mark},
  date = {2015-05-01},
  journaltitle = {Journal of the Association for Information Science and Technology},
  shortjournal = {Journal of the Association for Information Science and Technology},
  volume = {66},
  doi = {10.1002/asi.23349}
}

@inproceedings{wahle_are_2021,
  title = {Are {{Neural Language Models Good Plagiarists}}? {{A Benchmark}} for {{Neural Paraphrase Detection}}},
  shorttitle = {Are {{Neural Language Models Good Plagiarists}}?},
  booktitle = {2021 {{ACM}}/{{IEEE Joint Conference}} on {{Digital Libraries}} ({{JCDL}})},
  author = {Wahle, Jan Philip and Ruas, Terry and Meuschke, Norman and Gipp, Bela},
  date = {2021-09},
  pages = {226--229},
  doi = {10.1109/JCDL52503.2021.00065},
  eventtitle = {2021 {{ACM}}/{{IEEE Joint Conference}} on {{Digital Libraries}} ({{JCDL}})}
}

@inproceedings{wahle_d3_2022,
  title = {D3: {{A Massive Dataset}} of {{Scholarly Metadata}} for {{Analyzing}} the {{State}} of {{Computer Science Research}}},
  booktitle = {Proceedings of the {{Language Resources}} and {{Evaluation Conference}}},
  author = {Wahle, Jan Philip and Ruas, Terry and Mohammad, Saif and Gipp, Bela},
  date = {2022-06},
  pages = {2642--2651},
  publisher = {{European Language Resources Association}},
  location = {{Marseille, France}},
  url = {https://aclanthology.org/2022.lrec-1.283}
}

@inproceedings{wu_citeseerx_2019,
  title = {{{CiteSeerX}}: 20 Years of Service to Scholarly Big Data},
  shorttitle = {{{CiteSeerX}}},
  booktitle = {Proceedings of the {{Conference}} on {{Artificial Intelligence}} for {{Data Discovery}} and {{Reuse}}},
  author = {Wu, Jian and Kim, Kunho and Giles, C. Lee},
  date = {2019-05-13},
  pages = {1--4},
  publisher = {{ACM}},
  location = {{Pittsburgh Pennsylvania}},
  doi = {10.1145/3359115.3359119},
  url = {https://dl.acm.org/doi/10.1145/3359115.3359119},
  urldate = {2022-10-03},
  eventtitle = {{{AIDR}} '19: {{Artificial Intelligence}} for {{Data Discovery}} and {{Reuse}} 2019},
  isbn = {978-1-4503-7184-1},
  langid = {english}
}

@article{wu_population_2021,
  title = {A {{Population Model}} for {{Academia}}: {{Case Study}} of the {{Computer Science Community Using DBLP Bibliography}} 1960-2016},
  shorttitle = {A {{Population Model}} for {{Academia}}},
  author = {Wu, Yan and Venkatramanan, Srinivasan and Chiu, Dah Ming},
  date = {2021-01},
  journaltitle = {IEEE Transactions on Emerging Topics in Computing},
  volume = {9},
  number = {1},
  pages = {258--268},
  issn = {2168-6750},
  doi = {10.1109/TETC.2018.2855156},
  eventtitle = {{{IEEE Transactions}} on {{Emerging Topics}} in {{Computing}}}
}

@article{xia_research_2021,
  title = {Research {{Fronts}} of {{Computer Science}}: {{A Scientometric Analysis}}},
  shorttitle = {Research {{Fronts}} of {{Computer Science}}},
  author = {Xia, Wanjun and Jiang, Yanping and Zhu, Weifeng and Zhang, Shuang and Li, Tianrui},
  date = {2021-05-08},
  journaltitle = {Journal of Scientometric Research},
  shortjournal = {JSCIRES},
  volume = {10},
  number = {1},
  pages = {18--26},
  issn = {23216654, 23200057},
  doi = {10.5530/jscires.10.1.3},
  url = {http://www.jscires.org/article/397},
  urldate = {2022-09-11},
  langid = {english}
}

@article{zurita_bibliometrics_2020,
  title = {Bibliometrics in Computer Science: {{An}} Institution Ranking},
  shorttitle = {Bibliometrics in Computer Science},
  author = {Zurita, Gustavo and Merigo, Jose M. and Lobos, Valeria and Forteza, Carles},
  date = {2020-02-03},
  journaltitle = {Journal of Intelligent \& Fuzzy Systems},
  shortjournal = {Journal of Intelligent \& Fuzzy Systems},
  volume = {38},
  pages = {1--13},
  doi = {10.3233/JIFS-179636}
}
	}%
\endgroup
% Verzeichnisse (bitte nicht ändern) %%%%%%%%%%%%%
	%%%%%%%%%%%%%%%%%%%%%%%%%%%%%%%%%%%%%%%%%%%%%%%%%%%%%%%%%%%%%%%%%%%%%%%%%%%%

	% Anhänge
%   ...
	%\input{Kapitel/FAQ}              % Fragen und Antworten zur Vorlage und zu LaTeX
	%\input{Kapitel/LaTeX-Beispiele}  % Beispiele zu LaTeX

	%%%%%%%%%%%%%%%%%%%%%%%%%%%%%%%%%%%%%%%%%%%%%%%%%%%%%%%%%%%%%%%%%%%%%%%%%%%%
	%%% ENDE %%% ENDE %%% ENDE %%% ENDE %%% ENDE %%% ENDE %%% ENDE %%% ENDE %%%%
	%%%%%%%%%%%%%%%%%%%%%%%%%%%%%%%%%%%%%%%%%%%%%%%%%%%%%%%%%%%%%%%%%%%%%%%%%%%%
\end{document}